\documentclass[leterpaper,12pt]{ociamthesis}  

\usepackage{amssymb}

\usepackage{times}
\usepackage{amsmath}
\usepackage{setspace}
\usepackage{pstricks}
\usepackage{pst-grad}
\usepackage{pst-text}

\usepackage{graphicx}
\usepackage{epstopdf}

\usepackage{braket}
\usepackage{rotating}
\usepackage{booktabs}

\usepackage{subfig}

\usepackage{lipsum}

\usepackage{nomencl}
\usepackage{mathrsfs}

\makenomenclature

\newcommand{\oic}{(Original In Colour)}

\title{Physics in Regina\\[1ex]     
        for dummies}   

\author{Wenliang Li}             
\college{University of Regina}  

\degree{Doctor of Philosophy}     
\degreedate{2016}         

\begin{document}

\doublespacing

\baselineskip=18pt plus1pt

\setcounter{secnumdepth}{3}
\setcounter{tocdepth}{3}

\graphicspath{ {pics/} {pics/analysis_fig} }
\makeatletter
\let\insertdate\@date
\makeatother

{

\begin{titlepage}
\begin{center}
\thispagestyle{empty}
\renewcommand{\baselinestretch}{0}%
\textrm{\LARGE Exclusive Backward-Angle Omega Meson Electroproduction\\ } 
\vspace{1.5cm}
\renewcommand{\baselinestretch}{0}%
\textrm{\Large A Thesis \\[5mm]
Submitted to the Faculty of Graduate Studies and Research \\[5mm]
In Partial Fulfilment of the Requirements \\[5mm]
for the Degree of \\[5mm]
Doctor of Philosophy \\[5mm]
in Physics \\[5mm]
University of Regina}\\
\vspace{4cm}
\textrm{\large By \\[5mm]
Wenliang Li \\[5mm]
Regina, Saskatchewan\\[5mm]
October, 2017\\[1cm]
\copyright 2017: Wenliang Li
}
\end{center}
\end{titlepage}

}

\begin{figure}
 \centering 
	\includegraphics[scale=1]{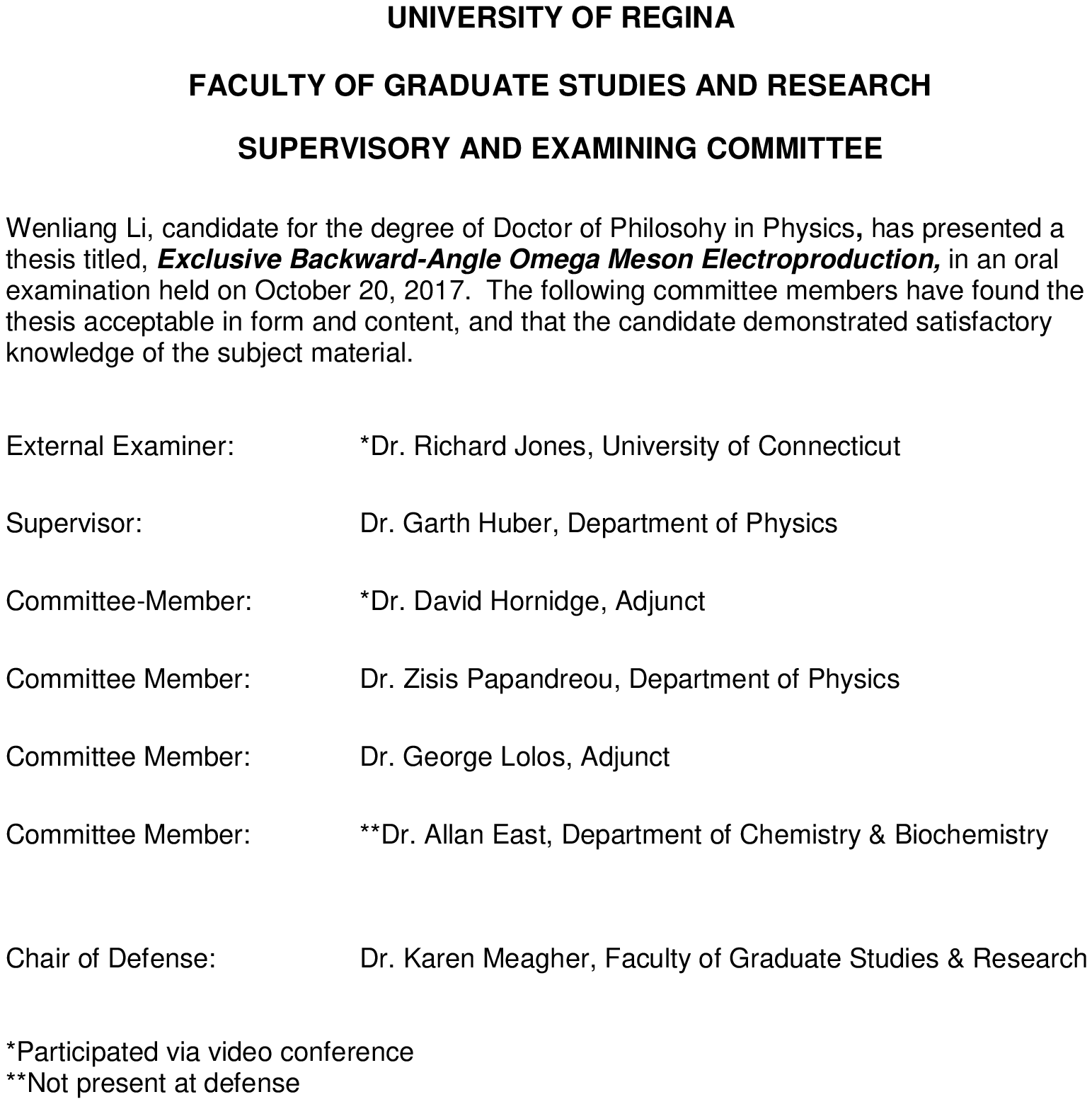}
\end{figure}

\doublespacing

\pagenumbering{gobble}

\chapter*{Abstract}



Exclusive meson electroproduction at different squared four-momenta of the exchanged virtual photon, $Q^2$, and at different four-momentum transfers, $t$ and $u$, can be used to probe QCD's transition from hadronic degrees of freedom at the long distance scale to quark-gluon degrees of freedom at the short distance scale. Backward-angle meson electroproduction was previously ignored, but is anticipated to offer complimentary information to conventional forward-angle meson electroproduction studies on nucleon structure.


This work is a pioneering study of backward-angle $\omega$ cross sections through the exclusive $^1$H$(e, e^{\prime}p)\omega$ reaction using the missing mass reconstruction technique. The extracted cross sections are separated into the transverse (T), longitudinal (L), and LT, TT interference terms.



The analyzed data were part of experiment E01-004 (F$_{\pi}$-2), which used 2.6-5.2~GeV electron beams and HMS+SOS spectrometers in Jefferson Lab Hall C. The primary objective was to detect coincidence $\pi$ in the forward-angle, where the backward-angle $\omega$ events were fortuitously detected. The experiment has central $Q^2$ values of 1.60 and 2.45~GeV$^2$, at $W$ = 2.21~GeV. There was significant coverage in $\phi$ and $\epsilon$, which allowed separation of $\sigma_{\rm T,L, LT, TT}$. The data set has a unique $u$ coverage of $-u\sim0$, which corresponds to $-t$ $>$ 4 GeV$^2$.





The separated $\sigma_{\rm T}$ result suggest a flat $\sim1/Q^{1.33 \pm 1.21}$ dependence, whereas $\sigma_{\rm L}$ seems to hold a stronger $1/Q^{9.43 \pm 6.28}$ dependence. The $\sigma_{\rm L}/\sigma_{\rm T}$ ratio indicate $\sigma_{\rm T}$ dominance at $Q^2$ = 2.45~GeV$^2$ at the $\sim$90\% confidence level.


After translating the results into the $-t$ space of the published CLAS data, our data show evidence of a backward-angle $\omega$ electroproduction peak at both $Q^2$ settings. Previously, this phenomenon showing both forward and backward-angle peaks was only observed in the meson photoproduction data.



Through comparison of our $\sigma_{\rm T}$ data with the prediction of the Transition Distribution Amplitude (TDA) model, and signs of $\sigma_{\rm T}$ dominance, promising indications of the applicability of the TDA factorization are demonstrated at a much lower $Q^2$ value than its preferred range of $Q^2$ $>$ 10~GeV$^2$.

These studies have opened a new means to study the transition of the nucleon wavefunction through backward-angle experimental observables.



\addcontentsline{toc}{chapter}{Abstract}

\chapter*{Acknowledgements}

I would like to express my sincere gratitude to my supervisor Prof. Garth Huber for encouraging me to undertake this Ph.D. project and for his continuous support through the development of this work, and for his meticulous and patient guidance. Working with him has been a deeply educational and challenging experience. I also cherish the personal bond that we manage to create along these years. I am also very grateful to Henk Blok and Dave Gaskell for their extremely valuable suggestions, comments and supports during this work. A special thanks to Tanja Horn for her great work on generating data Ntuples, this has significantly simplified the analysis. 

Great appreciation to Jean-Phillipe Lansberg, Bernard Pire, Krill Semenov and Lech Szymanowski, for providing the invaluable theoretical (TDA model) calculations. Visions offered by Christian Weiss and Mark Strikman have played a critical role throughout the thesis writing. 

Furthermore, I want to thank the research funding provided by NSERC of Canada and the FGSR of the University of Regina. Since the beginning, Department of Physics has provided amazing supports for my education and research. I would like to thank all members of the department, include Nader Mobed, George Lolos, Zisis Papandreou, Mauricio Barbi, Pierre Ouimet, Andrei Semenov, Cheryl Risling, Derek Gervais and others. I am also grateful to all the friends and colleagues met at the university: Dilli Paudyal, Ahmed Zafar, Ahmed Foda, Sameep Basnet, Tegan Beattie, Ryan Ambrose, Rory Evans, Nathanael Hogan and others. Together, we have created a pleasant working environment and they really had to put up with my loudness.

Finally I would like to dedicate this work to my parents, the Morrison family and the relatives for their kind support and hospitality throughout these years, and most importantly to my wife who cooks wonderfully.

\addcontentsline{toc}{chapter}{Acknowledgements}

\addcontentsline{toc}{chapter}{Dedication}
\begin{dedication}
This thesis is dedicated to\\
 my family\\
\end{dedication}

\begin{romanpages}          

\addcontentsline{toc}{chapter}{Table of Contents}
\tableofcontents            

\newpage

\addcontentsline{toc}{chapter}{List of Tables}
\listoftables              

\newpage

\addcontentsline{toc}{chapter}{List of Figures}
\listoffigures              
\end{romanpages}            

\doublespacing

\pagenumbering{arabic}



\graphicspath{{pics/1Introduction/}}

\chapter{Introduction}

\label{chap:intro}

The fundamental nature of matter in terms of elementary particles and their interactions is a central topic of research in subatomic physics. From the nuclear physics perspective, the atom consists of a cloud of electrons surrounding a positively charged core (nucleus), which contains protons and neutrons. The protons and neutrons are collectively called the nucleons and they are held together by the strong nuclear force via the exchange of mesons (the force charge carriers of the strong nuclear interaction). The strong nuclear force is described more fundamentally in terms of interactions between quarks and gluons. Hadrons, the strongly interacting particles such as nucleons and pions, are not considered elementary particles such as the electron (which is considered to be point-like), but instead contain a substructure based on fundamental particles, known as the partons. 

\begin{figure}[h!]
\centering
\includegraphics[width=0.75\textwidth]{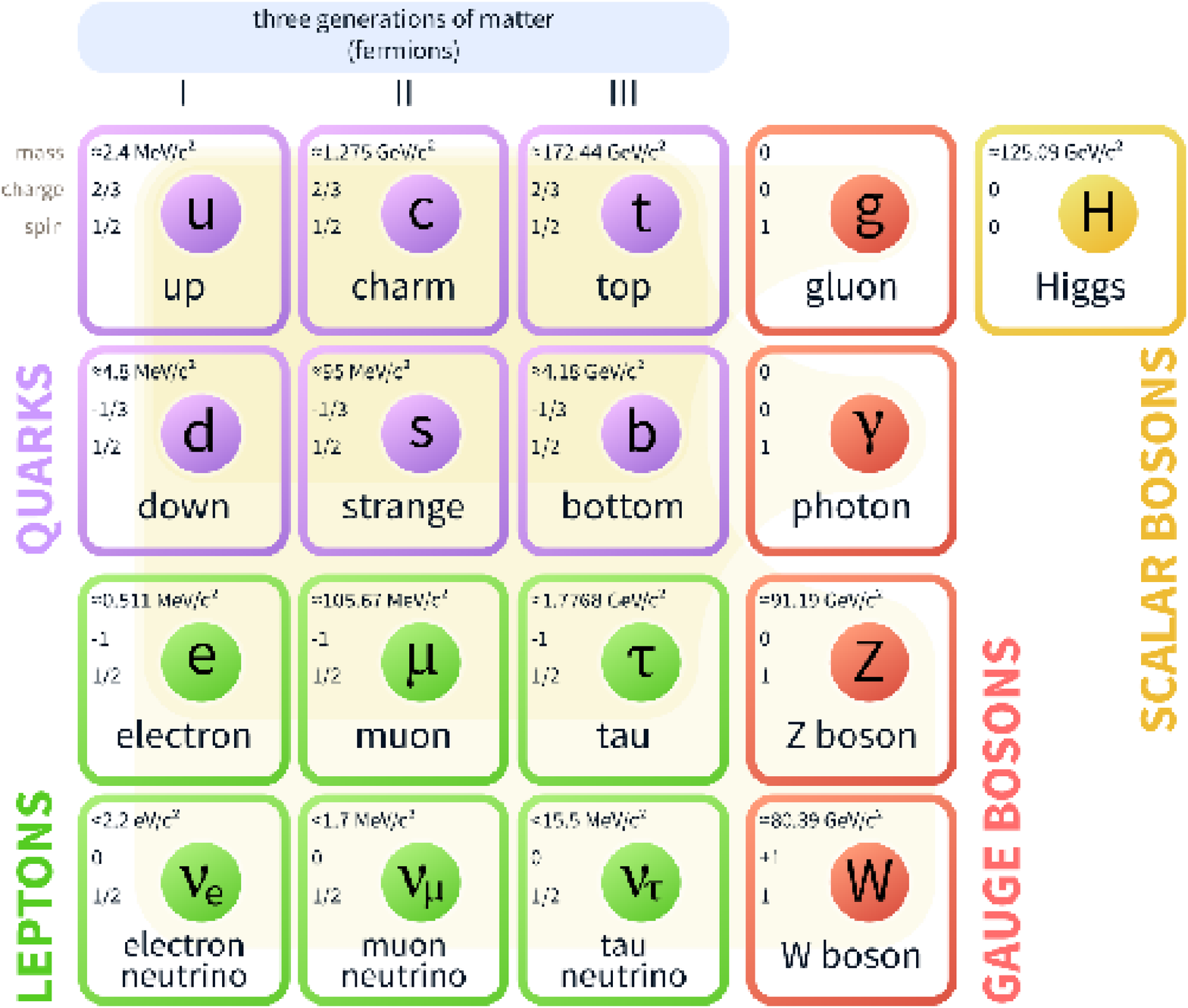}
\caption[Full list of the standard model fundemental particles]{Full list of the standard model fundamental particles~\cite{pdg}.\oic}
\label{fig:sm_particle}
\end{figure}

At the current stage, the most successful model (theory) available for the fundamental building blocks of matter is the Standard Model (SM\nomenclature{SM}{Standard Model}). According to the SM, there are four families of elementary particles, namely quarks ($q$), leptons (and their anti-particles), gauge bosons (the force charge carriers, also known as the quanta) and newly discovered Higgs boson. Examples of leptons include electrons and neutrinos. The quarks are identified as partons that are bound together by gluons to form hadrons. The forces between them are mediated via the exchanged gauge bosons, such as photons for the electromagnetic interaction and gluons for the strong interaction. The term `interaction', refers to the process of the gauge boson exchange. A complete list of standard model particles is shown in Fig.~\ref{fig:sm_particle}.

The field theory for the electromagnetic interaction is known as Quantum Electrodynamics (QED\nomenclature{QED}{Quantum Electrodynamics}). This theory has been developed into an instrument that allows high precision calculations for electromagnetic interactions, the intensity of these interactions is characterized by the electromagnetic coupling constant $\alpha_e \approx 1/137$. 

Analogously, the theory for the strong interaction between ``coloured'' quarks is known as Quantum Chromodynamics (QCD\nomenclature{QCD}{Quantum Chromodynamics}), where gluons are the field carriers that carry colour charges. In contrast to the QED field carriers (the photons), gluons can interact with other gluons. The intensity of the strong interaction is characterized by the strong coupling constant, $\alpha_s$, which has the particularity of being weak at short distance scales ($\sim 10^{-17}$~m) and strong at long distance scales ($\sim 10^{-15}$~m which is approximately the size of a nucleon). Experimentally, the long and short distance scales can be accessed through high and low energy interactions, respectively. Therefore, behavior of the strong interaction is significantly altered depending on the energy range of the reaction.

At a low energy scenario (corresponding to a long distance scale), where $\alpha_s \approx 1$ (dominates over other coupling constants such as $\alpha_e$), it is often difficult to detect all final state particles with great resolution to establish high quality data. These features make studying basic properties of hadrons very difficult.

This Ph.D. work is part of the general effort of studying the hadron structure (typical examples being protons and neutrons) in terms of $q$ and $g$ under the intermediate energy (under 10~GeV) scenario, where the proton target is probed by an accelerated electron beam. The thesis presents the extracted cross section of the $e+p \rightarrow e^{\prime} + p^{\prime} + \omega$ reaction from the experiment E01-004 (F$_\pi$-2) data taken at the Thomas Jefferson National Accelerator Facility (JLab).

This thesis consists of eight chapters:
\begin{itemize}

\item The first chapter gives a general introduction to subatomic physics, terminology and experimental methodology. 

\item The theoretical grounds for the interpretation of the extracted cross section observables are introduced in the second chapter. 

\item The experimental setup and apparatus at Jefferson Lab Hall C used in the experiment is presented in chapter three. 

\item Chapter four introduces the standard Monte Carlo simulation tool used for the  Hall C data analysis. The first part of the chapter describes the spectrometer models and various physics corrections which are taken into account, including: ionization energy loss, radiative corrections and multiple scattering. The second part of the chapter documents the development of the new C++ based software used in the analysis.

\item The analysis details regarding the elastic scattering events ($e+p\rightarrow e^{\prime}+p^{\prime}$) are introduced in the fifth chapter. The experimental conditions for the elastic scattering interaction resembles those for the $\omega$ production interaction, in both cases the scattered electrons and recoil protons are detected in coincidence mode. Thus, the study of elastic scattering events in greater detail significantly benefits the $\omega$ analysis in terms of particle identification (selection), experimental efficiency studies, dead time and experimental background subtractions (topics covered in the order as they are mentioned).

\item The detailed description of the $e + p \rightarrow e^{\prime} + p^{\prime} + \omega$ experimental data analysis is documented in chapter six. In the first part of this chapter, the particle selection and experimental kinematic coverage are discussed. The second part introduces the physics models used for simulating the $\omega$ and physics background processes, followed by the fitting methodology for the physics background subtraction. The chapter ends with a discussion of the statistical and systematic uncertainties in the extraction of the separated cross sections. 

\item In the seventh chapter, the experimental cross sections are presented. A comparison with past data from the CLAS collaboration, and the separated cross section ratios are presented to test the TDA predictions. Some general quantitative conclusions from the analysis are also discussed. 

\item In the last chapter, a brief overview is given to summarize the backward-angle meson production experiments expected in the near future.       

\end{itemize}

\section{Dynamical Properties of Hadrons}

Although the static properties of hadrons, like the total charge and magnetic moment, are explained by taking into account the quantum numbers (such as the total angular momentum $J$, and orbital momentum $l$ quantum numbers) of their constituent quarks, the dynamical properties of hadrons such as spin structure and parton distributions, particularly the gluon and sea quark contributions, are still not fully understood.

It is currently known that the dynamical properties of the nucleon constituents vary dramatically depending on the momentum scale at which the strong interaction is probed: at large momentum, the nucleon behavior is accurately described by its quarks and gluon fields, but at low momentum, it is necessary to use a description relying on effective hadronic degrees of freedom.

A complete understanding of nucleon properties requires an accurate description of the gluon interaction and sea quarks which directly contribute to the charge and current distributions. Particularly, as the fundamental part of the theory, topics of investigating the binding and confinement of quarks and gluons inside hadrons have been actively pursued, and prominent examples include the charged pion form factor experiments~\cite{jochen01, blok08} and GlueX experiment~\cite{gluex16}.

QCD is a fundamental theory, and is a part of the Standard Model of particle physics, which describes the interactions between quarks and gluons. QCD is a type of quantum field theory called a non-abelian gauge theory, with symmetry group SU(3)$_c$\footnote{SU(3)$_c$ represents the spacial unitary group that takes into account three colours of strong interaction.}, where subscript $c$ indicates the three colour charges: red, blue or green. The gluon is the strong force carrier, which plays the same role as the photons in the electromagnetic force described by QED, with the colour as analog of electric charge. By QCD description, the protons, neutrons and pions are made up as the lowest energy, colour neutral meson and baryon states. Since there are structural similarities between QCD and QED, it is assumed that the problems in hadron physics can be resolved using similiar perturbative methods (theory) what are successfully applied to QED~\cite{pdg}.

It is well understood that at the asymptotic (freedom) limit, where the exchange of momentum is large or interaction distance is sufficiently small (compared to the nucleon size), the experimental observables from a given physics process can be calculated from first principles via perturbative methods~\cite{gross73}. On the other hand, exact calculations are not yet possible at low momenta or long interaction length, since the binding of quarks is a long-distance effect, meaning that non-perturbative methods must play an important role.

A complete theory of QCD needs to take into account parton behavior at both interaction scales (perturbative and non-perturbative limits) to understand quark binding in hadrons. However, in the absence of a complete solution to QCD, the predictive power of the theory is limited, relying only on the extraction of related information from experimental data in the non-perturbative sector. Experimental data can be used to constrain effective models describing hadronic degrees of freedom in the strong interaction at larger distance scales, the QCD transition to quark-gluon degrees of freedom, to ultimately asymptotic freedom at progressively shorter scales.

The existence of partons inside hadrons is well established by scattering of energetic electrons off proton target~\cite{pdg}, such a process is often referred as the Deep Inelastic Scattering (DIS\nomenclature{DIS}{Deep Inelastic Scattering}). At sufficiently high electron energies, inelastic electron-proton scattering is viewed as elastic scattering of the electron from a free quark inside the proton. However, the internal structure of hadrons cannot treated as an simple constant structure consisting of three quarks.

The extrapolated mass of $u$ and $d$ quarks outside of any binding potential determined by DIS where quarks are only weakly bound, is 4-6~MeV~\cite{pdg}. These only account for $\sim$1\% of the nucleon mass. This is negligible compared to the gluon and sea quark contributions (virtual quark-antiquark pairs) to the nucleon mass. Contributions to the nucleon structure from the partons vary with the energy and momentum of interaction, i.e. asymptotic freedom versus confinement.

A reliable way to study the nucleon structure is to investigate collective observables of the bound systems. Electromagnetic (EM\nomenclature{EM}{Electromagnetism}) form factors of hadrons reflect the distribution of charge and current in the hadron. Therefore, the study of hadronic form factors can give insight into the internal structure of hadrons. 

Since no exact calculations can be done in the non-perturbative regime of QCD (soft QCD), it is extremely challenging to describe the strong interaction at small values of momentum transfer using an (non-perturbative QCD) effective model. Input from experimental data is needed to constrain those models.

\section{Electron Scattering: Access to Hadron Structure}
\label{sec:QED}

Electron scattering is a powerful tool, which gives clean access to study the structure of the nucleus. Because the electron-photon interaction is well described by QED, the point-like nature of the accelerated electron beam is a simple and well understood probe. Note the theory of QED been developed into an instrument that allows high precision calculations to describe the electromagnetic processes. 

Because the electromagnetic interaction is relatively weak compared to the strong interaction at the range comparable to the nucleon radius ($\sim$1~fm), it is well modeled by the exchange of a single virtual photon (force field carrier) between the incident electron and the hadron target. If the probed distance scale is sufficiently small, the virtual photon is able to resolve the structure inside of the proton, which is often referred to as the partonic structure (many partons). 

In terms of the spin and parity quantum numbers, the virtual photon is the same as the real photon. There are two kinds of virtual photons: the space-like virtual photon that carries more momentum than energy, and the time-like virtual photon that carries more energy than momentum. For the space-like virtual photon, since $E<p$, $E^2-p^2 < 0$. For the time-like virtual photon, since $E>p$, $E^2 - p^2 >0$. Unless otherwise specified, the virtual photon referred to in this thesis is the space-like virtual photon. Note that throughout this thesis work, all equations, parameters and experimental values are presented in the natural units where $\hbar = c = 1$. 

Another fundamental difference between real and virtual photons is that the real photon can only be transversely (perpendicular to the direction of propagation) polarized (as described by classical electrodynamics~\cite{jackson99}), while the virtual photon can be both longitudinally (parallel to the direction of the propagation) and transversely polarized. This property of virtual photon is directly related to principle of the L/T separation formalism, which is described in Sec.~\ref{sec:LT_sep}.

Even on the same target, the internal structure probed by a virtual photon can vary significantly, depending on the kinematics (such as the momentum transfer) of the scattering process. At extremely low energy transfers, the virtual photon interacts with the entire nucleus, scattering elastically or exciting a nuclear state or resonance. At higher energy and momentum transfers, scattering is dominated by quasielastic scattering, where the photon interacts with a single nucleon. As the energy and momentum transfer increase, and photon probes smaller distance scales, the interaction becomes sensitive to the quark and gluon degrees of freedom in the nucleus.

In addition to a clean separation of the scattering process from the structure of the target, electron scattering from a nucleus is well suited to the examination of the structure of the nucleus. Because electron scattering off a free nucleon is a well studied problem, one can separate the structure of the nucleon from the structure of the nucleus, and examine the nuclear structure, as well as modifications to the structure of the nucleons in the nuclear medium.

\section{Experimental Kinematics and Methodology}

\subsection{Interaction Reference Frame}

Conventionally, there are two frames of reference that are important for an experiment: the laboratory frame of reference (lab frame) and the center of mass frame of reference (CM\nomenclature{CM frame}{Center of Mass Reference Frame} frame). Intuitively, the lab frame is the frame of reference in which the experiment is performed, while the center of mass frame is that in which the total momentum of the system vanishes and the center of mass of the system remains at the origin. The connection between the two reference frames is through the Lorentz transformation (boost).

\subsection{Mandelstam Variables}

\begin{figure}[t]
\centering
\includegraphics[width=0.6\textwidth]{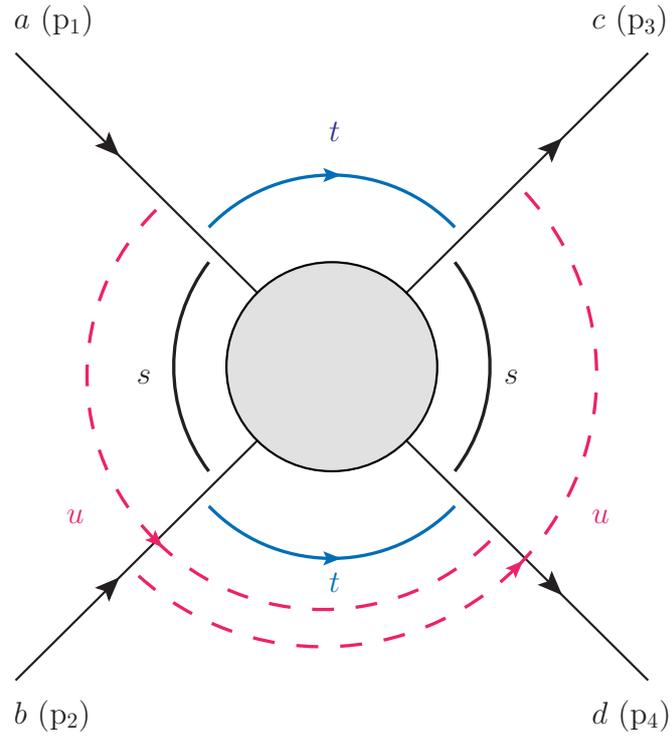}
\caption[Scattering diagram of a generic interaction]{Scattering diagram of a generic interaction: $a + b \rightarrow c + d$. The $s$, $t$ and $u$ cross relations between initial and final states of the interaction are indicated by black solid, blue solid and red dotted curves, respectively.~\oic}
\label{fig:man_va}
\end{figure}

Fig.~\ref{fig:man_va} shows the scattering diagram of the following interaction,
\begin{equation}
a({\rm p}_1) + b({\rm p}_2) \rightarrow c({\rm p}_3) + d({\rm p}_4)
\label{eqn:int}
\end{equation} 
neglecting $J$ and isospin ($I$) quantum numbers, $a$, $b$, $c$ and $d$ are the names of the particles; their four momenta are given as $${\rm p}_i=(E_i, -\vec{p}_i),~~~~~~~ {\rm p}_i^2= E_i^2 -\vec{p\,}_i^2 = m_i^2,$$ where $i= 1,\,2,\,3,\,4$; $E$ and $\vec{p}$ represent the energy and three momentum of the particle.

In this scattering process (Eqn.~\ref{eqn:int}), $E$, $\vec{p}$ and the scattering angle of the particles can be linked using the cross relations in a Lorentz invariant fashion (equal value in both lab frame and center of mass (CM) frame). These cross relations are known as the Mandelstam variables, and their definitions are given below, 
\begin{equation}
\begin{split}
s = &(\rm{p}_1+\rm{p}_2)^2=(\rm{p}_3+\rm{p}_4)^2 \\
t = &(\rm{p}_1-\rm{p}_3)^2=(\rm{p}_2-\rm{p}_4)^2 \\
u = &(\rm{p}_1-\rm{p}_4)^2=(\rm{p}_2-\rm{p}_3)^2.\\
\end{split}
\end{equation}
From the (three) momentum and energy conservation $$\vec{p}_1 + \vec{p}_2 = \vec{p}_3 + \vec{p}_4\,,$$ the following relation can be derived:
\begin{equation}
s + t + u ~ = m_1^2 + m_2^2 + m_3^2 + m_4^2.
\end{equation}

\subsection{Exclusive $\omega$ Meson Electroproduction}

The primary reaction studied in this thesis is the exclusive meson electroproduction reaction: $^1$H$(e, e^{\prime}p)\omega$. Meson electroproduction is a meson production process where the incoming projectile is the virtual photon ($\gamma^{*}$). Note that the $\gamma^{*}$ is induced by the incoming and scattered electron, a process well described by QED (introduced in Sec.~\ref{sec:QED}). 

Furthermore, a reaction is considered to be exclusive if all particles from the final states are detected or reconstructed, otherwise, the reaction is considered to be inclusive.  Note that all reactions analyzed in this thesis are exclusive reactions. 

Throughout the thesis, the interaction nomenclature such as $^1$H$(e, e^{\prime}p)\omega$ is used frequently. From the expression, the initial and final states of the interaction are separated by the comma (`,') symbol. The left hand side of the comma symbol: $^1$H and $e$ represents the liquid hydrogen target and incoming $e$ beam; on the right hand side:  $e^{\prime}$ is for the scattered electron beam, $p$ for recoil proton from the target and $\omega$ for produced omega meson. The energy and momentum information for particles inside of the bracket are measured directed using experimental hardware. Note that energy and momentum information of the $\omega$ are reconstructed using the missing mass technique (described below).

\begin{figure}[t]
  \centering
  \includegraphics[width=0.80\textwidth]{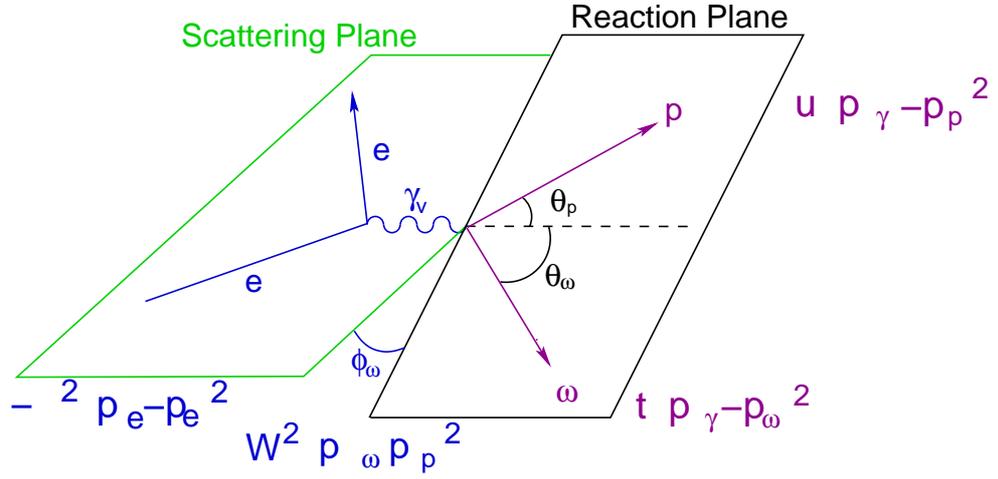}  
  \caption[The scattering and reaction planes for the $\omega$ production interaction]{The scattering and reaction planes for the $\omega$ production interaction: $^1$H$(e, e^{\prime}p)\omega$. The scattering plane is shown in green box and the reaction plane is shown in black box. Note that the recoil proton target after the interaction is labelled $p$; $\gamma_{\nu}$ represents the exchanged virtual photon and its direction defines the $q$-vector; the $\phi_p$ ($\phi_p = \phi_\omega + 180^\circ$) is defined as the angle between the scattering and reaction planes (the azimuthal angle around the $q$-vector); the $\theta_{p}$ and $\theta_{\omega}$ denote the angle of the $p$ and $\omega$ with respect to the $q$-vector, respectively. The definition of the Lorentz invariant variables such as $W$, $Q^2$, $t$ and $u$ are also shown.~\oic}
  \label{fig:plane}
\end{figure}

Fig.~\ref{fig:plane} shows a scattering schematic diagram of the exclusive meson electroproduction reaction: $^1$H$(e, e^{\prime}p)\omega$. The three-momentum vectors of the incoming and the scattered electrons are denoted as $\vec{p}_e$ and $\vec{p}_{e^\prime}$, respectively. Together they define the scattering plane, which is shown as a green box. The corresponding four momenta are p$_e$ and p$_e^{\prime}$. The electron scattering angle in the lab frame is labelled as $\theta_e$. The transferred four-momentum vector $q=(\nu,\vec{q})$ is defined as (p$_e-$p$_e^{\prime}$). In the one photon exchange approximation, the four-momentum of the virtual photon is taken as $q$. The square of the four momentum vector $q^2 =q_{\mu}q^{\mu}=\omega^2-|\vec{q}\,|^2$ = $-Q^2$ is always negative in the electron scattering process (for a space-like virtual photon). Note that the three momentum vector of the induced virtual photon is known as the $q$-vector\nomenclature{$q$-vector}{Three momentum vector of the induced virtual photon}.

The three-momentum vectors of the recoil proton target ($\vec{p}_p$) and produced $\omega$ ($\vec{p}_\omega$) define the reaction plane, which is shown as a black box. The azimuthal angle between the scattering plane and the reaction plane is denoted by the recoil proton angle $\phi_p$. From the perspective of standing at the entrance and looking downstream of the spectrometer, $\phi_p=0$ points to horizontal left of the $q$-vector, and it follows an anticlockwise rotation. The lab frame scattering angles between $\vec{p}_p$ (or $\vec{p}_{\omega}$) and $\vec{q}$ is labeled $\theta_p$ (or $\theta_\omega$). Unless otherwise specified, the symbols $\theta$ and $\phi$ without subscript are equivalent to $\theta_p$ and $\phi_p$, since the recoil protons were detected during the experiment. The parallel and antiparallel kinematics are unique circumstances, and occur at $\theta=0^\circ$ and $\theta=180^\circ$, respectively. Under the these circumstances, the interference (LT and TT) contributions from the virtual photon to the differential cross section are required to vanish. The implications of the parallel and antiparallel kinematics are further explained in Sec.~\ref{sec:LT_sep}.

In the $^1$H$(e, e^{\prime}p)\omega$ reaction, the missing energy and missing momentum are defined as:
\begin{equation}
\begin{split}
E_{m} =& E_e - E_{e^{\prime}} - E_{p}  \,,\\
\vec{p}_{m} =& \vec{p}_e - \vec{p}_{e^{\prime}} - \vec{p}_{p} = \vec{q} - \vec{p}_p \,.
\end{split}
\end{equation}
From these $E_m$ and $\vec{p}_{m}$, one can calculate the missing mass $M_m=\sqrt{E^2_m-\vec{p~}^2_m}$, which should correspond to the mass of the $\omega$ meson ($m_\omega$ = 0.738~GeV ~\cite{pdg}).

It is useful to describe the $^1$H$(e, e^{\prime}p)\omega$ reaction in terms of these Lorentz invariant quantities. In addition to $Q^2$, one can use the Mandelstam variables $s$, $t$ and $u$. In terms of the present reaction, these quantities can be defined as:  
\begin{equation}
\begin{split}
s = & ({\rm p}_{\rm H}  + {\rm q})^2          = ({\rm p}_p  + {\rm p}_\omega)^2  \,, \\
t = & ({\rm p}_{\rm H}  - {\rm p}_{p})^2      = ({\rm q}  -  {\rm p}_\omega)^2  \,, \\
u = & ({\rm p}_{\rm H}  - {\rm p}_{\omega})^2 = ({\rm q}  -  {\rm p}_p   )^2    \,.
\end{split}
\end{equation}
where ${\rm p}_{\rm H}$,  ${\rm q}$,  ${\rm p}_p$ and ${\rm p}_\omega$ are the four momenta of the liquid hydrogen nuclear target, virtual photon, recoil proton and $\omega$, respectively, and $\rm{q}$ is the equivalent to the $\rm{p}_{\gamma}$, defined in Fig.~\ref{fig:plane}. Instead of $s$, the invariant mass of the photon-target system, $W$, is often used here ($W=\sqrt{s}$), which can be expressed as $W=\sqrt{M^2_p+2M_p\nu-Q^2}$, where $M_p$ is the rest mass of the proton target and $\nu$ is the energy of the virtual photon. The quantities $t$ and $u$ are the squares of the four-momentum transfer to the nucleon system. They can be written as 
\begin{equation}
\begin{split}
t = & (E_p - \nu)^2 - |\vec{p}_p|^2 - |\vec{q}\,|^2 + 2|p_{p}||q|\cos{\theta_p}  \,, \\
u = & (E_\omega - \nu)^2 - |\vec{p}_\omega|^2 - |\vec{q}\,|^2 + 2|p_{\omega}||q|\cos{\theta_\omega}\,,
\end{split}
\end{equation}
respectively. In the present reaction, $t$ and $u$ are always negative. The minimum value $-t$ (or $-u$) known as $-t_{\rm min}$ (or $-u_{\rm min}$), is reached for $\theta=0^\circ$ (or $\theta=180^\circ$), respectively.  The minimum values of $-t$ and $-u$ increase as $Q^2$ increases, while $W$ is kept constant,
\begin{equation}
s + t + u ~ = ~ m_p^2 + q^2+ m_p^2 + m_\omega^2 = 2m_p^2 - Q^2 + m_\omega^2.
\label{eqn:constraints}
\end{equation}

In addition to the Mandelstam variables, the Lorentz invariant quantity Bjorken $x$ is also extremely important and detects the dynamical properties of nucleon. Bjorken $x$ is the fractional momentum carried by the struck parton and defined as $$x=\frac{Q^2}{2\,p\,q}.$$ Note that the $x$ in this thesis is defined as the Bjorken $x$ (often referred as $x_{B}$), and is not to be confused with the Feynman $x$ (often referred as $x_{F}$).

\subsection{L/T Separation}

\label{sec:LT_sep}

In the one-photon-exchange approximation, the $^1$H$(e, e^{\prime}p)X$ cross section of the $\omega$ and other meson production interactions ($X=\omega,\,\pi,\,\rho^{0}$, 2$\pi$, $\eta$ and $\eta^{\prime}$) can be written as the contraction of a lepton tensor $L_{\mu \nu}$ and a hadron tensor $W_{\mu \nu}$~\cite{mul90}. In the case of $\omega$ production:
\begin{equation}
\frac{d^6\sigma}{d\Omega_{e^{\prime}}\,dE_{e^{\prime}}\,d\Omega_p\,dE_p} = |p_p| \, E_{p} \, \frac{\alpha^2}{Q^4} \frac{E_{e^{\prime}}}{E_e} \, L_{\mu\nu}  \, W^{\mu\nu}\,,
\label{Xsection_6f}
\end{equation}
where the $L_{\mu\nu}$ can be calculated exactly in QED, and the explicit structure of the $W^{\mu\nu}$ is yet to be determined. Since the final states are over constrained (either detected or can be reconstructed), as in the case of the $^1$H$(e, e^{\prime}p)\omega$ reaction, the cross section can be reduced further to a five-fold differential form:
\begin{equation}
\frac{d^5\sigma}{dE^{\prime} d \Omega_{e^{\prime}} d \Omega_{p}^{*} } = \Gamma_v \, \frac{d^2\sigma}{d\Omega_{p}^*}\,,
\label{Xsection_5f}
\end{equation}
where the asterisks denote quantities in the center-of-mass frame of the virtual photon-nucleon system; $\Gamma_V$ is the virtual photon flux factor:
$$
\Gamma_v = \frac{\alpha}{2\pi^2} \frac{E_{e^{\prime}}}{E_e} \frac{q_{L}}{Q^2} \frac{1}{(1-\epsilon)}\,,
$$
where $\alpha$ is the fine structure constant, the factor $q_L = (W^2 - m_p^2)/(2M_p)$ is the equivalent real-photon energy, which is the laboratory energy a real photon would need to produce a system with invariant mass $W$; and $\epsilon$ is the polarization of the virtual photon which is defined as
$$
\epsilon = \left( 1 + \frac{2 |{\rm q}|^2 }{Q^2} \tan ^ 2 \frac{\theta_e}{2} \right)^{-1}\,.
$$
The two-fold differential cross section (Eqn.~\ref{Xsection_5f}) can be written in terms of an invariant cross section:
\begin{equation}
\frac{d^2\sigma}{d\Omega^{*}_\omega} = \frac{d^2\sigma}{dt~d\phi} \cdot \frac{dt}{d\cos\theta^*},
\end{equation}
where $$\frac{dt}{d\cos\theta^*} = 2|p^*||q^*|$$ is the Jacobian factor, and $p^*$ and $q^*$ are the three momentum of the proton and the virtual photon in the CM frame.

\begin{figure}[t]
\centering
\includegraphics[width=0.8\textwidth]{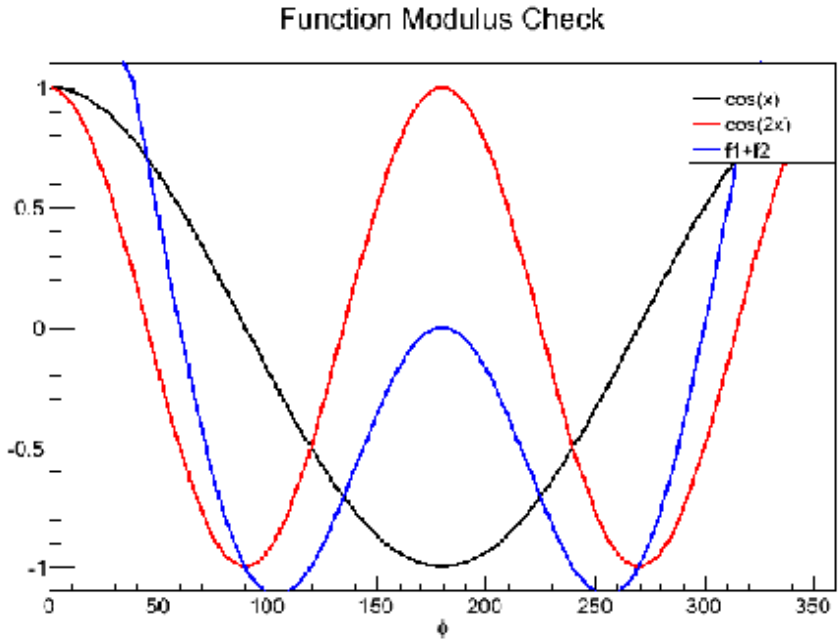}
\caption[Modulation of the interference terms versus $\phi$]{Modulation of the interference terms versus $\phi$ angle coverage. The $\phi$ modulation of $\sigma_{\rm LT}$ is shown as the black curve and the $\sigma_{\rm TT}$ is shown as the red curve. A possible interference modulation between the $\sigma_{\rm LT}$ and $\sigma_{\rm TT}$ is shown as the blue curve.~\oic}
\label{fig:modulus}
\end{figure}

The contraction of the lepton and the hadron tensor is decomposed into four structure functions corresponding to the polarization states of the virtual photon: a longitudinal (L), a transverse (T) and two interference terms (LT and TT). The general form of two-fold differential cross section in Eqn.~\ref{Xsection_5f} can be expressed in terms of structure functions as:

\begin{equation}
2 \pi \frac{d^2 \sigma}{dt ~ d\phi} = \frac{d \sigma_{\rm T}}{dt} + 
\epsilon ~ \frac{d \sigma_{\rm L}}{dt} + \sqrt{2\epsilon(1+\epsilon)}~ 
\frac{d\sigma_{\rm LT}}{dt} \cos \phi + \epsilon ~ 
\frac{d\sigma_{\rm TT}}{dt} \cos 2\phi \,.
\label{eqn:xsection_LT}
\end{equation}

The Rosenbluth separation, also known as the longitudinal/transverse (L/T) separation, is a unique method of isolating the longitudinal component of the differential cross section from the transverse component. The method requires at least two separate measurements with different experimental configurations, such as the spectrometer angles and electron beam energy, while fixing the Lorentz invariant kinematic parameters such as $x$ and $Q^2$. The only physical parameter that is different between the two measurements is $\epsilon$, which is directly dependent upon the incoming electron beam energy ($E_{e^{\prime}}$) and the scattering angle of the outgoing electron.

The two interference terms in Eqn.~\ref{eqn:xsection_LT} can be eliminated either by taking data parallel (or antiparallel) to the direction of the virtual photon ($\phi_{p}$), or by measuring those terms over the full angular $\phi$ range and integrating over the acceptance.

The former case is known as the parallel (or antiparallel) kinematics regime, where the recoil proton angle $\theta$ = 0$^{\circ}$ ( or $\theta$ = 180$^\circ$). As the result, $\phi$ coverage reduced to a single point and give no angular distributions (LT or TT interference contributions). Therefore, the Eqn.~\ref{eqn:xsection_LT} can be reduced to 
\begin{equation}
\frac{d\sigma}{dt} = \epsilon \frac{d\sigma_{\rm L}}{dt} + \frac{d\sigma_{\rm T}}{dt}\,.
\end{equation}
From the low and high $\epsilon$ measurements, the longitudinal and transverse components of the cross section can be written as
\begin{equation}
\frac{d\sigma_{\rm L}}{dt} = \frac{ \left(\frac{d\sigma}{dt}\right)_{\rm High}- \left(\frac{d\sigma}{dt} \right)_{\rm Low} }{\epsilon_{\rm High} -\epsilon_{\rm Low}}
\end{equation}
\begin{equation}
\frac{d\sigma_{\rm T}}{dt} = \frac{ \epsilon_{\rm High} \left(\frac{d\sigma}{dt}\right)_{\rm Low}- \epsilon_{\rm Low} \left(\frac{d\sigma}{dt} \right)_{\rm High} }{\epsilon_{\rm high} -\epsilon_{\rm Low}}\,.
\end{equation}

\section{The F$_{\pi}$-2 Experiment}
\label{sec:fpi2}

The data analysed in this thesis work were collected by the F$_\pi$-2 experiment, which was carried out at experimental Hall C of the Thomas Jefferson National Accelerator Facility (JLab), located in Newport News, Virginia, USA. An electron beam ($e$) was accelerated to an energy of 3.7-5.2~GeV before colliding with a liquid hydrogen ($p$) target. The scattered electrons ($e^{\prime}$) were detected by the Short Orbit Spectrometer (SOS\nomenclature{SOS}{Short Orbit Spectrometer}), and recoil protons $p^{\prime}$ were detected by the High Momentum Spectrometer (HMS\nomenclature{HMS}{High Momentum Spectrometer}) after the collision. The vector mesons such as $\omega$ were created as the result of the interaction. Since a large fraction of momentum was absorbed by the recoiled $p^{\prime}$, the $\omega$ was almost at rest in the lab frame. Therefore, the information needed to extract $\omega$ cross section must be reconstructed with the detected $e^{\prime}$ and $p^{\prime}$ data. A schematic diagram for backward angle $\omega$ production is shown in Fig.~\ref{fig:plane}.

Experiment E01-004 (F$_\pi$-2)~\cite{blok08} was the second charged pion form factor experiment undertaken at Jefferson Lab in 2003. The goal of the F$_\pi$-2 experiment was to extract the differential cross section of charged $\pi$ through the interactions $^1$H$(e,e^{\prime}\pi^{+})n$ and $^2$H$(e,e^{\prime}\pi^-)pp$, ($n$ represents neutron) at the intermediate energy level (few GeV), and further isolate the longitudinal part of the pion electro-production cross section for the purpose of extracting the charge pion form factor (F$_{\pi}$). These physical observables, allow study of the transition process from the non-perturbative QCD region to the perturbative QCD region to further understand hadron structure.

During the F$_\pi$-2 experimental data taking, a significant number of recoil protons were detected in coincidence with the scattered electrons. The missing mass distribution suggested strong evidence for the backward angle ($u$-channel) $\omega$ production. This thesis work used a similar technique as the earlier F$_\pi$-2 analyses, to extract the differential cross sections and perform a full Rosenbluth separation. Since these data offer unique backward angle kinematics, which have not been described by theory or studied by other experiments, the result from this research is expected to provide a new means to probe the quark component of the proton wavefunction.

\section{Past Exclusive $\omega$ Electroproduction Experiments}

\begin{figure}[t]
\centering
\includegraphics[width=0.8\textwidth]{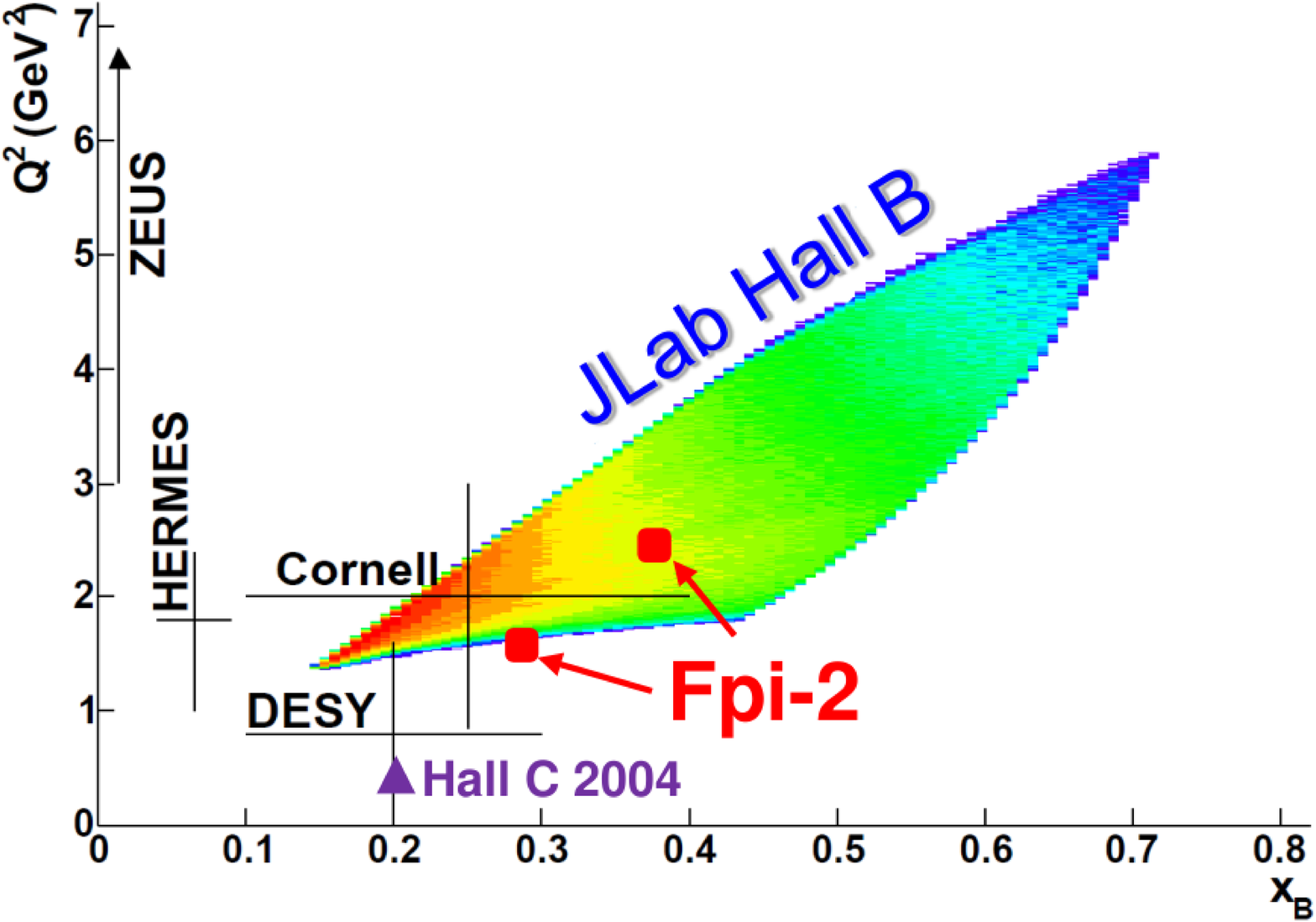}
\caption[$Q^2$ versus $x$ coverage for the world data on the $\omega$ electroproduction]{$Q^2$ versus $x$ coverage for the world data on $\omega$ electroproduction. Here, the x-axis variable $x_{\rm B}$, is equivalent to the $x$ (Bjorken $x$) used elsewhere in the thesis. Due to the limited acceptance of the spectrometer experiments, the kinematics coverages (settings) of the Hall C measurements are presented as individual points (red squares for F$_\pi$-2 and purple triangle for Ambrosewicz et al.~\cite{ambrosewicz04}, 2004). The full kinematics coverages in terms of $W$, $Q^2$, $x$ and $t$ for each of the experiments are listed in Table~\ref{tab:data}.}
\label{fig:coverage}
\end{figure}

The dynamical properties of nucleon greatly depend on the invariant mass of the probe-target system $W$, wavelength of the virtual photon probe ($\lambda\sim1/Q$) and fractional momentum of the struck parton $x$. Table~\ref{tab:data} shows the summary of the past exclusive $\omega$ electroproduction experiments, where each of the experiments has different coverages in terms of $W$, $Q^2$ and $x$, therefore not all data sets are suitable to compare to the result from the F$_{\pi}$-2 data. 

Fig.~\ref{fig:coverage} shows the $Q^2$ vs $x$ for all data sets. The exclusive $\omega$ meson data from ZEUS~\cite{zeus97}, HERMES~\cite{hermes14} and DESY~\cite{joo77} offer different coverages than the F$_{\pi}$-2 data, therefore cannot be used to perform any meaningful comparison. Any comparison study requiring significant extrapolation would introduce unavoidable bias to the physics observable, which can lead to the wrong conclusion. 

The Cornell~\cite{cassel81} data overlap F$_{\pi}$-2 kinematics coverage. The differential cross section $d\sigma_\omega/dt$ was extracted for 2.25 $<$ $W$ $<$ 3.7~GeV and 0.5 $<$ $Q^2$ $<$ 3~GeV$^2$. The $t$ coverage is given in terms of $t^{\prime}$, which is defined as $t^{\prime} = |t-t_{\rm min}|$ and ranges 0 $<$ $t^{\prime}$ $<$ 1~GeV$^2$. Despite similarity in the kinematics coverage, the Cornell data do not have sufficient statistics ($\omega$ events) to investigate the cross section evolution in terms of $t^{\prime}$ within a more constrained $Q^2$ and $W$ range. In addition, the large overall uncertainties (20-40\%) make the comparison much less meaningful.

Hall C experiment E91-016~\cite{ambrosewicz04, ambrosewicz} by Ambrosewicz et al., studied the $\omega$ electroproduction at low momentum transfer of $Q^2=0.5$~GeV$^2$ and $W\sim1.75$~GeV. Since the $W$ value is in the resonance region (excited states of baryons), the backward angle $\omega$ is due to the decay of a baryon resonance. This is a completely different physical mechanism compared to the $\omega$ created in F$_\pi$-2, whose $W$ is above the resonance region. Despite the differences in the physics objectives, the experimental methodologies used by the two experiments were extremely similar. In both experiments, the $\omega$ events were reconstructed using the detected final states information from the SOS and HMS (the missing mass reconstruction method), and the simulation method was used for the subtraction of the backward angle physics backgrounds.

The data published by Morand et al.~\cite{morand05}, from the CLAS collaboration, also overlaps the F$_\pi$-2 data. Different from the spectrometer setup at Hall C, the CLAS is a low luminosity, high precision detector with large solid angle acceptance. Thus, the methodology used to detect the $\omega$ mesons is completely different. The experiment measured $ep\rightarrow e^{\prime}p^{\prime}\omega$ reaction, where the $\omega$ decays through $\omega\rightarrow \pi^+\pi^-\pi^0$ channel. Since the detection of all three final state pions was extremely difficult, the $\omega$ event selection relied on the detection of one or two of the final state pions, which corresponds to $ep \rightarrow ep\pi^+ X$ and $ep \rightarrow e^{\prime}p^{\prime}\pi^+ \pi^- X$, respectively. After the events were selected, the missing mass $M_X$ distribution of $ep\rightarrow e^{\prime}p^{\prime}X$ was then reconstructed, where a distinctive peak corresponds to the $\omega$ is sitting on top of a smooth and
wide background. The CLAS data have extremely wide kinematics coverage, as shown in Fig.~\ref{fig:coverage}. The data set closest to the F$_{\pi}$-2 kinematics at $Q^2=2.35$~GeV$^2$, $W<2.47$~GeV and $0.21<-t<2.3$~GeV$^2$, is selected for comparison. Further details regarding the results comparison between CLAS and F$_{\pi}$-2 are given in Sec.~\ref{sec:u_channel_peak}.

\begin{table}[t]
\centering
\setlength{\tabcolsep}{0.4em}
\caption[Central kinematics for prior $\omega$ electroproduction data]{Central kinematics for prior $\omega$ electroproduction data. Data are arranged with respect to the published date (from past to present). Note that this work is expected to be published in late 2017.}
\label{tab:data}
\begin{tabular}{lcccccc}
\toprule
                        & Publication & $W$        & $Q^2$      & $x$         & $-t$ 	  &  Reference           \\
                        & Date        & GeV        & GeV$^2$    &             & GeV$^2$   &                      \\ \midrule
DESY                    & 1977        & 1.7-2.8    & 0.3-1.4    & 0.1-0.3     & $<$0.5    & \cite{joo77}         \\
Cornell                 & 1981        & 2.2-3.7    & 0.7-3      & 0.1-0.4     & $<$1      & \cite{cassel81}      \\
Zeus                    & 1997        & 40-120     & 3-20       & ~0.01       & $<$0.6    & \cite{zeus97}        \\
JLab Hall C Ambrosewicz & 2004        & $\sim$1.75 & $~\sim$0.5 & 0.2         & 0.7-1.2   & \cite{ambrosewicz04} \\
JLab Hall B Morand      & 2005        & 1.8-2.8    & 1.6-5.1    & 0.16-0.64   & $<$2.7    & \cite{morand05}      \\
HERMES                  & 2014        & 3-6.3      & $>$1       & 0.06-0.14   & $<$0.2    & \cite{hermes14}      \\
JLab Hall C F$_\pi$-2   & 2017        & 2.21       & 1.6,2.45   & 0.29, 0.38  & 4.0, 4.74 &                      \\
\bottomrule
\end{tabular}
\end{table}

\graphicspath{{pics/1Introduction/}}

\chapter{Literature review on Backward-Angle $\omega$ Meson Production}
\label{chap:omega_pro}

\section*{Production Mechanism of the Backward Angle $\omega$}

One of the key questions in this work is the production mechanism of the backward-angle $\omega$ meson. There are several possible interpretations (models) that can result in a backward-angle $\omega$ meson in the final state.

In one model, the $\omega$ is originated from the effect of vector meson dominance (VMD\nomenclature{VMD}{Vector Meson Dominance}), where the virtual photon produced by the incoming electron oscillates into one of the three vector mesons $\rho$, $\omega$ or $\phi$. Equivalent to the Rutherford scattering experiment, as a projectile, the $\omega$ meson recoils at 180$^{\circ}$ from the proton target.

\begin{figure}[t]
\centering
\includegraphics[width=0.8\textwidth]{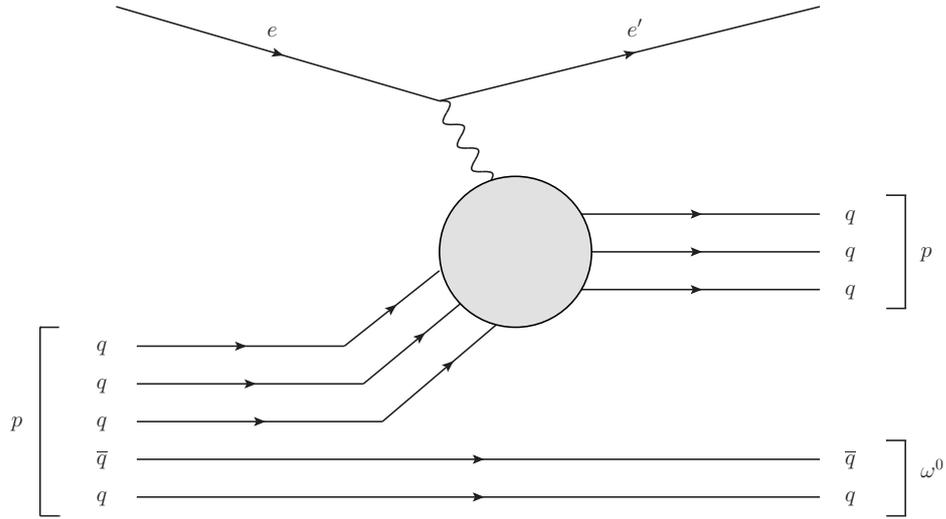}
\caption[Schematic diagram for backward-angle $\omega$ production]{Schematic diagram for backward-angle $\omega$ production. Note that the $q\overline{q}$ pair near the bottom of the plot are from the sea-quark or gluonic contribution from the nucleon structure. Figure created based on the description by Christian Weiss.~\cite{weiss14} }
\label{fig:omega_production}
\end{figure}

A second model is more complex: the $\omega$ is originated from the internal structure of the proton. An intuitive visualization of this interpretation is the following: the proton target consists of three valence quarks and an additional quark-antiquark ($q\overline{q}$) pair from the contribution of the quark sea. The incoming space-like virtual photon interacts with the proton target that includes three valence quarks, which results a ``new'' proton being pushed (at large momentum transfer) out of the target proton, the $q\overline{q}$ pair from the contribution of the quark sea remained target position. A schematic diagram of such interaction is shown in Fig.~\ref{fig:omega_production}. This unique $u$-channel meson interaction reaction is referred as a ``proton being knocked out of a proton process''~\cite{strikman14}. Other possible models, such as the $\omega$ is created by a decayed $N^*$ baryon resonance, are suppressed by the $W$ values of the F$_\pi$-2 data.

In this chapter, both models are examined using the currently available theoretical tools for the backward-angle meson production, to uncover the underlying mechanism for the $u$-channel physics.

\section{$u$-Channel Physics Overview}

\label{sec:u_channel_physics}

\begin{figure}[t]
\centering
\includegraphics[width=1.05\textwidth]{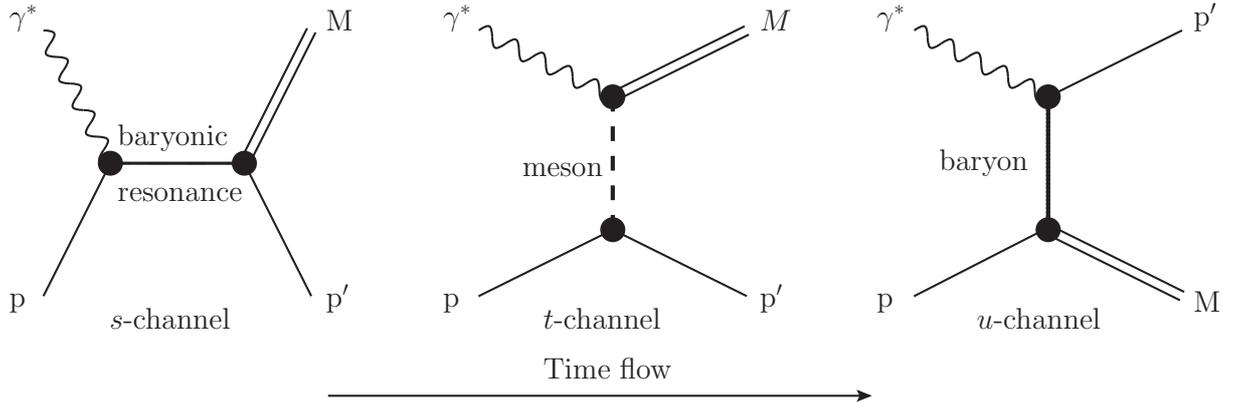}
\caption[Schematic diagrams for $s$, $t$ and $u$ channel scattering interactions]{Schematic diagrams for $s$, $t$ and $u$ channel scattering interactions: $\gamma^{*} + p\rightarrow  \omega + p^{\prime}$. Note that the virtual photon $\gamma^*$ is induced by the incoming electron $e$ and scattered outgoing electron $e^\prime$. Assuming the electron scattering in a fixed target experiment, the exchanged particle in the $s$-channel represents the excited baryonic resonance; the dashed line in the $t$-channel represents the meson exchange; the dashed line in the $u$-channel represents the baryon exchange. The direction of the time flow is from left to right.}
\label{fig:s_t_u_inter}
\end{figure}

In subatomic physics, a given reaction $a + b \rightarrow c + d$ (as shown in Fig.~\ref{fig:man_va}) is categorized as a $u$-channel interaction if the four momentum transfer squared $u=(q-p_p)^2$ approaches zero. $u$-channel interactions in the context of $p+p(\overline{p})$ collisions and the pion-nucleon ($\pi + N$) interaction have been studied for decades, since the 1960s~\cite{collins77,bransden73}, through the Regge theory~\cite{regge59}. These studies concentrate on the $u$-channel meson production processes through the creation of a resonance. One common feature of these early studies is that the $u$-channel interaction was only considered as a contribution (special case) of the $s$-channel interaction~\cite{collins77,bransden73}.

In the context of meson electroproduction, such as $\gamma^* + p \rightarrow \omega + p$, the conservation of the quantum numbers (charge, spin, isospin, parity and baryon number) suggests the exchange of a meson in the $t$-channel, and the exchange of a baryon in the $u$ and $s$ channels, as illustrated in Fig.~\ref{fig:s_t_u_inter}.

Derived from the original Regge theory formalism (described in Sec.~\ref{sec:regge}), the model developed by Vanderhaeghen, Guidal and Laget (VGL model)\nomenclature{VGL model}{Theory model developed by Vanderhaeghen, Guidal and Laget}~\cite{vgl96, guidal97} introduced the saturation of the Regge trajectory (explained in Sec.~\ref{sec:regge}) that allowed the smooth extrapolation of the scattering amplitude to the $-t$ $<$ 0 or $-u$ $<$ 0 regions, which led to the description of meson photoproduction ($\gamma N \rightarrow N \pi$) at low momentum transfer.

In the year 2000, the $Q^2$ dependence to the Regge based model was introduced by J. M. Laget (JML model\nomenclature{JML model}{Theory model developed by J. M. Laget})~\cite{laget04, laget00, laget02}. Currently, the JML model has the capability of describing meson photoproduction and electroproduction ($\gamma^* N \rightarrow \pi N$) data, even in the high momentum transfer range and the high $-t$ region. However, no $u$-channel electroproduction study by JML has been attempted~\cite{laget17}.

Despite its great success during 6 GeV era, the effectiveness of the Regge trajectory models can be further validated with experimental data in 12~GeV era of JLab. It is considered that, as the electron momentum transfer squared $Q^2$ (virtual photon resolving power) is increased to a sufficiently high level, the virtual photons are likely to couple with the partons directly. For this reason, it is beneficial to have a parton-based model that describes the nucleon structure in terms of the fundamental building blocks directly. In the past decade, one of the most important developments in hadronic physics has been the establishment of the theoretical framework of Generalized Parton Distributions (GPD) and Transverse Momentum Distributions (TMD)~\cite{ji97}, which offer the complete spatial and momentum information of the partons inside of a nucleon while fully taking into account the Heisenberg uncertainty principle. A complete understanding of the GPDs is equivalent to a full
spatial image of a nucleon. Currently, there is no known direct experimental access to measure GPDs.

Soon after the introduction of the GPD, a variant of the same framework known as the Transition Distribution Amplitude (TDA) was developed by a B. Pire, et al.~\cite{lansberg08b,lansberg14,pire05}. The TDA specifically describes the reaction of backward-angle meson production~\cite{pire15}, while GPDs are being actively studied through forward-angle meson production~\cite{kroll2016, liuti10}.

\subsection{Gateway to $u$-Channel Physics: $t$-Channel Physics}
\label{sec:link_t_u}

Developments in the Regge trajectory based models have created the linkage between physics kinematic quantities and the experimental observables. As a result, experimental observables at JLab physics are often parameterized in terms of $W$, $x$, $Q^2$ and $t$. By varying a particular parameter while fixing others, one can perform high precision studies to investigate the isolated dependence of the varied parameter for an particular interaction. During the JLab 6~GeV era, the $W$, $Q^2$ and $t$ dependences of exclusive meson photoproduction and electroproduction were actively pursued and resulted in extremely valuable conclusions~\cite{laget04}. Currently, this methodology remains the cleanest access to uncover the underlying mechanism.

In terms of experimental methods at JLab, the $t$-channel interactions are the most simple and straightforward approach, since they require the scattered $e^{\prime}$ (from the electron beam) and newly created particle to travel forward to be detected. In this picture, the recoil nucleon remains at the target station (recoiled 180$^\circ$ backward of the produced meson). The $u$-channel on the other hand, offers a unique and counter-intuitive scattering scenario, where the scattered $e^{\prime}$ and recoil nucleon move forward and the created meson remains at the target station (emitted 180$^\circ$ backward of the detected nucleon in the CM frame). The fact that the backward-angle emitted meson has smaller mass than the forward-going nucleon only makes the $u$-channel interaction more unconventional and interesting. 

The first step in gaining understanding of the $u$-channel interaction and uncovering the underlying physics mechanism is to understand the physical significance of $Q^2$ and $t$ (evolution of proton structure).  

Assuming the meson electroproduction interaction with a fixed $W$ value higher than the resonance region ($W>2$ GeV) and $x\sim0.3$, $Q^2$ can be visualized as the resolving power (wavelength of the virtual photon propagator $\lambda$) and $t$ is analogously linked to the impact parameter ($b$) of the interaction through $$ b \approx \frac{\hbar c}{\sqrt{-t}},$$ where the $\hbar$ is the Planck's constant and $c$ is the speed of light.

\begin{figure}[t]
\centering
\includegraphics[width=1\textwidth]{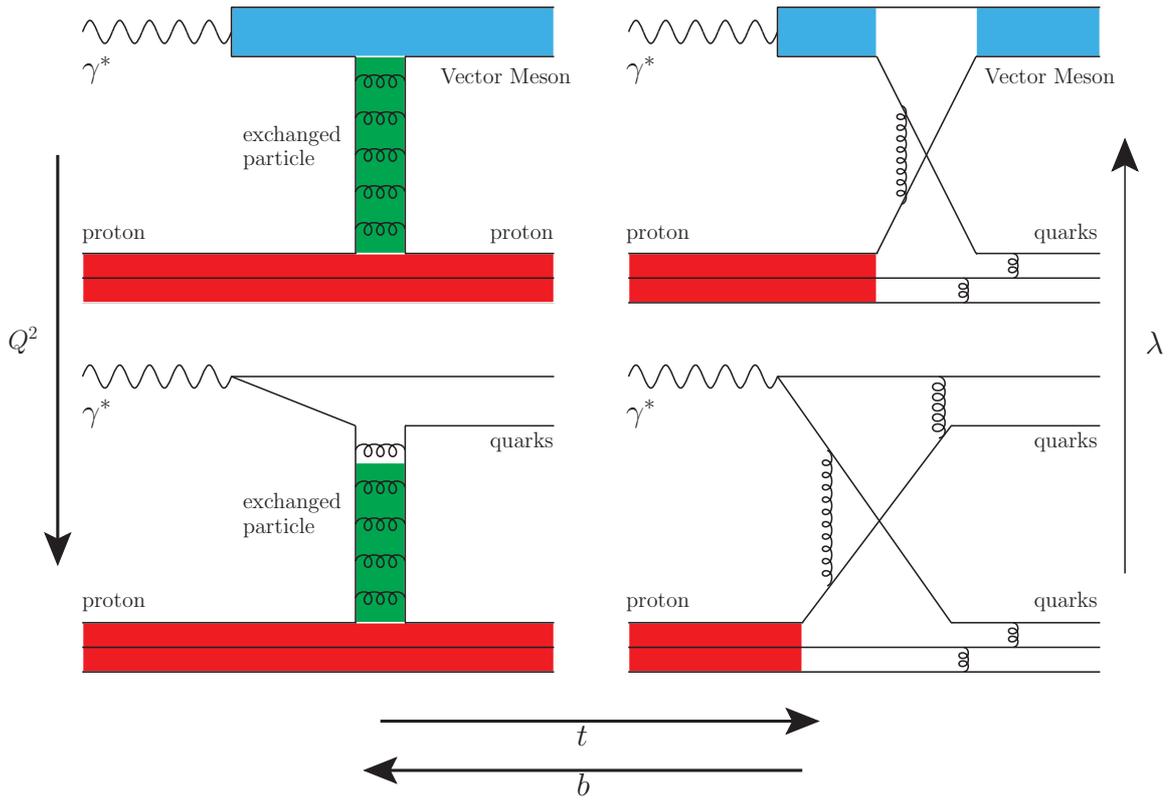}
\caption[A schematic view of the evolution of hard-scattering mechanisms in terms of $Q^2$ and $t$]{A schematic view of the evolution of hard-scattering mechanisms in terms of $Q^2$ and $t$. Plot created based the original from Ref.~\cite{laget04}.}
\label{fig:laget}
\end{figure}

Since the probe of the interaction is the virtual photon, two things will happen. First, as $Q^2$ increases the lifetime $\Delta \tau = 2\nu/(Q^2 + m^2_N)$ of its hadronic component decreases, therefore its coupling becomes more point-like. Second, the wavelength ($\lambda \sim 1/Q$) of the virtual photon decreases. This concept is illustrated in Fig.~\ref{fig:laget}~\cite{laget04}, in terms $Q^2$ and $-t$.

When both $Q^2$ and $-t$ are small (top left panel of Fig.~\ref{fig:laget}), the photon behaves as a beam of vector mesons which pass far away from the nucleon target (implying a large impact parameter $b$). The partons that may be exchanged have enough time to interact with each other and build various mesons. 

At low $Q^2$ and high $-t$ (top right panel of Fig.~\ref{fig:laget}), the small impact parameter $b$ corresponds to the hadronization length of the partons that are absorbed or recombined into the final state particles (within the interaction volume defined by $b$), before they hadronize. In simple terms, a pair of partons are exchanged between the meson and the nucleon and a gluon is exchanged between this pair of partons.

When $Q^2$ increases, the resolving power of the photon increases and begins to probe processes which occur at shorter and shorter distances and can couple to the constituents of the exchanged particles. When $-t$ is small (bottom left of Fig.~\ref{fig:laget}), the photon probes only the quarks inside the pion that is exchanged between the proton and the outgoing meson. When $-t$ and $Q^2$ are both large (bottom right of Fig.~\ref{fig:laget}), the quarks inside the proton are able to couple directly to the quarks inside the target because the wavelength $\lambda$ becomes comparable to the impact parameter $b$. The virtual photon sees the partons which are exchanged during the hard scattering.

This classical interpretation of the evolution of hard scattering was developed from Regge theory~\cite{vgl96} (further discussed in Sec.~\ref{sec:regge}), and has been successfully implemented to explain both meson photoproduction~\cite{vgl96, guidal97, laget04,laget00} and electroproduction~\cite{laget04}. Excellent agreement has been achieved between model and data. A extension to this interpretation using $u$ instead of $t$ is expected in the near future~\cite{laget17}.

\section{Regge Trajectory Model}
\label{sec:regge_chap}

\subsection{Regge Trajectory}
\label{sec:regge}

This section gives a brief summary on the concept of Regge trajectories and some of their most important features.

The partial-wave method introduced in Refs.~\cite{chew61,sakurai85} is a common methodology to analyze the scattering processes~\cite{tang00}. Consider the wavefunction in the form of
\begin{equation}
\psi({\bf r}) \simeq e^{i {{\bf k}}\cdot {\bf r}} + f({\bf k}, \cos\theta) ~ \frac{e^{i {\bf k}\cdot {\bf r}}}{\bf r}\,,
\end{equation}
where $\theta$ is the angle between the wave vector $\bf k$ and the position vector $\bf r$. In the case of bound states, the plane wave (first) term is absent. The form factor $f$ is written as a sum of partial waves as~\cite{regge59, tang00, regge60}
\begin{equation}
f(k^2, \cos \theta) = \sum^\infty_{l=0}~(2l+1)~a_l(k^2)~P_l(\cos \theta),
\end{equation}
if $$a_l (k^2) = \frac{1}{2}\int^{+1}_{-1} (2l+1)~f(k^2, \cos\theta)~P_l(\cos \theta) ~ d\cos\theta,$$
where $l$ is the orbital angular momentum quantum number and $P_l$ is the Legendre polynomial of degree $l$. In the initial introduction of Regge theory~\cite{regge59}, T. Regge generalized the solution for the solution of $f$ by treating $l$ as a complex variable. It was proven that for a wide class of potentials, the singularities of the scattering amplitude (simple poles $a_l(k^2)$ ) in the complex $l$ plane were poles, now known as the Regge poles~\cite{collins77, bransden73, chew62}.

For real values of $l$,  $Re(l) \ge -1/2$, the partial-wave components of the scattering amplitude have only simple poles and are functions of $k^2$,
\begin{equation}
a_l(k^2) \simeq \frac{\beta{k^2}}{l-\alpha(k^2)},
\end{equation}
where $\beta$ is the Regge residue and $\alpha$ is the position (Regge trajectory) of the poles. These poles correspond to the bound states or the resonances (baryons and mesons)~\cite{tang00}.

\begin{figure}[t]
\centering
\subfloat[][Meson exchange trajectory]{\includegraphics[width=0.515\textwidth]{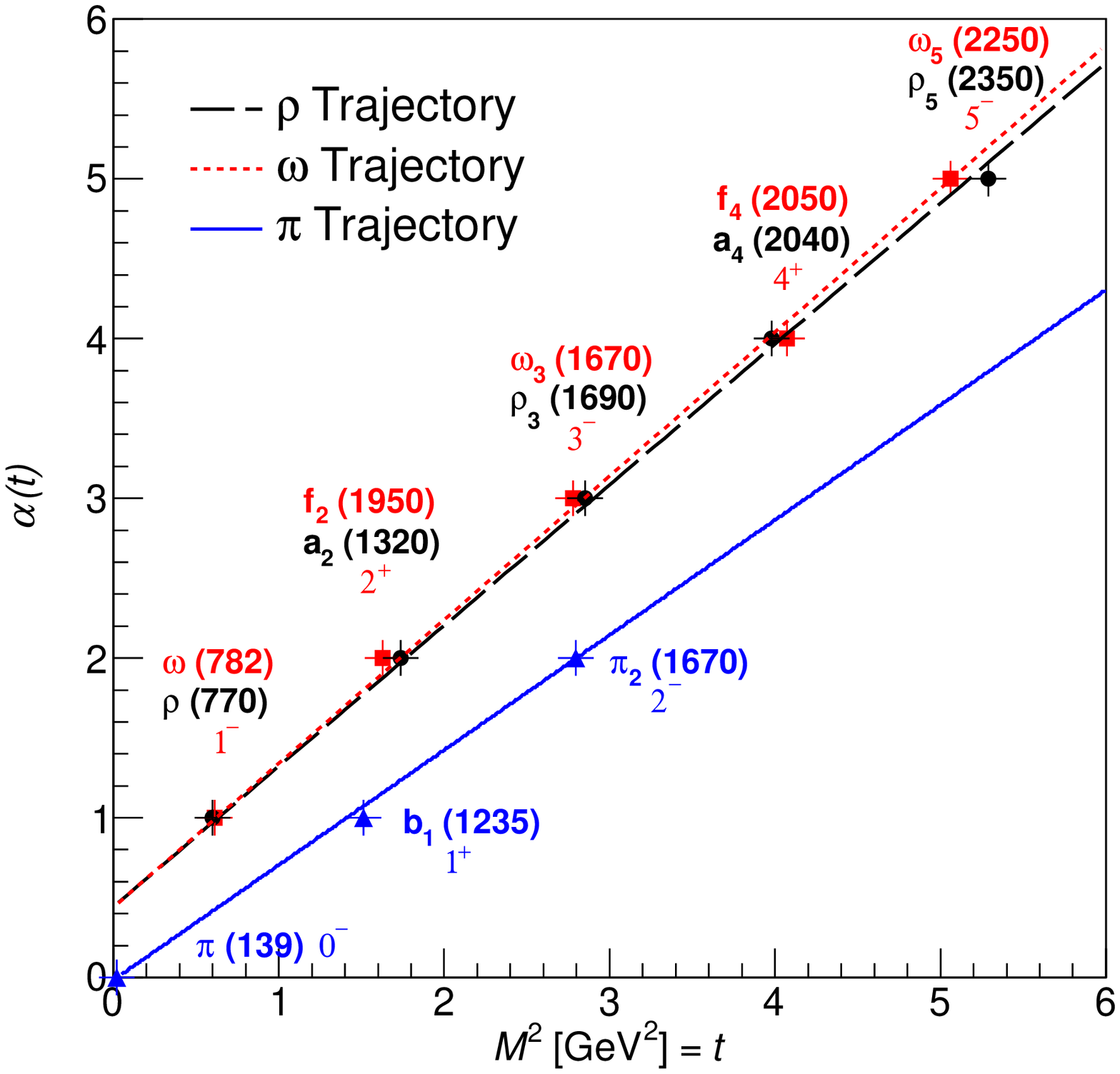}}
\subfloat[][Baryon exchange trajectory]{\includegraphics[width=0.515\textwidth]{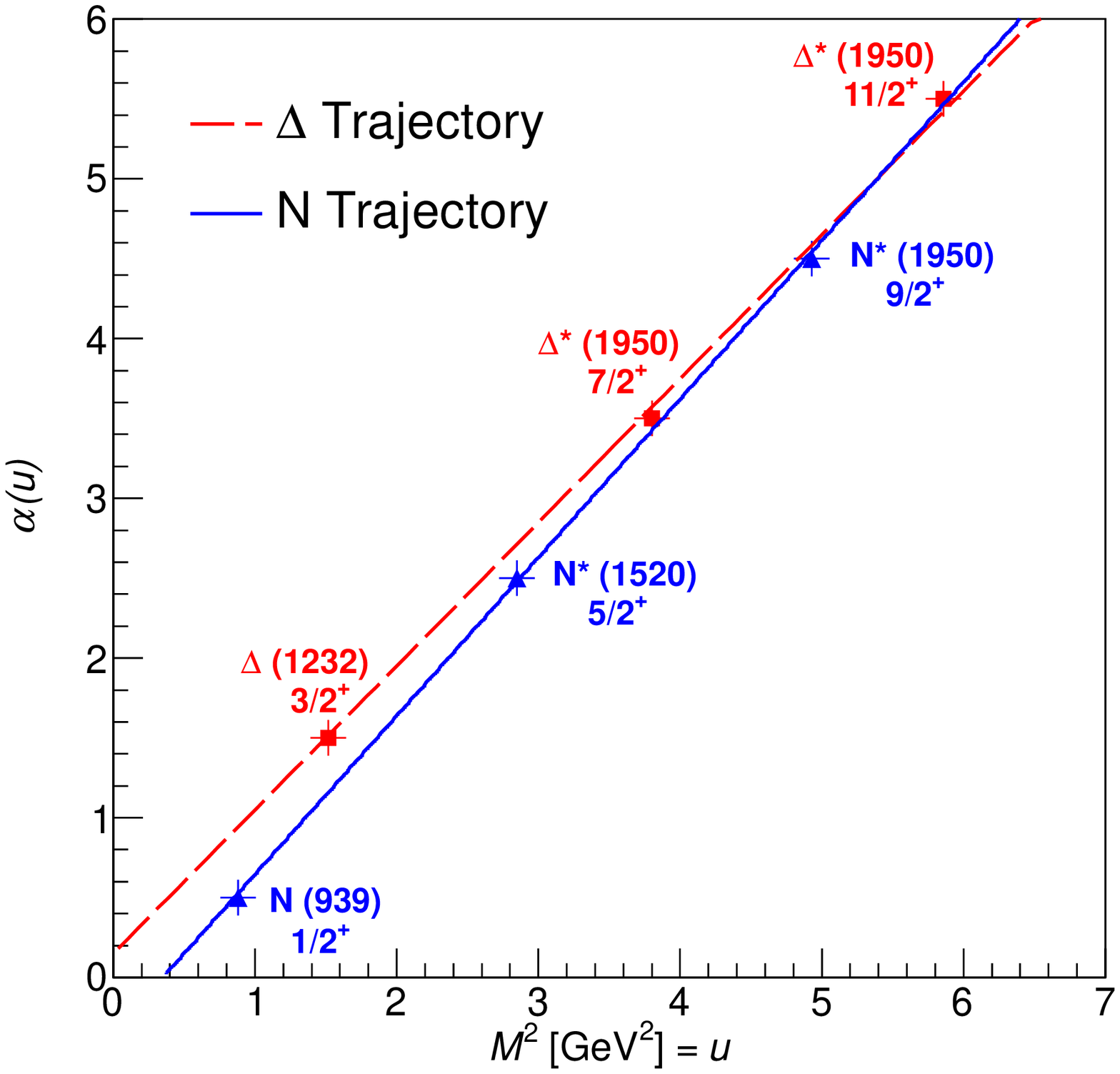}}
\caption[The meson and baryon exchange trajectories]{Figures (a) and (b) show the meson and baryon exchange trajectories, respectively. On the y-axis, $\alpha(t)$ and $\alpha(u)$, represents the real base of $\alpha$ trajectories, and are equivalent the total angular momentum quantum number ${\rm J}$ according to the Regge theory. On the x-axis, $M^2$ is interchangeable with the Mandelstam variable $t$ (or $u$) for the meson (or baryon) trajectories. Note that the shown trajectories are for demonstration of the linear relation between the $\rm J$ and $m^2$, therefore should not be used for as the actual Regge trajectory based calculation. }
\label{fig:Regge_traj}
\end{figure}

The Chew-Frautschi plots~\cite{chew62} that project the spin quantum number ($\rm J$) on $y$ axis and rest mass squared $M^2$ on the $x$ axis, for meson and baryon are shown in Figs.~\ref{fig:Regge_traj} (a) and (b), respectively. From the phenomenological point of view, the $\rm J$ values of the resonances seem to be linearly correlated to the $M^2$ values over a set of particles of a fixed radial node number $n$. Furthermore, Chew-Frautschi~\cite{chew61} were able to apply the Regge (pole) theory to investigate the properties of these linear trajectories ($\alpha(k^2)$) in the case of the strong interaction. This approach was a success, which allowed the Regge trajectory based models~\cite{collins77,vgl96,laget00,chew62} to predict the scattering amplitudes, the form factors and the experimental observables such as the cross sections which depend on the experimental kinematics variables such as $s$, $t$ and $u$.

In the Regge model, ${\rm J} = \alpha(k^2)$ is also sometimes expressed as ${\rm J}=\alpha(E)$, or more commonly in terms of the Mandelstam variable $t$ as ${\rm J}= \alpha(t)$, or $u$ as ${\rm J}=\alpha(u)$. In the $t$-channel (forward-angle) interaction, it is more convenient to use the ${\rm J}= \alpha(t)$ representation, which reflects the forward-angle meson production. Similarly, $\rm{J} = \alpha(u)$ is used for the $u$-channel (backward-angle) interaction. Note that the condition $\alpha(t)<0$ or $\alpha(u)<0$ does not correspond to any physical particles (pole) because $\rm{J}$ cannot be negative~\cite{forshaw97}.

\begin{figure}[t]
	\centering
	\subfloat[][$t$-channel]{\includegraphics[width=0.5\textwidth]{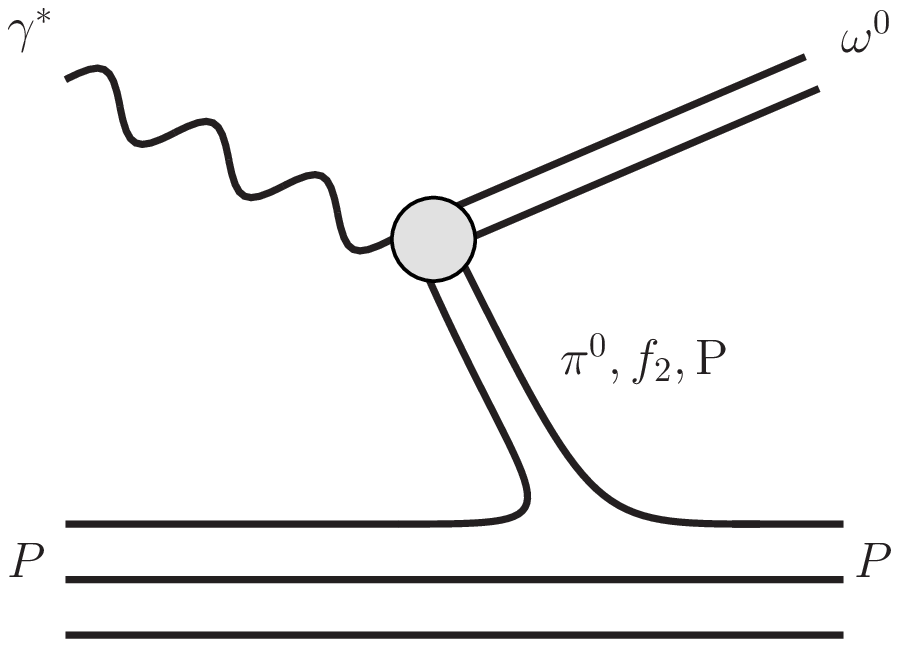}\label{fig:tchannel}}
	\subfloat[][$u$-channel]{\includegraphics[width=0.51\textwidth]{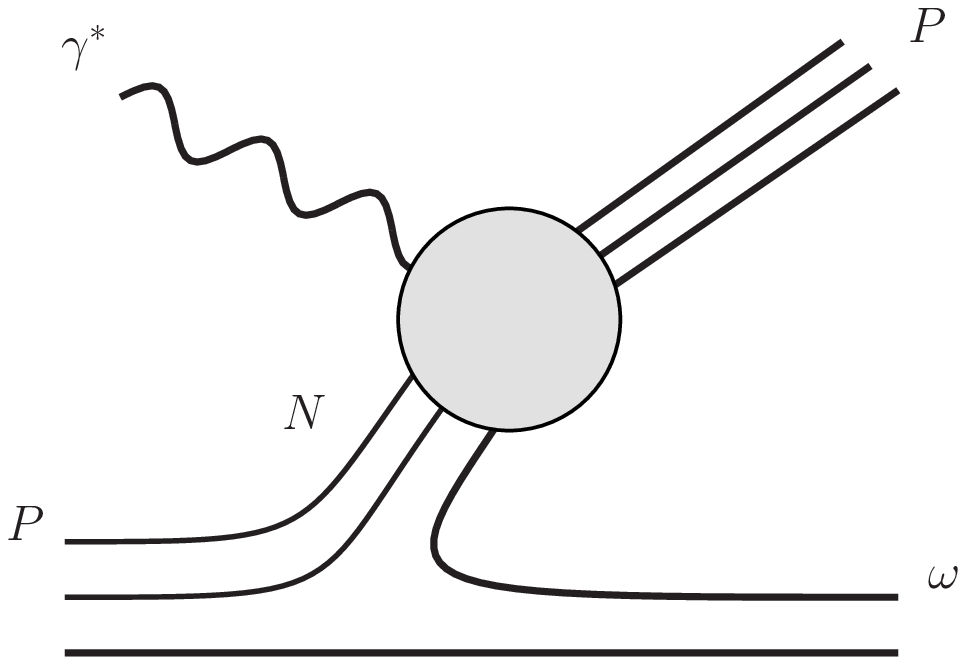}\label{fig:uchannel}}
	\caption[Regge model $t$ and $u$ channel interaction]{Regge model $t$ and $u$ channel interaction. (a) shows the $t$-channel $\omega$ production interaction, and the exchanged particles are based on the Regge trajectory model include $\pi^0$, $f_2$ and \textbf{P}. (b) shows the $u$-channel $\omega$ production interaction when a baryon is required to be exchanged.}
	\label{fig:regge_trajectory}
\end{figure}

Along with the existence of the primary Regge trajectory, there are also the daughter trajectories and the Regge residue. To reduce the level of complication, discussions of these are excluded. Complete discussions on these topics can be found in Refs.~\cite{collins77,bransden73,chew62}.

\subsection{$u$ and $t$ Kinematic Limits}
\label{sec:u_t_limits}

In pseudoscalar meson photoproduction ($E_\gamma>4$ GeV) reactions (such as $\gamma p \rightarrow n\pi^+$), the main feature involves a pair of strongly collimated peaks at forward ($|t|$ $\le$ 2~GeV$^2$), and backward-angles ($|u|$ $\le$ 1~GeV$^2$)~\cite{vgl96,guidal97}. Similar to the particle exchange diagram for $\omega$ production shown in Figs.~\ref{fig:regge_trajectory} (a) and (b), the $t$-channel interaction (peak) is dominated by meson exchange and $u$-channel interaction (peak) is dominated by baryon exchange. Note, this interpretation divides the interaction process into three separate regions with respect to $t$ (or $u$), and each of the three regions are dictated by different interaction mechanisms.

In the case of the electroproduction of the vector meson above the resonance production region ($W$ $>$ 2~GeV) and large momentum transfer ($Q^2$ $>$ 2~GeV$^2$), a similar feature in terms of the cross section behavior is expected as observed in pseudoscalar meson photoproduction. Currently, the strong forward-angle ($t$-channel) peak has been experimentally measured~\cite{morand05}. However, the expected existence of the backward-angle ($u$-channel) peak for vector meson electroproduction was not verified due to lack of experimental data until this Ph.D. work.

In this analysis work, the electroproduction of the vector mesons is divided in into three interaction regions with respect to $t$ (or $u$). The definition of the $t$ and $u$ limits are chosen based on similar definitions introduced by Ref.~\cite{vgl96}:
\begin{description}
\item[Low $-$\textit{t} Region:] $-t_{\rm min}$ $<$ $-t $ $<$ 1 GeV$^2$,
\item[Low $-$\textit{u} Region:] $-u_{\rm min}$ $<$ $-u$ $<$ 1 GeV$^2$,
\item[Large Emission Angle (LEA) Region:]  1 GeV$^2$ $<$ $-t$ $<$ $-t~(-u$ = 1 GeV$^2)$ or \\ 0.5 GeV$^2$ $<$ $-u$ $< -u~(-t=$ 1 GeV$^2)$.
\end{description}

Using the imposed momentum conservation constraints on the Mandelstam variables given by Eq.~\ref{eqn:constraints}, if the experiment has fixed $W$ and $Q^2$ values, the $t$ values can be converted into $u$. Thus, a small $t$ value corresponds to a large $u$ value, and vice versa.

\begin{figure}[p]
  \centering
  \includegraphics[width=0.9\textwidth]{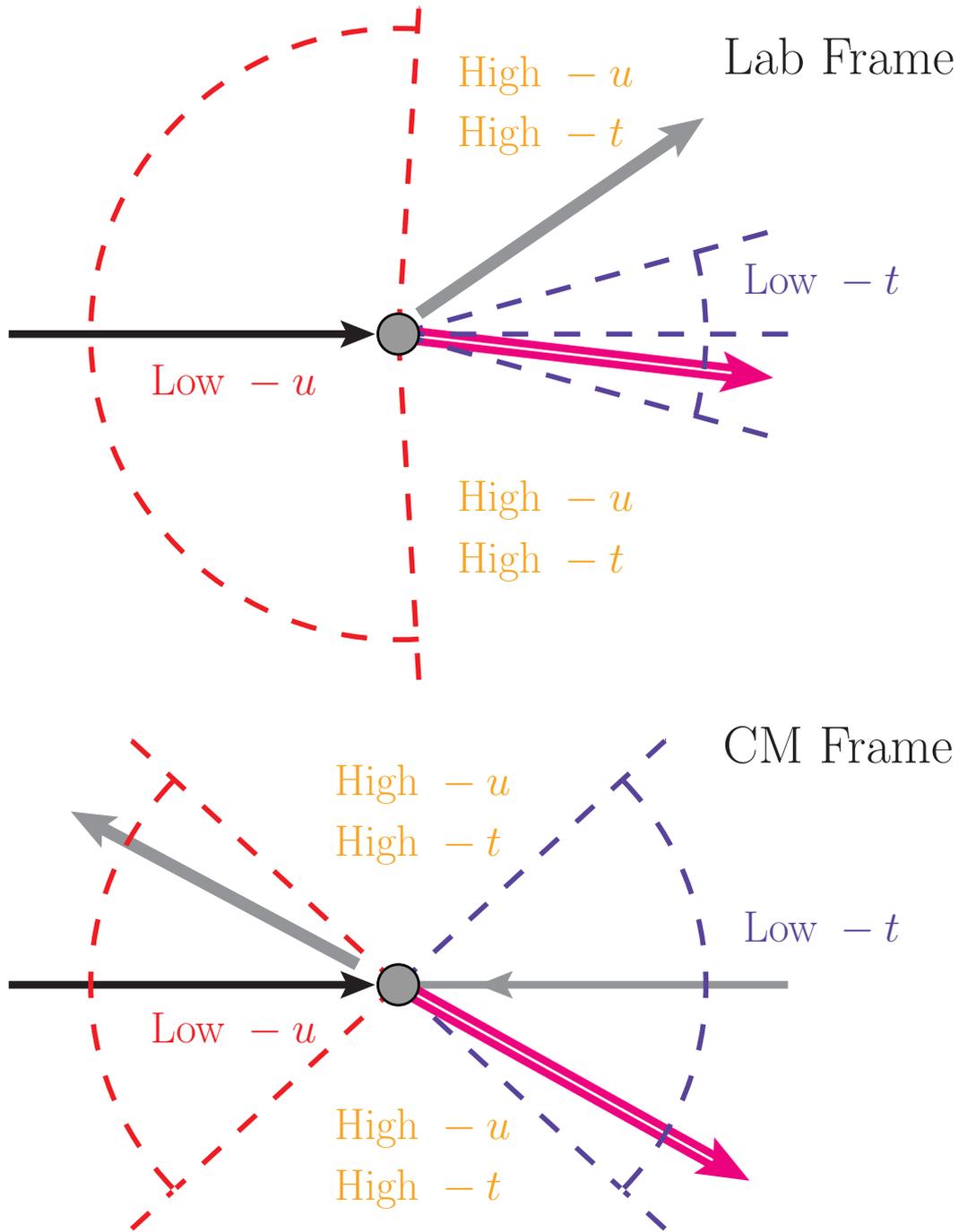}  
  \caption[Scattering angle distribution limits in $t$ and $u$ kinematics converage]{Scattering angle distribution limits for the low $-u$, low $-t$ and LEA (high $-u$ and high $-t$) regions, in both Lab frame (top) and CM frame (bottom). An $t$-channel meson production interaction through $\gamma^*p$ is shown as an example. In each diagram, the black thin arrow indicates the incoming virtual photon probe; thin grey arrow and circle describe the motion of proton target before the interaction; pink thick arrow is the produced meson; the thick grey represents the recoil proton after the interaction. Due to the Lorentz boost, the region limits in the Lab frame appear to be dramatically different from the limits in the CM frame.~\oic}
  \label{fig:scattering_u_t}
\end{figure}

In terms of $t$ coverage, the upper limit of LEA region does not correspond to the maximum possible $-t$ value: $-t_{\rm max}$. The $-t_{\rm max}$ value is inside of the low $-u$ region. The LEA upper limit is defined by the corresponding $-t$ value of $-u$ = 1~GeV$^2$ (low $-u$ upper boundary). Similarly, the upper limit of the LEA region in terms of $u$ coverage is defined by the corresponding $-u$ value of $-t$ = 1~GeV$^2$ (low $-t$ upper boundary). Depending on experimental kinematics, the boundaries between the three regions can vary, however, there is no overlap between the low $-t$ and low $-u$ regions.

Fig.~\ref{fig:scattering_u_t} shows three regions in terms of the scattering angle in both Lab and CM reference frames. A $t$-channel scattering process is used as an example.  Note the boundary lines between different regions are for illustration purposes only, which do not correspond to any particular $-t$ and $-u$ values.

\subsection{Regge Trajectory in Meson Production}

\label{sec:regge_model_intro}

In Regge-trajectory-based models, the standard treatment to take into account the exchange of high-spin, high-mass particles is to replace the pole-like Feynman propagator of a single particle (i.e. $\frac{1}{t-M^2}$) by the Regge (trajectory) propagator. Meanwhile, the exchange process involves a series of particles of the same quantum number (following the same Regge trajectory $\alpha(t)$), instead of single particle exchange. As an example, the Regge propagator for the pion trajectory is given as
\begin{equation}
\mathcal{P}_{Regge} = \frac{(\frac{s}{s_0})^{\alpha(t)}}{\sin(\pi\alpha(t))}  \frac{1+\zeta e^{-i \pi \alpha(t)}}{2} \frac{1}{\Gamma(1+\alpha(t))},
\end{equation}
where $\zeta=\pm1$ is the signature of the exchanged trajectory, and $\alpha(t)$ is the meson trajectory obtained from Chew-Frautschi plots such Fig.~\ref{fig:Regge_traj} (a). For vector meson ($\rho$, $\omega$ and $\phi$) production, the Regge propagator can be constructed in a similar form.

The required exchanged particles (trajectories) for vector meson production are listed in Table.~\ref{tab:regge_table}. For forward $\omega$ ($t$-channel) production the dominant trajectories are $\pi^0$ (plotted in Fig.~\ref{fig:Regge_traj} a), ${\rm f}_2$ and Pomeron ($\mathbb{P}$)~\cite{vgl96, guidal97, laget04, laget00, laget02}; in the backward-angle scenario ($u$-channel), the dominant baryon trajectory is $\Delta$~\cite{laget04, laget00, laget02}.

For the forward hard scattering process where $t<<0$, the meson trajectories are assumed to approach $-1$ (asymptotic limit)~\cite{vgl96}. This is known as the saturation of the Regge trajectory. Note that the saturation effect also applies to the backward hard scattering process where $u<<0$. Saturation is an extremely important and profound assumption, which allows a smooth transition and extrapolation from the soft scattering amplitude $\mathcal{M}_{soft}$ at $t>0$ (or $u>0$) to the hard scattering amplitude ($\mathcal{M}_{hard}$) at $t<0$ (or $u<0$)~\cite{vgl96}
\begin{equation}
\mathcal{M}_{hard} = \mathcal{M}_{soft}~F_3(t)~F_{4}(t)
\end{equation}
or
\begin{equation}
\mathcal{M}_{hard} = \mathcal{M}_{soft}~F_3(u)~F_{4}(u),
\end{equation}
where $F_3$ and $F_4$ are form factors of the two outgoing particles.

\begin{figure}[t!]
  \centering
  \includegraphics[width=0.75\textwidth]{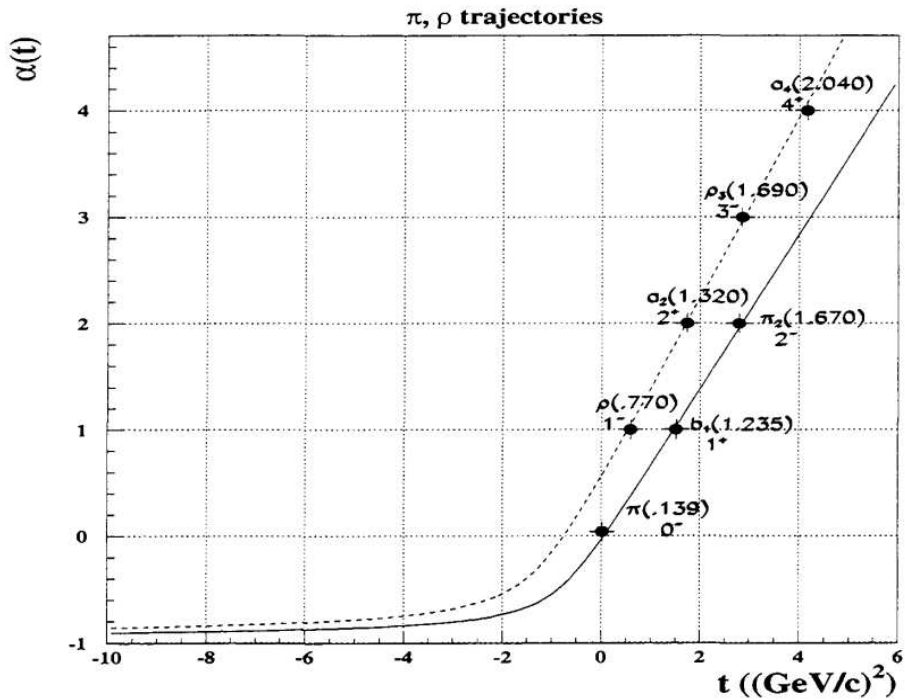}  
  \caption[Regge trajectory saturation for $\rho$ and $\pi$]{Regge trajectory saturation for $\rho$ and $\pi$. Original plot was from Ref.~\cite{vgl96, guidal97}.}
  \label{fig:saturation}
\end{figure}

Examples of the Regge trajectory saturation of $\alpha$ ($\alpha(t)\rightarrow-1$ or $\alpha(u)\rightarrow-1$) for $\pi$ and $\rho$ are shown in Fig.~\ref{fig:saturation}. As the result of the saturation effect, the differential cross sections will tend to a plateau in the LEA range since the exponential $t$-dependence ($e^{\alpha(t)}$) or $u$-dependence ($e^{\alpha(i)}$) vanishes. In potential models, the saturation of the Regge trajectories (approaching $-1$ when $-t\propto \infty$) is closely related to the one-gluon exchange interaction between two quarks~\cite{vgl96}.

\subsection{VGL and JML Models}

With the introduction of the saturation of the Regge trajectory~\cite{vgl96, guidal97}, Regge-based models such as VGL~\cite{vgl96, guidal97} and JML~\cite{laget04, laget02} have become effective methods to deal with hard-scattering mechanisms in the non-resonance region ($t<0$ and $u<0$) and have been successfully used to describe the meson photoproduction in $t$ and $u$-channels, and the meson electroproduction in $t$-channel.

\begin{figure}[t!]
  \centering
  \includegraphics[width=0.85\textwidth]{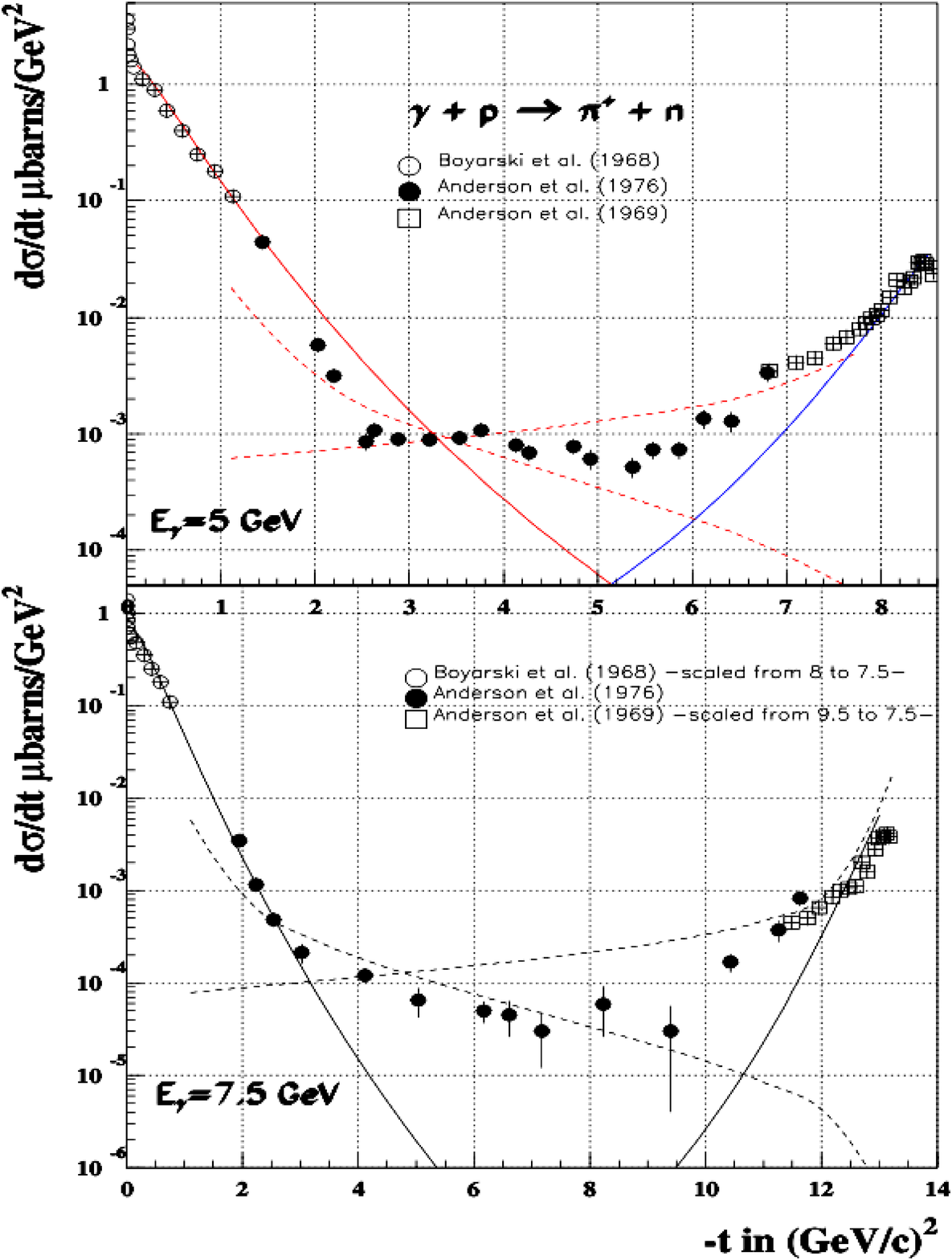}  

  \caption[Differential cross section charged pion photoproduction]{Differential cross section $\frac{d\sigma}{dt} (\gamma p \rightarrow n \pi^+)$ ~\cite{guidal97}. Solid lines represents the ``soft'' VGL model and dashed lines represents the ``hard''. The data are from Refs.~\cite{anderson69, anderson76, boyarski68}. Original plot was from Ref.~\cite{vgl96, guidal97}.}
  \label{fig:vgl_data}
\end{figure}

The VGL model has been validated with experimental data of pion photoproduction from Refs.~\cite{anderson69, anderson76, boyarski68}. Fig.~\ref{fig:vgl_data} shows the VGL model to data comparison. The peaks at $t$-channel ($t<1$~GeV$^2$) and $u$-channel ($t>13$~GeV$^2$ for $E_\gamma>7$~GeV or $E_\gamma>12$~GeV which corresponds to $u<1$~GeV$^2$) were successfully described by the model. The experimental data features three distinctive regions across the $t$ range (as described in Sec.~\ref{sec:u_t_limits}): the low $-t$ region, the $t$-channel peak dictated by the ``soft'' process of meson exchange; the LEA region, cross section plateau is the indication ``hard'' process; the $-u$ region, $u$-channel peak dictated by the ``soft'' process of hard baryon exchange. Here, the soft process refers to the photon probing the parton bound states (soft structure) inside of the nucleon; whereas the hard process describes the photon directly probing the point-like parton (hard structure). This
classic interpretation offered by the VGL model on the soft-hard-soft transition carries special significance in understanding the evolution of the scattering process with respect to $t$, and is elaborated in Sec.~\ref{sec:link_t_u}.

The $\rho^0$, $\omega$ and $\phi$ meson photoproduction~\cite{airapetian00,battaglieri03, battaglieri01} and the electroproduction of $\omega$ data from CLAS~\cite{morand05} have demonstrated the predictive power of the JML model in the low $-t$ and LEA regions. Compared to the VGL model, the JML model has included the $Q^2$ dependence. The validation of the $Q^2$-dependence extension of the JML model came from $\omega$ electroproduction for $Q^2\sim2.35$~GeV$^2$ data from CLAS~\cite{morand05}, and is further discussed in Sec.~\ref{chap:results}. The JML model was a successful milestone, significantly improving the knowledge regarding the hard scattering mechanism and establishing the direct linkage between kinematics variables (such as $t$) to the impact parameter~\cite{laget04, laget00, laget02} (described in Sec.~\ref{sec:link_t_u}).

Despite the great successes of Regge trajectory based models, there are some limitations that may require further research effort. 
\begin{itemize}

\item As introduced in Sec.~\ref{sec:link_t_u}, the classical interpretation of $Q^2$ is considered to be the resolving power of the probe and is inversely proportional to the virtual photon wavelength ($Q^2$ $\sim$ $1/\lambda$). As $Q^2$ increases beyond the effective $Q^2$ range of the Regge model, the virtual photon wavelength (interaction radius) would decrease and start to directly couple to the parton structure. Currently, the upper limit of effective range of $Q^2$ for the Regge theory has not been determined~\cite{morand05}. The study of the transition of the Regge theory in terms of $Q^2$ would be beneficial to the understanding off the proton structure in terms of the quarks and gluons, and their interaction mechanism,

\item In the case of the $\omega$ meson, the model predicts the dominance of the transverse component of the cross section $\sigma_{\rm T} >> \sigma_{\rm L}$ at large value of $Q^2$~\cite{morand05},

\item No calculation is available for $u$-channel electroproduction. Furthermore, the behaviour of the differential cross sections inside the LEA region is also unknown. See further discussion in Sec.~\ref{sec:u_channel_peak}.

\end{itemize}

\begin{table}
\centering
\caption[Main Regge trajectories for $t$ and $u$-channel interactions]{Table contains the main trajectories for the $t$-channel meson exchange and $u$-channel baryon exchange for $\gamma p \rightarrow p \rho^0$, $\gamma p \rightarrow p \omega$ and  $\gamma p \rightarrow p \phi$~\cite{laget04, laget00, laget02, morand_thesis}. Exchanged particle $\mathbb{P}$ represents the Pomeron. The Regge trajectories for listed exchanged meson and baryon are shown in Fig.~\ref{fig:regge_trajectory}. $^*$ The $u$-channel contribution of the $\phi$ production is unclear, currently, $\phi NN$ coupling is arbitrarily chosen based on ref.~\cite{jaffe89}.}
\label{tab:regge_table}
\begin{tabular}{lcccc}
\toprule
          &  Quark Composition                        &  $t$-channel         &  $u$-channel                            \\ \midrule
$\rho^0$  &  $\frac{u\overline{u}-d\overline{d}}{2}$  &  f$_2$, $\sigma$, $\mathbb{P}$ &   $\Delta$, $\Delta$-$N$ Interference   \\ 
$\omega$  &  $\frac{u\overline{u}+d\overline{d}}{2}$  &  $\pi$, f$_2$, $\mathbb{P}$    &  $N$                                    \\ 
$\phi$    &  $s\overline{s} $                         &  $\mathbb{P}$                  &  Unknown $\phi NN$ coupling$^*$         \\ 
\bottomrule
\end{tabular}
\end{table}

\section{GPD and TDA}
\label{sec:TDA}

As introduced in Sec.~\ref{sec:u_channel_physics}, Generalized parton distributions (GPDs) are an improved description of the complex internal structure of the nucleon, which provide access to the correlations between the transverse position and longitudinal momentum distribution of the partons in the nucleon. In addition, GPDs give access to the orbital momentum contribution of partons to the spin of the nucleon~\cite{ji97, jo12}.

In 1932, E. P. Wigner formulated a way to express quantum mechanical correlations using the language of classical statistical mechanics~\cite{wigner32}, which was later applied to describe the behaviour of quarks and gluons inside of the nucleon. 

Assuming a one-dimensional quantum mechanical system with wave function $\psi(x)$, the Wigner function is defined as~\cite{ji04}
\begin{equation}
W_{\Gamma} (x, p) = \int \psi^* (x-\eta/2)~\psi(x+\eta/2)~e^{ip\eta} d\eta,
\label{eqn:Wigner}
\end{equation}
where $\hbar$ is set to $\hbar = 1$; $x$ represents the position vector; $p$ is the momentum vector; $\eta$ represents the space-time separation. When integrating out the spatial information in $x$, one can obtain the momentum density $|\phi(p)|^2$; when integrating over the momentum space $p$, one can obtain the spatial density $|\phi(x)|^2$. This is a unique functionality that allows the Wigner distribution (derived from the Wigner function) to contain the most complete (spatial and momentum) information about a quantum system, while respecting the Heisenberg uncertainty principle~\cite{ji97}.

After constructing the ``rest-frame'' matrix element and averaging over all possible three-momentum transfer, the quantum phase-space quark distribution in a nucleon can be written as~\cite{ji04,dueren11}:
\begin{equation}
W_{\Gamma} (\vec{r}, k) = \dfrac{1}{2M} \int \dfrac{d^3\vec{q}}{(2\pi)^3}\langle\vec{q}/2 | \mathcal{\widehat{W}} |-\vec{q}/2\rangle\,,
\end{equation}
where $\mathcal{\widehat{W}}$ is the Wigner operator, $\vec{r}$ is the quark phase-space position; $k$ is the phase-space four momentum.

By integrating the transverse quark momentum information, the quark spatial structure of the proton is considered to be described by four independent leading twist helicity non-flip GPDs: $E$, $\tilde{E}$, $H$, $\tilde{H}$~\cite{ji04}. All of them are functions of longitudinal parton momentum $x$, of the momentum transfer squared $t$ and of the skewness parameter $\xi$, which is related to $x$ by $ \xi = x/(2 - x)\,$. By integrating over the GPDs across the nucleon radius, one can access the electric and magnetic distributions of the nucleon. Note that there are four additional GPDs associated with the helicity flip, which are not discussed in this thesis. Correspondingly, the eight gluon GPDs can be obtained following the same principle~\cite{ji04}.

\begin{figure}[t]
	\centering
	\subfloat[][$t$-channel]{\includegraphics[width=0.5\textwidth]{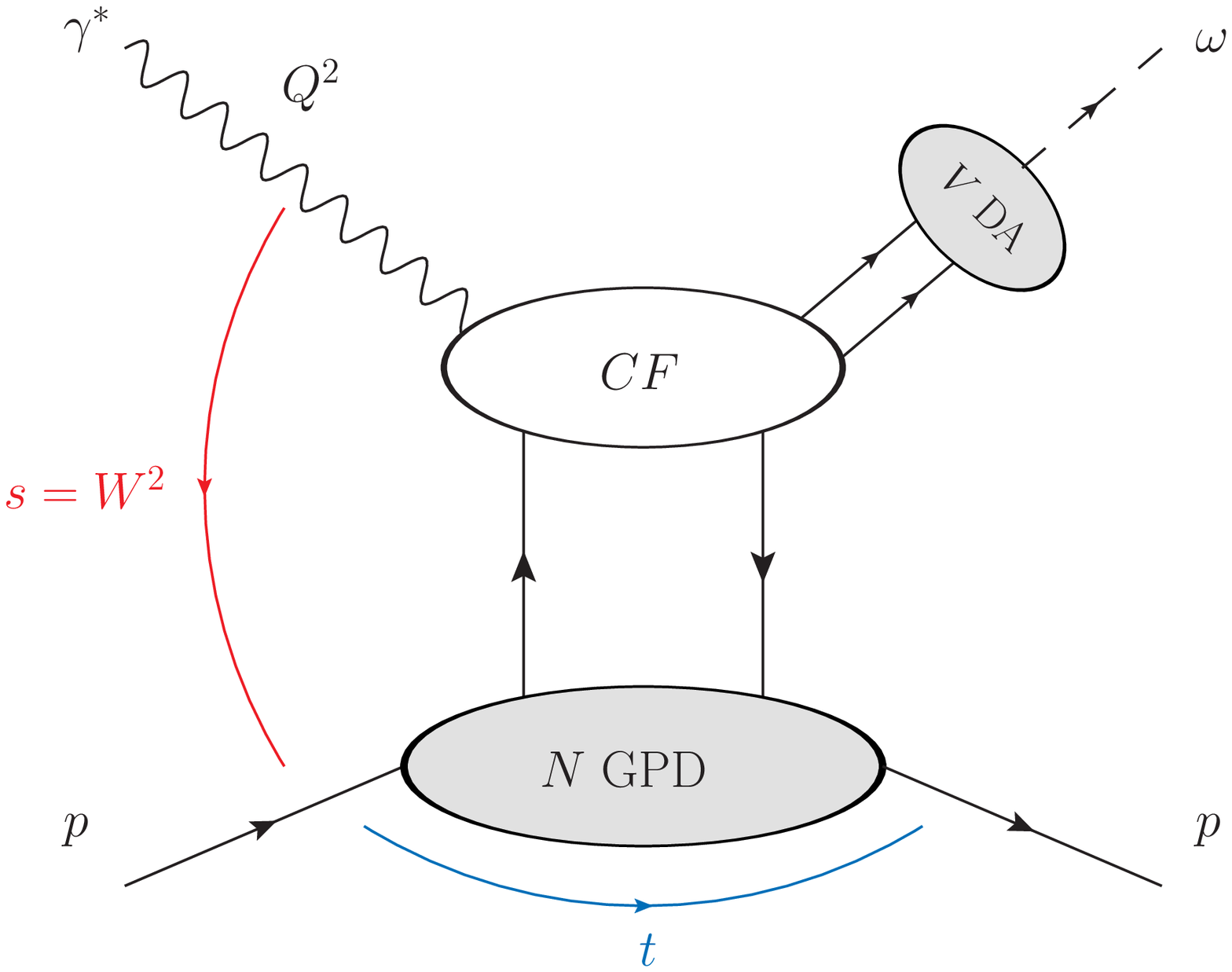}}
	\subfloat[][$u$-channel]{\includegraphics[width=0.5\textwidth]{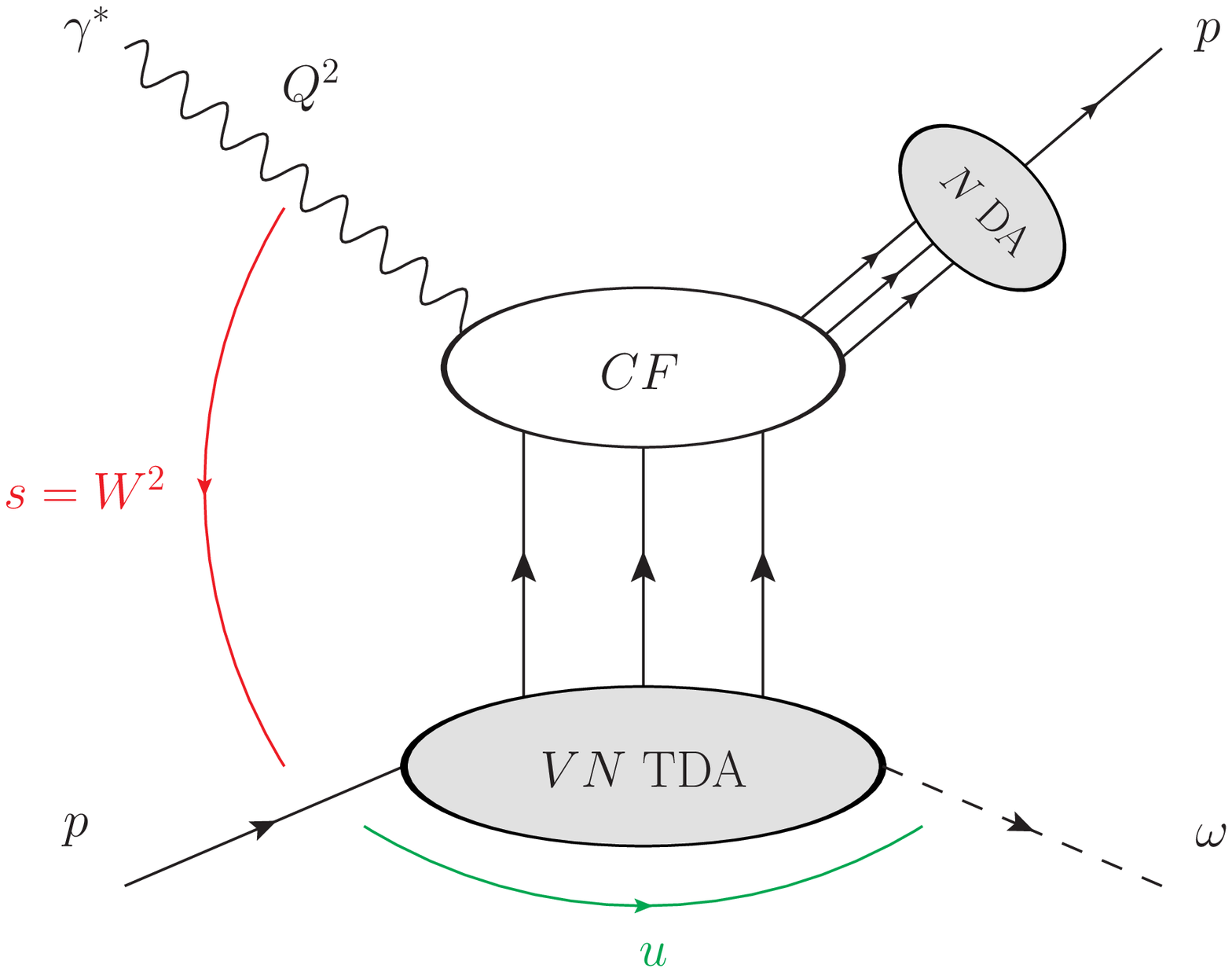}}
	\caption[Interaction diagrams for GPD and TDA collinear factorization]{(a) shows the $\omega$ electroproduction production interaction ($\gamma^*p\rightarrow p\omega$) diagram under the (forward-angle) GPD collinear factorization regime (large $Q^2$, large $s$, fixed $x$, $t\sim0$). $N$ GPD is the quark nucleon GPD (note that there are also gluon GPD that is not shown). $V$ DA stands for the vector meson distribution amplitude. The CF corresponds to the calculable hard process amplitude.  (b) shows the (backward-angle) TDA collinear factorization regime (large $Q^2$, large $s$, fixed $x$, $u\sim0$) for $\gamma^*p\rightarrow  p\omega$. The $VN$ TDA is the transition distribution amplitude from a nucleon to a vector meson.}
	\label{fig:GPD_TDA}
\end{figure}

Currently, there is no known direct experimental access to measure the GPDs~\cite{ji04}. The prime experimental channel to study the GPDs is through the Deep Virtual Compton Scattering (DVCS\nomenclature{DVCS}{Deep Virtual Compton Scattering}) and Deep Exclusive Meson Production (DEMP\nomenclature{DEMP}{Deep Exclusive Meson Production}) processes~\cite{ji97}. Both processes rely on the collinear factorization scheme; an example of DEMP reaction: $\gamma^*p\rightarrow p\omega$ is shown in Fig.~\ref{fig:GPD_TDA} (a). In order to access the forward-angle GPD collinear factorization (CF\nomenclature{CF}{Collinear factorization}) regime ($\gamma^*p\rightarrow \omega p$ interaction), the kinematics variables requirements are as follows: large $Q^2$, large $s$, fixed $x$ and $t\sim0$~\cite{ji04, pire15}.

Under the collinear factorization regime, a parton is emitted from the nucleon GPDs ($N$ GPDs) and interacts with the incoming virtual photon, then returns to the $N$ GPDs after the interaction~\cite{ji04}. Studies~\cite{kroll2016, liuti10} have shown that perturbation calculation methods can be used to calculate the CF process (top oval in Fig.~\ref{fig:GPD_TDA} (a)) and extract GPDs through factorization, while preserving the universal description of the hadronic structure in terms of QCD principles.

TDAs{\nomenclature{TDA}{Baryon-to-Meson Transition Distribution Amplitude}} are the backward analog of GPDs, with their full name being the baryon-to-meson transition distribution amplitude ($VN$ TDA). TDAs describe the underlying physics mechanism of how the target proton transitions into a $\omega$ meson in the final state, shown in the grey oval in Fig.~\ref{fig:GPD_TDA} (b). One fundamental difference between GPDs and TDAs is that the TDAs require three parton exchanges between $VN$ TDA and CF.

As introduced previously, the GPDs depend on $x$, $\xi$ and $t$. The $\omega$ production process through GPDs in the forward-angle ($t$-channel) and through TDAs in the backward-angle ($u$-channel) are schematically shown in Figs.~\ref{fig:GPD_TDA} (a) and (b), respectively. In terms of the formalism, TDAs are similar to the GPDs, except they require a switch from the impact parameter space ($t$ dependent) through Fourier transform to the large momentum transfer space ($u$ dependent), which brings a novel picture of the nucleon.

The backward-angle TDA collinear factorization has similar requirements: $x$ is fixed, the $u$-momentum transfer is required to be small compared to $Q^2$ and $s$; $u\equiv\Delta^2$, which implies the $Q^2$ and $s$ need to be sufficiently large. Based on these, the optimal $Q^2$ range of study for the TDA model is $Q^2$ $>$ 10~GeV$^2$. The parameter $\Delta$ is considered to encode new valuable complementary information on the hadronic 3-dimensional structure, whose detailed physical meaning still awaits clarification~\cite{pire15}.

In both the GPD and TDA collinear factorization interaction diagrams shown in Fig.~\ref{fig:GPD_TDA}, apart from the $N$ GPD, $VN$ TDA and collinear factorization, the parton structure (distribution amplitudes in terms of quarks) of the outgoing proton and meson have to be described. $V$ DA represents the $\omega$ meson distribution amplitude and $N$ DA is the proton distribution amplitude~\cite{pire15}.

\begin{figure}[t]
	\centering
	\includegraphics[width=0.65\textwidth,angle=90]{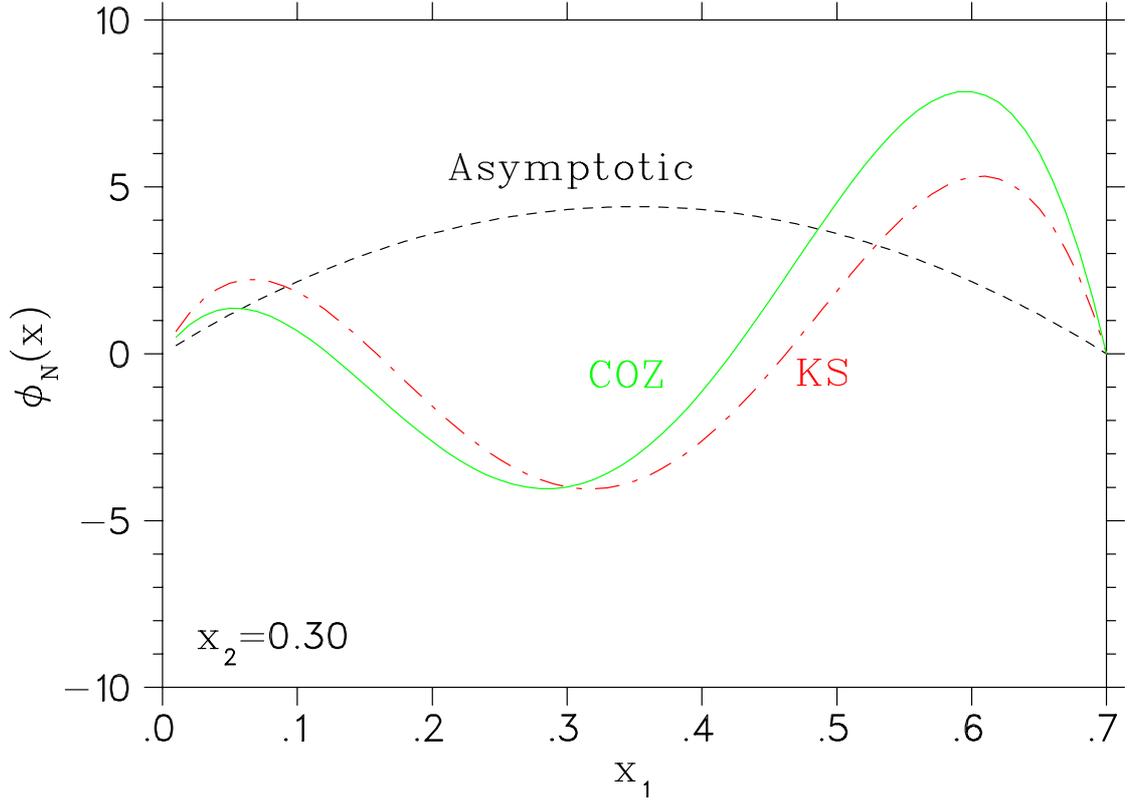}
	\caption[Nucleon DA model distributions]{Nucleon DA model distributions for $x_1$ assuming $x_2=0.3$. Black dotted line indicates the $N$ DA at asymptotic limits (large momentum transfer); green solid line is $N$ DA parameterized by COZ~\cite{chernyak89}; red dot-dashed is for the KS~\cite{king87}. Plot was created by G. Huber~\cite{huber17}.~\oic}
	\label{fig:NDA}
\end{figure}

The $V$ and $N$ DA are based on the choice of the phenomenological solution for the leading twist nucleon DA and the corresponding value of the strong coupling represents a complicated problem. In the TDA calculation made for this Ph.D. thesis, the Chernyak-Ogloblin-Zhitnitsky (COZ)~\cite{chernyak89} and King-Sachrajda (KS)~\cite{king87} $N$ DA models have been chosen. Both $N$ DA models have considerably different shapes from the $N$ DA asymptotic limit. Assuming the nucleon consists of three partons with momentum fractions, $x_1$, $x_2$ and $x_3$, the sum of the three distributions must equal to 1. If $x_2=0.3$, then $x_3 = 1-x_1-x_2$ and the distribution of $x_1$ predicted by COZ, KS and asymptotic $N$ DA ($\phi_N(x)$) are shown in Fig.\ref{fig:NDA}.  Note that both COZ and KS $N$ DA models are capable of providing a description of the nucleon electromagnetic form factors.

The $N$ DA model is an important part to the TDA model prediction, and the predicted experimental observable can change significantly depending on the choice of the $N$ DA model. Therefore, the improvement of TDA formalism would rely on an accurate nucleon spatial distribution parameterized by the $N$ DA models. As more experimental data are collected to constrain the $N$ DA model during the 12~GeV era at JLab, significant developments in the GPD and TDA pictures are expected in the coming decades.     

Due to its technical complexity, details regarding the nucleon distribution amplitude are excluded from this thesis, and further detail regarding the $N$ DA models can be found in Refs.~\cite{pire15, chernyak89, king87}.

\subsection{Two Predictions from TDA Collinear Factorization}

Through a private communication~\cite{pire15_priv}, a set of calculations matching the kinematics coverage of this Ph.D. work have been provided. Compared to the effective $Q^2$ range of the TDA formalism ($Q^2\approx$10~GeV$^2$~\cite{pire15}), the $Q^2$ range of this Ph.D. work is much lower $Q^2$ = 1.6, 2.45~GeV$^2$. A quantitative comparison between data and model may nevertheless provide an intuitive demonstration of the predictive power of the TDA model (see Sec.~\ref{chap:results}).

The TDA collinear factorization has made two specific qualitative predictions regarding backward vector meson electroproduction, which can be verified experimentally:
\begin{itemize}
\item The dominance of the transverse polarization of the virtual photon results in the suppression of the $\sigma_{\rm L}$ cross section by a least (1/$Q^2$): $\sigma_{\rm L}/\sigma_{\rm T}$ $>$ $1/Q^2$, 
\item The characteristic 1/Q$^8$-scaling behavior of the transverse cross section for fixed $x$, following the quark counting rules. 
\end{itemize}

The L/T separated differential cross section is directly relevant to the validation of the TDA frame work. However, due to the limited the $Q^2$ coverage, the second TDA prediction will be validated in the future studies.

\section*{Closing Remarks}

Recall the question regarding the mechanism of producing the backward-angle $\omega$ that was raised in the beginning of this chapter. By using the interaction demonstrated in Fig.~\ref{fig:laget} (top left panel), an answer to the question can be reached. In order to generate backward-angle $\omega$ from the photon probe through the VMD effect, the interaction requires a low resolution (low $Q^2$) and a high impact parameter (low $t$ or low $u$). Based on the kinematics of these data, in particular the $Q^2$ values, one needs to explore mechanisms beyond the VMD.

In the intermediate energy and momentum transfer scenario, such as in this thesis, the virtual photon wavelength is much smaller than the proton radius. Thus, the backward $\omega$ meson is originated from the nucleon target through the exchange (or knock out) of a baryon. Therefore, the study of backward-angle ($u$-channel) interactions at intermediate energy range contributes to the general understanding of dynamic properties of the nucleon.

\graphicspath{{pics/3Experimental/}}

\chapter{Experimental Apparatus}

\label{sec:exp}

\section{Overview}
\begin{figure}
\centering
\includegraphics[width=0.85\textwidth]{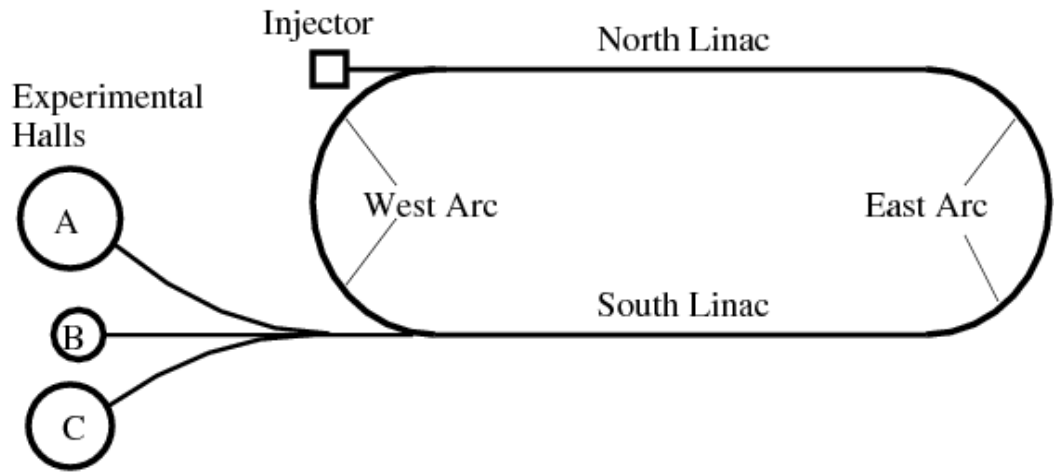}
\caption[Schematic of the CEBAF facility]{The schematic of the CEBAF facility at the Thomas Jefferson National Accelerator Facility.~\oic}
\label{fig:CEBAF}
\end{figure}

The Thomas Jefferson National Accelerator Facility\footnote{12000 Jefferson Avenue, Newport News, Virginia. https://www.jlab.org/} (JLab\nomenclature{JLab}{Thomas Jefferson National Accelerator Facility.}) is a U.S. Department of Energy (DOE) user facility for fundamental nuclear physics research. Started in 1984 as a dedicated laboratory to study hadronic structure and the fundamental properties of nuclear matter, it has since become one of the world's leading facilities for investigating the physics of quark-gluon interactions. JLab's main research facility is the Continuous Electron Beam Accelerator Facility (CEBAF)\nomenclature{CEBAF}{Continuous Electron Beam Accelerator Facility}, which consists of a polarized electron source, an injector and two anti-parallel superconducting RF linear accelerators (linacs\nomenclature{Linac}{Linear accelerator}), connected to each other by two arc sections which contain steering magnets. The electrons are kept in a racetrack configuration during
the acceleration process. A schematic diagram of the CEBAF is shown in Fig.~\ref{fig:CEBAF}. 

Since 2011, JLab has undertaken a major upgrade to double its maximum beam energy to 12~GeV. By 2014, the first 12 GeV beam was delivered to Hall D which started the 12 GeV era of JLab operation. It is important to note that the experimental details discussed in this section are applicable to the 6 GeV era of JLab operation (prior to 2011).

\section{Accelerator}

From a semiconductor photocathode, polarized electrons are excited by circularly polarized laser light and accelerated by the Radio-Frequency (RF\nomenclature{RF}{Radio frequency}) resonant cavities of the accelerators. One particular feature of JLab is the continuous nature of the electron beam, with a bunch length of less than 2~ps. In addition, a fundamental RF frequency of 1497 MHz allows for three sequential electron bunches serving three independent experimental halls, each bunch having independent current amplitude during the 6~GeV operation period.

Conceptually, CEBAF is a linear accelerator that has been folded up in a design similar to that of a racetrack. Recirculation of the beam is possible up to five times to achieve the maximum beam energy: electrons are accelerated by the injector to an energy of 45~MeV and sent to the North Linac, where they gain an additional energy up to 600~MeV through acceleration on superconducting RF resonant cavities. From the North Linac, the electron beam is bent through the east arc and guided through the South Linac, where it gains up to another 600~MeV.

After the electron beam exits the South Linac for a given pass, the Beam Switch Yard (BSY\nomenclature{BSY}{Beam switch yard}) alternately delivers one out of every three bunches of electrons to each of the three experimental halls, or recirculates them through the west arc for an additional pass through the linacs.

During the JLab 6~GeV era operation, the maximum energy gain of the CEBAF was 1.2~GeV per pass, corresponding to a nominal energy of 6~GeV.  Each linac consisted of 20 cryomodules, each of which contained eight superconducting niobium cavities cooled by liquid helium at 2~K. The same linacs were used for the acceleration in each circulation. Nonetheless, the beams from different passes were split into different vacuum pipes before being steered by the steering magnets and traversing through the recirculating arcs. Before the entering the linac, the beams from different passes were recombined. This unique configuration allowed the experimental halls to run simultaneously at different energies.

The CEBAF accelerator produces beams in bunch lengths of less than 2~ps, which occur at a frequency of 1497~MHz as a result of the RF power used in the resonating cavities. During the 6~GeV operation period, every third pulse was delivered to each of the experimental halls resulting in one pulse every 2~ns, which corresponded to a beam frequency of 499~MHz. The RF separators at the BSY separated the beam pulses after each linac pass. It should be noted that at this rate, the beam delivery can be effectively considered continuous. The continuous beam property is critical for a coincidence experiment such as F$_\pi$-2, which requires a high precision and high luminosity to insure reliable extraction of the cross section with acceptable statistical uncertainty.

To achieve the same luminosity, a non-continuous (pulsed) linac such as SLAC\footnote{Stanford Linear Accelerator Center (SLAC) National Accelerator Laboratory, 2575 Sand Hill Rd. Menlo Park, CA 94025. https://www6.slac.stanford.edu/}\nomenclature{SLAC}{Stanford Linear Accelerator Center} would require a higher electron density within a bunch and longer bunch width within the operation window. This would significantly increase the random coincidental backgrounds and reduce the timing separation. Conceptually, the real coincident events would be diluted by the random coincident events and raise the statistical uncertainty for the cross section. Thus, performing a coincidence measurement is not feasible for with a non-continuous linac.

\section{Hall C}
\begin{figure}[t]
\centering
\includegraphics[width=0.65\textwidth]{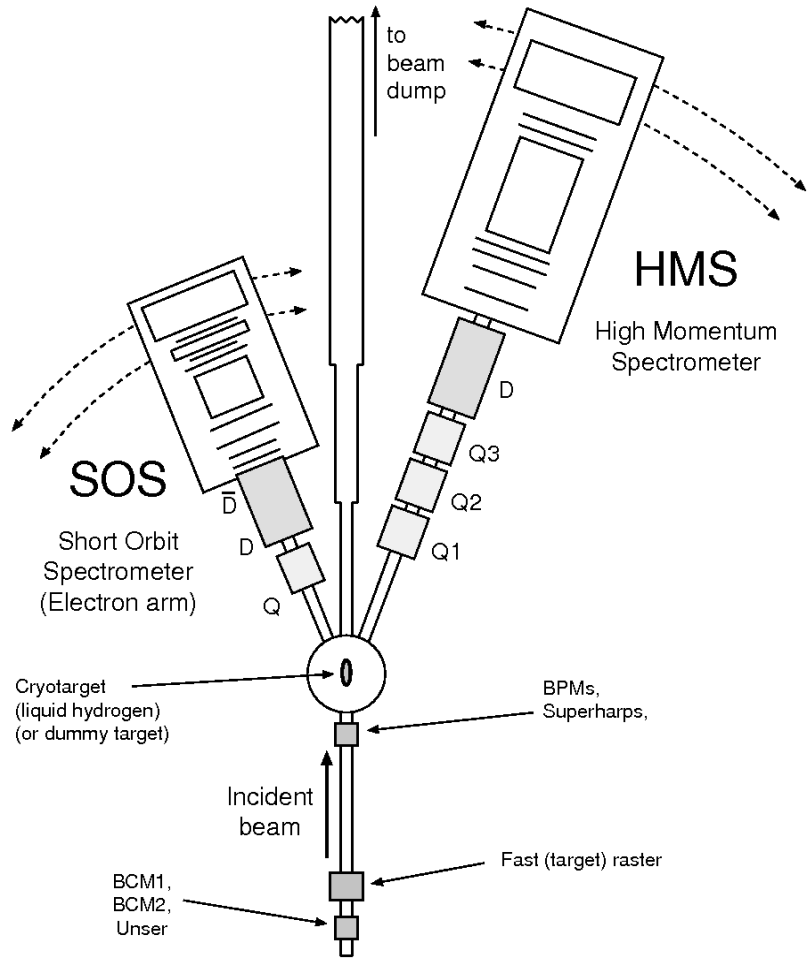}
\caption[Schematic top view of the Hall C spectrometers]{Schematic top view of the Hall C spectrometers relative to the target and beam line~\cite{horn}.}
\label{hallc_top}
\end{figure}

Fig.~\ref{hallc_top} shows an overhead schematic layout of experimental Hall C during the JLab 6~GeV operation. The hall has a nearly circular geometry with a diameter of 32~m. A large fraction of the experimental hall is located underground and it is well shielded to contain the hazardous level of radiation. 

\begin{figure}
\centering
\includegraphics[width=0.85\textwidth]{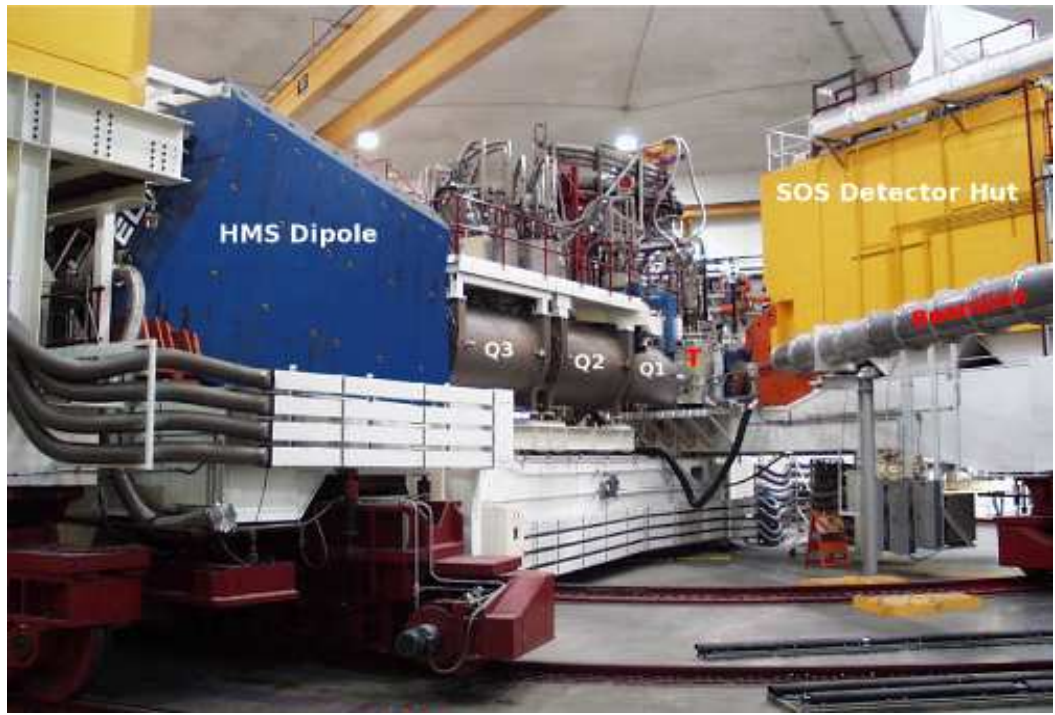}
\caption[Hall C image taken during the F$_\pi$-2 experiment]{Hall C image taken during the F$_\pi$-2 experiment. The critical spectrometer and beamline components are labeled. The red ``T'' symbol indicates location of the target chamber. Note that the image is taken from a location between the HMS spectrometer and the beamline (downstream from the target chamber).~\oic}
\label{fig:hall_c}
\end{figure}

\begin{figure}[p]
\centering
\includegraphics[width=0.8\textwidth]{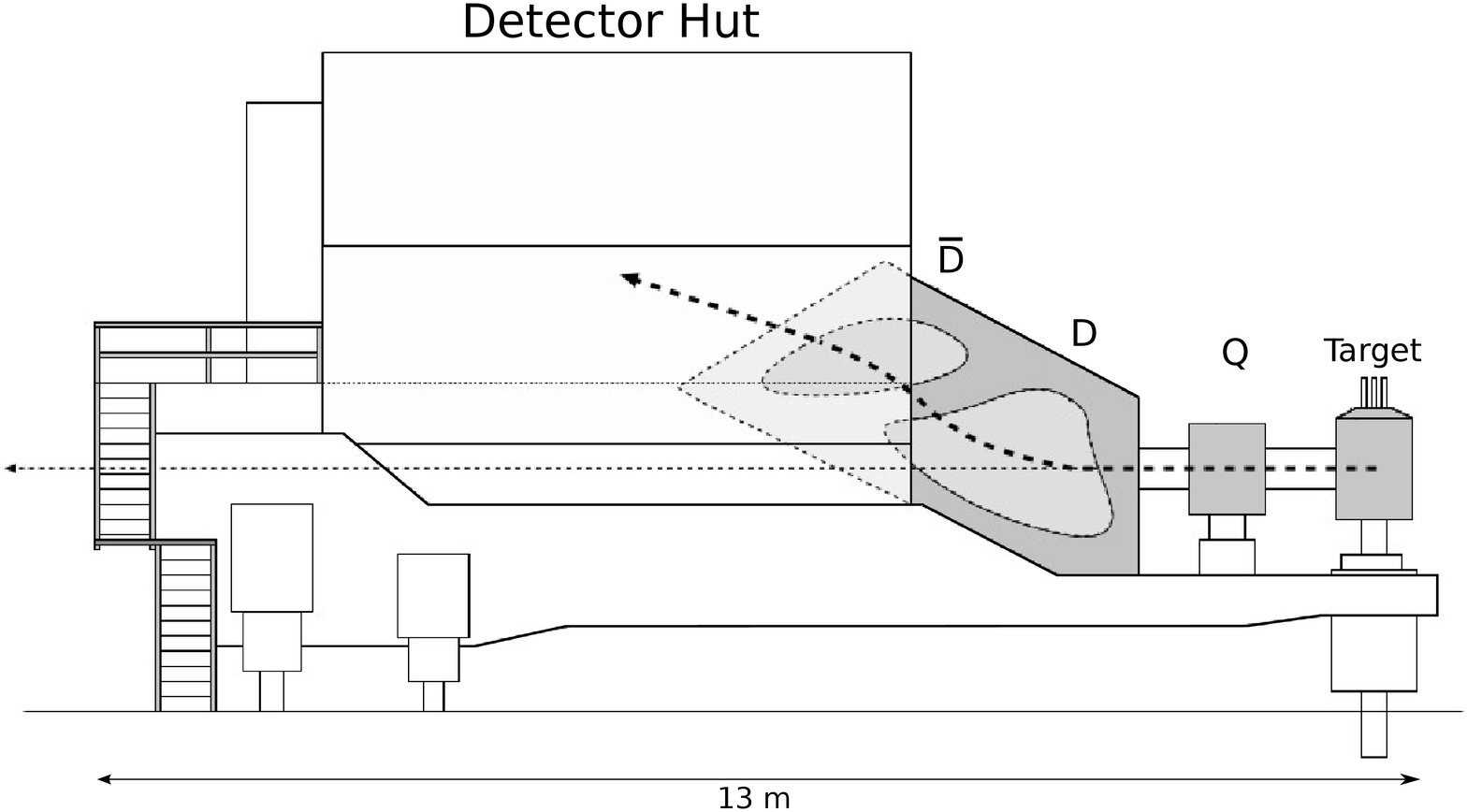}
\caption[Schematic drawing of SOS spectrometer]{Schematic drawing of the SOS spectrometer. Quadrupole $Q$ and the dipoles (D and $\overline{\textrm{D}}$) are used as the optical elements to focus and select particles, before they reach the detector hut. This figure is modified based on the original from Ref.~\cite{mohring99}.}
\label{fig:sos_spec}
~\\
~\\
\includegraphics[width=1\textwidth]{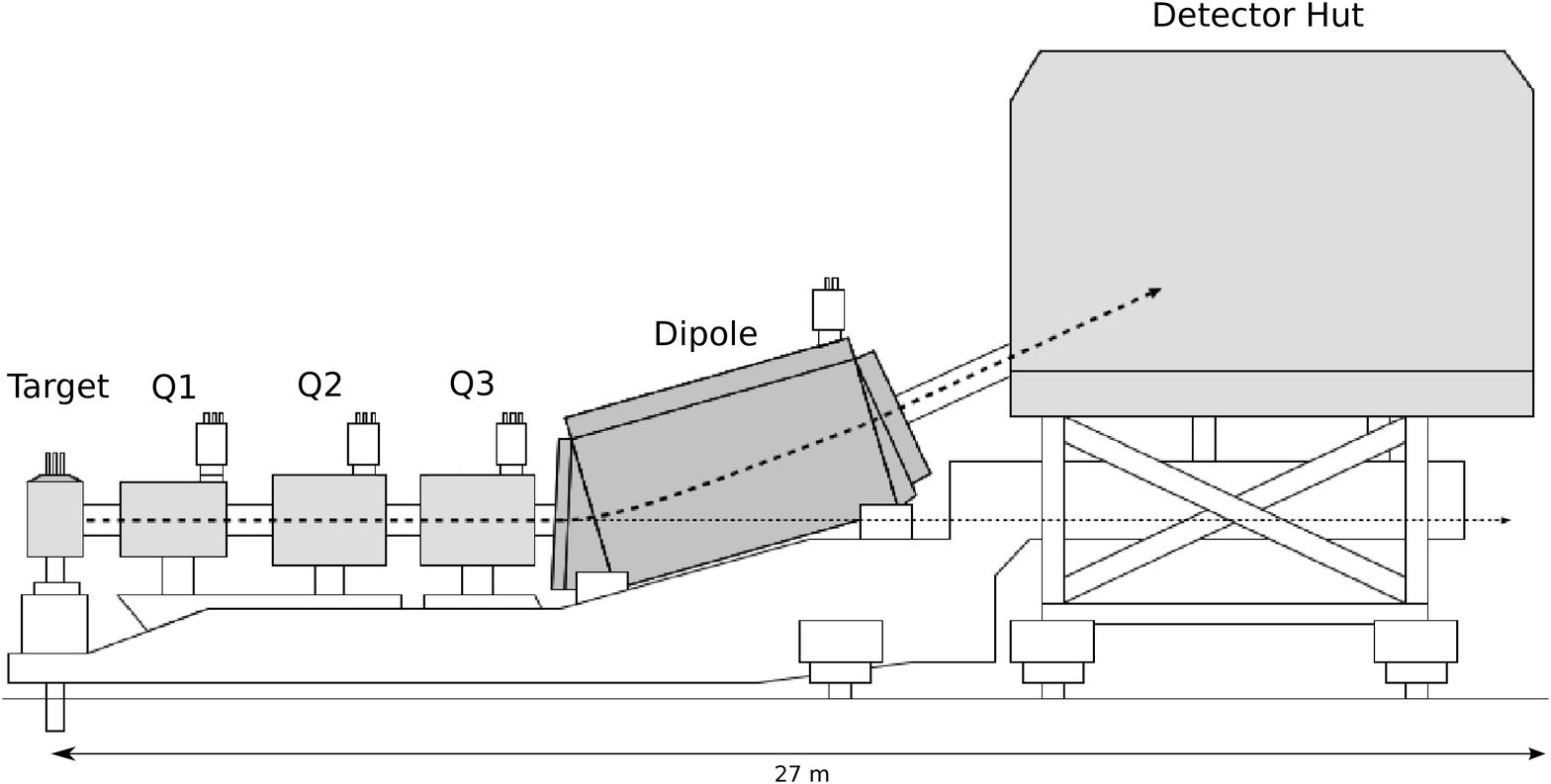}
\caption[Schematic drawing of HMS spectrometer]{Schematic drawing of the HMS spectrometer. Quadrupoles ($Q1$, $Q2$ and $Q1$), and dipole (D) are used as the optical elements to focus and select particles, before they reach the detector hut. This figure is modified based on the original from Ref.~\cite{mohring99}.}
\label{fig:hms_spec}
\end{figure}

The standard Hall C apparatus consists of two magnetic focusing spectrometers: the High Momentum Spectrometer (HMS) shown in Fig.~\ref{fig:hms_spec},  and the Short Orbit Spectrometer (SOS) shown in Fig.~\ref{fig:sos_spec}. Fig.~\ref{fig:hall_c} shows an image of Hall C during the F$_{\pi}$-2 experiment, where the critical spectrometer and beamline components are labelled.

The HMS optics configuration consists of three superconducting quadrupoles followed by a dipole and has a path length of approximately 26~m from the target to the focal plane. In contrast, the SOS optics consists of three resistive magnets and has a path length of 10~m, which is adequate for the detection of short-lived particles at low momentum. The momentum resolutions of the HMS and SOS are better than 10$^{-3}$~m and the horizontal angular resolutions are better than 2~mrad. The designed maximum central momenta for the HMS and SOS are 7 and 1.74~GeV/c, respectively. The standard instrumentation in Hall C has been used successfully for a variety of experiments requiring the full CEBAF beam current of 200~$\mu$A.
 
\section{Beamline}
\label{sec:beam}

For a precision L/T-separation experiment such as F$_\pi$-2, the characteristic (profile) of the electron beam is an important factor that needs to be monitored throughout the experiment. In this section, the techniques and apparatus for determining the beam position, current and energy information are briefly introduced.

\begin{figure}[t]
\centering 
\includegraphics[width=0.8\textwidth]{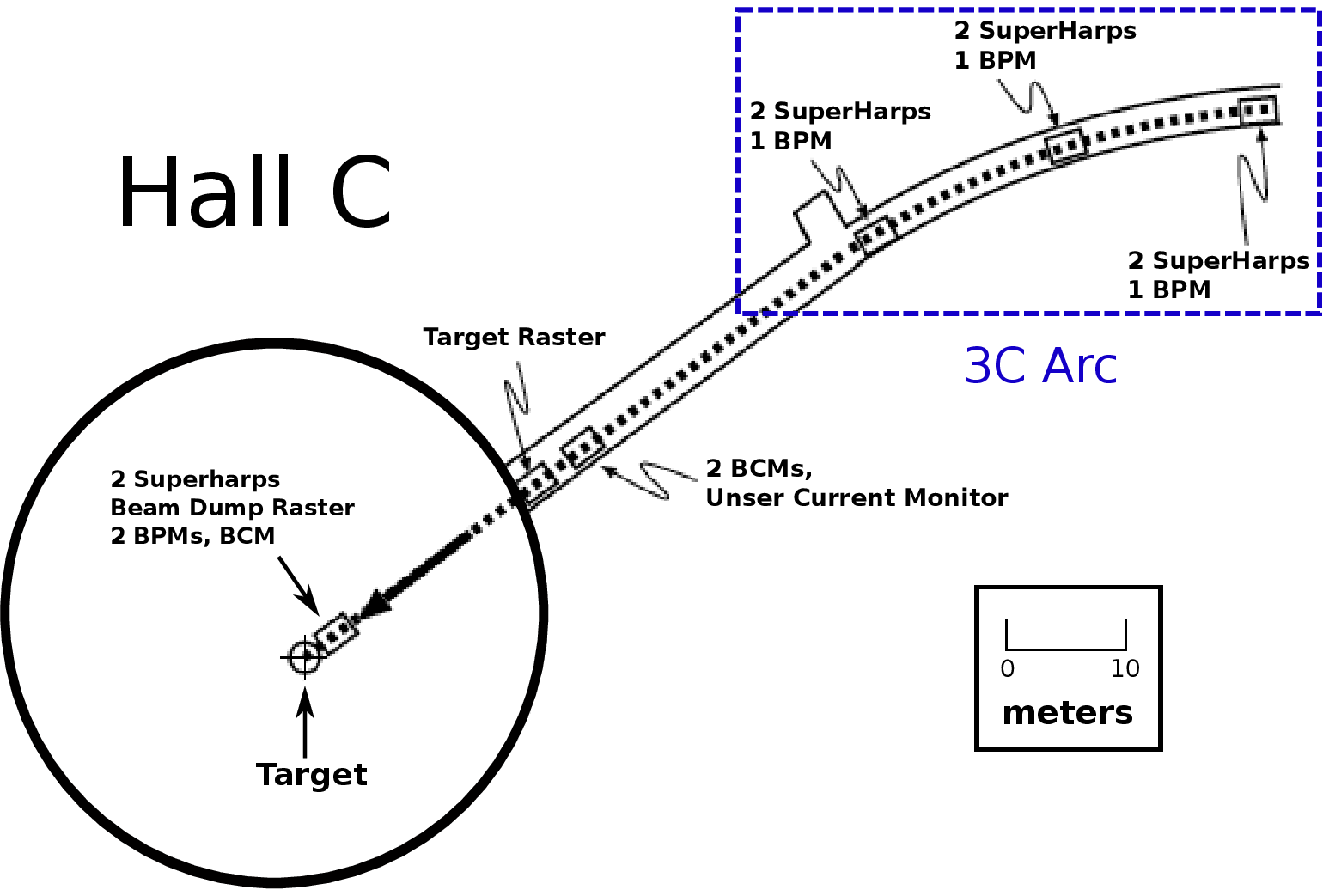}
\caption[Schematic representation of the instrumentation of the Hall C beamline]{Schematic representation of the instrumentation of the Hall C beamline. The critical (beamline) components are labelled in the diagram. The arc section  (3C Arc) of the beamline that guide the electron beam (from the BSY) into Hall C is indicated by the blue dashed box. This figure is recreated based on the original from Ref.~\cite{horn}.~\oic}
\label{fig:instrumentation}
\end{figure}

\subsection{Beam Position Monitors}

The monitoring of the position of the beam in the Hall C arc and beamline is accomplished with Beam Position Monitors (BPM\nomenclature{BPM}{Beam position monitors}, as shown in Fig.~\ref{fig:instrumentation}. The BPM monitors consist of resonating cavities with a fundamental frequency to match both the $1497$~MHz accelerator beam pulse frequency and the $499$~MHz pulse frequency into Hall C. Each cavity has four antennas which are rotated by $45^\circ$ with respect to the horizontal and vertical axes to minimize damage caused by the synchrotron radiation. The $45^\circ$ angle was chosen is due to the beam being focused in horizontal and vertical directions by quadrupoles along the beamline. The amplitude of the signal picked up from the fundamental frequency by each antenna allows for the determination of the relative position of the beam~\cite{niculescu}. 

The primary beam steering is guided by the BPMs located in Hall C, and additionally, the BPMs closest to the target (H00A, H00B, H00C) were also monitored to ensure precision. The beam position was set based on information from spectrometer optics data and it varied for each of the four beam energies used during the F$_\pi$-2 experiment. Note that the BPM coordinates do not represent the absolute position of the beam and are chosen based on the requirement of simultaneous mid-plane symmetry in both spectrometers.

The beam position at the target location can be determined by combining the projection of any pair of BPMs. During the experiment, BPM C was determined to be unreliable, so that for all subsequent calculations only BPM A and BPM B were used. Note that the typical size of the position variation at the target was less than 0.5~mm.

The beam position and direction at the entrance, middle and the exit of the Hall C arc are measured using the high resolution wire sensors (harps\nomenclature{Harps}{High resolution wire sensors}) system. The harps system consists of two vertically and one horizontally oriented  wires in a non-stationary frame. During a `harps scan', the vertical wires move across a low current beam at the same time, then followed by the same action from the horizontal wires. The signals generated at each wire as they are intercepted by the beam are recorded by an Analog to Digital Converter (ADC) unit. The corresponding position of the wire intercepted is then determined by a position encoder.

The superharps system is an upgrade of the harps system, including absolute position readout electronics, a dual beam profile detection system with two analog pick-up channels and a vibration-free support system. The harps system and its operation are described in more detail in Ref.~\cite{yan95}.

\subsection{Beam Energy Measurement}

\begin{figure}[t]
\centering 
\includegraphics[width=0.7\textwidth]{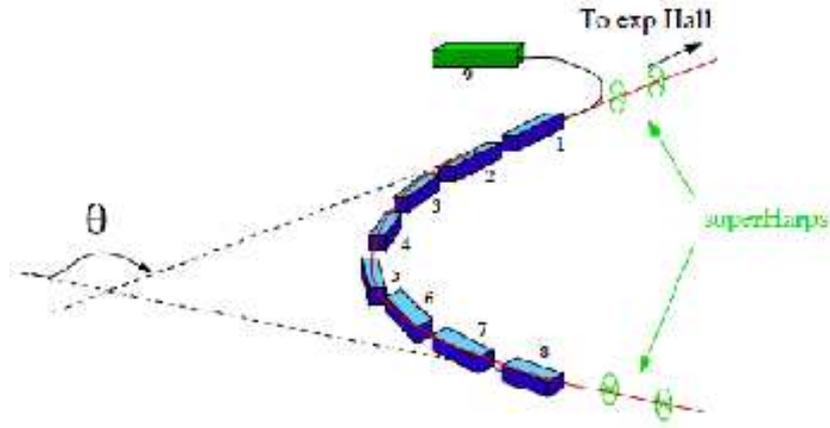}
\caption[Conceptual drawing for an arc energy measurement instrumentation]{Conceptual drawing for an arc energy measurement instrumentation used for Hall C. Not all superharps are shown. This figure modified based on the original from Ref.~\cite{horn}.~\oic}
\label{fig:arc_energy}
\end{figure}

The energy of the electron beam in Hall C is measured using the deflection of the beam in a known magnetic field in the Hall C arc. The technique makes use of the fact that an electron traversing a constant magnetic field moves in a circular trajectory, where its radius depends on the strength of the magnetic field and the electron momentum. The arc method uses the arc magnets as a type of spectrometer and the beam position measurement to determine the deflection of the beam in the section of the beamline between the BSY and the hall entrance. The conceptual drawing of such instrumentation is shown in Fig.~\ref{fig:arc_energy}. The blue dashed box in Fig.~\ref{fig:instrumentation} (top right corner), shows the Hall C arc (3C Arc).

This measurement cannot be performed simultaneously with regular data taking because it requires all the focussed elements to be turned off and degaussed (neutralizing the residual magnetic field). The beam position and direction at the entrance, middle and the exit of the arc are measured using the superharps system. The bend angle of the Hall C arc ($\theta_{arc}$) was measured to be $34.3^{\circ}$. The beam is then steered so that the central trajectory and the beam energy can be determined from the electron momentum using:
\begin{equation}
p= \frac{e}{\theta_{arc}} \int{B\,dl}\,,
\label{eq:eq_p}
\end{equation}
where $e$ is the electron charge and $B$ is the magnetic field in the dispersive elements. The extraction of the beam energy from the field integral requires the knowledge of the magnetic fields in the arc dipoles. For this reason, one of the dipoles in the Hall C arc has been field-mapped as a function of current. 

The remaining eight dipoles are calibrated relative to the reference dipole assuming similar field maps. Using the value of the field integral, the beam energy can be determined with a precision of $\frac {\delta p}{p}\approx5\times10^{-4}$ ~\cite{horn}. A more detailed description of the energy measurement of the beam using the arc method is documented in Ref.~\cite{yan93}.

\subsection{Beam Current Monitors}
The F$_\pi$-2 experiment uses two Beam Current Monitors (BCM\nomenclature{BCM}{Beam current monitor}) that measure the electron beam current delivered to Hall C. The primary BCMs (BCM1 and BCM2) are cylindrically shaped waveguides tuned to the frequency of the beam. The geometry of these cavities was designed to be excited by the $TEM_{010}$\footnote{Transverse electromagnetic (TEM) is a mode of propagation where the electric and magnetic field lines are all restricted to directions normal (transverse) to the direction of propagation. The subscript $010$ refers to the resonating mode of the standing (EM) wave created inside of the superconducting cavity.} mode of the electron beam pulse frequency~\cite{horn}. This mode has the particular advantage that its magnitude changes slowly with respect to the position of the beam within the cavities. The output voltage levels of the waveguides are proportional to the beam current when the waveguides are tuned to the frequency of the beam.

The resonant frequency of the cavities is sensitive to the temperature fluctuations, since the current monitor cavities can thermally expand or contract due to temperature changes. To minimize these effects, the temperature is stabilized by thermally insulating the beam monitor cavities at a constant value of $43.3^{\circ}$C. The cavity temperature was checked during each shift and found to be oscillating within the range of $\pm0.2^{\circ}$C. Note that the temperature of the readout electronics can also affect the current measurement. In order to minimize this effect, the electronics room was maintained at a nearly constant temperature throughout the experiment.

Both BCM1 and BCM2 exhibit reasonable gain stability as a function of time. Nonetheless, to minimize drifts in the gain, both BCMs are calibrated to an absolute standard device at regular intervals. The calibration is performed using an Unser current monitor~\cite{Unser}, which is a parametric DC current transformer. The Unser monitor has an extremely stable gain, but suffers from large drifts in the offset on short time scales. Thus, the Unser monitor cannot be used alone to measure the beam current reliably on a run-to-run basis. The resonant cavity BCMs were calibrated by taking dedicated runs with periods of no beam (the purpose was to monitor the Unser zero/baseline) interspersed with periods of beam at various currents.

During the F$_\pi$-2 experiment, the currents ranged from 10 to 110~$\mu$A, and the actual current values were continuously adjusted. The BCMs are generally stable enough so that calibrations have to be performed only infrequently during the experiment. The run-to-run uncertainty in the current, as measured by BCM1 and BCM2, is estimated from a combined analysis. The averaged current drift between calibrations was found to be on the order of $0.2\%$ at 100~$\mu$A~\cite{horn}. Considering in addition the normalization uncertainty from the Unser monitor, which is estimated to be $0.4\%$, results in an absolute uncertainty for the charge measurement of $\pm0.5\%$.

\subsection{Modification to Beamline}
The beamline of Hall C was modified for the F$_\pi$-2 experiment by adding a small diameter beam pipe installation downstream of the target, to allow for data taking at the smallest possible angle between the beam line and the spectrometers in particular the HMS. With this particular geometry (at small SOS central angles), the beam pipe is susceptible to magnetic fields from an unshielded edge of the SOS dipole magnet. The presence of these magnetic fields was confirmed prior to the experiment from measurements at a momentum setting of 1.74~GeV/c~\cite{huber17, gaskell16}. The dominant fields are parallel to the dipole yoke and oriented along and perpendicular to the spectrometer axis. The contribution from magnetic fields vertical to the magnet yoke and perpendicular to the spectrometer axis are $20\%$ smaller~\cite{MackMuMe1}.

Since a beam deflection that exceeds the upstream beamline aperture can cause damage to one of the flanges of the Hall C beam dump due to an excessive deposition of energy, there was a concern about the beam deflection at the diffuser at the exit of the hall. The deflection of the beam was calculated for different kinematic settings using a magnetic field map data. In the calculations, a SOS momentum of 1.74~GeV/c and a beam energy of 5~GeV were assumed. The deflection at the smallest angle for F$_\pi$-2 experiment was determined to be $\pm$4~mrad from the target centre~\cite{MackMuMe2,MackMuMe3}. The vertical deflection of the beam at the diffuser was addressed with magnetic shielding of the downstream beam pipe. Two layers of magnetic shielding foil were also installed around the beam pipe in order to reduce the value of the field integral and its corresponding beam deflection.

Detailed tests of the beam deflection with the modified beam pipe entailed measurements at SOS angles between $22^{\circ}$ and $30^{\circ}$. Furthermore, beam deflection under the SOS full saturation mode was confirmed to be adequately suppressed within acceptable boundaries~\cite{MackMuMe2,MackMuMe3}. 
 
\section{Targets}
\label{sec:experiment_target}

\begin{figure}[t]
\centering
\subfloat[][Hall C target ladder.]{\includegraphics[width=0.5\textwidth]{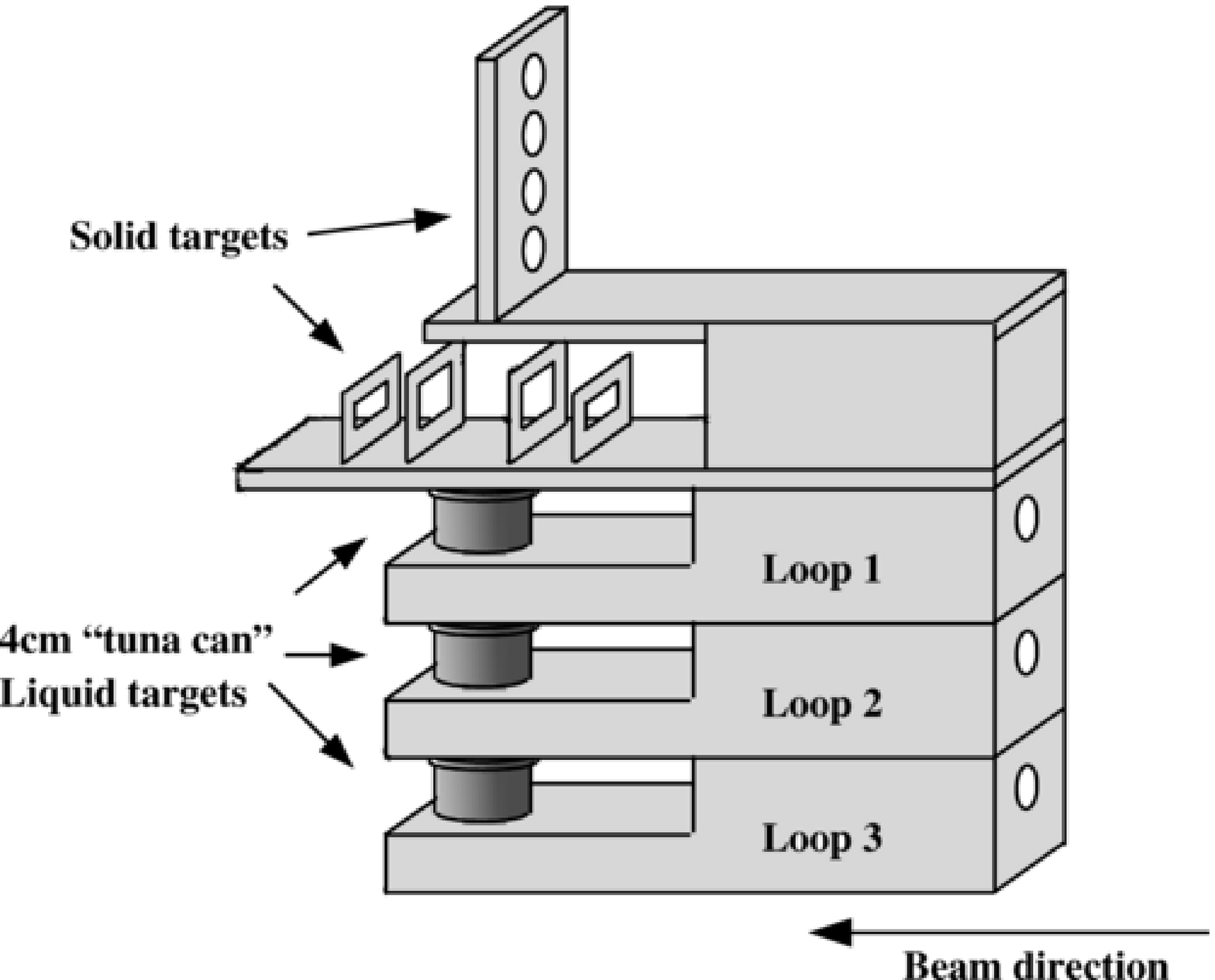}}
\subfloat[][Cross section of the cryotarget loop.]{\includegraphics[width=0.5\textwidth]{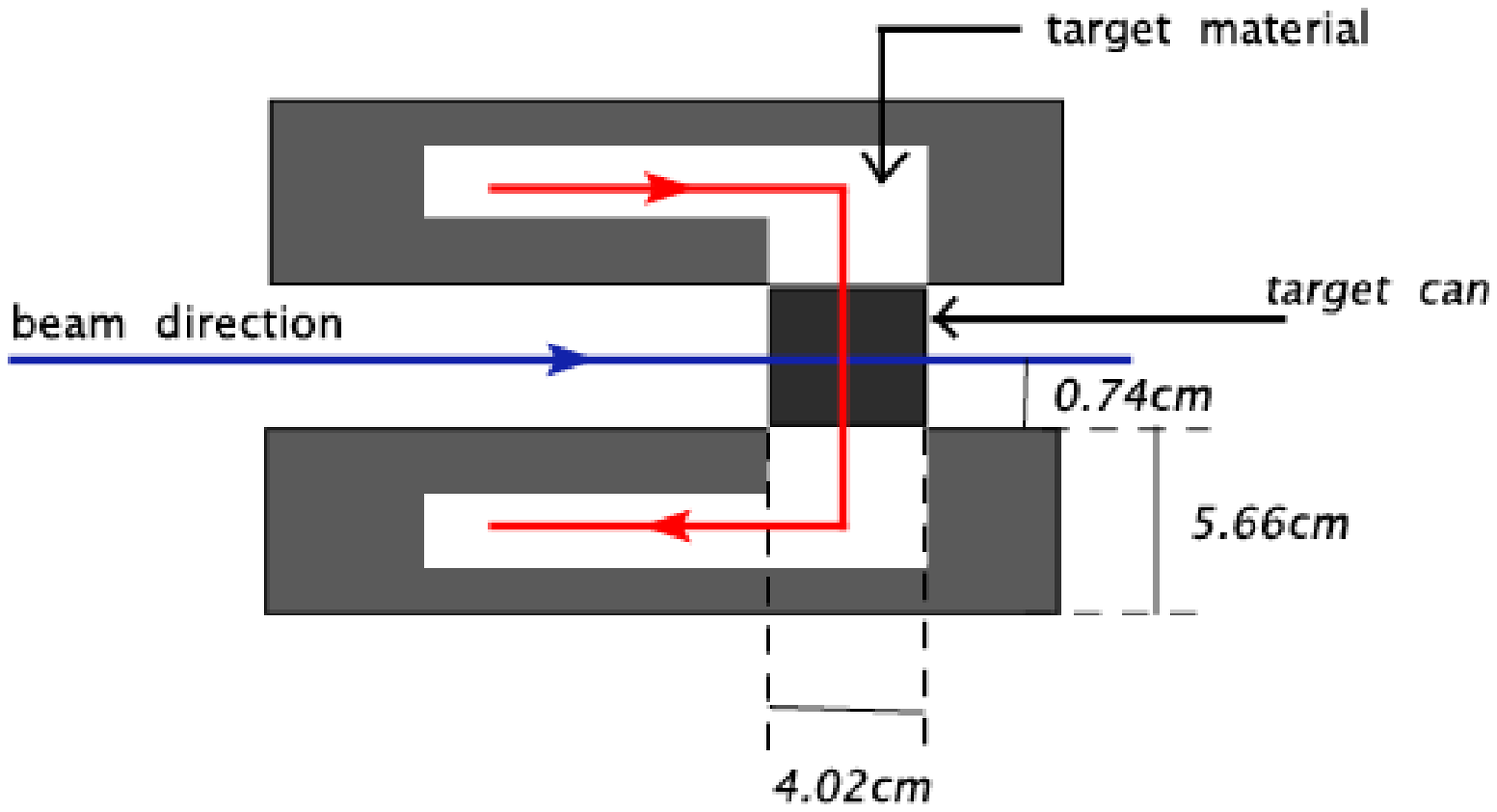}} 
\caption[Schematic diagrams for the target configuration]{Schematic diagrams for the target configuration in Hall C~\cite{horn}. Figure (a) shows the five layer configuration for the target ladder, which includes (top to bottom): carbon target layer, dummy target layer, and three layers (loops) of cryogenic targets. Figure (b) shows the cross section view for each of the cryotarget loop. Note that the red arrow (line) indicates the cryogenic fluid flow vertically through tuna can target. The incoming electron beam direction is indicated in both diagrams. Theses figures are recreated based on the originals from Ref.~\cite{horn}.~\oic}
\label{fig:target}
\end{figure}

The Hall C target system contains a three-loop cryogenic target (cryotarget) stack mounted together with optics and ``dummy'' targets on a target ladder enclosed in a high vacuum scattering chamber. Fig.~\ref{fig:target}(a) shows a schematic drawing of the target stack configuration inside of the scattering chamber. The solid target ladder consists of five carbon and two aluminum foils at different positions ($z$=0~cm, $z$=$\pm$2~cm, $z$=$\pm$7~cm) along the beam direction~\cite{horn, meekins}, here $z$ is along the beam direction from the BSY to the beam dump, and the center of the target station is at $z$ = 0~cm. The two aluminum foils situated at $z$ = $\pm$2~cm constitute the ``dummy target'', which is used to quantify the experimental yield from the aluminum cryotarget cell wall. Note that the dummy target is 7.022 times thicker than the nominal thickness of the cryotarget cell walls. The remaining solid carbon foils are used with beam incident on two or five (``quintar'') foils simultaneously for the purpose of calibrating the spectrometer optics properties. 

The average energy deposition in the cryotarget is relatively large ($\sim$4~MeV~cm$^2$g$^{-1}$), while the diameter of the incident electron beam is relatively small ($<0.5$~mm). The electron beam needs to be rastered\footnote{Beam rastering technique is a standard procedure to uniformly distribute the electron beam onto the cryogenic target, thus, minimizing the localized heat deposit. This is generally achieved by using a combination small dipole magnets upstream of the cryogenic target.} to a $2\times2$~mm$^2$ profile in order to distribute the energy in a more uniform manner across the cryotarget volume, since the local heating can lead to a target density fluctuation (i.e. a target boiling effect). The rastering profile used during the F$_\pi$-2 experiment consisted of a constant and uniform pattern in contrast to the sinusoidal pattern used in previous experiments~\cite{horn}. This system is described more fully in Refs.~\cite{Wojcik2002,Yan95raster}.

\subsection{Cryogenic Targets: LH$_2$}
The cryogenic targets were each held inside of a cylindrical ``tuna-can'' cell, of 4~cm in diameter, oriented vertically as shown in Fig.~\ref{fig:target}(b). Each target cell occupies one of the three available loops. During the F$_\pi$-2 experiment, the loop 1 was empty, while loops 2 and 3 contained liquid hydrogen (LH$_2$) and liquid deuterium (LD$_2$), respectively. Both cryotargets used the same coolant supply (liquefied helium) and were cooled on the cryotarget ladder simultaneously. The End Station Refrigerator (ESR) supplied the helium at 15~K and the coolant flow to the individual loop was controlled by the target operator using Joule Thompson (JT) valves. 

The cryogenic coolant is circulated continuously through the heat exchanger from the target cell. Low and high power heaters are controlled by a Proportion, Integral and Derivative feedback system, keeping the LH$_2$ at 19~K. During low current or beam-off periods, the target control system regulates the cryotarget temperature by replicating the power deposition of the electron beam using high power heaters, while the target fluid moves continuously through the heat exchanger around the target cell. From Fig.~\ref{fig:target}(b), each target cell is 3.95-4.02~cm long in the beam direction, with cell walls made from aluminum alloy T6061 and of a thickness of 0.013-0.014~cm. The alloy used in manufacturing the aluminum dummy targets is AI-T7075, a higher strength alloy. More information regarding the cryotargets' mechanical structure, composition and design can be found in Refs.~\cite{Dunne,MeekinsThesis,TerburgThesis}.

During the experiment, conditions such as the flow rate and temperature of the cryogenic fluid, thermal expansion (contraction) and boiling effects can affect the target density and volume. In order to minimize these effects, the cryotarget (such as the LH$_2$ target) at a density of 0.0723$\pm$0.0005~mg/cm$^3$ is kept at a nominal operating temperature of 19~K, which is around 2~K below the boiling point. Measuring the target length at room temperature and doing the offset corrections for the target from the center of the beamline and thermal contraction ($0.4\%\pm0.2\%$), the real length of the cell in the cooled-down state is calculated to be 3.98$\pm$0.01~cm~\cite{horn,meekins}.

As the electron beam traverses the target, significant energy per unit area is deposited. The energy deposition from the electron beam in the target is predominately due to the ionization process and can be estimated by the Bethe-Bloch formula~\cite{pdg}. Assuming a 100~$\mu$A current electron beam accelerated to 6~GeV of energy, the estimated stopping power in the LH$_2$ target is around 4~MeV$\,$cm$^2$/g~\cite{triumf}. The LH$_2$ target density and length are 70.8~mg/cm$^3$ and 3.98~cm, which yield energy loss of 1.1 MeV. The power loss is equivalent to a 100~Watt light bulb (assuming most of the energy is converted into heat). In order to keep the LH$_2$ target below the 20.28~K boiling temperature, and avoid localized density fluctuation,  a large amount of cooling power ($>$ 100~Watts) is required. In addition, the electron beam is rastered in a uniform pattern to distribute the heat evenly across the tuna can target (if electron beam radius
is less than 0.5~mm, the power per unit area can reach $\sim10^8$~Watts/m$^2$).

A 100~$\mu$A current 6~GeV electron beam on a carbon target does not require target density (temperature) correction or beam rastering. Since the carbon material has a lower stopping power of $\sim$2~MeV/g/cm$^2$~\cite{triumf}, the energy loss in the material is estimated to be 0.34~MeV, which corresponds to a power output of 34~Watt. Note that the carbon target thickness is taken as 0.173~g/cm$^2$~\cite{meekins}, in combination with the high melting temperature of the carbon material (3550$^\circ$C)~\cite{carbonmelt}. The carbon target data are perfect to study the rate dependent efficiencies such as tracking efficiency under high trigger rate environment ($>$ 500~kHz).

\subsection{Target Thickness}
\label{sec:target_thick_mess}
The cryotarget thickness and associated uncertainties are listed in Table~\ref{tab:cryotarget}, where the cryotarget length at room temperature is corrected for thermal contraction of the aluminum cell walls and the offset of the cryotarget from the surveyed position (3.42~mm). The actual target thicknesses for these targets were also corrected for the beam offset from the target center at each kinematic setting, the target thickness uncertainty is the quadrature sum of $0.6\%$ uncertainty on the target length and $0.5\%$ on the target density. The total target thickness is determined using the target cell geometry at operating temperatures in combination with the target density derived from cell temperature and pressure. 

For the cryotarget, the cell temperature was kept constant to 100~mK within the operation temperature (19~K) during the F$_\pi$-2 experiment. The dominant uncertainty in target density is due to the thermal expansion and contraction, and it is about $0.5\%$. Note that the uncertainty contributions from the measured temperature are negligible. The uncertainty for the outer diameter of the target cell at room temperature was measured to be $\pm0.3\%$. The uncertainty for thickness of the cell walls was determined to be $\pm0.0013$~mm~\cite{horn}. 

 \begin{table}[t]
   \centering  
   \begin{tabular}{l|c|c}
     Target & Target length (cm) & Target thickness (g/cm$^2$)\\
     \hline
     LH$_{2}$ & $3.918\pm0.01$ & $0.283\pm0.002$\\
   \end{tabular}
    \caption[Cryotarget thicknesses not corrected for beam offset]{Cryotarget thicknesses not corrected for beam offset. The target length is given by $L=2\sqrt{R^{2}-{dx}^{2}}$, and the target cell radius $R$ is corrected for thermal contraction, and $dx$ is the beam offset from the target center.~\cite{horn}}
   \label{tab:cryotarget}
 \end{table}

The target length is sensitive to the size and the form of the raster pattern and the central position of the beam from the target center. The reduction of effective target length due to the constant raster pattern was determined to be $\approx 0.005\%$~\cite{horn}. Initial target survey results and measurements of thermal effects, like vacuum motion and target cool-down motion, indicate that the target cells were on average located at 3.42$\pm$0.50~mm (with the beam right facing downstream) relative to the nominal beamline. Optics data and information from the beam position monitors suggest that the beam was offset between 0.15 and 2.00~mm in the same direction for the four different kinematic settings. In the worst case deviation of the beam from the target center, the correction of the effective target length is $1.50\pm0.05\%$. The variation in target thickness due to the central beam position between high and low $\epsilon$ settings is $0.2\%$. Additional uncertainties to the target
thickness are given by the purity of the target gas and dynamic effects such as target heating due to energy deposited by the electron beam. To determine the target purity, samples of the target materials were examined after the experiment. Both cryotarget purities were found to be $>99.9\%$, so no correction was assigned~\cite{horn}.

Localized target density fluctuations due to heating effects can have a significant effect on the average density of cryotargets. The rastering of the beam reduces local density fluctuations of the liquid targets, but cannot eliminate them entirely. The change in luminosity due to beam heating was measured by comparing yields at fixed kinematics as a function of beam current. To account for the net reduction in measured target density due to localized target boiling, a correction factor is applied. Taking into account the uncertainty in the beam current, the total uncertainty in the target density is on the order of $0.5\%$. Based on the target boiling study in Sec.~\ref{sec:target_boil_study}, no significant target density reduction due to localized heating was found. Thus, no target boiling correction factor was applied.

In order to understand and subtract the thin aluminum wall (target chamber) contribution to the experimental yield, data runs with a dummy target were used to correct experimental data. The dummy target thickness was designed to be greater than the thickness of the wall of the target cell, thus reducing the dummy target data taking period. 

According to the information documented in the F$_\pi$-2 target configuration technical report ~\cite{meekins}, the normalized dummy target experimental yield has to be corrected by the following factor:
\begin{equation}
\dfrac{\textrm{Hydrogen~Target~Wall~Thickness}}{\textrm{Dummy~Target~Thickness}} = \dfrac{0.0746~\textrm{g/cm}^3}{0.5237~\textrm{g/cm}^3} = \dfrac{1}{7.022} = 0.142.
\end{equation}
Note that the percentage uncertainty for the dummy-target ratio is the quadratic sum of the percentage uncertainties in LH$_2$ target cell wall thickness and dummy target thickness, and is calculated to be 1\%.

\section{Detectors}

\begin{figure}[t]
\centering
\includegraphics[width=1\textwidth]{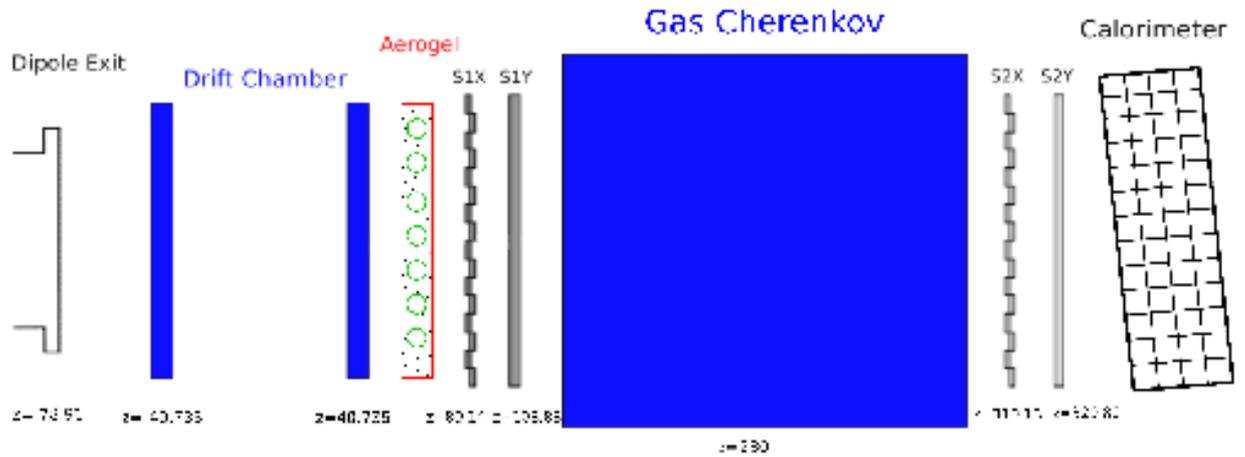}
\caption[Schematic drawing of the HMS detector stack]{Schematic drawing of the HMS detector stack inside of the detector hut. The detector stack provides full capability particle tracking and identification. The conceptual design of the SOS detector stack is very similar to HMS design, except that SOS does not have an aerogel Cherenkov detector during the F$_{\pi}$-2 experiment. The charged particle enters the detector hut from the dipole exit window and travels towards the calorimeter (left to right). This figure is modified based on the original from Ref.~\cite{horn}.~\oic}
\label{fig:hms_det_stack}
\end{figure}

The detector package layout in the HMS and SOS are very similar. As an example, the conceptual drawing of the HMS layout is shown in Fig.~\ref{fig:hms_det_stack}. The detector package consists of two horizontal drift chambers for the track reconstruction, four scintillating hodoscopes used for generating the triggers and measuring the time-of-flight (TOF), the threshold gas Cherenkov detectors and lead-glass calorimeters used for particle identification. The HMS detector package also includes an aerogel Cherenkov detector used for separating protons from pions. $\pi$-$e$ separation is performed using the gas Chereknov detectors in both spectrometers. A complete review of the detector packages, including detailed geometry and performance, can be found in Ref.~\cite{ArringtonThesis}.

\subsection{Drift Chamber}
\label{sec:drift_chamber}

The drift chambers are used to measure the horizontal and vertical angles and positions of the charged particles before and after the focal plane, in order to determine their momenta and trajectories. The basic operation principle is as follows: charged particles induce ionization of the atoms inside the gas chambers and the free electrons are produced due to the ionization process. These free electrons are eventually captured by the sense wires. A good spatial resolution is achieved by measuring the deviation in the electron drift time. The electric field inside of the gas chamber is required to be a very specific configuration, which is achieved by surrounding the sense wires with non-sensed wires at high voltage. The trajectory information of the two sets of chambers are combined to determine the trajectory of the charged particles through the focal plane.

The HMS spectrometer is equipped with a pair of drift chambers and each consists of six wire planes. For each chamber, the wire planes are ordered $x$, $y$, $u$, $v$, $y^\prime$, $x^{\prime}$. The $x$ and $x^\prime$ planes determine the dispersive (vertical) coordinates of the particle trajectory, while two $y$ and $y^\prime$ planes determine the non-dispersive (horizontal) track position. The $u$ and $v$ plane wires are at $\pm15^\circ$ with respect to the $x$ plane, the purpose of these wires is to enhance the tracking resolution in the vertical direction. The cell spacing is 1~cm and the position resolution is approximately 150~$\mu$m per plane. Note that the HMS has better resolution in the dispersive direction due to its wire plane configuration. The two drift chambers are placed at a distance of 40~cm before and after the focal plane. The ionizing medium in the HMS drift chambers is an argon-ethane mixture of 1:1 ratio, which is controlled by a gas handling system located on the outside of the experimental hall.

The design of the SOS drift chamber is similar in design to the HMS drift chambers. The SOS is also equipped with a pair of drift chambers consisting of six planes of wires. The wire planes are ordered $u$, $u^\prime$, $x$, $x^\prime$, $v$, $v^\prime$. In the same fashion as the HMS, the $x$ and $x^\prime$ planes determine the vertical particle trajectory. The $u$ and $u^\prime$ planes are rotated 60$^\circ$ clockwise with respect to the $y$ coordinate determined by the $x$ and $x^\prime$, while the $v$ and $v^\prime$ planes are rotated 60$^\circ$ counterclockwise. Similarly, in the HMS the matched planes are offset by 0.5~cm perpendicular to the sense wire direction to resolve the left-right ambiguity in the case of multiple hits in both planes. An argon-ethane mixture of 1:1 ratio is also used for this chamber.

\subsection{Hodoscopes}

Both HMS and SOS are equipped with four planes of scintillator hodoscopes divided into pairs of $x$-$y$ planes. Each pair contains one plane segmented in the vertical and one plane segmented in the non-dispersive direction (horizontal plane). Each plane is composed of several components: the detector paddles made of long narrow strips of scintillator material with PMTs attached to both ends. The scintillator paddles are arranged in an overlapping configuration to eliminate gaps between the elements.

The principle of scintillation detectors can be summarized as follows: charged particles travelling through the scintillator material ionize atoms in the medium. The emitted electrons interact with the scintillating material, exciting molecules to higher energy levels. The excited molecules return back to the ground state (de-excitation) by emitting photon energy. The emitted photons propagate through the material via total internal reflection and are detected by PMTs attached to either ends of the paddle.

The reflecting material is aluminum foil for the HMS and aluminized mylar for the SOS scintillator elements. The HMS scintillator paddles have dimension of $8.0\times1.0$~cm$^2$ (width $\times$ thickness) and the dimensions in the SOS are $7.5\times1.0$~cm$^2$ for the $x$ planes and $7.5\times1.1$~cm$^2$ for the $y$ planes. The length of the paddles depends on the spectrometer and their location and orientation in the detector hut. It should be noted that the scintillator paddles are shorter in the SOS, thus, resulting in a generally better time resolution per plane than in the HMS, due to reduced attenuation. However, the overall TOF resolution remains similar because of smaller separation distance between the front and back pair of planes in the SOS.

The arrangement of the two pairs of planes is similar in both spectrometers. However, the separation between the front and the back planes and the order of the four planes is different. In the HMS, the first plane is segmented vertically ($x$ planes) and the second plane segmented horizontally ($y$ planes), with separation of 220~cm between the front and back pair. The plane order is reversed in the SOS and the pair separation is 180~cm.

The main purpose of the scintillator hodoscopes is to provide the raw trigger for the data acquisition system and to determine the particle velocity by measuring the TOF between the front and back planes. The hodoscope signals are read out through a combination of Analog to Digital Converters (ADCs), discriminators and Time to Digital Converters (TDCs), and signal logic modules. Note that the electronics use leading-edge discriminators, which will result in timing shifts due to the difference in energy deposited (signal pulse height), also known as the time-walk effect. The timing information from each scintillator paddle is corrected for the time-walk effect and (timing) offset using a software calibration routine. The detailed description of this routine can be found in Ref.~\cite{ArringtonThesis}.

\subsection{Cherenkov Detectors}
\label{sec:gas_cherenkov}

As an important part of the standard particle identification (PID\nomenclature{PID}{Particle identification}) package, both spectrometers are equipped with gas threshold Cherenkov detectors. The primary objective for both detectors is to perform $e$-$\pi$ separation.

The basic working principle of the threshold Cherenkov detector relies on the Cherenkov effect, which is described by classical electrodynamics~\cite{jackson99}. Cherenkov radiation is emitted when a charged particle traverses a dielectric medium of index of refraction $n$ with velocity ratio $\beta$ that is faster than the light speed inside of the medium ($c/n$). The Cherenkov radiation angle can be calculated as $$\cos\theta_c =  \frac{c}{v n} = \frac{1}{\beta n}\,.$$ The emitted Cherenkov light is reflected from parabolic mirrors inside of the detector and focused onto the sensitive area of the photon multiplier tubes (PMT\nomenclature{PMT}{Photon multiplier tube}).

The ($\beta>1/n$) threshold property of the Cherenkov radiation allows the possibility to adjust the dielectric (gas) medium in the detector to allow identification of electrons and pions over a wide range of momentum settings. Although the separation of electrons and pions is highly efficient, the rejection of pion events is complicated by the presence of knock-on (secondary) electron due to multiple scattering inside of the detector. The secondary electrons ($\delta$-rays) are produced when a proton or pion interacts with the material in front (or inside) of the Cherenkov gas and subsequently results in a hit in the Cherenkov detector. The mis-identification of proton due to $\delta$-ray is a significant effect during the analysis, and further detail can be found in Sec.~\ref{sec:p_int_corr}.

The HMS Cherenkov detector is a cylindrical tank holding two mirrors and two PMTs. The detector design allows for gas pressure in the tank above and below atmospheric pressure. Therefore, the detector can be used for $e$-$\pi$ separation at atmospheric pressure or below, but it can also be used to separate pions from protons using Freon-12 (CCl$_2$F$_2$) above the atmospheric pressure. During the F$_{\pi}$-2 experiment, the HMS Cherenkov was filled with C$_4$F$_{10}$ gas at a pressure of 0.47~atm. The corresponding index of refraction at this pressure is 1.00066, which yields an electron Cherenkov radiation (momentum) threshold of 14~MeV/c and a pion threshold of 3.8~GeV/c.
  
The SOS Cherenkov detector design is similar to the that of the HMS, but contains four mirrors and four PMTs. The Cherenkov medium was Freon-12 at a pressure of 1~atm. The corresponding refractive index at this pressure is 1.00108, which results an electron Cherenkov radiation momentum threshold of 11~MeV/c and a pion threshold of 3~GeV/c. Note that such momentum thresholds exceed the SOS maximum central momentum by about a factor of two.

\subsection{Lead Glass Calorimeter}

\label{sec:calorimeter}

The primary objective of the lead glass calorimeter is to provide an additional means of selecting and separating electrons from pions. The lead glass calorimeter is positioned at the back of the detector hut for both spectrometers. The calorimeters use $10\times10\times70$~cm$^3$ blocks arranged in four planes, and stand 13 and 11 blocks in both height and width in HMS and SOS, respectively. The entire detector is tilted by 5$^\circ$ relative to the central ray of the spectrometer to minimize losses due to particles passing thought the gaps between the blocks. To ensure light tightness, each block is wrapped in aluminized mylar and tedlar film. The calorimeter signal from each block is read out by PMTs attached at one side.

Particle detection using electromagnetic calorimeters is based on the production of electromagnetic showers in the lead glass material. As particles enter the calorimeter, they interact with nuclei inside the lead glass material and radiate photons via the bremsstrahlung process. The bremsstrahlung photons produce electron-positron pairs that also radiate photons (by either secondary bremsstrahlung or Cherenkov processes). The particular choice of the calorimeter thickness ensures that incident electrons or positrons deposit all their energy in the particle shower. The light radiated by the charged particle is collected by PMTs through internal reflection, and the amplitude of the signal is proportional to the incident momentum of the primary charged particles. Pions and muons entering the calorimeter do not produce bremsstrahlung showers, and instead they deposit a constant amount of EM shower ($\approx$300~MeV) in the calorimeter. Similar to pions, protons will not generate bremsstrahlung showers, and deposit even less EM shower. However, pion, muons and protons can undergo nuclear interactions in the lead glass and produce particle showers similar to the electron-positron induced particle showers.

The separation of electrons from other particles, such as decayed pion events, is based on the normalized energy deposited in all layers in the calorimeter. During the F$_\pi$-2 experiment, the SOS calorimeter was used in combination with the SOS gas Cherenkov detector to select electrons and exclude $\pi$ events. 

The first layer of the calorimeter stack carries a unique significance, since it contributes two important trigger conditions: \textrm{PRHI} and \textrm{PRLO}, which are defined and explained in Sec.~\ref{sec:trigger}.

\subsection{HMS Aerogel Cherenkov Detector}
\label{sec:ACD}
In addition to the standard PID detectors, the HMS has an aerogel threshold Cherenkov detector that provides adequate hadron identification for the spectrometer central momenta above 3~GeV/c. The primary objective of the HMS aerogel Cherenkov detector (ACD\nomenclature{ACD}{Aerogel Cherenkov detector}) is to separate pions from protons at high momentum ($>$3~GeV/c). Different from the low momentum region, the $\pi$-$p$ separation at high momenta is not effective by using the standard TOF method (examine the velocity of the particle) due to the decreased timing separation: $\Delta t \propto \frac{1}{p^2_{\rm HMS}}$, where $p_{\rm HMS}$ is the central momentum setting of the HMS. Similar to the gas Cherenkov detector, the principle of the ACD is based on the threshold Cherenkov radiation, which depends on the refractive index of the dielectric medium.

The dielectric medium in the HMS ACD was specifically chosen to perform $\pi$-$p$ separation from 3.0 - 4.6~GeV/c. Aerogel is a hydrated silicon oxide of molecular structure: n(SiO$_2$)+2n(H$_2$), and its density ranges between 0.04 - 0.20~g/cm$^3$. The hydrate surrounding the aerogel molecule yields an average refractive index between gases and liquids. During the F$_\pi$-2 experiment, aerogel with refractive index of $n=1.030$ was used ($n=1.015$ was also available), which yields a $\pi$ Cherenkov radiation threshold of 0.565~GeV/c and a proton threshold of 3.802~GeV/c. The threshold momenta for muons and kaons are 0.428 and 2.000~GeV/c, respectively. The highest HMS momentum setting during the F$_\pi$-2 experiment was 3.336~GeV/c, so that $\pi$-$p$ separation can be done adequately.

The HMS ACD consists of 650 tiles ($110\times110\times10$~cm$^3$) arranged into nine 5~mm honeycomb sheets stacked in a $117\times67$~cm$^2$ tray. The individual layers were offset with respect to each other by 2-3~cm to minimize the loss of particles passing through the detector without hitting any aerogel material. The Cherenkov radiation generated by charged particles passing through the aerogel is collected by 16$\times$5 inch Photonis XP4572B PMTs\footnote{https://www.photonis.com/} mounted on each side of the reflecting diffusion box.

The reflective surface results in multiple internal reflections of the produced Cherenkov photons before detection. The aerogel tiles were made light-tight by wrapping them in reflective material, except for the surface facing the diffusion box with Millipore paper. To ensure high reflectivity from the internal walls, the inside of the diffusion box was covered with Membrane GSWP-0010 Millipore paper\footnote{http://www.emdmillipore.com}. The entire assembly of tiles was held in place by a 100~$\mu$m stainless steel wire. Further details on the design and testing of the HMS ACD can be found in Ref.~\cite{asaturyan}.

\section{Trigger System}
\label{sec:trigger}

\begin{figure}[t]
\centering
\includegraphics[width=1.0\linewidth]{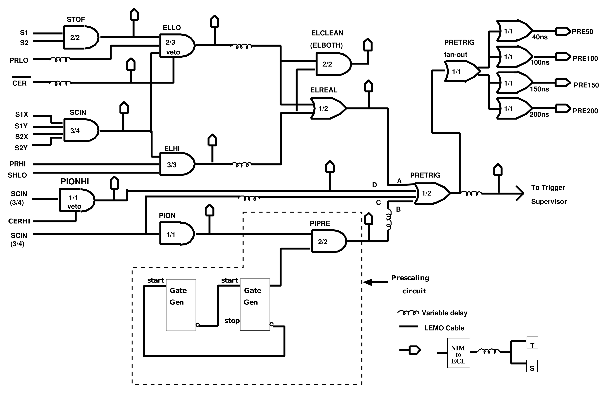}
\caption[Schematic diagram for the HMS electronic pre-trigger]{Schematic diagram for the HMS (single arm logic) electronic pre-trigger during the F$_{\pi}$-2 experiment. The four options for pre-trigger are described in the text. This diagram is modified based on the original from Ref.~\cite{horn}}
\label{fig:trigger}
\end{figure}

The purpose of this section is to provide a basic introduction to the trigger system used during the F$_\pi$-2 experiment, and also introduce some terminology such as pre-scale factor (PS\nomenclature{PS}{Pre-scale factor}) and pre-trigger.

A schematic diagram for the single-arm trigger logic for the HMS (the SOS is similar) is shown in Fig.~\ref{fig:trigger}. The purpose of the single arm trigger logic is to generate a pre-trigger signal when a particle arrives. The pre-trigger signals from both spectrometers are fed into a trigger supervisor circuit (TS\nomenclature{TS}{Trigger supervisor circuit}). The single arm (HMS or SOS) and two arm coincidence trigger signals are formed depending on the state of the TS (GO, ENABLE or BUSY). Both triggers and pre-triggers are fed into scaler modules, providing information such as the electronic dead time (EDT\nomenclature{EDT}{Electronic dead time}) and computer dead time (CDT\nomenclature{CDT}{Computer dead time}). The TS takes in all the pre-trigger and trigger signals and effectively controls the readout of all detector ADCs\nomenclature{ADC}{Analog-to-digital converter} and TDCs\nomenclature{TDC}{Time-to-digital converter} for the events. In order to reduce EDT and CDT, especially at
high event rates, a pre-scaling circuit is introduced to control how frequently an event type is selected to proceed to the TS. One can adjust the event selection frequency by changing the pre-scale factor, i.e. a PS factor set to 1000 means a given event type is forwarded to the TS once every 1000 events.

\subsection{HMS Pre-trigger}
Fig.~\ref{fig:trigger} shows the schematic diagram of the HMS pre-trigger that is composed of signals from different HMS detectors. The main component of the HMS trigger are the signals (or information) from the hodoscopes (scintillators) that are indicated as SCIN or STOF. 

The SCIN (``3/4'') signal requires a signal from three out of four layers from the hodoscope scintillator planes within a timing window of 50~ns. The advantage of using this trigger configuration is to minimize the impact of the single layer of hodoscope efficiency on the trigger efficiency. The practical experiences gathered from the previous experiments~\cite{ambrosewicz, volmer} have demonstrated the effectiveness (a consistent reliable high efficiency) of this trigger configuration. In addition, 3/4 signals is a good methodology to monitor the performance of the hodoscopes during the experiment. 

The signal condition STOF is satisfied when two of the scintillator planes independently give a signal, with one signal from the front and other signal from the back hodoscope plane. This is the minimum condition for the computation of the TOF  information of a detected particle. Note that satisfaction of the signal condition SCIN would imply the automatic satisfaction of STOF.

If the scintillator signal is present, the pre-trigger signal can be formed in one of the two different configurations: ELLO and ELHI. The ELHI configuration is formed if all three of the following signals are present: the SCIN signal, the PRHI signal and SHLO signal. Both PRHI and SHLO signals are formed at the calorimeter, where the former is satisfied when the signal from the first layer of the calorimeter exceeds a particular ``high'' threshold and the latter is formed when the total energy deposited in the calorimeter is above a particular ``low'' threshold. The ELLO pre-trigger requires a two out of three coincidence of SCIN, PRLO and STOF, where PRLO is defined as a signal from the calorimeter and the energy deposited in the first layer of the calorimeter exceeds a particular ``low'' threshold. In addition, absent of the Cherenkov signal: CER is
used as the signal veto for the ELLO, meaning if CER is not present the ELLO signal will be vetoed. Further clarification regarding the PRHI and PRLO thresholds, can be found in Ref.~\cite{horn}.

As shown in Fig.~\ref{fig:trigger}, there are four different pre-trigger options: 1. Standard ELREAL, 2. prescaled 3/4 pion trigger PIPRE, 3. ``Open'' 3/4 trigger SINC and 4. Pion trigger, 3/4 with Cherenkov veto PIONHI.

The logic-OR of ELHI and ELLO forms the electron pre-trigger or ELREAL signal. The advantage of using a two path electron pre-trigger is to reduce sensitivity due to particular hardware in either the Cherenkov detectors or the calorimeter. Two copies of ELREAL are formed, of which one signal is sent to the HMS PRETRIG module where it is logic-ORed with the pre-scaled pion signal, PIPRE. The PIPRE signal is effectively a pre-scaled 3/4 SCIN and is formed to ensure that a sample of pions is recorded by the data acquisition system to allow determination of PID efficiencies for Cherenkov and calorimeters. The PRETRIG signal is split into four copies after the PRETRIG module, called PRE50, PRE100, PRE150 and PRE200. These four copies of PRETRIG are used for determination of the electronic live time (ELT). Detailed explanation of these four different pre-trigger modules and ELT analysis can be found in Sec.~\ref{sec:ELT}.

The fourth pre-trigger option PIONHI was implemented during the F$_\pi$-2 experiment. PIONHI is satisfied by the presence of the 3/4 SCIN signal in absence of CERHI, which is the HMS Cherenkov signal with a high threshold for the detected number of photo-electrons N$_{p.e.}\approx4$ \nomenclature{N$_{p.e.}$}{Number of photo-electrons} in order to reject a larger fraction of electrons. The PIONHI is sent to the HMS PRETRIG module and is read out in the scalers and TDCs. During the first part of the experiment, the pion trigger condition was implemented as PIONHI logic-ORed with the SCIN signal to allow for cautious monitoring of the Cherenkov veto signal. Later in the experiment, the pion condition was reduced to PIONHI. The F$_{\pi}$-2-$\pi^+$\nomenclature{F$_{\pi}$-2-$\pi^+$}{Second charged pion form factor experiment $\pi^+$ analysis: $^1$H$(e, e^\prime\pi^+)n$ } analysis by T. Horn~\cite{horn} shown no significant difference in terms of data quality between these triggers.

\subsection{SOS Pre-trigger}
The configuration of the SOS pre-trigger is similar to that of the HMS. Analogous to the HMS, ELLO is formed from the SCIN, STOF, PRLO given the presence of the SOS Cherenkov signal in the trigger. The ELHI signal is formed by SCIN, PRHI and PRLO. The ELLO signal is then sent to the ELREAL module and two duplicated signals are sent to the SOS PRETRIG module. The PRETRIG signal is split into four copies after the PRETRIG module PRE50, PRE100, PRE150 and PRE200 for determination of the ELT. Similar to the HMS trigger, the first part of the experiment required ELREAL logic-ORed with PIPRE. This requirement was reduced to ELREAL only at the same time as the pion trigger (in HMS) was relaxed. The F$_{\pi}$-2-$\pi^+$ analysis by T. Horn~\cite{horn} shown no significant difference in terms of data quality between these triggers.

\section{Spectrometer Acceptance and Optics}
\label{sec:secptrometer_acc} 

The two magnetic spectrometers, SOS and HMS, were used to detect electrons and protons for the $^1$H$(e,ep)X$ reaction, respectively. There are two main coordinate systems: the beam (experiment) coordinate system and the spectrometer coordinate system.

The three components of the beam coordinate system are defined as follows:
\begin{itemize}
\item{$z$} points along the beam direction (downstream),
\item{$x$} points to the right of the beam (looking downstream), in the horizontal plane,
\item{$y$} points down towards the floor.													
\end{itemize}

The three components of the spectrometer coordinate system are defined as follows:
\begin{itemize}
\item{$z$} points along the optical axis at any point inside the spectrometer,
\item{$x$} points outwards in the direction of increasing the spectrometer momentum (the dispersive direction),
\item{$y$} points in the corresponding non-dispersive direction as required to complete a right-handed system. In the case of HMS and SOS, $y$ points to the left of the spectrometers.
\end{itemize}

A central trajectory (ray), also known as the nominal trajectory, is defined as the trajectory of a particle entering the spectrometer through the center of the entrance aperture, or in the case of the HMS and SOS, along the optical axis of the first quadrupole magnet. The detection plane is defined as the plane in the middle between two consecutive drift chambers detecting the charged particles. Defining $\vec{p}$ as the particle momentum and $B$ as the central magnetic field of the dipole magnet, central ray particles of different $p/B$ values would reach the detector plane at different positions. The optical axis is defined as the central ray that passes through a chosen point: the center of the detector plane (dispersive direction). The momentum of the particles traveling along the optical axis is called the central momentum $p_0$ or the ``excitation'' of the spectrometer.

Two reference frames in the spectrometer coordinate system are commonly used. One has the origin in the center of the detection plane. For historical reasons, the subscript $fp$ (focal plane) is used. However, in the case of the two spectrometers of Hall C, the focal plane and the detection plane do not coincide (see below). Thus, by definition $z_{fp}=0$ is at the detection plane. 

\begin{figure}[t]
\centering
\includegraphics[width=0.7\linewidth]{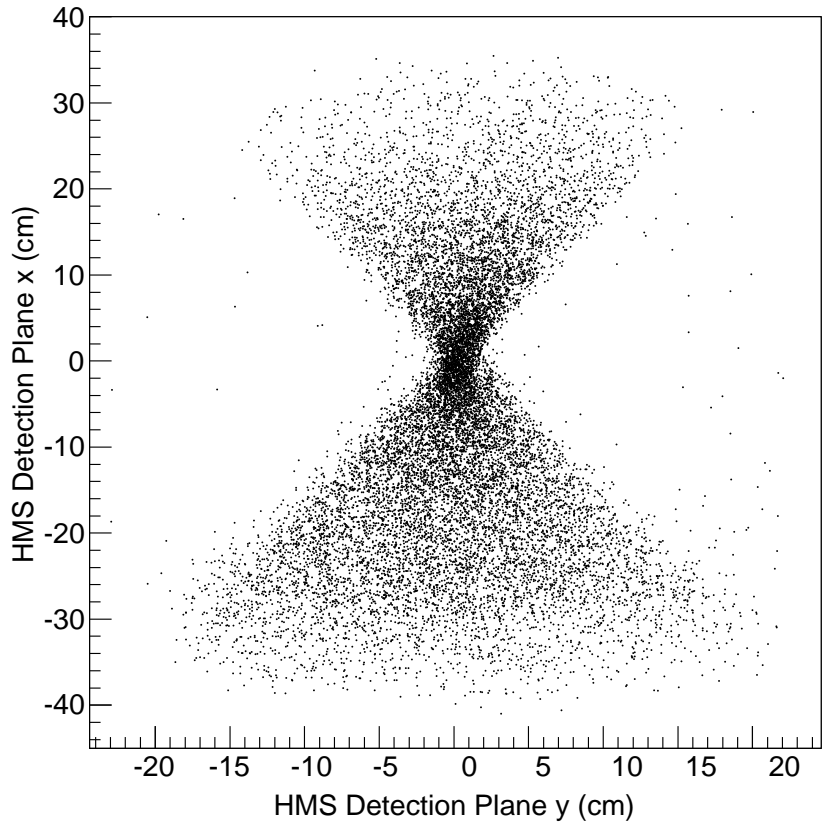}
\caption[Hourglass distribution ($x_{fp}$ versus $y_{fp}$) of coincidence proton event]{Example hourglass distribution ($x_{fp}$ versus $y_{fp}$) of coincidence proton events in the HMS detection plane for a given F$_\pi$-2 data run with the LH$_2$ target.}
\label{fig:hourglass}
\end{figure}

The other reference frame is directly related to the target, which are written with subscript $tar$. The trajectory of a particle is characterized by the two $x^\prime_{tar}$ and $y^{\prime}_{tar}$ (defined as above), and the point of origin $y_{tar}$. It is assumed that $x_{tar}=0$. The particle momentum $p$ is expressed in terms of the fractional difference compared to the central momentum $p_0$
\begin{equation}
\delta=\frac{p-p_0}{p_0}.
\end{equation}
The strength of the quadrupole fields for a given field $B$ of the dipole magnet is called the tune of the spectrometer. The field strengths have been chosen such that, for both HMS and SOS, there is a point-to-point focus in both directions ($x$ and $y$) for particles travelling along the optical axis, i.e., with $p=p_0$, or $\delta=0$. For trajectories with other values of $\delta$, the $x$ focus of the HMS behaves in such a way that for $\delta>0$ or $\delta<0$ it is at $x>0$ or $x<0$. The optical focusing in $y$ is more complicated: it is moved from positive $z_{fp}$ to $-\infty$ and then from $+\infty$ to negative $z_{fp}$, depending on the $\delta$. Fig.~\ref{fig:hourglass} shows the resulting $x_{fp}$ versus $y_{fp}$ distribution in the detection plane. The waist of the hourglass distribution at $x_{fp}=0$, $y_{fp}=0$ is the point where $x$ and $y$ focal planes and the detection plane coincide.

\begin{table}
\centering
\begin{tabular}{lcc}
\toprule
Quantity                  &   HMS         &  SOS            \\ \midrule
In-plane angle ($\theta$) &   -           &  -              \\ 
Out-of-Plane ($\phi$)     &  $+$1.1~mrad  &  $+$3.2~mrad    \\ 
Central Momentum ($p_0$)  &  $-$0.13\%    &  0.0-1.4\%      \\ 
\bottomrule
\end{tabular}
\caption[Kinematic offsets determined in the F$_\pi$-2 experiments]{Kinematic offsets determined in the F$_\pi$-2 experiments. This table was originally documented in Ref.~\cite{horn}.}
\label{tab:kin_offset}
\end{table}

\section{Determination of the Spectrometer Kinematic Offsets}

The discrepancies in the extracted cross section can be corrected by taking into account the kinematic offsets, particularly the deviations of the spectrometer central angles ($\theta$ and $\phi$) and momentum ($p_0$) from the nominal values.

The kinematic offsets can be determined and verified by reconstructing the physics quantities for the overdetermined (all final states particles are directed detected) elastic reaction ($e+p \rightarrow e+p$), such as the invariant mass $W$ and missing momentum components. The determination procedure is usually a two-step process. First, elastic-electron singles data (described in Sec.~\ref{sec:singles}) for beam energies between 1-3~GeV were used to fit the $W$ deviations, thus one can extract the spectrometer angles ($\theta$ and $\phi$) and the momentum ($p_0$) offsets. The effect of radiative corrections, energy loss and multiple scattering were taken into account using the Monte Carlo Simulation (described in Sec.~\ref{sec:simc}). Any vertical beam position offset from the center position must be determined and included, since such an offset would resemble a momentum offset. The second step is to verify the determined experimental offsets.

If the kinematic offsets are taken into account properly, the reconstructed invariant mass $W$ must be consistent with the proton mass within uncertainty. The full list of the kinematics offsets for the F$_{\pi}$-2 experiment was determined by T. Horn during the F$_{\pi}$-2-$\pi^+$ analysis, and is presented in Table~\ref{tab:kin_offset}. Further details regarding the determination of the kinematics offsets can be found in Ref.~\cite{horn, volmer}.

\graphicspath{{pics/4Simulation/}}

\chapter{SIMC and New Analysis Software}

The chapter starts with a brief overview of the Monte Carlo simulation software used for the analysis. The physics parameterizations used for the Heep reaction, $^1$H$(e^{\prime},ep)$, and the $\omega$ production reaction, $^1$H$(e^{\prime},ep)\omega$, are documented in later chapters. Finally, the new C++ based analysis software used to extract the experimental yields is introduced towards the end of the chapter.

\section{SIMC}
\label{sec:simc}

The Single Arm Monte Carlo package, SIMC\nomenclature{SIMC}{Single Arm Monte Carlo Simulation Software Package}, is the standard simulation package for Hall~C experimental data. It was used for the similar analyses of several previous experiments including F$_{\pi}$-1\footnote{Hall C experiment E93-021}\nomenclature{F$_\pi$-1}{First charged pion form factor experiment (E93-021)}~\cite{volmer}, F$_{\pi}$-2-$\pi^+$~\cite{horn} and F$_{\pi}$-2-$\pi^-$\cite{piminus}. A detailed description of SIMC can be found in Ref.~\cite{ArringtonThesis}, and therefore only an brief overview is given in this thesis.

For each event, the Monte Carlo generates both the initial coordinates of the interaction vertex $(x, y, z)$ and the kinematic quantities such as the energy ($E$) and three-momentum ($\vec{p}$) of the particles of interest. The kinematic offsets determined from the experimental data are required as the input parameters to the SIMC to correctly match the data and simulation. These kinematic offsets can be found in Table~\ref{tab:kin_offset}. The initial values for the generation limits in angle and momentum are fixed by the input files to the simulation and generally chosen to be larger than the physics acceptance of the spectrometers. If the kinematic quantities of an event are allowed (within the limitations of the acceptance), the outgoing event is followed through the target while the effects of ionization energy loss and multiple scattering are taken into account.

After the event generation process is completed, the events are sent to the single arm spectrometer modules, which simulate the optics as the result of combining multiple magnetic fields inside of each spectrometer. The propagation of the particles is monitored as they exit the target station, pass through the spectrometer aperture and magnetic field, and eventually into the spectrometer hut. Note that in SIMC, all angles are generated in the coordinate systems of the respective spectrometers.

A physics model that parameterizes the event production cross section in terms of Lorentz invariant physics quantities such as $W$, $Q^2$, $x$, $t$ and $u$, is required to weight the distribution of the generated events. A variety of effects, such as spectrometer acceptance and radiative correction are taken into account in the SIMC. Over the years, a large amount of effort~\cite{horn, volmer} was spent on customizing and refining the SIMC's capabilities.

Inside of each spectrometer hut, the particle trajectories are examined at each detector aperture. Events that are within all apertures and cross the minimum number of detectors in the huts are considered to generate a valid trigger. Only particles with a valid trigger have their trajectories fully simulated. Since detector apertures are simulated, no inefficiencies are assigned in the event selection. However, each event is weighted by the relevant model cross section, which is corrected for radiative processes, a luminosity factor and a Jacobian transformation that converts between the spectrometer coordinate and the physics coordinate.~\cite{horn}. The advantages and disadvantages of the event handling by SIMC are further elaborated in Sec.~\ref{sec:mm_me}.

\subsection{Spectrometer Models}

After the angle and momentum information for each event is generated at the event vertex, the events are sent to the single arm subroutines, which transport the particles through the magnetic field in the spectrometers using a COSY INFINITY model~\cite{cosy91}. In short, the COSY model consists of matrix elements that transport the particle sequentially through the magnetic optics in the spectrometer. The sequential implementation of the COSY model is advantageous in terms of allowing for the modeling of hadron decay.

By comparing simulated reconstructed quantities to the experimental data (particularly with exclusive interactions such as the Heep reaction), one can verify the measured experimental cross section and spectrometer optics models. Since a cross section weight is applied to each event, the agreement of the distributions of physics quantities, such as $Q^2$, $W$ or $t$, give information about the description of the kinematic dependence of the cross section model used. In addition, a comparison of the reconstructed spectrometer quantities, such as $hsytar$ and $hsyptar$ (target frame $y$ position and angle as viewed by the HMS, see Sec.~\ref{sec:secptrometer_acc}), provide a good check of the reconstructed optics matrix elements. Determination the optical matrix is documented in Ref.~\cite{horn}.

\subsection{Ionization Energy Loss}
The ionization energy loss of the incoming/outgoing electrons and the produced (recoiled) hadrons can be estimated by using the Bethe-Bloch formula~\cite{pdg}. In the scenario of low absorber thickness and high momentum, the mean energy loss distribution is better described with a Landau distribution, due to its asymmetrical feature~\cite{pdg}. Therefore in SIMC, the ionization energy loss is simulated using this type of distribution function.

The energy loss function is determined by two parameters: the most probable energy loss ($E_{prob}$), and full width at half maximum of the distribution ($\xi$). The most probable energy loss can be calculated from a random number $\lambda$, obtained from a Landau distribution, which can be written as
\begin{equation}
E_{prob} = \lambda \xi + E_{true} \,,
\end{equation}
and $$\xi = \frac{2\pi N_A z^2 e^4}{m_e c^2}\,,$$ where $N_A$ is Avogadro's number; $ze$ is the charge of the incident particle; $m_e$ denotes the electron mass; $t$ is the material thickness in g/cm$^2$; $Z$ and $A$ are the atomic number and mass of the material; and $\beta$ denotes the velocity of the incident particle in units of $c$.

In SIMC, the incident electrons are tracked as they travel through the target cell (cryogenic target and exit window) and their energy losses are calculated. The energy losses of the outgoing electrons and hadrons after travelling through various materials from the target to the spectrometer exit windows are also determined. Further detail regarding the general procedure on the electrons and hadrons energy loss correction can be found in Refs.~\cite{horn,volmer}.

\subsection{Multiple Scattering}
The experimental resolution determined by the wire chambers is modeled in SIMC, which includes the multiple scattering of the charged particles inside the target and spectrometers. A Gaussian distribution can be used to describe the deflection of the charged particles from their original scattering angle as they pass through a medium. The width of the Gaussian distribution describing multiple scattering is given as~\cite{horn}:
\begin{equation}
\theta_{0} = \frac{13.6~\textrm{MeV} \cdot \sqrt{t}~[\,1 + 0.088 \, \log(t/\beta^2)\,]}{\beta \, p \, c}\,,
\end{equation}
where $p$ denotes the momentum of the incident particle in MeV/c; $t$ is the thickness of the scattering medium in radiation length (the unit of radiation length is given by $1/X_0$). The angles defining the direction of a particle traversing a material with thickness $t$ are changed by a factor $g\cdot\theta_{0}$ , where $g$ is a random number following a Gaussian distribution centered at zero and with unit width. Note that the multiple scattering in horizontal and vertical directions are simulated independently.

The effect of multiple scattering is calculated in SIMC for both the incident and scattered electrons and also for the produced (recoiled) hadrons. After including the multiple scattering, the experimental and simulated resolutions agree to a level of 30\%. Although this deviation appears to be significant, the effect of changing the simulated resolution to match the experimental resolution has been tested with elastic electron singles data, and only a relatively small effect was observed on the simulated acceptance~\cite{horn, volmer}.

Compared to the early commissioning Hall C experiments (such as the F$_\pi$-1 experiment~\cite{volmer}), the multiple scattering in the F$_\pi$-2 experiment has increased in the HMS due to the thicker titanium spectrometer exit window. Further detail regarding the correction for multiple scattering in SIMC can be found in Ref.~\cite{horn}.

\subsection{Radiative Process}
\label{sec:rad_process}

The radiative process describes the emission of photons (Bremsstrahlung radiation) by charged particles involved in the reaction, meaning that the reconstruction of the missing mass and missing energy spectra would appear to be wider (corresponding to a poorer resolution) and the central value deviates away from the expectation. Therefore, the understanding the radiative process is an important part of the analysis for the electron scattering experimental data. Traditionally, the radiative (process) correction of the experimental data involves computation of a correction factor in terms of missing energy or missing mass distributions to account for any redistribution in the cross section. However, such a correction factor is only capable of correcting redistributions from the nominal experimental setting, while ignoring the variation across the experimental acceptance. One way to address this short coming is to directly calculate cross section spectra with SIMC,
which takes into account the variation across the full spectrometer acceptance as described in detail in Ref.~\cite{ent01}.

The radiative correction algorithm used in SIMC is based on a formalism originally derived to apply the radiative corrections to the inclusive electron scattering off a proton target~\cite{mo69}, and was extended to take into account coincidence $(e, e^\prime p)$ reactions~\cite{ent01, mo69, makins94} before being implemented in SIMC.

In SIMC, the radiative correction of meson production processes is based on the assumption that the target particle is treated as a stationary proton and the final state meson is treated as an offshell (virtual) proton. This radiative correction for meson production was implemented in Ref.~\cite{horn}, but this default assumption is for the meson to travel forward after the interaction and is detected by one of the two spectrometers. However, in the $u$-channel meson production reaction, the proton target travels forward and is detected, replicating the exact scenario of the coincidence Heep reaction. Since the $u$-channel meson ($\omega$ in this analysis) is reconstructed with the detected proton information, the radiative correction used the coincidence $(e, e^\prime p)$ reactions is sufficient to correct the $u$-channel reaction $^1$H$(e, e^\prime p)\omega$.

The radiative effect (emission) in the electron scattering reaction is a result of the acceleration of the charged particle in the presence of an electric field. Under the external radiation emission scenario, one of the charged particles involved in the reaction emits a real photon upon interacting with the electric field of the encountered nuclei while traversing through a medium. This external radiative correction is relatively straightforward, since the particles radiates independently without inference effect.

In the case of internal radiation, the charged particles radiate in the field of the primary nucleon target. The correction is complicated by various interference effects. The internal radiative correction contains second order QED diagrams such as vacuum polarization and self energy diagrams~\cite{horn}. Further explanation regarding the higher order correction terms of the internal radiation correction can be found in Ref.~\cite{horn, ent01}.

The radiative correction implementation in SIMC includes an approximation to the photon energy and angular distributions of the radiated photon. The radiated photon energy is restricted to be much less than the energies of the initial and final state particles, and this is referred to as the soft photon approximation. Under this limit, the fundamental one photon exchange amplitude can be factorized from the radiative process. In addition, the extended peaking approximation provides an important simplification for the calculation of radiative effects in the coincidence framework. With this approximation, the single photon bremsstrahlung radiation can be divided into three discrete photon directions (along the direction of incoming electron, scattered electron and meson momentum). The total radiated strength in this limit is preserved by dividing the non-peaked terms of the angular distribution evenly between the electron peaks.

The radiative correction is part of the overall weighting factor, which is directly multiplied by the cross section. The generation of radiative correction factors and further discussion on the two-photon exchange diagrams are extremely complicated topics, and more information regarding these topics can be found in Refs.~\cite{horn,ent01,makins94}.

The Heep reaction provides a good validation of the radiative correction factor; the missing energy and missing mass distributions are compared between data and simulation in Sec.~\ref{sec:mm_me}.

During the F-$_\pi$-2-$\pi^+$~\cite{horn} and F-$_\pi$-2-$\pi^-$~\cite{piminus} analyses, a standard 2\% correlated systematic uncertainty were used. Since the radiative correction for coincidence $(e, e^\prime p)$ is better understood than for the coincidence $(e, e^\prime \pi)$ process, a slightly reduced correlated systematic uncertainty of 1.75\% is used for this analysis.

\subsection{Monte Carlo Yield}
In order to extract the experimental cross section by comparing absolute normalized data to Monte Carlo, the equivalent SIMC yield has to be determined. The data yield is calculated in counts per unit of integrated luminosity. The Monte Carlo luminosity can be written as
\begin{equation}
L = \frac{\rho \, t N_A N_e}{M}\,,
\end{equation}
where $\rho$ is the target density in g/cm$^3$, $t$ is the target thickness in cm, $N_A$ is Avogadro's number of the target, $N_e$ is the number of electrons in 1~mC of beam charge (10$^{-3}$/1.60218$\cdot$10$^{-19}$) and $M$ is the target mass.

The SIMC yields are calculated differently for Heep and meson production reactions. For the Heep reaction, $^1$H$(e^{\prime}, ep)$, the SIMC yield can be written as
\begin{equation} 
Y_{\rm SIMC} = L \int_V{\left( \frac{d^5 \sigma}{d\Omega_{e^{\prime}} \, dE_{e^{\prime}} \, d\Omega_{p} }\right)^{model} A(V)R(V)J(dX^{\prime}) \, dX_{e^{\prime}} \, dE_{e^{\prime}} \, dX^{\prime}_p } \,,
\end{equation} 
where $A$ is the coincidence acceptance function including energy loss and other relevant effects, $R$ is the radiative correction factor, $dX^{\prime}= dx^{\prime}dy^{\prime}$ is the differential solid angle in spectrometer coordinates and $J(dX^{\prime})$ is the Jacobian transforming the model cross section from spherical to spectrometer coordinates which are used in event generation.

For the meson production reactions, $^1$H$(e^{\prime}, ep)X$, where $X$ is $X=\omega$, $\rho$, or other background final states, and the SIMC yield can be written as
\begin{equation} 
Y_{\rm SIMC} = L \int_V{\left( \frac{d^6 \sigma}{d\Omega_{e^{\prime}} \, dE_{e^{\prime}} \, d\Omega_{p} \, dM_R}\right)^{model} A(V)R(V)J(dX^{\prime}) \, dX_{e^{\prime}} \, dE_{e^{\prime}} \, dX^{\prime}_p \, dM_R}\,,
\end{equation} 
where $M_R$ is the recoil mass of the system. In the case of the $\omega$ production, the choice of $M_R$ is determined from the mass and width of the $\omega$. When analyzing the simulation data, it is extremely important to scale the simulated distributions (i.e. missing mass) by the weight factor. The weight factor is generated on an event-by-event basis and is a variable in the simulation ntuple. The complete normalization of the simulation data is discussed in the next section. Further detail regarding the SIMC model for the meson productions can be found in Sec.~\ref{sec:sim_physics}.

\section{The New C++ Analysis Platform}
\label{sec:ana_platform}

\begin{figure}[t]
  \centering
  \includegraphics[width=0.85\textwidth]{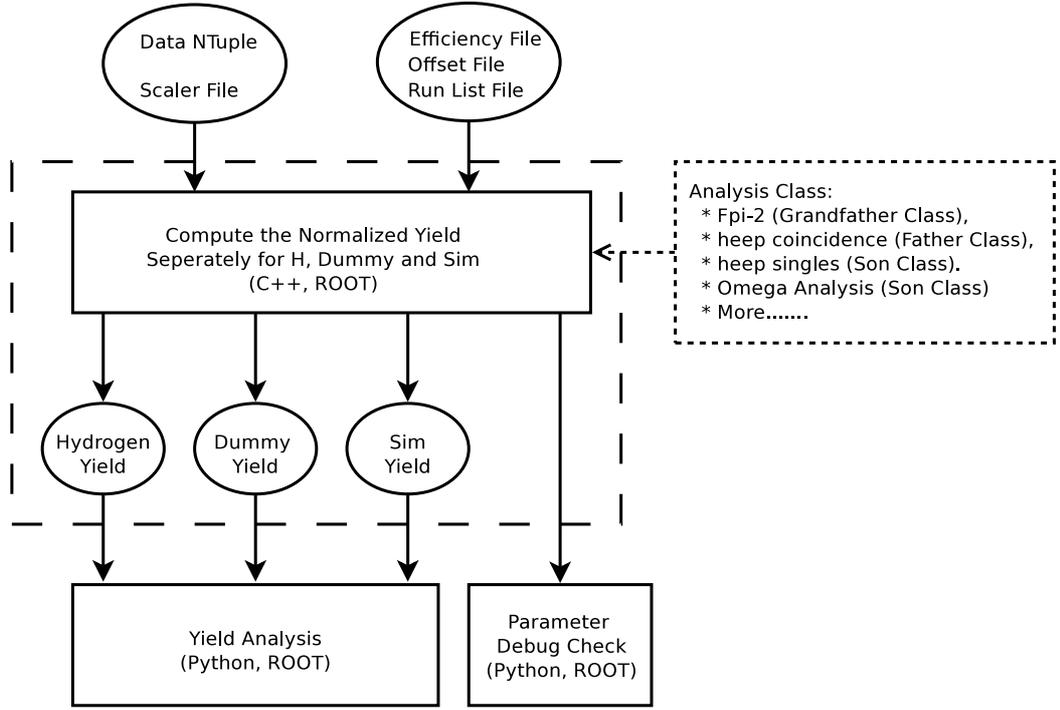}
  \caption[Flow chart of the data analysis procedure]{A simple flow chart of the data analysis procedure.}
  \label{fig:code_pro}
\end{figure}

\begin{figure}[t!]
  \centering
  \includegraphics[width=0.85\textwidth]{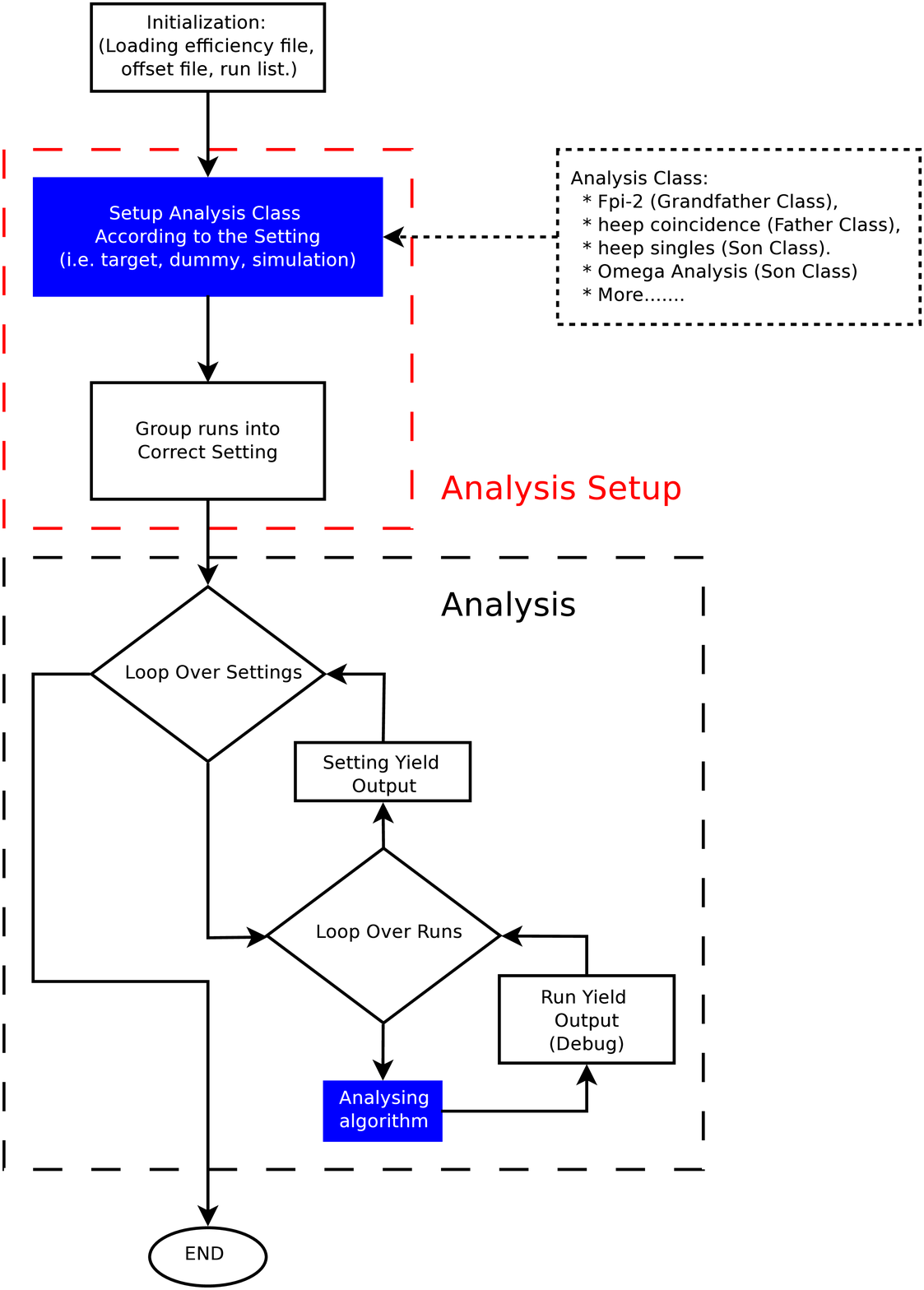}
  \caption[Flowchart of the normalized yield analysis code]{Flowchart of the normalized yield analysis code. Different analysis classes are required when performing different types of analysis, i.e. Heep coincidence ($^1$H$(e,e^{\prime}p)$), Heep singles ($^1$H$(e,e^{\prime})X$). When a new analysis created, the only part of the code that requires modification is colored in blue. \oic}
  \label{fig:flowchart}
\end{figure}

For the previous F$_\pi$-2 analyses, the PAW code (for the yield computation) was inherited from the F$_\pi$-1~\cite{volmer} analysis. One of the major objectives of this work is to translate the FORTRAN based PAW code to a C++ based analysis platform, before extracting the $\omega$ yield. The new code is designed to use the libraries from ROOT~\cite{root}, while maintaining the same functionality and algorithm structure.

Similar to the PAW macros, when computing the normalized experimental and Monte Carlo simulation yields, efficiency tables and offset files are needed as input as well as the data Ntuples. The computed yield and distributions (for cross-checking) are saved into a ROOT file. The schematics for computing the normalized yield are shown in Fig.~\ref{fig:code_pro}. A small portion of the F$_\pi$-2 data were used to test the new analysis platform and were compared with the PAW macro. The computed yields and cross-checking distributions ($hsxptar$, $hsyptar$, $hsdelta$,) agree 100\%. Note that the C++ code takes one third of the time to run in comparison with the PAW code.

Fig.~\ref{fig:flowchart} shows the flow chart of the new C++ platform for extracting the normalized yield. In the analysis setup (red dashed box) section of the code, lists are created according to different experimental settings; here the experimental settings are categorized with respect to $\epsilon$ (virtual photon polarization), $Q^2$, $\theta_{\rm HMS}$ and target type (hydrogen or dummy). Depending on the type of data analysis (F$_\pi$-2, Heep singles, Heep coincidence), the number of settings can vary significantly. The constructor for each analysis class is initialized for each setting. The analysis class is specific to different types of analysis, where the earlier analysis class can be inherited and their functions can be used. The last part of the analysis setup is to loop over the data runs to associate each run with the correct efficiencies.

In the analysis (black dashed box) section of the code, the program will first loop over the setting list, then analyze each run in the list. The analysis (yield computation) takes into account the efficiencies, cuts and offsets on a run-by-run basis, and yields are accumulated over the setting. For debugging and cross-checking purposes, kinematic variables ($Q^2$, $M_m$), spectrometer acceptance parameters ($hsxptar$, $ssyptar$) and absolute yield are saved to a ROOT file. After each setting ends, the yield sum and error from each run are normalized to 1~mC of beam charge. The normalized yield and yield error are saved, as well as the normalized distribution of kinematic variables and spectrometer acceptance parameters for the each settings.

For the simulation, the analysis procedure is much simpler since no particle identification (PID) cut is required. The efficiency files are replaced with a normalization factor file. Since SIMC generates events with unequal weighting, but uniform phase-space coverage, the normalization of the simulated data requires the additional weight applied to each event; the overall distribution needs an additional scale factor which is given by:  
\begin{equation}
\textrm{scale factor} = \frac{\textrm{normalization factor}}{\textrm{number of events}}\,,
\label{eqn:normfac}
\end{equation}
where the `normalization factor' takes into account the luminosity ($L$) and simulated phase-space.

The coding philosophy is to maximize the customizability of the individual analysis while maintaining the standardized analysis setup procedure. For example, the differences in terms of the analysis codes for analysing the F$_\pi$-2 test data and the Heep coincidence data are shown in the blue boxed region, where the main structure of the code remains identical. 

In order to avoid repetitive coding, the earlier analysis classes such as F$_\pi$-2 and heep\_coin, can be directly inherited by any later class. Shared cuts and general utility functions can be easily accessed. It is the author's hope that this code will be used to save time and effort by more and more students performing similar analysis. The final version of the code is located at the following GitHub\footnote{https://github.com} \textit{repository: https://github.com/billlee77/omega\_analysis}.

After the normalized yield for the hydrogen target, dummy target and simulation is computed and saved into a ROOT file, the experiment-simulation ratio is then generated by a python based script, as shown in Fig.~\ref{fig:code_pro}. As a consistency check, the dummy target-subtracted distribution for all kinematic variables and acceptance parameters should not generate negative peaks.

From the flow chart given, one can develop a general structural picture regarding the iterative L/T separation procedure. The further details and justification regarding the iterative L/T separation procedure are given in Chap.~\ref{chap:omega_ana}.

\graphicspath{{pics/5Heep/}}

\chapter{Heep Data Analysis} 
\label{chap:heep}

The elastic reaction $^1$H$(e, e^{\prime}p)$ is often referred to as the Heep\nomenclature{Heep}{$e$-$p$ Elastic Scattering Reaction: $^1$H$(e, e^{\prime}p)$} process. In this reaction, the recoil electrons are detected by the Short Orbit Spectrometer (SOS) and the protons are tracked by the High Momentum Spectrometer (HMS). Fig.~\ref{hallc_top} shows an overhead view of the standard Hall C experimental setup, which shows the SOS and HMS locations with respect to the target. These elastic scattering data provide a good check for the spectrometers, as well as various effects on reconstruction, such as radiative processes, multiple scattering and energy loss that were simulated by the Monte Carlo simulation (SIMC).

The data for the elastic $^1$H$(e, e^{\prime}p)$ reaction were taken in four different kinematic settings, see Table~\ref{tab:heep}. These kinematic conditions were modelled in SIMC, then compared to the data. For a detailed description regarding the cross section parameterization used for the Heep model, see Sec.~\ref{sec:heep_model}. 

During the Heep data runs, the data acquisition was operated in the coincidence mode for the $^1$H$(e,e^{\prime}p)$ interaction. The coincident Heep study relies on the acceptance, tracking and PID information from both spectrometers to reconstruct variables such as missing mass, missing energy and $Q^2$. In addition, one can also perform the Heep study by examining only the recoil electron information in SOS (electron singles Heep mode) for cross checking purposes.

This coincidence Heep measurement almost replicates the exact experimental condition of the backward-angle production interaction: $^1$H$(e,e^{\prime}p)\omega$. The coincidence trigger modes are identical between the two data sets. Furthermore, the event selection criteria and detector efficiencies used for both data sets are almost identical. Since the Heep data set has much less pion contamination than the $\omega$ data set, it is the optimal choice to study the proton detector efficiencies. 

In this chapter, brief introductions on event selection criteria and background subtractions are given in Secs.~\ref{sec:heep_selection} and \ref{sec:heep_bg_sub}. A variety of efficiencies specifically related to the proton selection in the HMS are described in Sec.~\ref{sec:heep_eff}. Finally, the simulation to experiment yield ratio (comparison) is presented in Sec.~\ref{sec:heep_yield}.

\begin{table}[t]
\centering
\caption[Nominal experimental kinematic values for the $^1$H$(e,e^{\prime}p)$ coincidence runs]{Nominal experimental kinematic values for the $^1$H$(e,e^{\prime}p)$ coincidence runs. $E_e$ is the electron beam energy; $Q^2$ represents the four momentum transfer square between the electron before and after the interaction; $p_{e^{\prime}}$ is the nominal momentum setting of the electron arm (SOS); $\theta_{e^\prime}$ is the electron arm (SOS) angle with respect to the incident electron beam; $p_{p}$ is the nominal momentum setting of the proton arm (HMS); $\theta_{p}$ is the proton arm (HMS) angle with respect to the incident electron beam.}
\label{tab:heep}
\setlength{\tabcolsep}{1.2em}
\begin{tabular}{cccccc}
\toprule
$E_e$   &   $Q^2$         & $p_{e^{\prime}} $ & $\theta_{e^\prime}$  &  $p_{p}$  &  $\theta_{p}$  \\ 
GeV     &   GeV$^2$       &  GeV/c            &  deg                 &  GeV/c    &  deg           \\ \midrule
3.778   &   4.44          &  1.442            &  54.02               &  3.154    &  21.40         \\ 
4.210   &   2.41          &  1.582            &  51.03               &  3.437    &  20.90         \\ 
4.709   &   5.42          &  1.726            &  48.06               &  3.756    &  20.50         \\ 
5.248   &   6.53          &  1.726            &  50.07               &  4.335    &  18.00         \\ 
\bottomrule
\end{tabular}
\end{table}

\section{Data Selection and Correction}
\label{sec:heep_selection}

The first step in the identification of $^1$H$(e,e^{\prime}p)$ events depends on the correct identification of electrons and protons in the SOS and HMS spectrometers, and on the precise coincidence timing information for the separation of the true and random coincidence events. 

\subsection{Particle Identification in SOS and HMS }
\label{sec:PID_cuts}

In the SOS, due to its negative polarity setting, negatively charged pions are detected along with the recoil electrons. The pion contamination that was not rejected by correct coincidence time cut and $E_m$ cut, is less than 3$\%$~\cite{horn}. Electrons were detected and identified in the SOS using a combination of the Heavy Gas Cherenkov detector (HGC)\nomenclature{HGC}{Heavy Gas Cherenkov Detector} and calorimeter. The HGC detector was used as a threshold detector with a mean SOS signal of 7 photo-electrons (pe\nomenclature{pe}{Photo-electron}) for one individual electron event. Good electron events were selected for number of pe threshold of N$_{photoelectron} > $ 0.5. A cut was also placed on the SOS calorimeter. In previous data analyses~\cite{horn, piminus}, a threshold of $E_{cal}/E^{\prime} > $ 0.7 was in place, which is $>$99\% efficient for selecting electrons.

In the HMS, where the proton events are selected, the background particles are pions and positrons. The rejection of the positrons is done via the signal from the Cherenkov detector. The positrons that were not rejected by the HMS pion trigger contribute 2.2\% of all events with the correct coincidence timing and reconstructed missing mass. The limit of 0.5 photo-electrons in the Cherenkov detector provides positron rejection better than 99.5\%~\cite{horn}. The remaining positron contamination is negligible (much less than 0.1\%). In addition, there is a nonzero probability for a proton to produce a knock-on electron ($\delta$ radiation) while passing through the detector, which will result in a false signal. The contamination of the random electron-proton coincidence events that have the correct missing mass value is $\sim3\%$, and is sufficiently corrected by the random coincidence background subtraction (described in Sec.\ref{sec:heep_rand}). For documentation purpose, the complete set of particle identification (PID) cuts are given below: 

\begin{description}
\item[PID Cut: ] $hsbeta > 0.1$ \&\& $hsbeta < 1.5$ \&\& $hcer\_npe < 0.5$ \&\& $abs(haero\_su) < 4$ \&\&\\
 $ssshtrk > 0.70$ \&\& $scer\_npe > 0.50$
\end{description}

\subsection{Coincidence Timing vs.Particle Speed in the HMS}
\label{sec:cointime_heep}

\begin{figure}[t]
\centering
\includegraphics[width=0.85\textwidth]{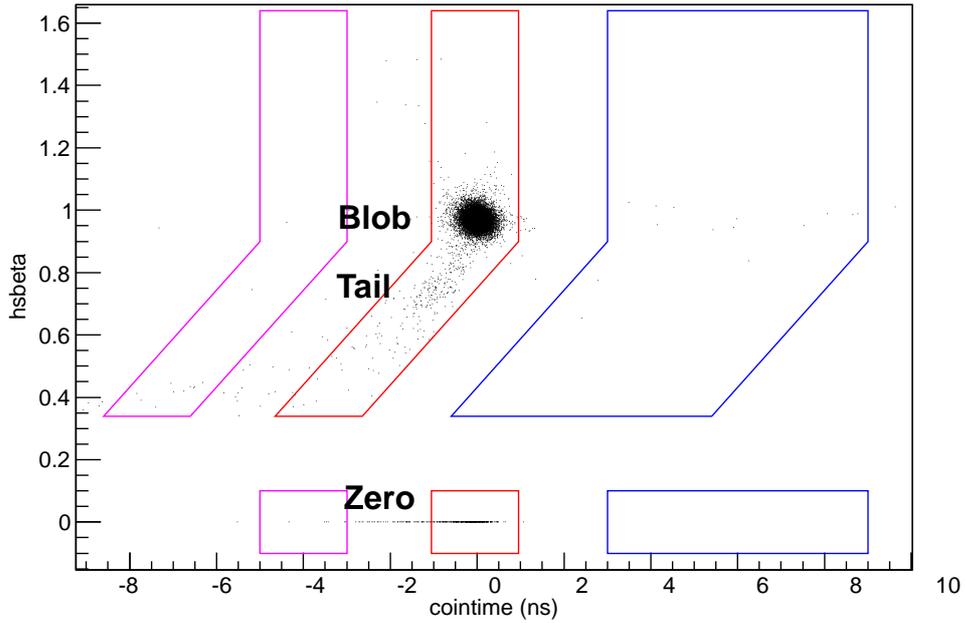}
\caption[$hsbeta$ versus $cointime$ distribution for the Heep data set]{Example $hsbeta$ versus $cointime$ distribution for the Heep data set. Acceptance and PID cuts are applied. Red box shows the real coincidence box with width of 2.1~ns; blue box shows the early random coincidence box with width of 6.3~ns; magenta box shows the late random coincidence box with width of 2.1~ns. The box boundary positions are fixed across all settings. The blob, tail and zero events are indicated in the figure. Note that the criteria of the blob, tail and zero events are defined in the corresponding text. \oic}
\label{fig:hsbeta_cointime_heep}
\end{figure}

\begin{figure}[t]
\centering
\includegraphics[width=0.8\textwidth]{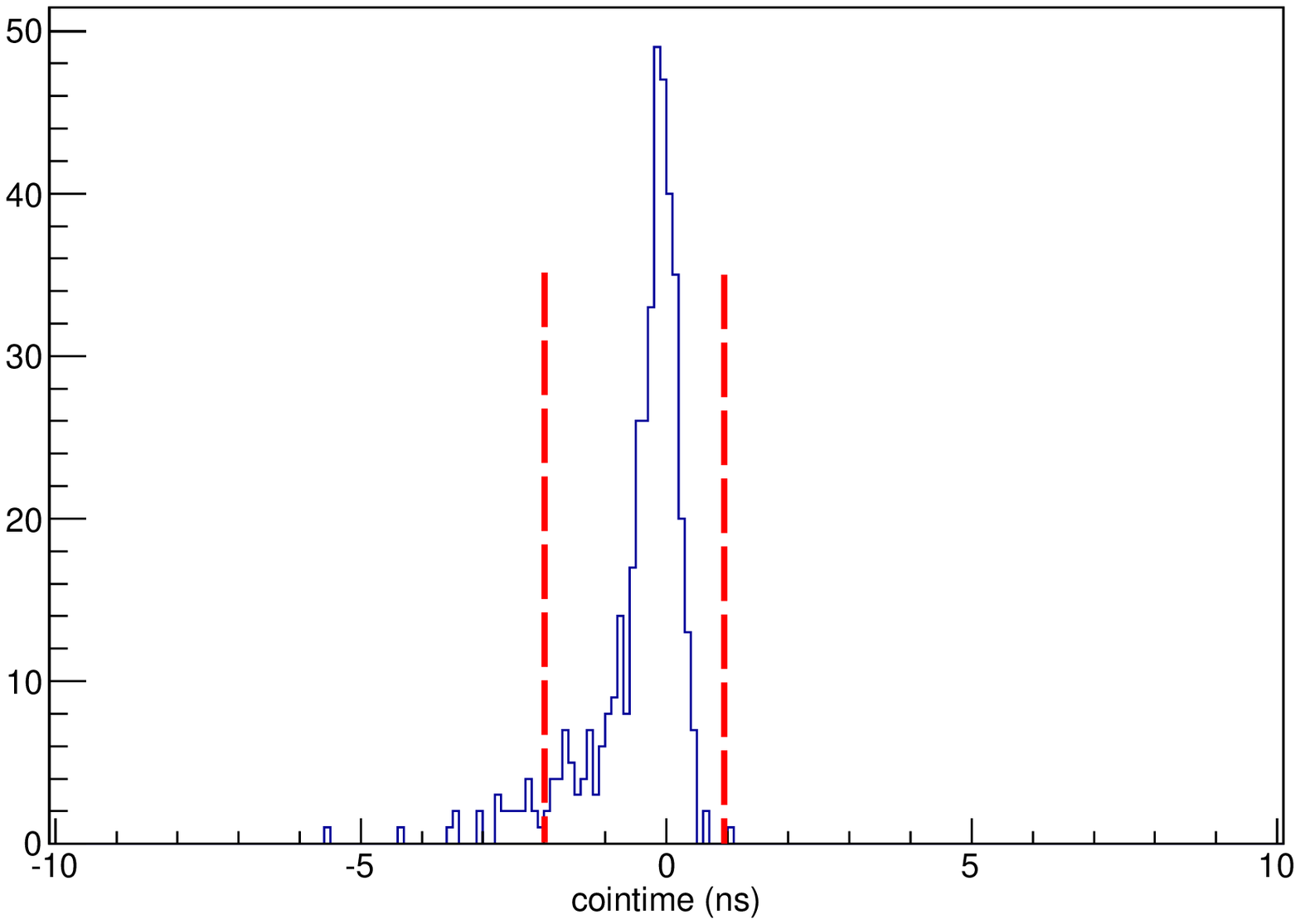}
\caption[$cointime$ distribution projection for the zero ($hsbeta=0$) events]{Example $cointime$ distribution projection for the zero ($hsbeta=0$) events from Fig.~\ref{fig:hsbeta_cointime_heep}. Same cuts applied as in Fig.~\ref{fig:hsbeta_cointime_heep}. Red box shows the real coincidence box with width of 2.1~ns. \oic}
\label{fig:cointime_zero_check}
\end{figure}

The most effective criterion for selecting the proton coincidence events is formed by examining the correlation between the relative particle velocity ratio $\beta$ (to the speed of light) inside of the HMS focal plane($hsbeta$) versus the coincidence timing information. The particle velocity determined from the time of flight (TOF\nomenclature{TOF}{Time of Flight}) information is generally an important element in the selection of events from the reaction of interest. The particle velocity in each spectrometer is calculated from the TOF information provided by the four scintillator element hodoscopes in the SOS and HMS. An example $hsbeta$ versus coincidence time distribution for the Heep data is plotted in Fig.~\ref{fig:hsbeta_cointime_heep}.

Note that small offsets in the location of the $cointime$\nomenclature{$cointime$}{Coincidence timing window is started by HMS pre-trigger trigger and stopped by SOS pre-trigger} blobs are observed (among data runs). For a given run, this offset is much smaller than the 2.1~ns timing window, therefore would not result any significant discrepancies. In this analysis, the blob positions have to be corrected on a run-by-run basis, before the $hsbeta$ versus $cointime$ distributions are summed over the same $Q^2$ setting.

The HMS-SOS coincidence trigger TDC is timed by a HMS pre-trigger signal starting the TDC and stopped by a delayed SOS coincidence trigger signal. The time difference between the two triggers is the raw coincidence time. The raw coincidence time is corrected for time differences resulting from the variation in particle velocity and path distance traveled through each spectrometer. The difference in path length is estimated from the difference of the particle trajectory compared to the central trajectory. The corrected coincidence time allows for a resolution of 200~ps, which is sufficient to resolve the beam structure of the accelerator. Further details regarding the path length correction are given in Ref.~\cite{ArringtonThesis}.

From Fig.~\ref{fig:hsbeta_cointime_heep}, a single `blob' represents the coincidence proton events at $cointime=0$~ns. There is a `tail'-like structure (towards low $hsbeta$) attached to the blob and in addition there is a cluster of `zero' events with $hsbeta=0$. These `tail' and `zero' events are the effects due to the proton undergoing multiple scattering inside the scintillator material, HGC window and other material in their path inside of the HMS focal plane stack.

Fig.~\ref{fig:cointime_zero_check} shows the $cointime$ distribution for the zero ($hsbeta=0$) events. Note that the zero events contribute less than 3\% of the random subtracted yield. 85\% of the zero events are included by the real coincidence time window, as indicated by the red dashed lines, despite the fact that the most appropriate location of the coincidence window boundary may not correspond to the $\pm1$~ns window from the location of the blob events. A vague correlation can be observed between the locations of the tail and zero events. Due to low statistical significance, the effect of the position of the real coincidence for the zero events has a negligible contribution (less than 0.4\%) to the real experimental yield.

Based on further investigation, the `zero' and `tail' events have valid acceptance information (such as $hsxfp$), which can be treated as the `blob' events. One of the possible causes for the `zero' and `tail' events are the interactions between proton and detector material downstream of the wire chamber. Since the detailed tracking and TOF information requires a fiducial cut on the hit location and signal strength from the hodoscopes, the deflected events due to multiple scattering can easily fail the fiducial cut and result an TOF overflow ($hsbeta=0$).

During the data analysis, the `blob' events correspond to $hsbeta\ge0.9$, the `tail' events have $hsbeta<0.9$ and the `zero' events have $hsbeta=0$.

\subsection{Event Selection Criteria}
\label{sec:heep_cuts}

The event selection criteria (cuts) used to analyze the Heep data are listed below, many of these conditions are similar to the ones used in the previous analysis efforts~\cite{horn, volmer, piminus}. Note that the same cuts are used for the $\omega$ analysis. 
\begin{description}
\item[HMS Acceptance Cut: ] $abs(hsytar) \le 1.75$ \&\& $abs(hsdelta) \le 8.0$ \&\& $abs(hsxptar) \le 0.080$ \&\& $abs(hsyptar) \le 0.035$.

\item[SOS Acceptance Cut: ] $ssytar \le 1.5$ \&\& $abs(ssdelta) \le 15.$ \&\& $abs(ssxfp) \le 20.$ \&\& \\ 
 $abs(ssxptar) \le 0.04$ \&\& $abs(ssyptar) \le 0.065$.

\item[Partial PID Cut: ] $hsbeta > -0.1$ \&\& $hsbeta < 1.5$ \&\& $hcer\_npe < 2$ \&\&
 $ssshtrk > 0.70$ \&\& $scer\_npe > 0.50$.

\item[ACD Threshold Cut:] Depending on the HMS central momentum setting, a different Aerogel Cherenkov threshold is required. See detail in Sec.~\ref{sec:aero}.

\item[Full PID Cut:] Partial PID Cut \&\& ACD Threshold Cut.

\item[Missing Mass ($M_m$\nomenclature{$M_m$}{Missing Mass}) and Energy Cut ($E_m$\nomenclature{$E_m$}{Missing Energy}): ] $Em < 0.10$ \&\& $M_m>-0.032$ \&\& $M_m< 0.018$.
 
\end{description}

Note that depending on the HMS central momentum setting, different ACD threshold cuts were applied, this is explained in detail in Sec.~\ref{sec:aero}. The Full PID cut combines the Partial PID cut and ACD threshold cut. From this point onwards, unless specified, the term PID cut refers to the Full PID cut. Furthermore, a missing energy ($E_m$) cut of $E_m < 0.1$~GeV and a missing mass cut were used. These cuts are further explained in Sec.~\ref{sec:mm_me}.

\section{Background Subtraction}
\label{sec:heep_bg_sub}

The experimental data contain two types of non-physics background: random coincidences from unrelated electrons, pions, and protons in the two spectrometers, and coincident electrons and protons originating from the aluminum walls of the target cell. They are described in Secs.~\ref{sec:heep_rand} and \ref{sec:heep_dummy}, respectively. Note that the same techniques were applied to the $\omega$ analysis.

\subsection{Random Coincidence Background Subtraction}
\label{sec:heep_rand}

Random $e$-$p$ coincidence events constitute a background and have to be subtracted from the data sample. The estimation of the random background includes two separate random windows, one before and one after the real $e$-$p$ window. The `early' random coincidence window (blue boxes right side of proton blob) was 6.3~ns wide (three times the real window), and the `late' random coincidence window (magenta boxes left side of the real peak) was 2.1~ns wide (one real window). The real $e$-$\pi$ peak is avoided in the placement of random coincidence cuts. The number of random events within the real window can be estimated as the total of random events over the number of the random windows (four random windows), and it is subtracted from the total number of events in the real window.

\subsection{Cell Wall Contribution and Dummy Target Data Subtraction}
\label{sec:heep_dummy}

Another type of background that needs to be removed from the sample of good events is the background due to the scattering from the aluminum target cell walls enclosing the liquid hydrogen. The target cell wall contributes a relatively small percentage of the total yield (2-4.5\%).

\begin{figure}[t]
\centering
\includegraphics[width=0.8\textwidth]{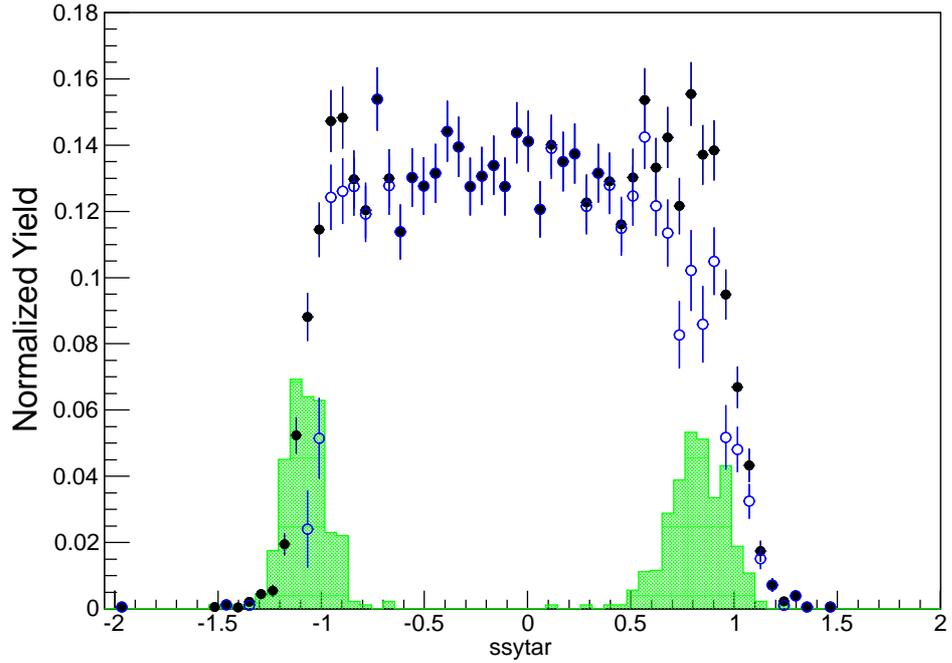}
\caption[normalized $ssytar$ distributions for $\omega$ data at $Q_2$ = 2.45 GeV$^2$]{Example of normalized $ssytar$ (events horizontal position information in the field of view of SOS) distributions for $\omega$ data at $Q^2$ = 2.45 GeV$^2$, $\epsilon$ = 0.55, central HMS angle setting. The LH$_2$ target data is shown in solid black dots; green shaded distribution is for the dummy target; the dummy target subtracted distribution is shown in blue circles (black$-$green).~\oic}
\label{fig:ssytar}
\end{figure}

The target cell wall contribution to the background events can be estimated by taking dummy target data and subtracting them from the data with the LH$_2$ target. The dummy target consists of two aluminum foils 4~cm apart and centered at the target station. Note that the dummy target is intentionally made thicker, thus maxing the yield while minimizing the run time. Fig.~\ref{fig:ssytar} shows the $ssytar$ distribution for the LH$_2$ target for the $Q^2$ = 2.45~GeV$^2$ setting in black dots and the cell wall contribution in green. Since the SOS angle is 50$^{\circ}$ with respect to the electron beam, the separation between the two green bumps is not 4~cm.

The dummy target data are analyzed in the same way as the regular data, including the same method of random coincidence subtraction and applying the same event selection criteria (cuts). The extracted experimental yields (number of events which pass the event selection criteria) are then subtracted from the real data yields, taking into account the additional weight of 7.022 to account for the difference in wall thickness between target cell and dummy target. When compared to other experimental uncertainties, the uncertainty in the target thickness ratio between target cell wall and dummy target is negligible.

\section{Efficiency Study}
\label{sec:heep_eff}

\subsection{Analysis Information}
\label{sec:yield_cal}

In computing the normalized yield, one must apply corrections for inefficiencies such as trigger, track reconstruction and data acquisition deadtime. The total experimental yield can be written as 
\begin{equation}
Y_{exp} = \frac{N}{\epsilon_{tot} Q_{tot}}\,,
\end{equation}
where $N$ is the total number of selected good events, $\epsilon_{tot}$ is the efficiency correction factor (all detector efficiencies and electronic live times combined) and $Q_{tot}$ is the total accumulated electron beam charge. Details regarding efficiency studies, such as event tracking efficiencies and electronic live time, are documented in this section. These efficiencies are also applied to the $\omega$ analysis. 

Note that the analysis presented in this thesis does not include the raw data replay (calibration and conversion). The $e$-$p$ data ntuples used were created during the F$_\pi$-2-$\pi^+$ analysis by T. Horn~\cite{horn}.

\subsection{Computer and Electronic Live Time}
\label{sec:live_time}

The data acquisition (DAQ) system efficiency for experiments is rarely perfect (100\%). During the experiment, the DAQ which consists of electronics and computers, has a number of rate-dependent efficiencies (live time) due to their processing speed and the level of the logic complexity. These rate-dependent efficiencies need to be carefully studied in order to obtain an accurate absolute physics measurement.

The computer live time (CLT) can be calculated from the ratio between the recorded events and total events (triggers). The DAQ system used during F$_\pi$-2 recorded data on a single-event basis, meaning once the recording process was initiated, no more event could be recorded until the first event processing is completed. This would inevitably cause event loss at high event rate. Note that there is also an efficiency associated with the performance of the trigger supervisor, whose effect is negligible when compared to the CLT.

\begin{table}[p]
\centering
\setlength{\tabcolsep}{.46em}
\caption[Data for the Computer and Electronic Life Time study for the HMS]{Example data for the Computer and Electronic Life Time study for the HMS. Four runs from each data type are selected. PS1 is the singles pre-scale factor of the HMS; $atrig$ is the total number of triggers generated by the trigger supervisor, which can also be calculated using Eqn.\ref{eqn:atrig}; $hpre$ is the total number of HMS pre-trigger; BoT and hS1X are described in Table~\ref{carbon_tab}; $ctrig$ is the number of coincidence trigger;  $htr\_ccut$ is the HMS tracking efficiencies which is explained in detail in Sec.~\ref{sec:tr_eff}. $hcomp$ and $helec$ are the computer and electronic live times of the spectrometer.}
\label{sample_data_HMS}
\small
\begin{tabular}{cccccccccc}
\toprule
Run   & PS1 &   $atrig$ & $hpre$    &   BoT  &    hS1X    & $ctrig$  & $htr\_ccut$ & $hcomp$  & $helec$  \\ 
\midrule
\multicolumn{10}{c}{$\omega$ Production Runs} \\
\midrule
47055 & 1300 &   82386 &  46476348 & 1108.5 &  143246000 & 39774  & 0.9646    & 0.9926 & 0.9973 \\ 
47056 & 1300 &   33544 &  18895112 &  448.5 &   58256076 & 16225  & 0.9643    & 0.9926 & 0.9973 \\ 
47057 & 4000 &   26115 &  19616695 &  475.5 &   60404320 & 16775  & 0.9676    & 0.9944 & 0.9973 \\ 
47062 & 4000 &  137852 & 103301496 & 2485.5 &  318552993 & 88744  & 0.9662    & 0.9939 & 0.9974 \\ 
\midrule
\multicolumn{10}{c}{Carbon Data Runs} \\
\midrule
47012 &  700 &  360222 & 204144117 & 1106.5 & 279010493  &     5  & 0.9655    & 0.9671 & 0.9879 \\   
47017 &  300 &  351032 &  90054730 &  608.5 & 123248274  &     2  & 0.9000    & 0.9425 & 0.9903 \\   
47018 &  200 &  369808 &  63678174 &  600.5 &  87321171  &     1  & 0.8696    & 0.9394 & 0.9931 \\   
47023 &  200 &  377130 &  58634905 &  923.5 &  80547701  &     0  & 0.8919    & 0.9607 & 0.9959 \\
\midrule
\multicolumn{10}{c}{Heep Runs} \\
\midrule
47049 &    1 & 1000087 &    882351 & 1167.5 &  38251650  &  4833  & 0.9615    & 0.9254 & 0.9999 \\   
47050 &    1 & 1002450 &    882596 & 1183.5 &  38263209  &  4915  & 0.9618    & 0.9268 & 0.9999 \\   
47051 &    1 & 1000300 &    881914 & 1152.3 &  38243611  &  4817  & 0.9624    & 0.9264 & 0.9999 \\   
47054 &  100 &  855628 &    747156 &  784.5 &  34215154  &   366  & 0.9638    & 0.9072 & 0.9999 \\   
\bottomrule
\end{tabular}

\caption[Data for the Computer and Electronic Life Time study for the SOS]{Example data for the Computer and Electronic Life Time study for the SOS. The description of the parameters are similar to Table~\ref{sample_data_HMS}.}
\label{sample_data_SOS}

\small
\setlength{\tabcolsep}{.5em}
\begin{tabular}{cccccccccc}
\toprule
Run   &   PS2 &  $atrig$ &    $spre$   &   BoT  &    sS1X    & $ctrig$ &  $str$ & $scomp$ &   $selec$ \\ 
\midrule
\multicolumn{10}{c}{$\omega$ Production Runs} \\
\midrule
47055 &  1100 &   82386  &   7931280  &  1108.5 &  53758202 &   39774 & 0.9926 &  0.9924 &  0.9991  \\  
47056 &  1100 &   33544  &   3219778  &   448.5 &  21869817 &   16225 & 0.9961 &  0.9924 &  0.9991  \\  
47057 &   750 &   26115  &   3361715  &   475.5 &  22677390 &   16775 & 0.9955 &  0.9941 &  0.9991  \\  
47059 &   750 &   31564  &   4049601  &   562.5 &  27328976 &   20347 & 0.9907 &  0.9941 &  0.9991  \\

\midrule
\multicolumn{10}{c}{Carbon Data Runs} \\
\midrule

47012 &   300 &  360222  &  23832512  &  1106.5 & 166164337 &       5 & 0.9925 &  0.9665 &  0.9982  \\  
47017 &   150 &  351032  &  10578904  &   608.5 &  73265713 &       2 & 0.9882 &  0.9413 &  0.9985  \\  
47018 &   100 &  369808  &   7408278  &   600.5 &  51820714 &       1 & 0.9890 &  0.9381 &  0.9990  \\  
47023 &    70 &  377130  &   6918378  &   923.5 &  47723650 &       0 & 0.9904 &  0.9594 &  0.9993  \\  

\midrule
\multicolumn{10}{c}{Heep Runs} \\
\midrule

47049 &     1 & 1000087  &    203454  &  1167.5 &  18730307 &    4833 & 0.9939 &  0.9261 &  1.00  \\ 
47050 &     1 & 1002450  &    203584  &  1183.5 &  18717954 &    4915 & 0.9949 &  0.9266 &  1.00  \\ 
47051 &     1 & 1000300  &    202534  &  1152.3 &  18711655 &    4817 & 0.9935 &  0.9254 &  1.00  \\ 
47052 &     1 &  556253  &    112095  &   668.5 &  10388704 &    2682 & 0.9940 &  0.9269 &  1.00  \\ 

\bottomrule
\end{tabular}

\end{table}

The probability of $n$ events occurring in an interval $\tau$ for a certain event rate $x$ can be described by the Poisson distribution:
\begin{equation}
P(n) =  (\tau x)^n \frac{e^{-\tau x}}{n!}.
\end{equation}
The probability of zero events occurring in the interval $\tau$ is thus 
\begin{equation} 
P(0) = e^{-\tau x}.
\label{eqn:eff_comp_cor}
\end{equation} 
For small $x$, the probability can be estimated as $P(0) = 1-\tau x$. In this analysis, $P(0)$ is the live time, $x$ is the event rate and $\tau$ is the time needed to process one event (computer or electronic processing time).

In this section, the studies of the CLT and electronic live time (ELT) as functions of event rate are presented. The data runs used to perform both studies include: LH$_2$ target $\omega$ production runs, carbon target luminosity runs and $e$-$p$ elastic scattering (Heep) runs. Note that carbon runs are selected to extend the event range, since Heep and $\omega$ runs have relatively low event rate. Four data run examples were selected from each data type and are listed in Table~\ref{sample_data_HMS}. The actual correction applied was determined by using the scaler information of each run. Eqn.~\ref{eqn:eff_comp_cor} was only used to check for consistency, to be sure the live time values make sense.

\subsubsection{Electronic Live Time}
\label{sec:ELT}

\begin{figure}
\centering
\subfloat[][HMS ELT]{\includegraphics[width=0.52\linewidth]{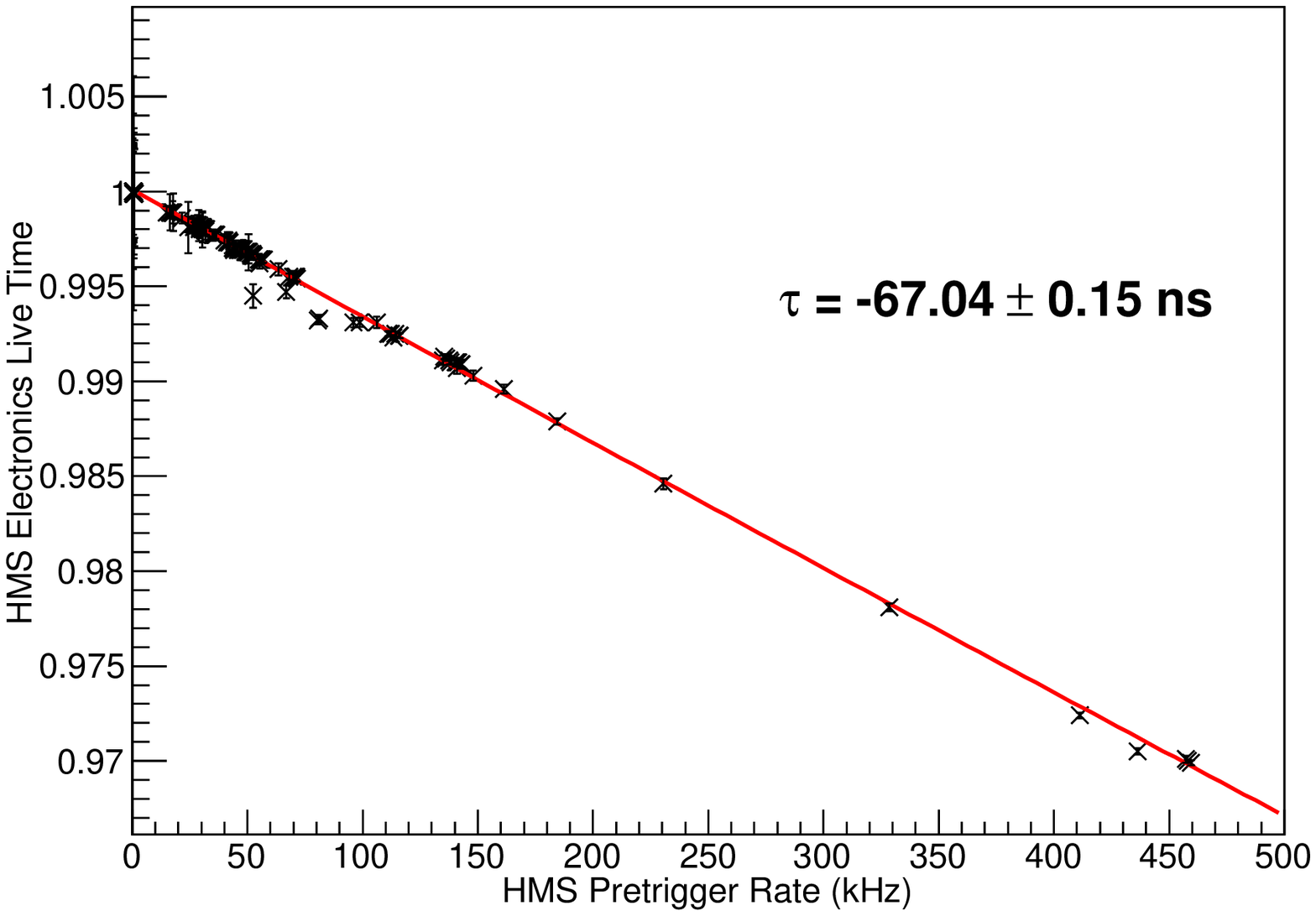}}
\subfloat[][SOS ELT]{\includegraphics[width=0.52\linewidth]{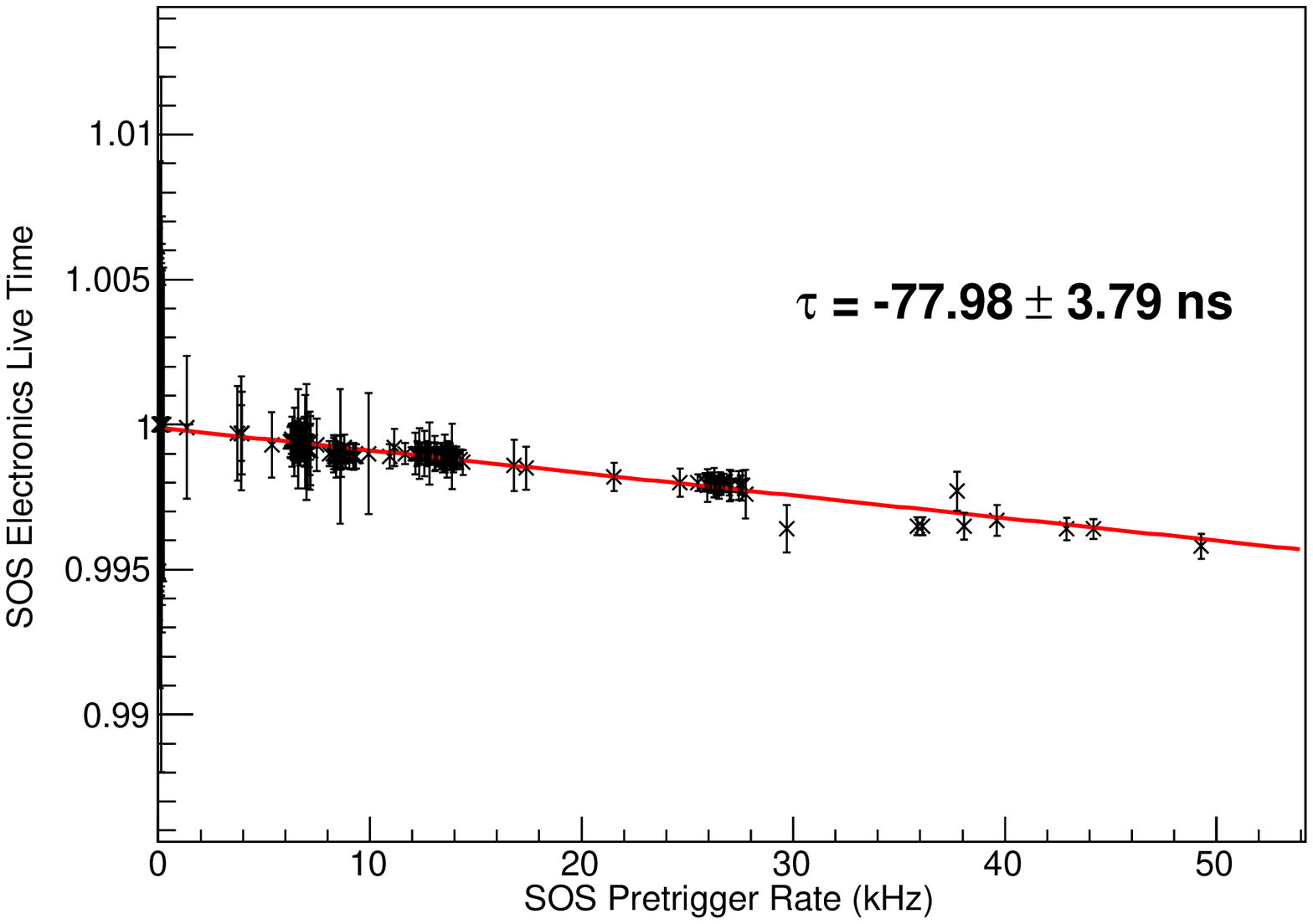}}
\caption[HMS and SOS Electronics Live Time versus pre-trigger rate]{HMS and SOS Electronics Live Time (ELT) versus pre-trigger rate. The ELT plot for HMS is on the left and SOS is on the right. The curves are fits using Eqn.~\ref{eqn:eff_comp_cor}. Note that these fitted curves are only for the better visualization of the general trend. \oic}
\label{hms_sos_elec_live_time}
\end{figure}

The electronic dead time (EDT) is normally estimated by observing the variation on pre-trigger scaler counts of various gate widths. Apart from the pre-trigger (PRE), there are PRE50, PRE100, PRE150 and PRE200 scalers corresponding to pre-trigger gate widths of 40~ns, 100~ns, 150~ns and 200~ns. Note that the PRE gate width is around 60~ns. PRE50 is intentionally set to be 40~ns to help understand the relationship between EDT and the pre-trigger gate width. 

The real number of pre-triggers is calculated as
\begin{equation}
N_{true} = N_{measured} + N_{correction},
\end{equation}
where $N_{measured}$ is the measured pre-trigger scaler counts with 60 ns gate width; $N_{correction}$ is the pre-trigger correction computed by the pre-trigger scaler counts of other gate widths, i.e. 40~ns and 100~ns. Since the EDT is expected to scale linearly with the gate width, there are a number of ways to compute the $N_{correction}$:
\begin{equation}
N_{correction} = N_{PRE50} - N_{PRE100}  = \left( \frac{N_{PRE100} - N_{PRE150} }{50~\textrm{ns}}\right) \times 60~\textrm{ns}
\end{equation}
where $N_{PRE50}$, $N_{PRE100}$ and $N_{PRE150}$ are pre-trigger scaler counts of PRE50, PRE100 and PRE150, respectively. An additional factor of 6/5 is needed as the gate width difference between PRE100 and PRE150 is only 50~ns instead of 60~ns. The chosen methodology in this analysis involves PRE100 and PRE150 to compute $N_{correction}$.

The EDT can be calculated as
\begin{equation}
EDT \approx \frac{6}{5} \times \frac {N_{PRE100} - N_{PRE150} }{ N_{PRE100} } \,,
\end{equation}
where the approximation $N_{PRE100}$ = $N_{true}$, where $N_{true}$ corresponds to the actual number of events collected during the experiment since the coincidence ggate width is set to 100~ns. Thus, ELT is given by
\begin{equation}
ELT = 1 - EDT  = 1 - \frac{6}{5} \times \frac {N_{PRE100} - N_{PRE150} }{ N_{PRE100} }\,,
\end{equation}
and its uncertainty
\begin{equation}
\delta (ELT) = \frac{6}{5} \times \frac{ \sqrt{N_{PRE100}} + \sqrt{N_{PRE150}} }{ N_{PRE100} \times ELT }\,.
\end{equation}
Note the equation of $\delta(ELT)$ assumes binomial statistics due to the fact that EDT values are very close to zero. The function form of the binomial statistics is different from the standard Poison statistics.

The HMS and SOS ELT versus the pre-trigger rate are plotted separately in Fig.~\ref{hms_sos_elec_live_time}. The fitting results suggest a time constant $\tau_{\rm HMS}~\approx$ 67 $\pm$ 0.15~ns for the HMS and $\tau_{\rm SOS}~\approx$ 77 $\pm$ 4~ns for SOS. These values differed slightly from the previously reported time constants during the F$_\pi$-2 studies: $\tau_{\rm HMS}~\approx$ 63.9~ns and $\tau_{\rm SOS}~\approx$ 72.5~ns~\cite{horn}. The differences might be contributed by two major sources: 1) Different data runs were used for this and the previous study; 2) In this study, since all data points from the $\omega$ production, Heep and carbon luminosity runs were included, therefore the number of data points was significantly more than the previous ELT study. The ELT time constants extracted from this study are accurate, since they were obtained directly from the $\omega$ production data.

\subsubsection{Computer Live Time }
\label{sec:CLT}

\begin{figure}
\centering
\includegraphics[width=0.75\linewidth]{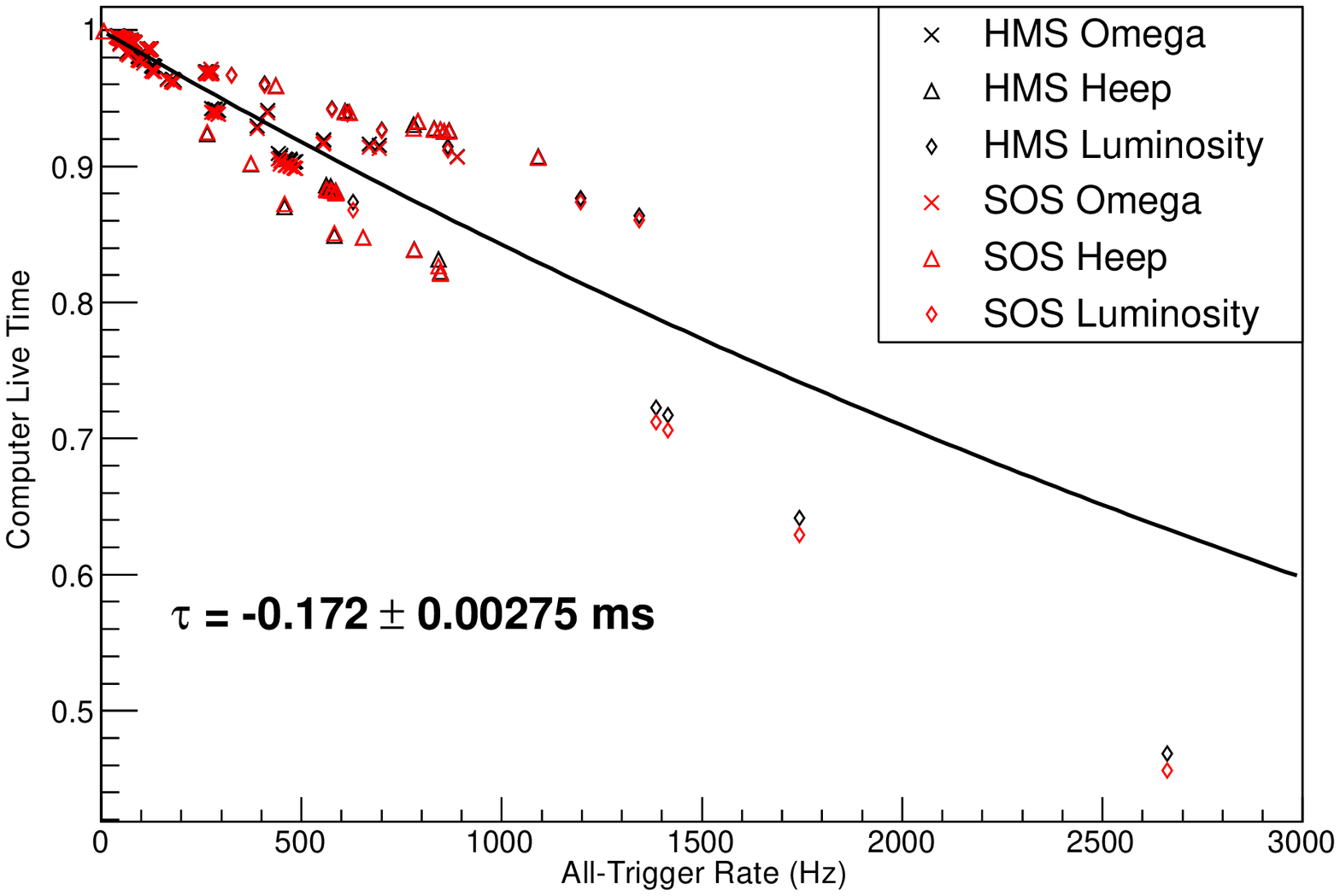}
\caption[HMS and SOS CLTs against the HMS+SOS (all) trigger rate]{HMS and SOS CLTs are plotted separately against the HMS+SOS (all) trigger rate. The two measurements should agree within the statistical uncertainties. The plot includes data points from $\omega$ production (crosses), Heep (triangles) and luminosity (diamonds) runs. The zoomed-in plots that contain the CLT from the $\omega$ data only, are separately for the HMS and SOS shown in Fig.~\ref{hms_sos_clt} (a) and (b). The data fitting curve only takes into account the data points from the $\omega$ production runs. Error bars are smaller than the plotting symbols. The fitted curve is only for the better visualization of the general trend. Note that the systematic uncertainty is between the red and black points, not between the points and the curve.~\oic}
\label{hms_sos_all}
\includegraphics[width=0.75\linewidth]{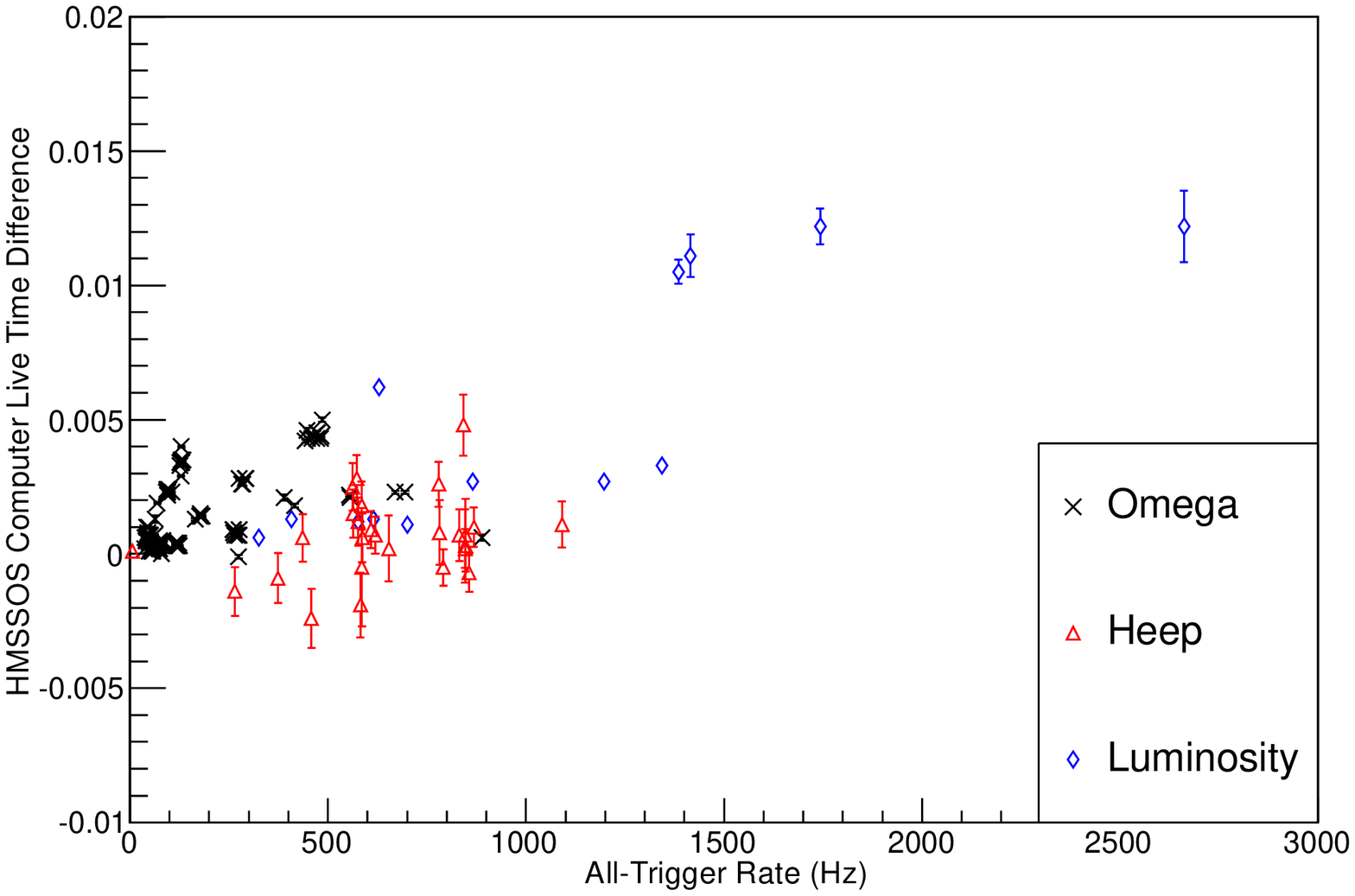}
\caption[CLT difference between HMS and SOS]{CLT difference between HMS and SOS (HMS$-$SOS from Fig.~\ref{hms_sos_all}). The plot includes data points from $\omega$ production (black crosses), Heep (red triangle) and luminosity (blue diamond) runs.~\oic}
\label{hms_sos_diff}
\end{figure}

Fig.~\ref{hms_sos_all} shows the CLT\nomenclature{CLT}{Computer Live Time} versus All Trigger Rate (ATR). All Trigger ($atrig$ from Table~\ref{sample_data_HMS}) is the total number of triggers over HMS, SOS singles and coincidences, which can be calculated as:
\begin{equation}
\rm atrig \approx \frac{hpre}{PS1}+\frac{spre}{PS2}+\frac{ctrig}{PS3} \,,
\label{eqn:atrig}
\end{equation}
where $hpre$ and $spre$ are the number of HMS and SOS single arm pre-triggers; PS1 and PS2 are the HMS and SOS singles pre-scale factors, and PS3 is the HMS+SOS coincidence pre-scale factor, which is typically set PS3=1; the ATR is calculated as
\begin{equation}
\rm ATR = \frac{atrig}{BoT}
\end{equation}
where BoT is the beam on target time.

The data points in Fig.~\ref{hms_sos_all} are from $\omega$ production, Heep and carbon target luminosity runs. The fitting curve only takes into account the data points from the $\omega$ production runs, and yields a CLT time constant of: $\tau$ = 0.172 $\pm$ 0.003~ms. It seems that the Heep data points are spread much wider in rate than those of the $\omega$ data for both HMS and SOS, and the Heep spread subsequently forms two (top and bottom) trails. It suggests that there may exist more than one effective value for the CLT constant.

Since there was only one data acquisition computer for both spectrometers, the CLT for HMS and SOS should be identical within statistical uncertainties. However, a small deviation is observed and the difference (HMS$-$SOS) is shown in Fig.~\ref{hms_sos_diff}. The difference seems to form a increasing trend as the all trigger rate increases; the CLT difference for Heep and $\omega$ is within $\pm$0.5\%.

\begin{figure}
\centering
\subfloat[][HMS CLT]{\includegraphics[width=0.5\linewidth]{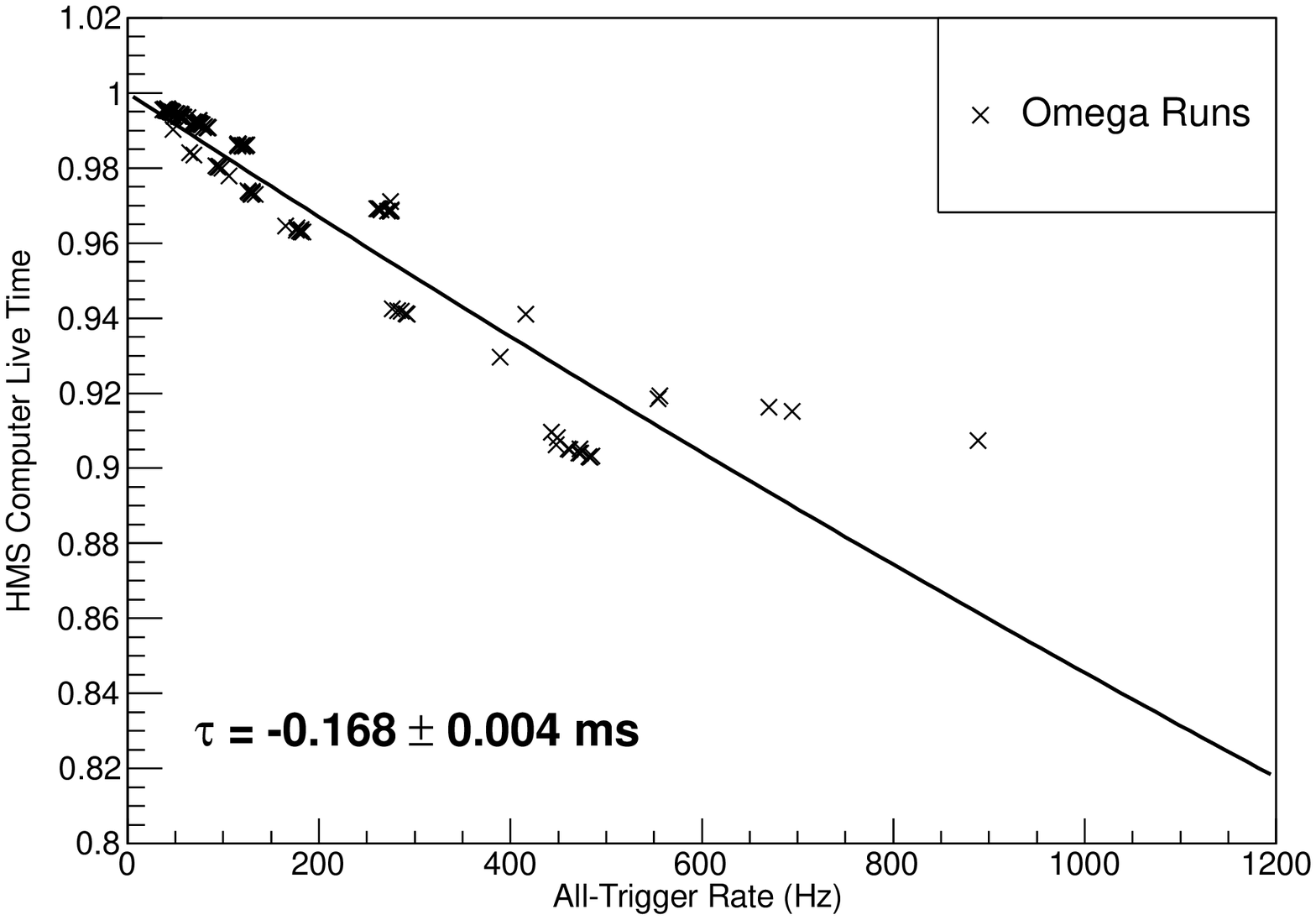}}
\subfloat[][SOS CLT]{\includegraphics[width=0.5\linewidth]{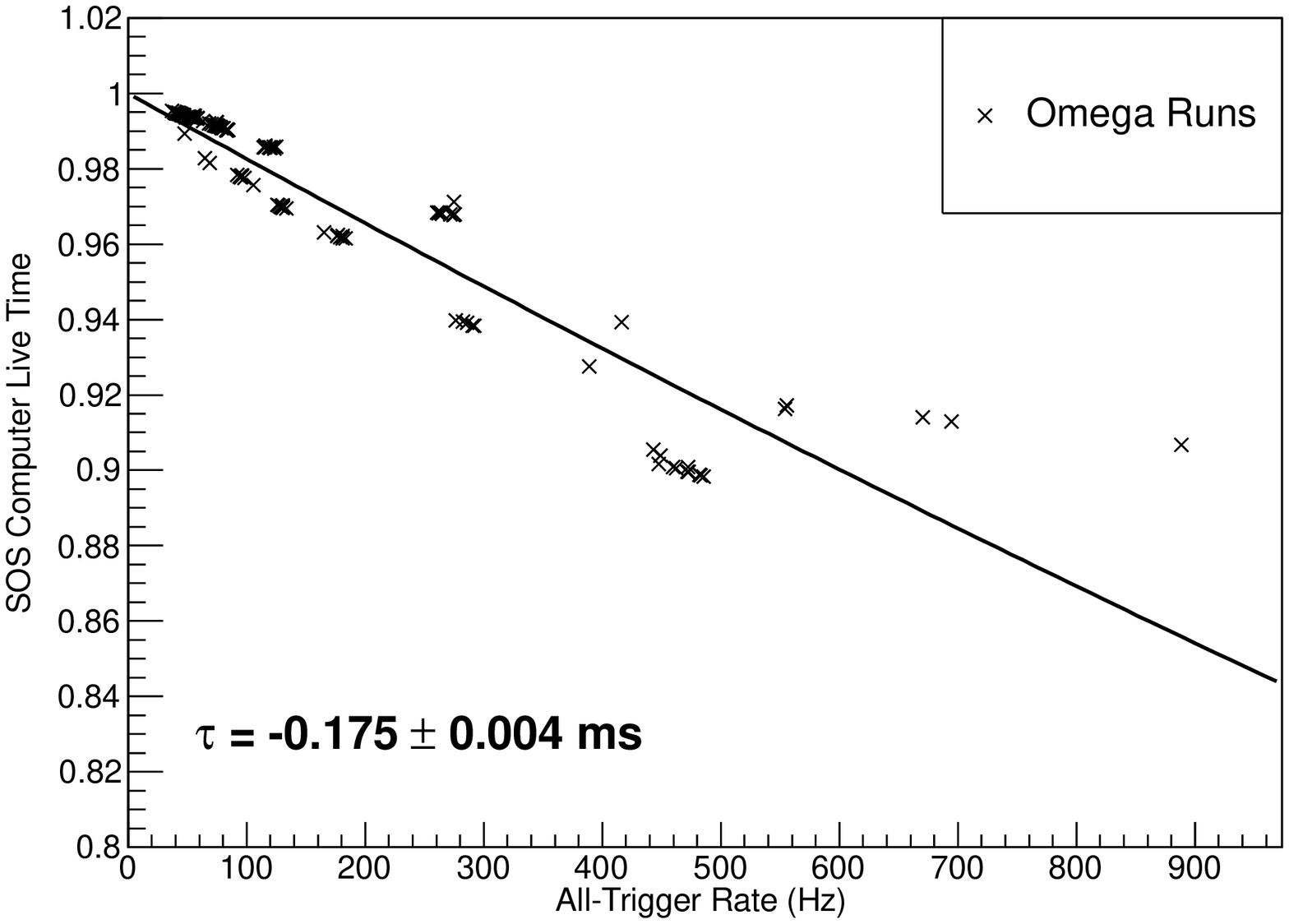}}
\caption[HMS and SOS Computer Live Time plotted versus all-trigger rate]{Figures (a) and (b) shown the HMS and SOS Computer Live Time plotted versus all-trigger rate, respectively. The data points from $\omega$ runs are used.  Error bars are smaller than the plotting symbols. The fitted curve is only for the better visualization of the general trend.}
\label{hms_sos_clt}
\end{figure}

To study further the CLT at low rate, the HMS and SOS CLT versus ATR are plotted separately in Fig.~\ref{hms_sos_clt} with Heep and $\omega$ runs. The Heep data seems to split into top-bottom trails similar to Fig.~\ref{hms_sos_all}.

Besides the small difference between HMS and SOS CLT, a difference is observed in the fitted HMS and SOS CLT time constants as shown in Fig.~\ref{hms_sos_clt}.  Note that only the $\omega$ production runs are used to extract the $\tau$ values, since these data scatter much less. The extracted HMS $\tau_{\rm HMS}$ = 0.168 $\pm$ 0.004~ms and SOS $\tau_{\rm SOS}$ =  0.175 $\pm$ 0.004~ms, where the combined HMS and SOS $\omega$ CLT time constant from Fig.~\ref{hms_sos_all} is $\tau_{all}$ = 0.172 $\pm$ 0.004~ms.

The CLT was previously reported as $\tau$ = 0.49~ms from the F$_{\pi}$-1 data analysis~\cite{volmer}, which suggests the computer processing speed for F$_\pi$-2 is 3.5 times faster than that for F$_\pi$-1, due to DAQ upgrades that occurred between F$_\pi$-1 and F$_\pi$-2.

Runs \#47141 and \#47183 are identified as bad runs. The relevant information are listed in Table~\ref{bad_data}. Both runs have large HMS pre-trigger number ($\times$10 higher than runs with similar Beam On Time) and over 90\% computer dead time.   

Note that the purpose of this analysis is to understand the trends of the CLT; the applied rate dependent correction to the experimental data used the actual efficiency value determined for each run.

\begin{table}[t]
\centering
\footnotesize
\setlength{\tabcolsep}{.23em}
\caption[Identified bad $\omega$ production runs]{Identified bad $\omega$ production runs. Both runs have very large HMS pre-trigger ($hpre$) values and more than 90\% computer dead time. The description of the parameters are the same as in Table~\ref{sample_data_HMS}.}
\begin{tabular}{ccccccccccccc}
\toprule
\multicolumn{13}{c}{$Q^2$ = 2.45~GeV$^2$, $W$ = 2.21~GeV, $E_{Beam}$ = 4.21~GeV, $\epsilon$ = 0.27} \\
\midrule
Run   &   PS1 & $atrig$ &    $hpre$    &    BoT  &    hS1X    & $ctrig$   &   $htr\_p$ & $htr\_ct$  & $htr\_ccut$ & $htr\_ccc$ &  $hcomp$ &   $helec$ \\ 
\midrule
47141 &  3000 &  150902 & 4397010598 &   3510.5&  343335387 &   84215  & 0.9536  & 0.9551 &  0.9548 &  0.9549 &  0.0231 &  0.9982 \\   
47183 &  3000 &  142913 & 4386158993 &   3212.5&  306009605 &   80627  & 0.9476  & 0.9489 &  0.9495 &  0.9486 &  0.0207 &  0.9982 \\   
\bottomrule
\end{tabular}
\label{bad_data}
\end{table}

\subsection{Spectrometer Tracking Efficiencies}
\label{sec:tr_eff}

The tracking efficiency is defined as the probability that a particle, identified as an electron or proton, is associated with a valid track from the wire chambers.

In the HMS spectrometer, an good proton event is determined if it satisfies the proton identification criterion. The HMS proton PID criterion requires signals in at least three of the four hodoscope planes (a valid SCIN signal, see Sec.~\ref{sec:trigger}), in addition to the PID information from TOF, both HGC and ACD detectors, and the calorimeter (with the fiducial and EM shower cuts). Except for the $Q^2$ = 6.52~GeV$^2$ Heep setting (proton momentum exceeds ACD threshold momentum), the proton momentum is below the threshold for generating the Cherenkov radiation in both HGC and ACD, therefore absence of signal or sub-threshold signals (from both Cherenkov detectors) are expected. 

For the electron selection in the SOS, the similar levels of information are required (such as a valid SCIN signal) by the electron identification criterion. The electrons are expected to generate an over-threshold signal in the SOS HGC, an EM shower in the calorimetry and velocity much closer to the speed of light. 

The failure of the tracking algorithm to identify a valid event can be caused by a wire chamber inefficiency, or a failure of the tracking algorithm itself. While the raw data are processed, the replay engine keeps count of the number of events for a given type with, and without, a track.

The scaler output from the data analysis engine generally includes several tracking efficiencies for different particle types at different thresholds and trigger conditions. 

Carbon target luminosity runs are excluded from this study. Since luminosity runs have few coincidence triggers, the corresponding coincidence tracking efficiencies are unavailable. Also, the HMS was configured to operate with negative polarity during the carbon target luminosity runs, therefore no proton events are expected at the wire chamber.

\subsubsection{Choice of HMS Tracking Efficiency}

\begin{figure}
\centering
\includegraphics[width=0.8\linewidth]{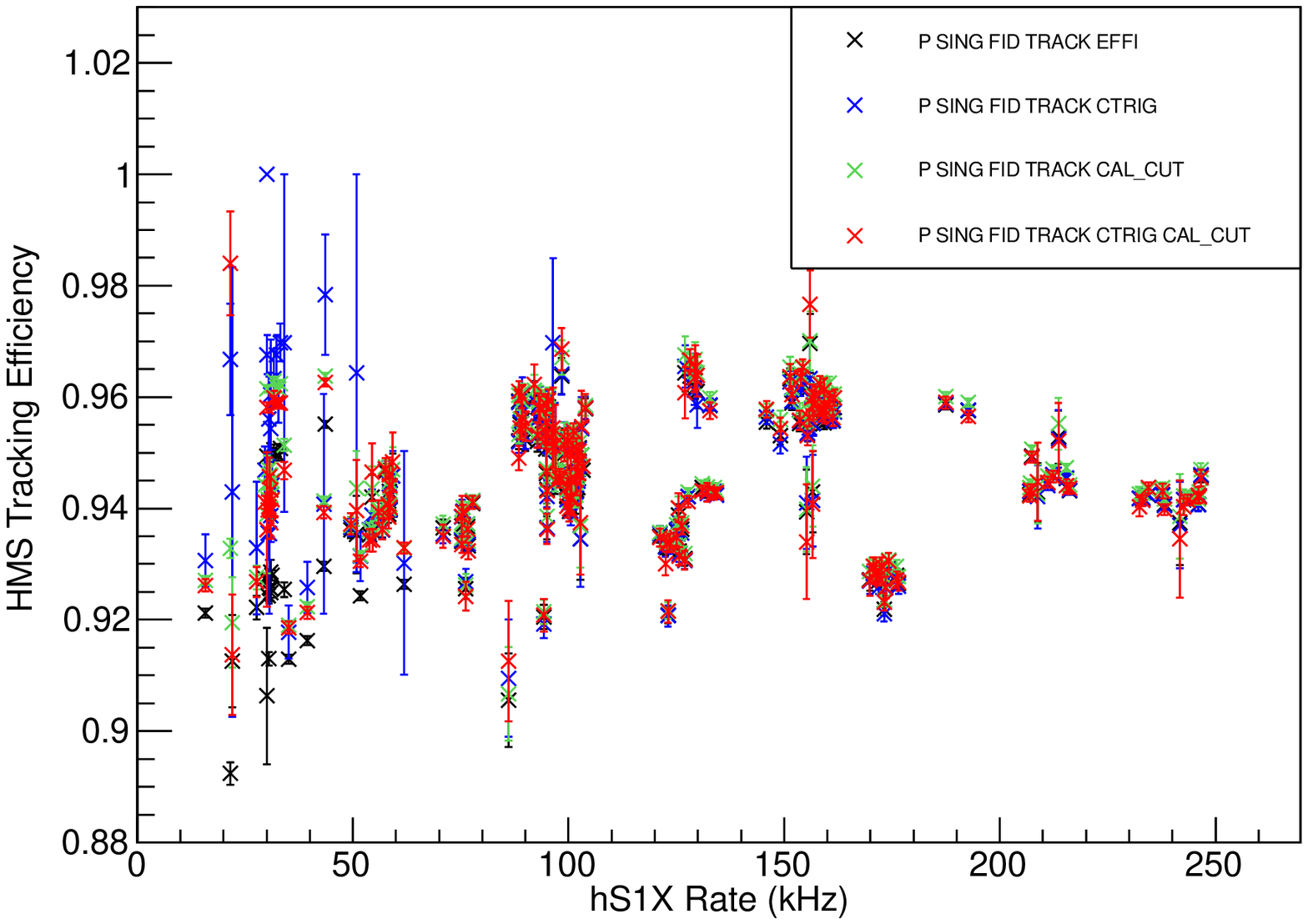}
\caption[HMS tracking efficiencies plotted as a function hS1X]{HMS tracking efficiencies plotted as a function of first hodoscope plane rate of the HMS (hS1X). All efficiencies have been corrected according to Eqn.~\ref{eqn_correction}. \oic}
\label{hms_tacking}
\includegraphics[width=0.8\linewidth]{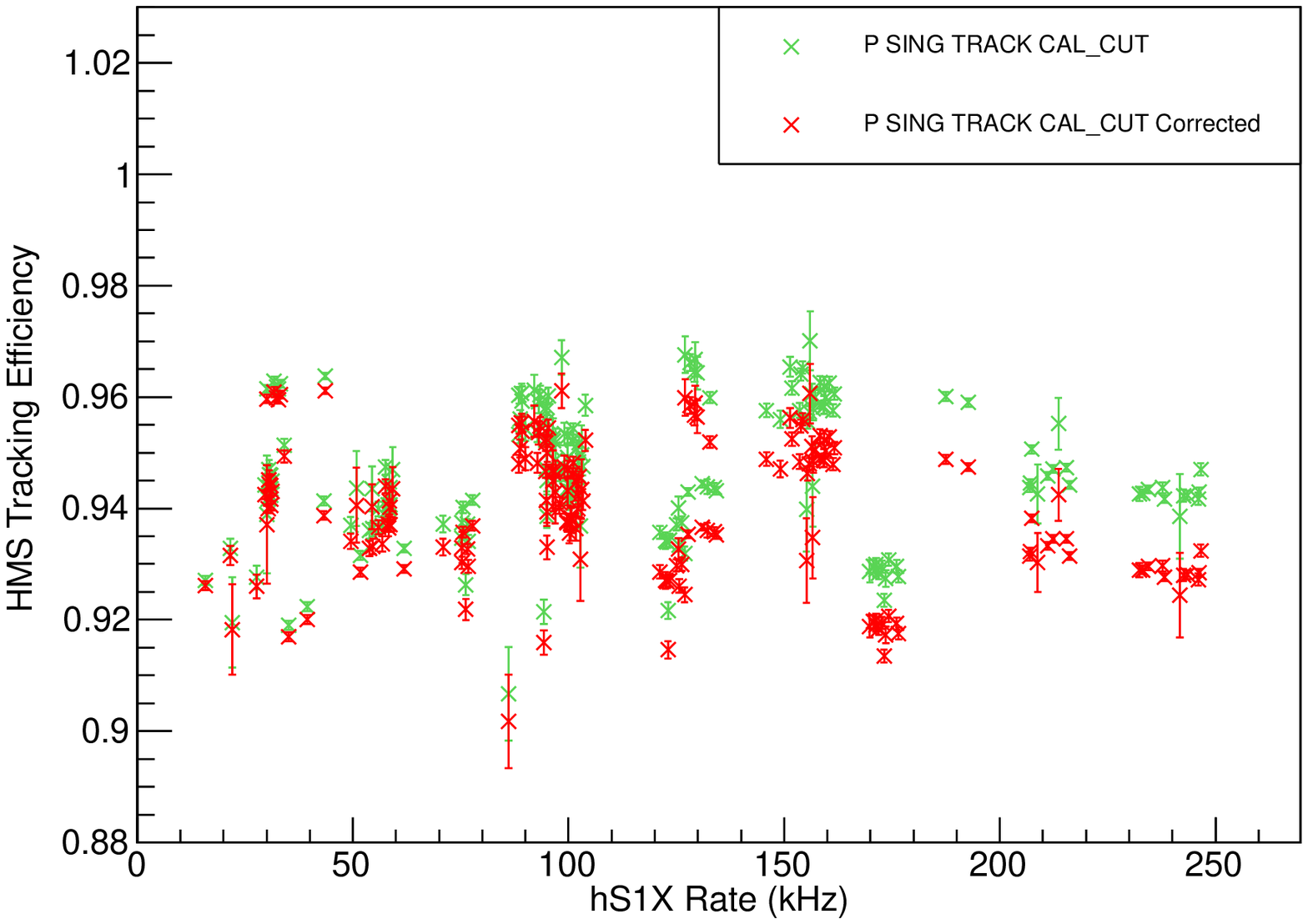}
\caption[Commparison between different sets of tracking efficiencies]{P SING FID TRACK CAL\_CUT (green) and corrected P SING FID TRACK CAL\_CUT versus first hodoscope plane rate of the HMS (hS1X). All efficiencies have been corrected according to Eqn.~\ref{eqn_correction}. \oic}
\label{hms_tacking_corrected}
\end{figure}

There were four sets of tracking efficiencies generated by the replay engine and stored in the HMS scaler files: 
\begin{description}
\item[P SING FID TRACK EFFI ($htr\_p$)] Proton tracking efficiency for HMS singles events. Protons were selected by ACD cut less than 4 pe and a HGC cut less than 0.5~pe.
  
Cut limits: $haero\_npe\_sum < 4$ \&\& $hcer\_npe\_sum < 0.5$.
 
\item[P SING FID TRACK CTRIG ($htr\_ct$)] Proton tracking efficiency for HMS+SOS singles and coincidence events. Protons were selected by the same ACD and HGC cuts. Note that, despite the name (`CTRIG'), there was no cut on the coincidence trigger.

Cut limits: $haero\_npe\_sum < 4$ \&\&  $hcer\_npe\_sum < 0.5$ \&\& $ctrig$.

\item[P SING FID TRACK CAL\_CUT ($htr\_cct$)] Proton tracking efficiency for HMS singles events. Protons were selected with the same ACD and HGC cuts, in addition to lower and upper limits on the energy deposited in the calorimeter. 

Cut limits: $haero\_npe\_sum < 4$ \&\& $hcer\_npe\_sum < 0.5$ \&\& $hcal\_et > 0.02*hpcentral$ \&\& $hcal\_et < 0.70*hpcentral$.

\item[P SING FID TRACK CTRIG CAL\_CUT ($htr\_ccc$)] Proton tracking efficiency for HMS singles. Protons were selected with the same ACD and HGC cuts, in addition to lower and upper limits on the energy deposited in the calorimeter. The cut on calorimeter is tighter than that of P SING FID TRACK CAL\_CUT.

Cut limits: $haero\_npe\_sum < 4$ \&\&  $hcer\_npe\_sum < 0.5$ \&\& $hcal\_et > (0.02*hpcentral)$ \&\& $hcal\_et < (0.125*hpcentral)$.

\end{description}
Note that a fiducial cut on the spectrometer focal plane is applied to all four sets of efficiencies to reject events which come close to the edges of the focal plane.

Fig.\ref{hms_tacking} shows all four tracking efficiencies versus the HMS S1X hodoscope plane (hS1X) rate. The differences between all sets of tracking efficiencies are surprisingly small, especially for event rate $>$ 100~kHz. The differences at low event rate are more substantial, however, the error bars are also dramatically larger. As to be explained in Sec.~\ref{sec:rate_study}, the selected HMS tracking efficiencies will be corrected using Eqn.~\ref{eqn_correction}. Fig.\ref{hms_tacking_corrected} shows the P SING FID TRACK CAL\_CUT and its corrected value.

P SING FID TRACK CAL CUT was selected for the Heep and $\omega$ analyses for the following reasons:
\begin{itemize}
\item $htr\_cct$ is the only tracking efficiency that requires a ``positive'' signal in PID detectors, thus eliminating ``junk'' hits. $htr\_ccc$ also has the same condition, but the cut appears to be too aggressive (over constraint)~\cite{gaskell16}, 
\item $htr\_cct$ gives a consistent experiment-to-simulation yield ratio of 1 using different parameterizations and has a small uncertainty shown in Fig.~\ref{hms_tacking}; this is further explained in Sec.~\ref{sec:heep_yield},
\item $htr\_cct$ was used as HMS tracking efficiency during the F$_\pi$-2-$\pi^+$ analysis.
\end{itemize}

Further information regarding on the choice of the tracking efficiency is documented in Ref.~\cite{wenliang_heep_1}.

\subsubsection{Choice of SOS Tracking Efficiency}

\begin{figure}
\centering
\includegraphics[width=0.8\linewidth]{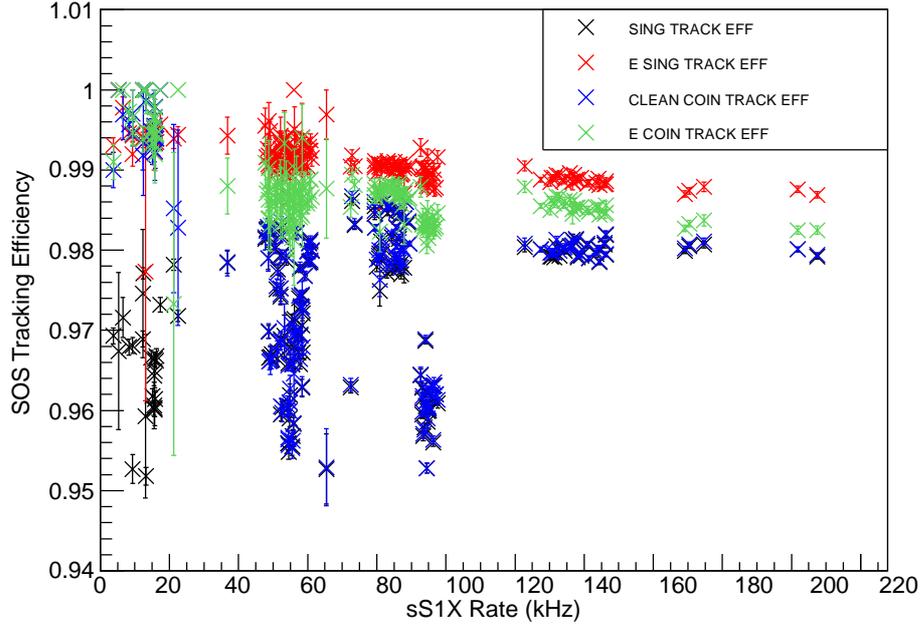}
\caption[SOS tracking efficiencies plotted as a function sS1X]{SOS tracking efficiencies plotted as a function of first hodoscope plane rate of the SOS (sS1X). \oic}
\label{sos_tacking}
\end{figure}

For electron tracking inside the SOS, there are also four tracking efficiencies from the SOS scaler files can be used to perform the data analysis as listed below: 
\begin{description}
\item[SING TRACK EFF] Tracking efficiency for all SOS singles events including pion, hadron and electron events,

\item[E SING TRACK EFF] Electron tracking efficiency for SOS singles events,

\item[CLEAN COIN TRACK EFF] Tracking efficiency for all clean HMS+SOS coincidence pion, hadron and electron events,

\item[E COIN TRACK EFF] Electron tracking efficiencies for HMS+SOS coincidence events.

\end{description}

Fig.\ref{sos_tacking} shows all four tracking efficiencies versus the SOS sS1X rate. Data points for SING TRACK EFF and CLEAN COIN TRACK EFF show a surprising amount of scatter, particularly for sS1X rate below 100 kHz. E SING TRACK EFF and E COIN TRACK EFF seem to be consistently following a linear relation with the rate. The average difference between the two sets of efficiencies is around 0.5\%. The error bars for both sets of efficiencies are around 0.2-0.4\% below 100~kHz and less than 0.2\% above 100~kHz. E SING TRACK EFF was selected for the Heep and $\omega$ analyses, since it has higher statistics. There is no additional correction applied to the SOS tracking efficiency due to its lower event rate (compared to the HMS rate) during the data taking, which implies the SOS tracking efficiency is not so sensitive to the rate.

\subsection{Carbon Target Rate Study}
\label{sec:rate_study}

In theory, physical observables (measurements) such as differential cross-section should not depend on the luminosity. In reality, detector efficiencies are often affected by high event rates which will directly influence the experimental results.

Particular to the L/T separation studies, two separate measurements are required at two different beam energies. Due to the difference in scattering cross sections and experimental acceptances, the event rate for low beam energy setting can be dramatically different from the rate for the high beam energy setting. In order to accurately correct the rate dependence in the measured experimental data, it is critical to carefully study the relationship between the overall HMS efficiency versus event rate. The standard technique is to take deep inelastic scattering measurements using a carbon target at a range of electron beam currents.

In the HMS spectrometer, the charged particle trajectories are measured by two drift chambers, each with six planes of wires, as described in Sec.~\ref{sec:drift_chamber}. Note that a ``good track'' requires 5 out of 6 wire planes to fire in each drift chamber for both spectrometers. The effectiveness of the overall tracking algorithm in the software analyzer is expected to have a rate dependence, as the detection efficiency and multiplicity drop with rate. The tracking algorithm is capable of taking into account multiple track events, which are more probable at high rate.

There are two separate methods to calculate the spectrometer tracking efficiency with the same tracking algorithm. The F$_\pi$-1 method~\cite{piminus} (which is the default method for the data analyzer) and the F$_\pi$-2-$\pi^+$ method~\cite{horn}.

Comparing to the F$_\pi$-2-$\pi^+$ method~\cite{piminus}, the F$_\pi$-1 method applies additional fiducial cuts on the scintillator planes. In the case of multiple track events, these cuts place a bias on the event sample used to calculate the HMS tracking efficiency. Since 2-track events have lower efficiencies than 1-track events, the resulting bias caused the tracking efficiencies to be overestimated. The experience from the F$_{\pi}$-1 experiment~\cite{piminus} has suggested the HMS normalized yield from carbon target (for electrons) computed using the F$_{\pi}$-1 method efficiencies fall linearly with rates, due to the presence of multiple track events in the drift chambers at high rate environment and particles hitting near the edge of the wire chambers that fail the fiducial cuts. This observation implies the tracking efficiency can not adequately correct the rate dependent effect in the tracking algorithm and an additional (linear) correction was required.

At low rate ($< 150$~kHz), the tracking efficiency computed using the F$_\pi$-2-$\pi^+$ method shows no additional dependence on rates~\cite{horn, piminus}. However, the tracking efficiency is determined to be unreliable at a high rate environment ($> 500$~kHz)~\cite{piminus} due to the looser event selection used (demonstrated during the F$_{\pi}$-2-$\pi^-$ analysis~\cite{piminus}).   

In the $\omega$ analysis, the F$_\pi$-1~\cite{volmer} method is used due to its greater reliability over a wider range of rates. Therefore, an additional rate dependent correction on the tracking efficiency study is required. A similar approach was successfully used in the F$_{\pi}$-2-$\pi^-$ analysis~\cite{piminus}.

In order to rigorously study the tracking efficiency, a study of yields from carbon target versus rate  was performed. The study requires the extraction of the inclusive experimental yields, plotted against the trigger rate of the first hodoscope plane for the HMS (hS1X). The single arm experimental yield ($N_{yield}$) is calculated as
\begin{equation}
N_{yield} = \frac{N_e \times {PS}}{Q_{e} \times EL_{lt} \times CPU_{lt} \times Tr_{eff}} \,,
\label{eqn:yield}
\end{equation}
where $N_e$ is the number of electrons obtained using a loose cut to the ntuples; $PS$ is the HMS singles pre-scale factor applied during the data acquisition; $Q_{e}$ is the accumulated electron beam charge; $EL_{lt}$ is the Electronic Live Time (ELT); $CPU_{lt}$ is the Computer Live Time (CLT); $Tr_{eff}$ is the tracking efficiency obtained using the scaler information (from the $\omega$ data replay using the F$\pi$-1 tracking efficiency computation method). Note that all scaler information except $Tr_{eff}$ came from the F$_\pi$-2-$\pi^+$ replay, since it uses the F$_\pi$-2 method to compute the tracking efficiency.

The scaler information for all the carbon luminosity runs are listed in Table~\ref{carbon_tab}. $N_e$ is listed as SING in the table. $N_e$ refers to the number of events passed the selection criteria after applying a loose cut to various parameters, and the actual cuts are listed below,

\begin{description}
\item[Event selection criteria:] $evtype < 3$ \&\& $hcer\_npe > 1$ \&\& $abs(hsdelta) < 8.5$ \\ 
\&\& $abs(hsytar) < 5$ \&\& $abs(hsxptar) < 0.08$ \&\& $abs(hsyptar) < 0.05$. 
\end{description}

\begin{table}
\centering
\caption[HMS carbon target luminosity study]{HMS carbon target luminosity study data taken during F$_\pi$-2 measurement. $Q_{tot}$ is the total accumulated beam charge in mC; hELCLEAN is the number of generated HMS ELCLEAN triggers (see Fig.~\ref{fig:trigger}); hS1X is the number of triggers from the first HMS hodoscope plane. BoT is the average of beam on time 1 \& 2 (threshold cuts: 5$\mu$A for BCM1 (Beam Current Monitor 1) and 1 $\mu$A for BCM2; PS1 is the HMS pre-scale factor; htr is the tracking efficiency; $hcomp$ is the HMS Computer Live Time (CLT); $helec$ is the HMS Electronics Live Time (ELT). All scaler information except $htr$ is from F$_\pi$-2-$\pi^+$ data replay; $htr$ is from the $\omega$ data replay using the F$\pi$-1 tracking efficiency computation method; SING refers to the number of events passed the selection criteria from the F$_\pi$-2 replay.}
\label{carbon_tab}
\small
\setlength{\tabcolsep}{.25em}
\begin{tabular}{ccccccccccc}

\toprule

Run   & \multicolumn{1}{c}{$Q_{tot}$}  &  \multicolumn{1}{c}{hELCLEAN}  &  \multicolumn{1}{c}{hS1X}  &  
		\multicolumn{1}{c}{BoT}  & \multicolumn{1}{c}{hS1X/BoT} &  \multicolumn{1}{c}{PS1}  &  
		\multicolumn{1}{c}{$htr$}  &  \multicolumn{1}{c}{$hcomp$}  &  \multicolumn{1}{c}{$helec$}  &  
		\multicolumn{1}{c}{SING} \\
\midrule
\multicolumn{11}{c}{$E_e$ = 4.210~GeV, $\theta_{\rm HMS}$=12.00$^\circ$, $P_{\rm HMS}$=$-$3.000~GeV/c} \\
\midrule
47012 & 100809 & 181614321 & 279010493 & 1106.5 & 252 kHz &  700 & 0.9797 & 0.9671 & 0.9879 & 165047 \\ 
47017 &  44330 &  80184449 & 123248274 &  608.5 & 202 kHz &  300 & 0.9810 & 0.9425 & 0.9903 & 166750 \\
47018 &  31249 &  56737629 &  87321171 &  600.5 & 145 kHz &  200 & 0.9815 & 0.9394 & 0.9931 & 176610 \\
47023 &  28692 &  52272440 &  80547701 &  923.5 &  87 kHz &  200 & 0.9834 & 0.9607 & 0.9959 & 167320 \\
\midrule
\multicolumn{11}{c}{$E_e$ = 4.702 GeV, $\theta_{\rm HMS}$=10.57$^\circ$, $P_{\rm HMS}$=$-$4.050~GeV/c} \\
\midrule
47757 &  42974 & 233316130 & 303675199 &  575.5 & 527 kHz &  500 & 0.9743 & 0.6415 & 0.9705 & 222339 \\
47759 & 124775 & 675461151 & 880596706 & 1590.5 & 553 kHz & 2000 & 0.9742 & 0.8738 & 0.9700 & 219433 \\
47760 &  19126 & 107791771 & 138777445 &  710.5 & 195 kHz &  250 & 0.9807 & 0.7171 & 0.9896 & 232082 \\
47763 &  56962 & 308207310 & 402136284 &  724.5 & 555 kHz &  750 & 0.9749 & 0.7225 & 0.9699 & 220876 \\
47764 &  29473 & 159412019 & 208067731 &  376.5 & 552 kHz &  250 & 0.9746 & 0.4685 & 0.9701 & 222754 \\
\bottomrule
\end{tabular}

\end{table}

\begin{figure}
\centering
\includegraphics[width=0.75\linewidth]{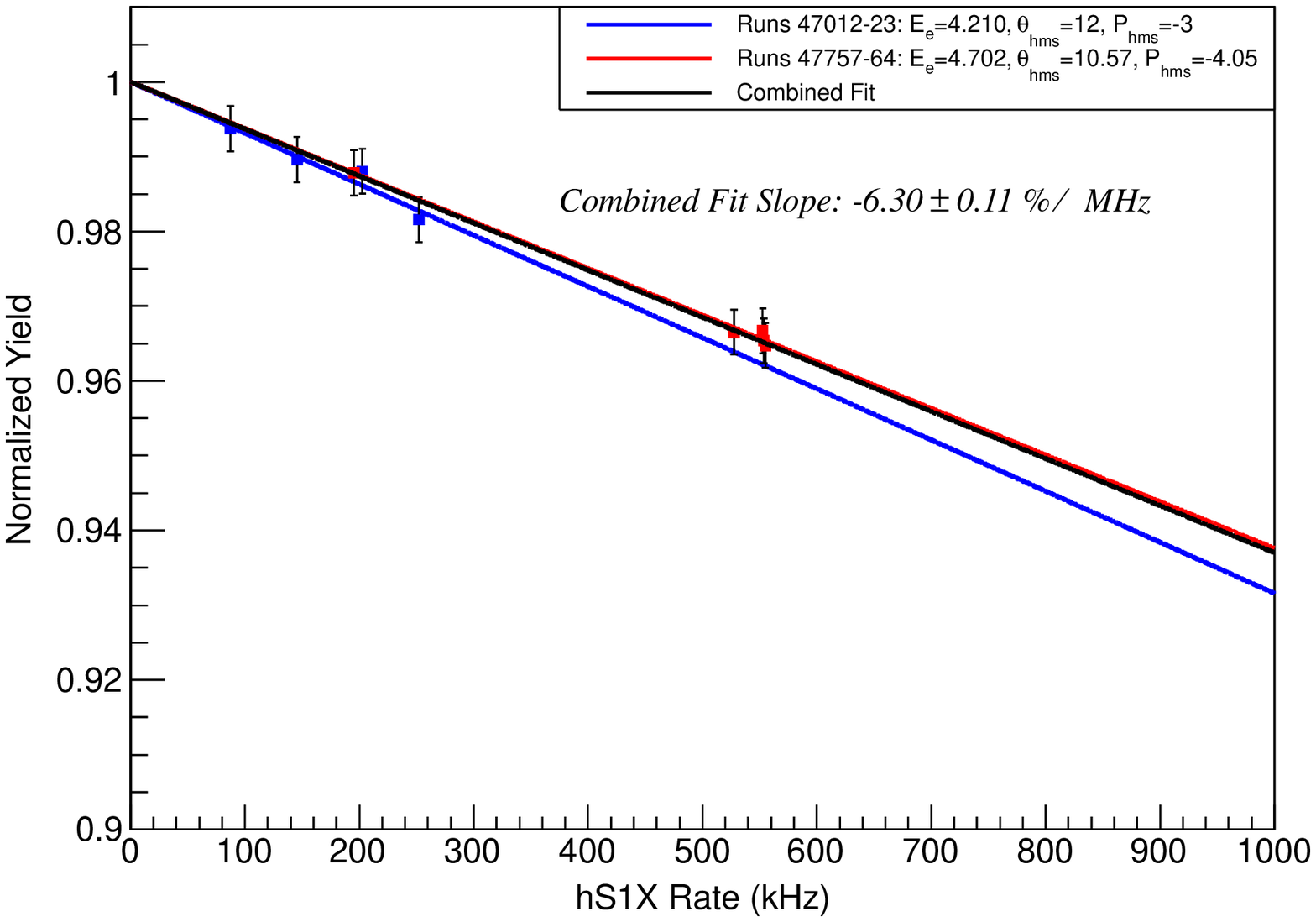}
\caption[Normalized yields from carbon target versus HMS S1X event rate]{Normalized yields from carbon target versus HMS S1X event rate. The error bars include the statistical uncertainty and an estimated systematic uncertainty of 0.3\% added in quadrature. Red and blue lines represents the fitting results from data runs of different kinematic settings as indicated in the legend. The black curve is the overall fitting result. \oic}
\label{rate_check}
\includegraphics[width=0.75\linewidth]{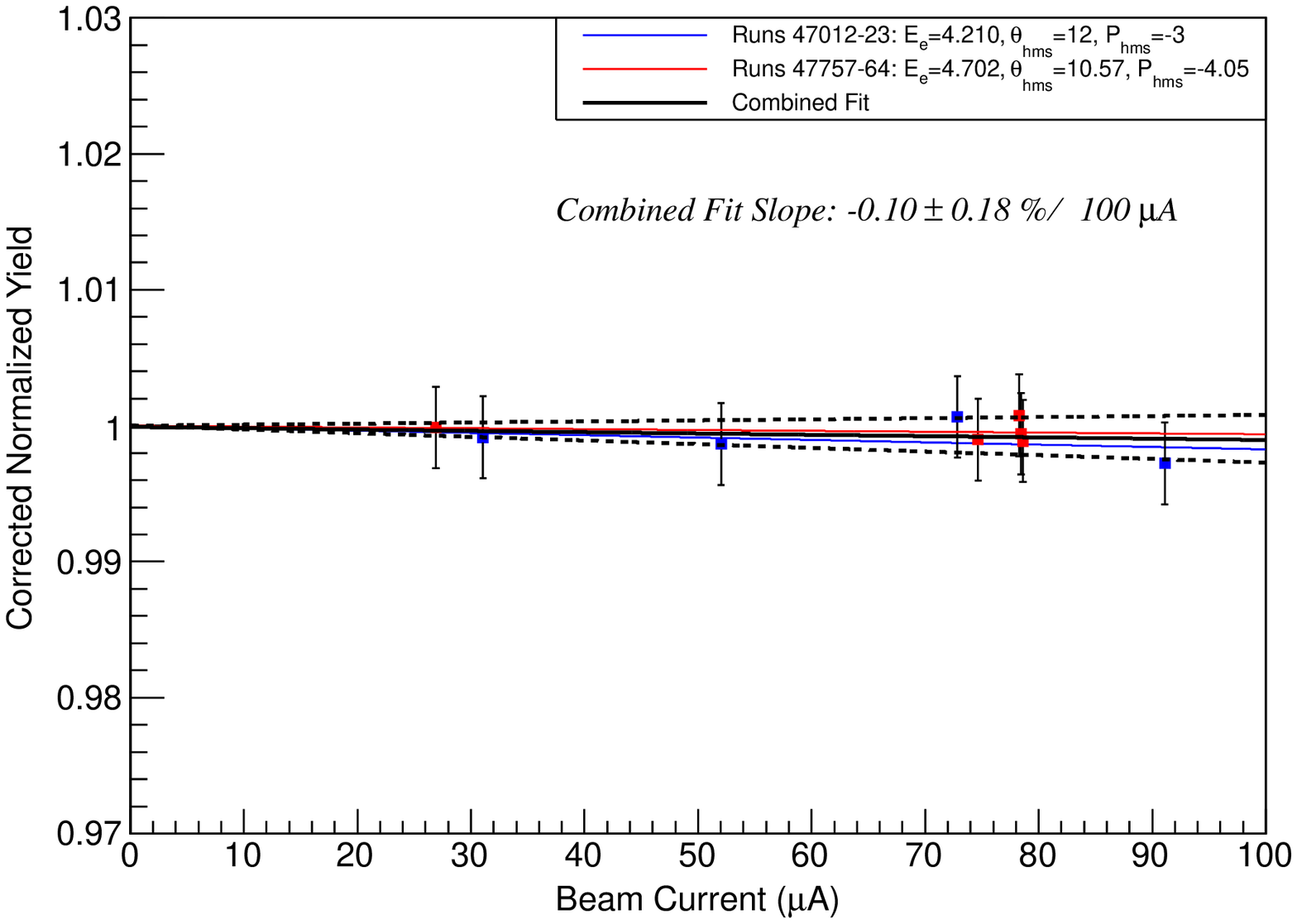}
\caption[Corrected normalized yields from the carbon target versus the beam current]{Corrected normalized yields from the carbon target versus the beam current. The correction to the HMS tracking efficiency (htr) is according to Eqn.~\ref{eqn_correction}. Red and blue lines represent the fitting results from data runs of the different kinematic settings as indicated in the legend. The black curve is the overall fitting result. Dashed lines indicate the 1$\sigma$ error band for the overall fitting result. \oic}
\label{current_check}
\end{figure}

The experimental yields were computed using Eqn.~\ref{eqn:yield} and information from Table~\ref{carbon_tab}, then normalized to unity at hS1X rate of 0~Hz. The normalized yields versus hS1X rate are plotted in Fig.~\ref{rate_check}. The error bars include the statistical uncertainty and an estimated systematic uncertainty of 0.3\% ~\cite{piminus} added in quadrature, to take into account beam steering on the target and other sensitive effects. Data from the two kinematic settings were separately fit versus rate (blue and red curves in the figure), and they are combined to yield the black curve.

The reason for the rate dependent tracking efficiency correction has been given in the earlier text, and the tracking efficiency correction as a function of rate is given by
\begin{equation}
htr\_corrected = htr \times e^{\rm -hS1X/BoT \, \times \, 6.30 \times10^{-5}/kHz } \,.
\label{eqn_correction}
\end{equation}

Fig.~\ref{current_check} shows the corrected normalized yields plotted against the beam current. The fitting result seems to be consistent with 0 within the 1$\sigma$ error band (dashed line). This confirms that the converted efficiencies now have the correct dependence on event rate, which will produce a normalized yield that is independent of luminosity.

Furthermore, carbon luminosity data runs \#47758 and \#47761 are not used for the target study since their normalized yields are dramatically far away from the fitted slope.

\subsection{LH$_{2}$ Target Boiling Study}

\label{sec:target_boil_study}

\begin{figure}
\centering
\includegraphics[width=0.8\linewidth]{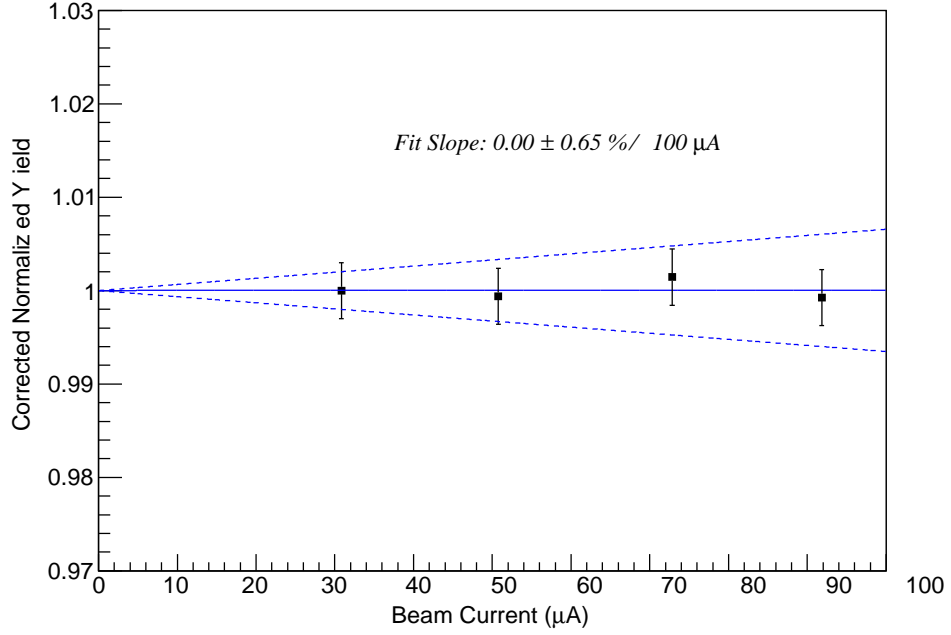}
\caption[LH$_2$ target boiling study]{Corrected normalized HMS yields from LH$_2$ target luminosity run data plotted as a function of beam current. The dashed lines indicate the 1$\sigma$ error band for the fitting result. \oic}
\label{LH2_check}
\end{figure}

When the electron beam hits the liquid cryogenic target, the energy deposit is equivalent to a 100~Watt bulb (based on the estimation from Sec.~\ref{sec:experiment_target}) across a small area. This consequently induces localized density fluctuation often referred to as ``target boiling'', more detail was given in Sec.~\ref{sec:experiment_target}. In order to minimize the target density fluctuations, the beam was rastered over an area of $2\times2$~mm$^2$, rather than being focussed to a single point on the cryotarget. The target boiling effect can be measured by comparing the yields at fixed kinematics and varied beam current. 

The F$_\pi$-2 measurement used the ``tuna can'' cryotarget geometry and circular beam raster design, which are expected to result in boiling corrections $<$ 1\% due to better flow of the cryogenic fluids. In order to make sure the LH$_2$ target boiling contributes no additional dependence to the normalized yields, the LH$_2$ target study was repeated and compared with the previous LH$_2$ target studies~\cite{piminus}. 

\begin{table}[t]
\centering
\footnotesize
\setlength{\tabcolsep}{.3em}
\caption[HMS LH$_2$ target luminosity study]{HMS LH$_2$ target luminosity study data taken during F$_\pi$-2. The variables in the table are the same as in Table~\ref{carbon_tab} with the exception of htr\_ct, which has been corrected via Eqn.~\ref{eqn_correction}.}
\label{LH2_tab}

\begin{tabular}{cccccccccccc}

\toprule

Run   & \multicolumn{1}{c}{$Q_{tot}$}  &  \multicolumn{1}{c}{hELCLEAN}  &  \multicolumn{1}{c}{hS1X}  &  
		\multicolumn{1}{c}{BoT}  & \multicolumn{1}{c}{hS1X/BoT} &  \multicolumn{1}{c}{PS1}  &  
		\multicolumn{1}{c}{$htr$}  & \multicolumn{1}{c}{$htr\_ct$}      &  \multicolumn{1}{c}{$hcomp$}  &  
		\multicolumn{1}{c}{$helec$}  &  \multicolumn{1}{c}{SING} \\
\midrule
\multicolumn{11}{c}{$E_e$ = 4.210~GeV, $\theta_{\rm HMS}$=12.00$^\circ$, $P_{\rm HMS}$=$-$3.000~GeV/c} 	   \\
\midrule
47010 & 55703 & 219259338 & 346569464 & 606.5 & 571kHz & 700.0 & 0.9738 & 0.9394 & 0.9268 & 0.9724 & 183051 \\
47014 & 37921 & 150584896 & 238130720 & 520.5 & 457kHz & 300.0 & 0.9762 & 0.9485 & 0.8765 & 0.9781 & 279893 \\
47019 & 15462 &  61940123 &  98109699 & 304.5 & 322kHz & 200.0 & 0.9792 & 0.9595 & 0.8638 & 0.9846 & 171455 \\
47022 & 13647 &  55126254 &  87477847 & 442.5 & 197kHz & 200.0 & 0.9812 & 0.9691 & 0.9145 & 0.9907 & 162904 \\
\bottomrule
\end{tabular}

\end{table}

The information for the LH$_2$ target study runs are listed in Table~\ref{LH2_tab}. The cuts of
\begin{description}
\item[Event selection criteria:] $hcer\_npe > 1$ \&\& $abs(hsdelta) < 8.0$ \&\& $abs(hsxptar) < 0.09$ \&\& $abs(hsyptar) < 0.055$,
\end{description}
are applied to the data Ntuples for each of these runs to extract the number of events that passed the selection selection criteria (SING). 

The experimental yields were calculated using Eqns.~\ref{eqn:yield} and \ref{eqn_correction}. The normalized yields are plotted versus current in Fig.~\ref{LH2_check}. The error bars include statistical uncertainties and an estimated systematic uncertainty of 0.3\% is added in quadrature. The fitting curve is consistent with 0 across the measured beam current range, thus confirming no additional correction is needed for the effect of target boiling. This is consistent with the study presented in Ref.~\cite{horn}.

\subsection{SOS Coincidence Blocking}
\label{sec:coin_blocking}

\begin{figure}[t]
\centering
\subfloat[][Heep data]{\includegraphics[width=0.52\linewidth]{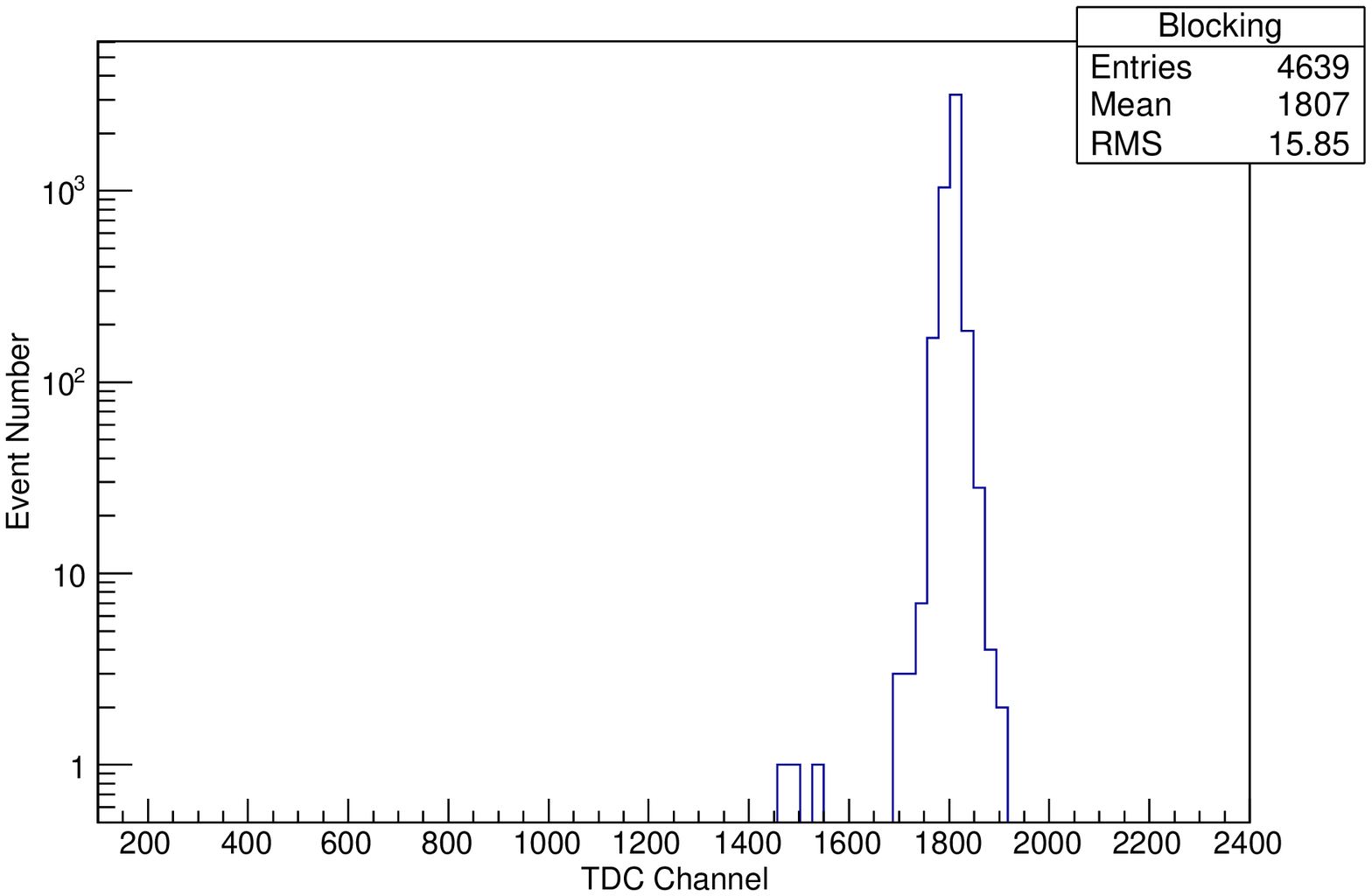}}
\subfloat[][$\omega$ data]{\includegraphics[width=0.52\linewidth]{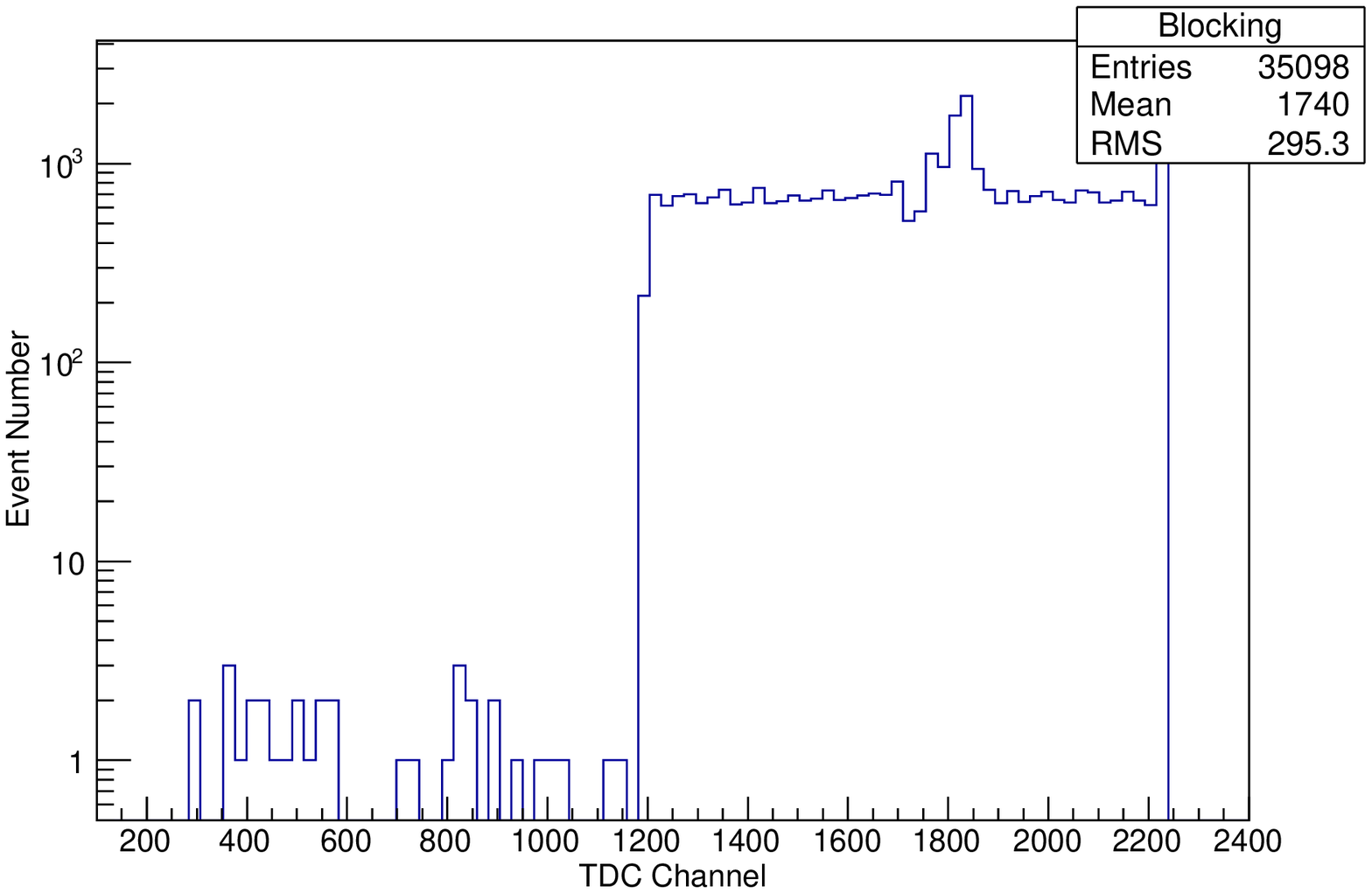}}
\caption[Uncorrected coincidence time spectra for Heep data]{Uncorrected coincidence time spectra for Heep data on the left and $\omega$ data on the right. Note the spectrometer acceptance and PID cuts are applied.}
\label{fig:uncorrected_coin_time}
\end{figure}

A coincidence event will normally be started at the TDC with a delayed HMS trigger and stopped by the SOS trigger. Due to interference between random coincidence and real coincidence events, a fraction of events are recorded with the coincidence time outside the main timing window, as defined by the pre-trigger signal width. The ``coincidence blocking'' events will be lost from the data due to the coincidence time cuts used in the analysis, therefore a study is needed to correct for data loss. The $\omega$ and Heep data are used for this study.

Examples for the coincidence time spectra of $\omega$ and Heep data runs are shown in Fig.~\ref{fig:uncorrected_coin_time}. The main coincidence window corresponds to the region between TDC channel 1186 to 2240 for $\omega$ runs and 1756 to 1910 for Heep runs. The conversion between picoseconds and TDC channels is approximately 120~ps per channel. The events left of the main timing windows are the coincidence blocking events due to the SOS singles triggers arriving earlier than the SOS coincidence trigger. Thus, the TDC is stopped too early, and the resulting events fall outside of the main coincidence time window (early triggers). Note that due to the trigger setup (started by the HMS pre-trigger and stop by the SOS pre-trigger), there is no early HMS event therefore no need for the coincidence correction.

\begin{figure}
\centering
\includegraphics[width=0.8\linewidth]{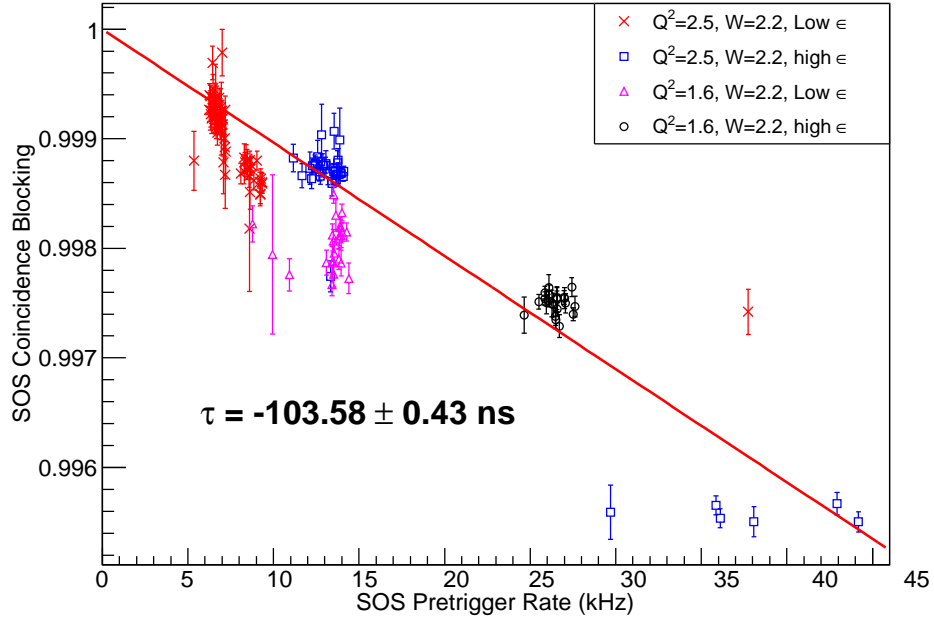}
\caption[The SOS coincidence blocking correction study]{The SOS coincidence blocking correction due to SOS early triggers as a function of pre-trigger rate. Eqn.~\ref{eqn:blocking_fit} is used to fit the data, and the fitted curve is only for the better visualization of the general trend. \oic} 
\label{fig:blocking}
\end{figure}

The coincidence blocking correction can be estimated from the rate dependence of the number of blocked events, similar to the deadtime correction in Sec.~\ref{sec:live_time}. The comparison of the number of events outside of the main coincidence time window, and the total number of events, yields the coincidence blocking rate:
\begin{equation}
Coin~block~rate = \frac{N_{block}}{N_{total}} \,,
\end{equation}
where $N_{block}$ is the number of ``early'' SOS events (triggers) in the measured coincidence time spectrum and the $N_{total}$ is the total number of events independent of the coincidence time. The good coincidence rate takes into account events within the main coincidence time window, and can be written as a function of the coincidence blocking constant $\tau_{b}$,
\begin{equation}
Good~coin~rate = \frac{N_{good}}{N_{total}} = 1 - Coin~block~rate = e^{-\tau_{b} x},
\label{eqn:blocking_fit}
\end{equation}
where $x$ is the SOS pre-trigger rate. Binomial statistics are used to calculate the uncertainty for the good coincidence rate~\cite{paterno}
\begin{equation}
\delta(Good~coin~rate) = \frac{1}{N_{total}} \sqrt{N_{good}\left(1-\frac{N_{good}}{N_{total}}\right)}.
\end{equation}

Fig.~\ref{fig:blocking} shows the SOS blocking correction plot, where the good coincidence rate is plotted as a function of SOS pre-trigger rate. From the fitting result, the coincidence blocking time constant ($\tau_b$) in Eqn.~\ref{eqn:blocking_fit} is determined, $\tau_{\rm SOS} \approx$ 103.58$\pm$0.43~ns.

From the F$_{\pi}$-2-$\pi^+$ analysis~\cite{horn}, the $\tau_{b}$ constant was reported as $\tau_{b} = 92$~ns~\cite{horn}. The difference between two $\tau_{b}$ values, is due to different data set used for each studies. In the $\omega$ (this) analysis, the study only includes the $\omega$ data set which is a small subset of the F$_{\pi}$-2-$\pi^+$ data. The previous coincidence blocking study included the whole data set. 

This Eqn.~\ref{eqn:blocking_fit} with newly determined $\tau_{b}$ constant was applied to the data as the Cherenkov coincidence correction.

\subsection{HMS Aerogel Cherenkov Detector Threshold Cuts}
\label{sec:aero}

\begin{table}
\centering
\renewcommand{\arraystretch}{1.4}
\caption[Proton interaction correction study]{Proton interaction correction study for four Heep and two $\omega$ settings. $P_{\rm HMS}$ is the HMS central momentum; $f_1$ and $f_2$ are defined in the text of relevant section.}
\label{aerogel_tab}

\begin{tabular}{ccccccc}
\toprule
$Q^2$   & $P_{\rm HMS}$ & $f_1$  & $f_2$ &  Inter. Cor. & Data Set  & haero\_su Cut         \\       
GeV$^2$ & GeV/c         &  \%    &  \%   &   \%         &           & pe                   \\ \hline
2.41    & 3.44          & 54.14  &  80   &  4.58        &  Heep     & $-50<haero\_su<95$    \\ 
4.42    & 3.15          & 50.82  &  80   &  4.36        &  Heep     & $haero\_su<29$        \\
5.42    & 3.76          & 56.82  &  80   &  4.75        &  Heep     & $haero\_su<29$        \\
6.53    & 4.34          & 62.89  &  80   &  5.15        &  Heep     & $haero\_su<29$        \\ \hline
1.60    & 2.93          & 37.77  &  80   &  3.51        &  $\omega$ & $-2.5<haero\_su<2.5$  \\
2.45    & 3.33          & 70.65  &  80   &  5.65        &  $\omega$ & $-2.5<haero\_su<2.5$  \\ 
\bottomrule
\end{tabular}

\end{table}

\begin{figure}[t]
\centering
\subfloat[][$Q^2=2.41$ GeV$^2$, $P_{\rm HMS}=3.44$~GeV/c]{\includegraphics[width=0.52\textwidth]{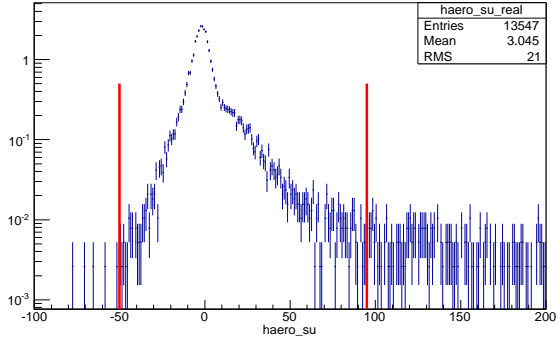}}
\subfloat[][$Q^2=4.42$ GeV$^2$, $P_{\rm HMS}=3.15$~GeV/c]{\includegraphics[width=0.52\textwidth]{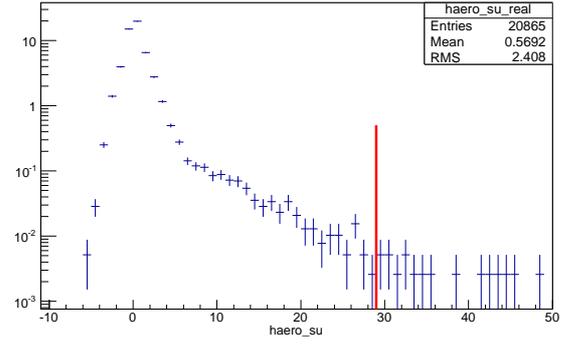}} \\
\subfloat[][$Q^2=5.42$ GeV$^2$, $P_{\rm HMS}=3.76$~GeV/c]{\includegraphics[width=0.52\textwidth]{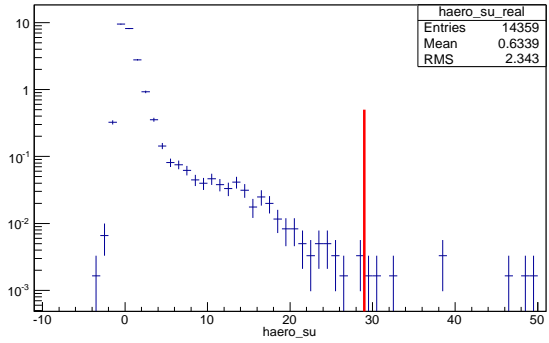}}
\subfloat[][$Q^2$ = 6.53 GeV$^2$, $P_{\rm HMS}$ = 4.34~GeV/c]{\includegraphics[width=0.52\textwidth]{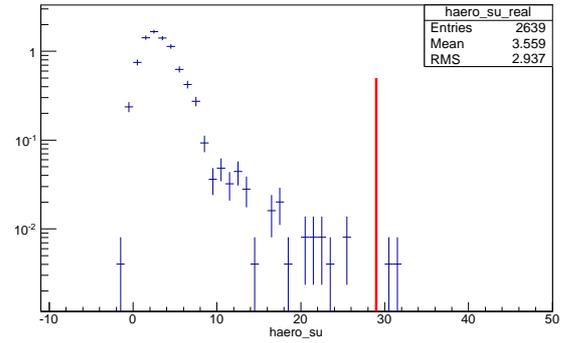}}
\caption[ACD distributions for four Heep settings]{ACD ($haero\_su$) distributions for four Heep settings. $Q^2$ values and HMS central momentum ($P_{\rm HMS}$) are listed under each plot. Acceptance, Partial PID cuts, $hsbeta$-coincidence, missing mass and missing energy cuts are applied. Random coincidence is also subtracted.  Note that the Cherenkov radiation threshold momentum for a proton inside the ACD with an index of refraction $n$=1.030 is 3.80~GeV/c. \oic}
\label{fig:aerogel}
\end{figure}

During the F$_\pi$-2 experiment, the primary objective of the ACD was to perform a clean $\pi/p$ separation; see Sec.~\ref{sec:ACD} for further detail regarding the ACD.

Table~\ref{aerogel_tab} shows the HMS central momentum values ($P_{\rm HMS}$), ACD cuts and other relevant information for four Heep settings and two $\omega$ settings. For the Heep data,  ACD cuts were used to exclude events beyond the applied threshold. For the $\omega$ data, ACD cuts were used to ensure selections of clean coincidence proton events, while events beyond the cuts were corrected by an ACD cut efficiency factor (described in Sec.~\ref{sec:aero_eff}). 

The HMS ACD cuts for the Heep data were determined using the HMS ACD ($haero\_su$) distributions (logarithmic scale), shown in Fig.~\ref{fig:aerogel}. The red boundary lines were drawn as the distributions started to plateau. 

The $Q^2$=2.41~GeV$^2$ Heep setting is a sub-threshold HMS central momentum setting, however, its ACD distribution (see Fig.\ref{fig:aerogel}(a)) is unusually wide, which may be due to mis-calibration; an  $haero\_su$ cut of $-50<haero\_su<95$ is applied. The $Q^2$ = 6.53~GeV$^2$ Heep setting has a HMS central momentum of 4.34~GeV/c that is above the Cherenkov radiation threshold momentum for a proton, and its ACD distribution seems to allow the same cut of $haero\_su<29$ as the other two well-calibrated sub-threshold settings. 

\subsection{HMS ACD Cut Study for the $\omega$ Analysis}
\label{sec:aero_eff}

\begin{figure}[t]
\centering
\subfloat[][Without $hsbeta$-$cointime$ cut]{\includegraphics[width=0.52\textwidth]{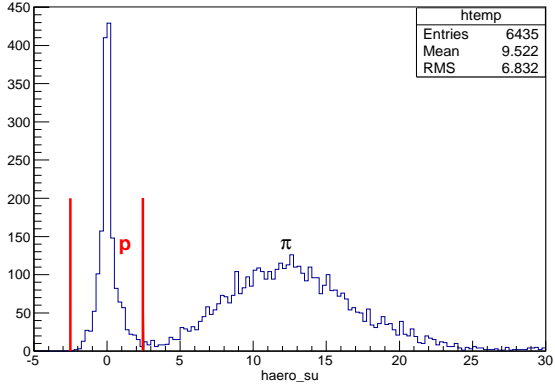}}
\subfloat[][With $hsbeta$-$cointime$ cut]{\includegraphics[width=0.52\textwidth]{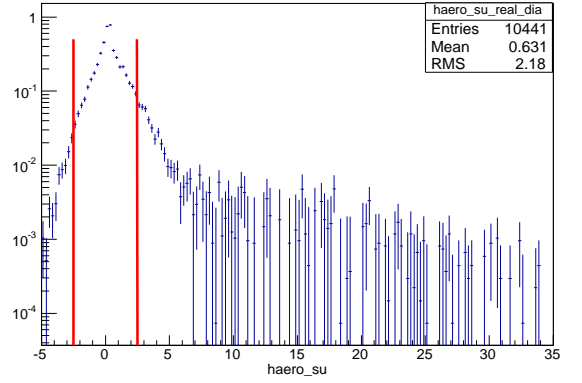}}
\caption[Example HMS Aerogel Cherenkov detector signal distributions]{(a) and (b) show example HMS Aerogel Cherenkov detector signal distributions ($\omega$ data $P_{\rm HMS}=2.93$~GeV/c) without and with $hsbeta$-$cointime$ cut, respectively; (b) is normalized and random coincidence subtracted. The units of both distributions are in Number of detected pe. Both distributions have the same acceptance and Partial PID cuts. The contribution from the random coincidence is not removed in (a). \oic}
\label{fig:haero_omega}
\end{figure}

\begin{figure}[t!]
\centering
\subfloat[][$Q^2=4.42$~GeV$^2$]{\includegraphics[width=0.52\textwidth]{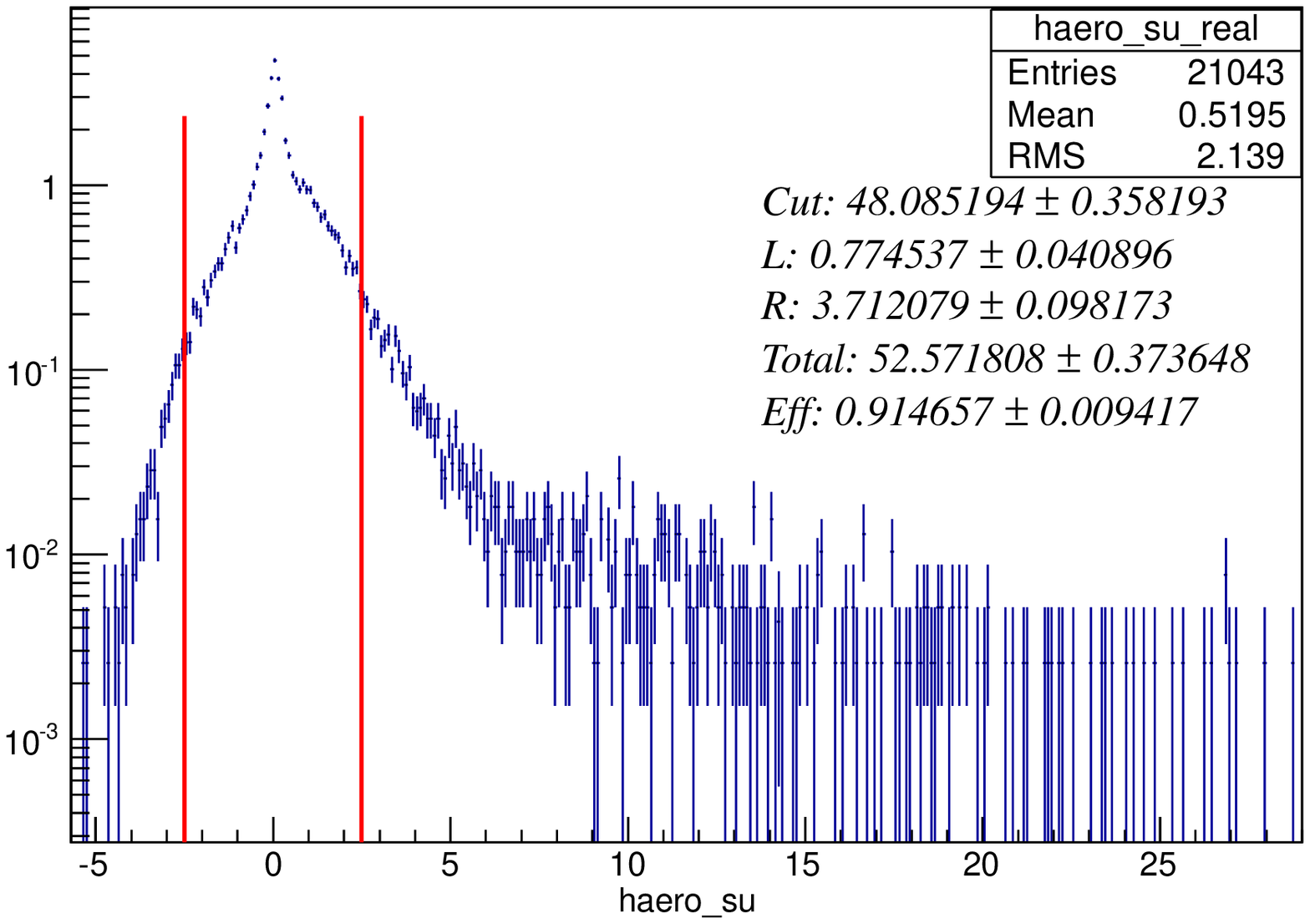}}
\subfloat[][$Q^2=5.42$~GeV$^2$]{\includegraphics[width=0.52\textwidth]{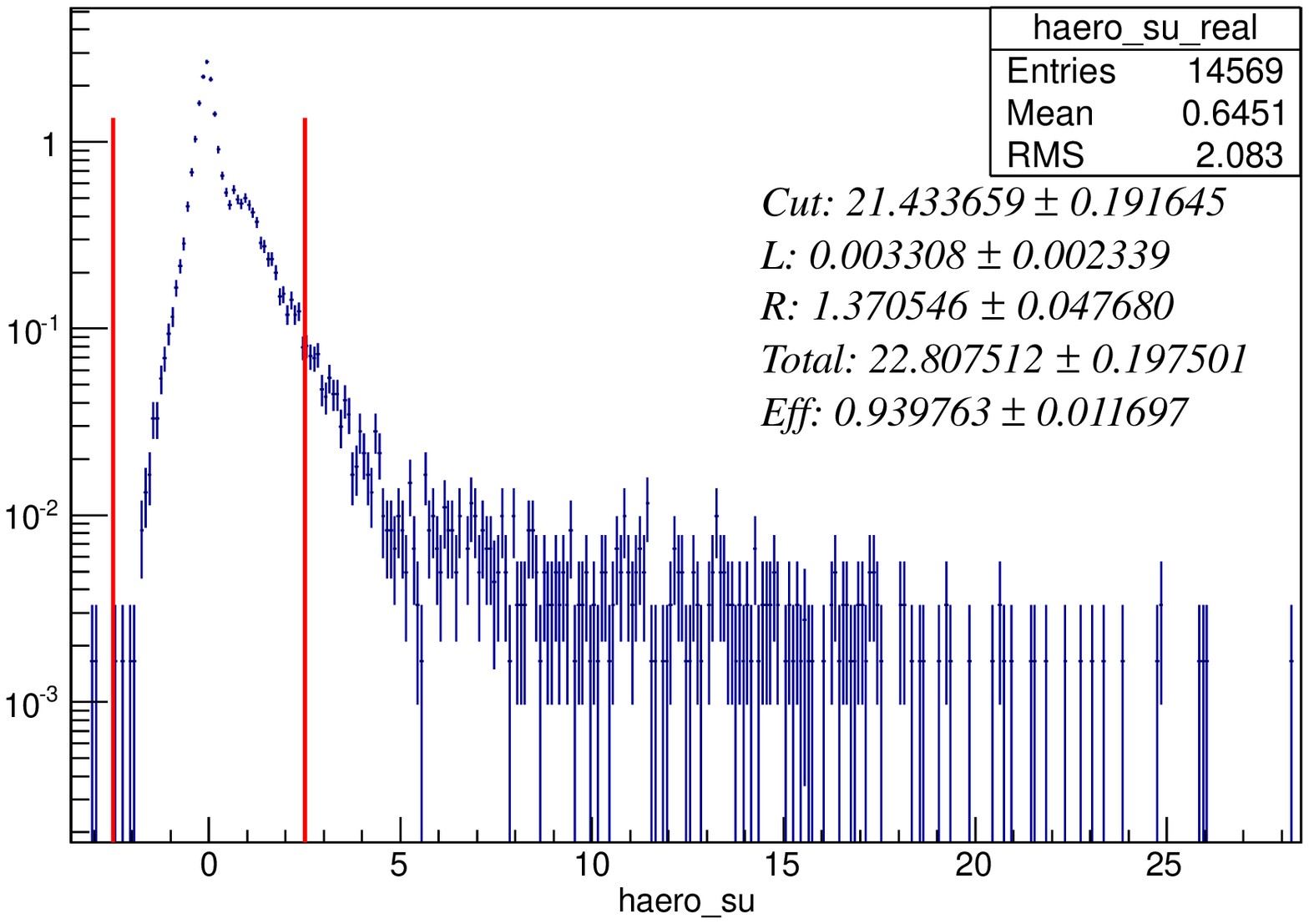}}
\caption[$hsbeta$ distributions for Heep setting $Q^2$=4.42 and 5.42~GeV$^2$]{$haero\_su$ distributions for Heep setting $Q^2$=4.42 and 5.42~GeV$^2$. The red lines indicate the boundary of $|hsaero\_su|< 2.5$. The averaged efficiency for HMS ACD cut of $|hsaero\_su| < 2.5$ is 92.7$\pm$1.2\%. Acceptance, Partial PID, $hsbeta$-$cointime$, missing mass and missing energy cut is applied. The contribution of the random coincidence is subtracted. \oic}
\label{fig:haero_eff}
\end{figure}

Unlike the Heep runs, the $\omega$ data have a much higher rate of pion contamination in the coincidence trigger that requires both the $hsbeta$-$cointime$ cut and the HMS ACD cut ($|haero\_su|<2.5$) to cleanly select the proton coincidence events.

The $hsbeta$-$cointime$ distribution for Heep data in Fig.~\ref{fig:p_abs_hebeta}(b) shows only the prompt proton coincidence bunch with scattered random coincidence events. The $hsbeta$-$cointime$ distribution for the $\omega$ data looks dramatically different, where the separate $p$ and $\pi$ coincidence bunches are visible (described in Sec.~\ref{sec:cointime_omega}). Since $\pi$ have a higher velocity due to its lighter mass (at the nominal HMS momentum), it arrives the at HMS detector package (hodoscopes) 4-5~ns earlier. Thus, by applying the $hsbeta$-$cointime$ cut, the main proton bunch can be separated from the pions.

Fig.~\ref{fig:haero_omega}(a) and (b) show the $haero\_su$ distributions without, and with, the $hsbeta$-$cointime$ cut, respectively. The red boundary indicates the HMS ACD cut of $|haero\_su|<2.5$~pe. Both distributions have the same acceptance and PID cuts. (b) is normalized (to 1~mC of beam charge) and random coincidence contribution has been subtracted, where (a) is not. Fig.~\ref{fig:haero_omega}(b) is visually clean, 90.9\% the events are within the cut region (red boundary). Beyond the $haero\_su>2.5$~pe limit, the tail contains predominantly proton events with a small $\pi$ contamination in the tail region ($haero\_su>5$~pe), since the spectrometer setting is optimized for $\pi$ detection. The level of pion contamination is difficult to estimate, since coincidence proton events beyond the haero\_su cut cannot be accurately counted.

As indicated in Sec.~\ref{sec:aero}, the Heep data have much less pion contamination than the $\omega$ data. Thus, two sub-threshold (HMS ACD) Heep settings were used to estimate the proton coincidence events beyond the $haero\_su$ cut ($|haero\_su|>2.5$~pe). The ACD distributions of Heep settings of $Q^2$=4.42 and 5.42~GeV$^2$ are plotted in Fig.~\ref{fig:haero_eff}(a) and (b). The $haero\_su$ cut efficiency is the ratio between events within the red boundary and the total. The averaged efficiency between the two Heep settings is 92.7$\pm$1.2\%; the normalized experimental yield will be divided by this efficiency for the $\omega$ analysis. Since very few coincidence proton events from the dummy target are able to survive the Heep analysis cuts, the contributions from the Dummy target runs are negligible.

For the pion contamination within the $haero\_su$ cut boundary, the random subtraction process is sufficient assuming the random proton and $\pi$ contamination is identical for each bunch; therefore, no additional correction is necessary.

\subsection{\textit{hsbeta} Distribution and Proton Interaction Correction}

\label{sec:p_int_corr}

\begin{figure}[t]
\centering
\includegraphics[trim=0cm 0.4cm 0cm 0mm, clip, width=0.8\textwidth]{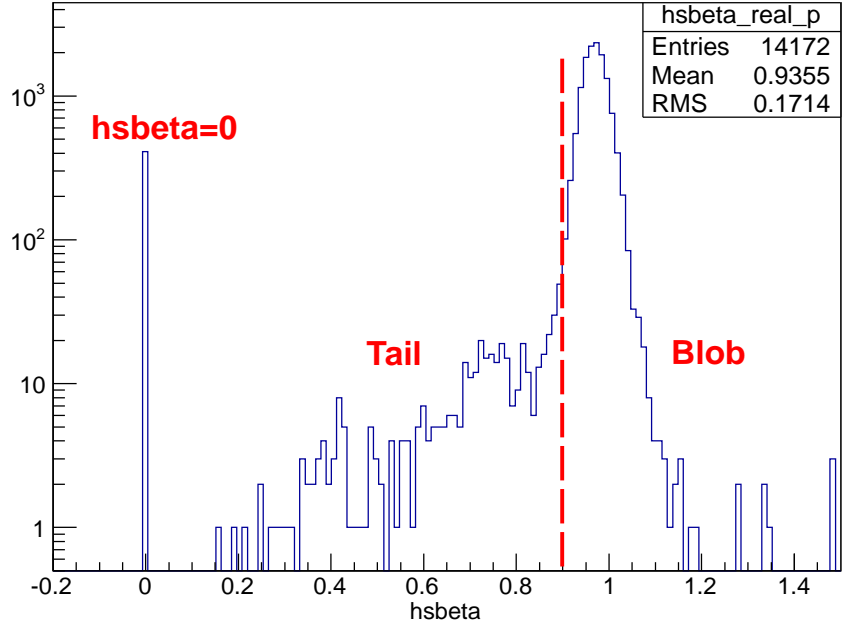}
\caption[$hsbeta$ distribution of Heep $Q^2$=2.41~GeV$^2$]{$hsbeta$ distribution of Heep $Q^2$=2.41~GeV$^2$. Note that the $hsbeta$ boundary between blob and tail events was set to 0.9. Since Heep runs have exceptionally good coincidence to random ratio, only few events appear in the random boxes.  Acceptance, PID cuts, missing mass and missing energy cuts are applied. From the further investigation, the `zero' and `tail' events have valid acceptance information (such as $hsyfp$ and $hxtar$) which can be treated as the `blob' events. One of the possible cause for the `zero' and `tail' events are the interaction between proton and detector material down stream of the wire chamber. This is further explained in the text. \oic}
\label{fig:p_abs_hebeta}
\end{figure}

By taking a closer look at the $hsbeta$ distribution and the $hsbeta$ versus $cointime$ spectrum in Fig.~\ref{fig:p_abs_hebeta}, a cluster of `zero' events ($hsbeta=0$) can be seen. After performing studies, such as those described in Sec.~\ref{sec:PID_cuts}, and through private communication with Hall C experts~\cite{gaskell16}, the `zero' events have been included along with `tail' ($hsbeta<0.9$) and `blob' ($hsbeta>0.9$) events for the Heep and $\omega$ analysis. In addition, a proton interaction correction study is performed using the same methodology used for determining the pion absorption correction during F$_\pi$-2-$\pi^{-}$  \nomenclature{F$_\pi$-2-$\pi^-$}{Second charged pion form factor experiment (E01-004) $\pi^-$ analysis: $^2$H$(e,e^{\prime}\pi^-)p$} analysis~\cite{piminus}.

Note that in the F$_\pi$-2-$\pi^{-}$ analysis, the `tail' events were corrected by a correction factor, whereas both `zero' and `tail' events are included in the yield computation for the Heep and $\omega$ analysis.

From a recoil proton (from the target chamber) traversing through the HMS entrance window, to the generation of a valid trigger, a proton can interact with a variety of detector materials along its path through the spectrometer. 

The dominant proton reaction for the recoil protons is inelastic scattering (mainly pion production), elastic and (quasi) elastic scattering (with heavier elements than $^1$H). In the case of pion production and (quasi) elastic scattering, a secondary pion, proton or neutron is emitted along the path of the recoil proton momentum, therefore has a probability to generate a valid trigger. From Table~\ref{aerogel_tab}, the mean value of the HMS central momentum settings is 3.5~GeV/c. The $pp$ and $pn$ total cross sections are dependent on the proton momentum, and are estimated to be 43~mb at 3.5~GeV/c~\cite{pdg}, where the elastic cross section is 1/3 of the total cross section. The proportion of protons lost due to these interactions must be correctly accounted for.

The situation is complicated by the fact that the proton interaction in the scintillators, ACD and HGC detector material leading to the emission of energetic nucleons can generate valid triggers or tracks. These events are part of the $hsbeta$ distribution and are already included in the analysis. Subsequently, if one applies a simple proton transmission correction based on the scattering cross section and material properties, it would result in an overcorrection. The proton interaction correction study is intended to account for the HMS triggers that are lost due to proton interactions in the material upstream of the drift chambers, or interaction in the detector stack, such as large angle deflection or leading to the emission of low momentum nucleons, which do not give enough signal in the scintillators providing a valid trigger.

\begin{table}[p]
\small
\centering
\setlength{\tabcolsep}{.45em}
\renewcommand{\arraystretch}{1.4}

\caption[HMS spectrometer material table ]{HMS spectrometer material table modified from similar table recreated by Henk Blok, which was originally produced during the F$_\pi$-2 analysis~\cite{horn}. Original version of the table was documented in Ref.~\cite{westrum}. $t$ shows the material thickness; $\rho$ is the material density; $\lambda$ is the nuclear collision length at $\sigma=38.4$~mb; $X=t\times\rho$; rescaled nuclear collision length at $\sigma=43$~mb: $\lambda^\prime = \lambda \times 43/38.4$; $X/\lambda^\prime$ denotes the proton interaction probability as it travel through the each spectrometer component.}
\label{tab:material}

\begin{tabular}{ll*{5}{c}c}
\hline
Absorber         & Material    &  $t$                 & $\rho$     &  $\lambda$  &  $X$      & $X/\lambda^\prime$ & Partial Sums \\
                 &             &  cm                  & g/cm$^3$   &  g/cm$^2$   &  g/cm$^2$ &    \%              & \%           \\ \hline
Target           & LH$_2$      &  1984                & 0.072      &  43.3       & 0.143     &  0.370             &              \\
Target Window    & Al          &  0.013               & 2.700      &  70.6       & 0.035     &  0.056             &              \\
Chamber Window   & Al          &  0.0406              & 2.700      &  70.6       & 0.110     &  0.174             &              \\
Chamber Gap      & Air         &  15                  & 0.001      &  62.0       & 0.018     &  0.033             &              \\
Entrance Window  & Kevlar      &  0.0381              & 0.740      &  60.0       & 0.028     &  0.052             &              \\
Idem             & Mylar       &  0.0127              & 1.390      &  60.2       & 0.017     &  0.032             &              \\
Exit Window      & Titanium    &  0.0508              & 4.540      &  79.9       & 0.231     &  0.324             &              \\ \hline
\multicolumn{5}{c}{Target - Exit Window Sum}                                     &           &                    & 1.04         \\ \hline
Dipole-DCGap     & Air         &  35                  & 0.001      &  62.0       & 0.042     &  0.076             &              \\
DC Windows       & Mylar       &  4$\times$(0.0025)   & 1.390      &  60.2       & 0.014     &  0.026             &              \\
DC Gas           & Ar/C6H6     &  12$\times$(1.8)     & 0.002      &  65.0       & 0.033     &  0.057             &              \\
DC Sensewires    & W           &  2$\times$(5.89E-06) & 19.30      &  110.3      & 0.001     &  0.001             &              \\
DC Fieldwires    & Be/Cu       &  36$\times$(0.00018) & 5.400      &  70.0       & 0.035     &  0.056             &              \\
Airgap DC-S2X    & Air         &  83.87               & 0.001      &  62.0       & 0.101     &  0.182             &              \\
ACD Entrance & Al          &  0.15                & 2.700      &  70.6       & 0.405     &  0.642             &              \\
Aerogel          & SiO2        &  9                   & 0.04-0.06  &  66.5       & 0.450     &  0.758             &              \\
ACD Airgap   & Air         &  16                  & 0.001      &  62.0       & 0.019     &  0.034             &              \\
ACD Exit     & Al          &  0.1                 & 2.700      &  62.0       & 0.270     &  0.488             &              \\
S1X              & polystyrene &  1067                & 1.030      &  58.5       & 1.100     &  2.106             &              \\
S1Y              & polystyrene &  1067                & 1.030      &  58.5       & 1.100     &  2.106             &              \\ \hline
\multicolumn{5}{c}{Dipole-DCGap - S1 Sum}                                              &           &                    & 6.53         \\ \hline
Cer Windows      & Al          &  2$\times$(0.102)    & 2.700      &  70.6       & 0.550     &  0.872             &              \\
Cer Gas          & C4F10       &  135                 & 0.002      &  63.0       & 0.332     &  0.590             &              \\
Cer Mirror       & Support     &  1.8                 & 0.050      &  53.0       & 0.090     &  0.190             &              \\
Cer Mirror       & SiO2        &  0.3                 & 2.200      &  66.5       & 0.660     &  1.111             &              \\
S2X              & polystyrene &  1.067/4             & 1.030      &  58.5       & 0.275     &  0.526             &              \\ \hline
\multicolumn{5}{c}{Cer Windows - S2X Sum}                                        &           &                    & 3.29        \\ \hline
\end{tabular}

\end{table}

To avoid any possible overcorrection for the proton interaction, the proton transmission from the target through to S2X was calculated and used to estimate which fraction of these events end in the parts of the $\beta$ versus coincidence time spectrum, see Fig.~\ref{fig:p_abs_hebeta}. The proton transmission for each material was calculated by making use of their known  areal densities and the nuclear collision lengths $\lambda$, as listed in Table~\ref{tab:material}. It was assumed that all proton interactions from the target to the spectrometer exit window resulted in lost triggers (1.04\%). 

For the protons interacting from the drift chambers to S1 (6.53\%), it was assumed that a fraction $f_1$ of protons were lost triggers, while the remaining fraction ($1-f_1$) of protons would successfully generate a trigger. These non-lost protons would either end up in the `zero'  or in the `tail' section of the $hsbeta$ distribution. Note that $f_1$ is a parameter to be determined later in this section.

Finally, for the interactions from the front window of the HGC detector through the first 1/4 thickness of S2 (corresponding to approximately the deposition which is necessary to generate a trigger), it was assumed that a fraction $f_2$ resulted in a low $\beta$ value (`zero' and `tail'), while the remaining ($1-f_2$) were indistinguishable from those protons that did not undergo nuclear interactions (`blob').

To determine the fractions $f_1$, $f_2$ appropriate for the Heep and $\omega$ data, a similar procedure to the F$_\pi$-2-$\pi^{-}$ analysis~\cite{piminus} was followed. The fractions of `zero' ($hsbeta=0$), `tail' ($hsbeta<0.9$) and `blob' ($hsbeta\ge0.9$) events, indicated in Fig.~\ref{fig:p_abs_hebeta}a, were determined for each data setting (four Heep setting and two $\omega$ settings) in Table~\ref{aerogel_tab}. Note that for this study, acceptance and PID cuts were applied.

Since low $\beta$ values can be due to instrumental timing effects, the `zero' and `tail' contributions were also determined using runs with an electron in the HMS (i.e. the HMS is set to negative polarity). The electron `zero' and `tail' fractions used in the study are the same as those determined from the F$_\pi$-2-$\pi^{-}$ analysis, and are 0.17\% and 0.66\%, respectively. The electron fractions were then subtracted from the proton fractions, yielding typical `zero+tail' values  of 5.8\%, with the reminder in the `blob' (model). $f_1$ and $f_2$ were then inferred by comparison to the observed `zero+tail' and `blob' values (experiment) to the calculated interaction probabilities. Note that this comparison is carried out on a setting-by-setting basis.  Both model and experiment `blob' fractions for all settings are determined to be 89.8$\% ~ \pm$ 1$\%$.

The proton interaction probability from the HMS HGC to 1/4-S2 is calculated as 3.29\%. Due to the close distance to the scintillator plane S2 (a valid trigger requires a particle to reach at least 1/4 thickness of S2), the forward-going energetic nucleons through $pp$ and $pn$ interactions can generate valid triggers. Most of these events (70-90\%) are likely to end up in the `zero+tail' section of the $hsbeta$ distribution. Based on this assumption, the $f_2$ factor is assumed to be around 80\%, subsequently, $f_1$=37-70\% resulted in good agreement with the data. Table~\ref{aerogel_tab} lists the $f_1$, $f_2$ and proton interaction correction determined for each of the data settings.

\begin{figure}[t]
\centering
\includegraphics[width=0.8\textwidth]{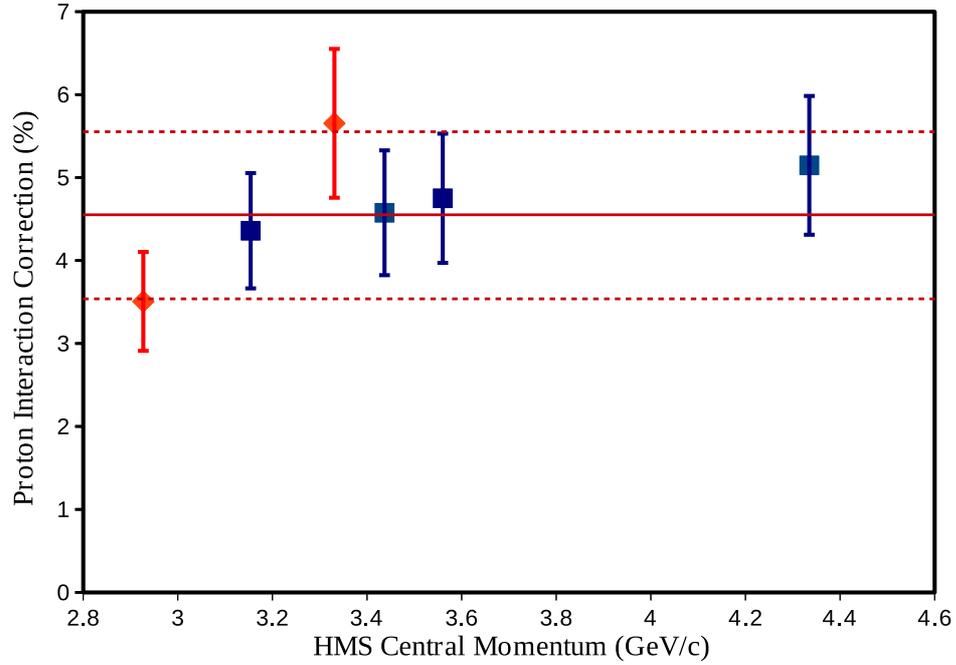}
\caption[Proton interaction correction study]{Proton interaction correction (\%) versus HMS central momentum setting for a proton $P_{\rm HMS}$. Red dots are for the $\omega$ data at $Q^2=$1.60 and 2.45~GeV$^2$;  where black dots are for the Heep data at $Q^2=$2.41, 4.42, 5.42 and 6.53 GeV$^2$. The error bars are the quadratic sum of 10\% model uncertainties for $f_1$ and for $f_2$. The averaged proton interaction correction factor of 4.7\% is indicated by the red solid line and the $\pm1\%$ point-to-point error bands are indicated by the red dashed lines. \oic}
\label{fig:proton_abs}
\end{figure}

The overall proton interaction correction factor consists of the calculated interaction probability from the target to the exit window (1.04\%) and the lost proton fraction from dipole and S1 (dictated by $f_1$). Fig.~\ref{fig:proton_abs} shows the proton interaction correction versus the HMS central momentum calculated for each of the Heep and $\omega$ settings. The plotted proton interaction correction values were calculated with $f_2=$80\%. An estimate of 10\% uncorrelated uncertainties were assigned to the $f_1$ and $f_2$ factors. The uncorrelated uncertainties (for $f_1$ and $f_2$) were then added in quadrature to calculate an overall uncertainty. 

Other uncertainties, such as the statistical uncertainty (from data) and the estimated scattering cross section uncertainty (from Table~\ref{tab:material}), were negligible compared to the dominant uncertainties described above, therefore not included in the quadratic sum for the overall uncertainty.

The averaged proton interaction, indicated by the red horizontal line in Fig.~\ref{fig:proton_abs}, implies an averaged $f_1$ factor of 55.5\%. The variation of $f_1$ factors shown in Table~\ref{aerogel_tab} becomes insignificant given the large assigned uncertainty of 10\%. The $f_1$ variation is also not visually noticeable by comparing $hsbeta$ distributions (such as Fig.~\ref{fig:p_abs_hebeta}(a)) from one setting to another.

Due to the large uncertainties, an average HMS central momentum-independent proton interaction correction of 4.7\%$\pm$1\% was applied to all settings. Note that the 1\% uncertainty is the point-to-point deviation in Fig.~\ref{fig:proton_abs}. The proton interaction correction is a higher correction than the pion absorption applied in the F$_\pi$-2-$\pi^+$ analysis~\cite{horn}, due to the larger $pp$ and $pn$ total cross section ($\sim$43~mb). The implementation of the proton interaction correction is to divide the yield by 0.953$\pm$0.01, and is combined with other corrections when computing scaler information for each data run. In the early stage of HMS commissioning, a proton interaction study was performed and a correction of 0.945$\pm$0.02 was determined~\cite{westrum}. Despite the two proton correction values agreeing within the error bar, the `old' correction is considered to have overestimated the proton interaction, since more detector materials were added (thicker HMS dipole exit window and presence of ACD) in the path of the proton in F-$_\pi$-2 compared to the early commissioning experiments of Ref.~\cite{volmer}.

\section{Missing mass and Energy Distributions}
\label{sec:mm_me}

\begin{figure}[t]
\centering
\includegraphics[width=0.85\textwidth]{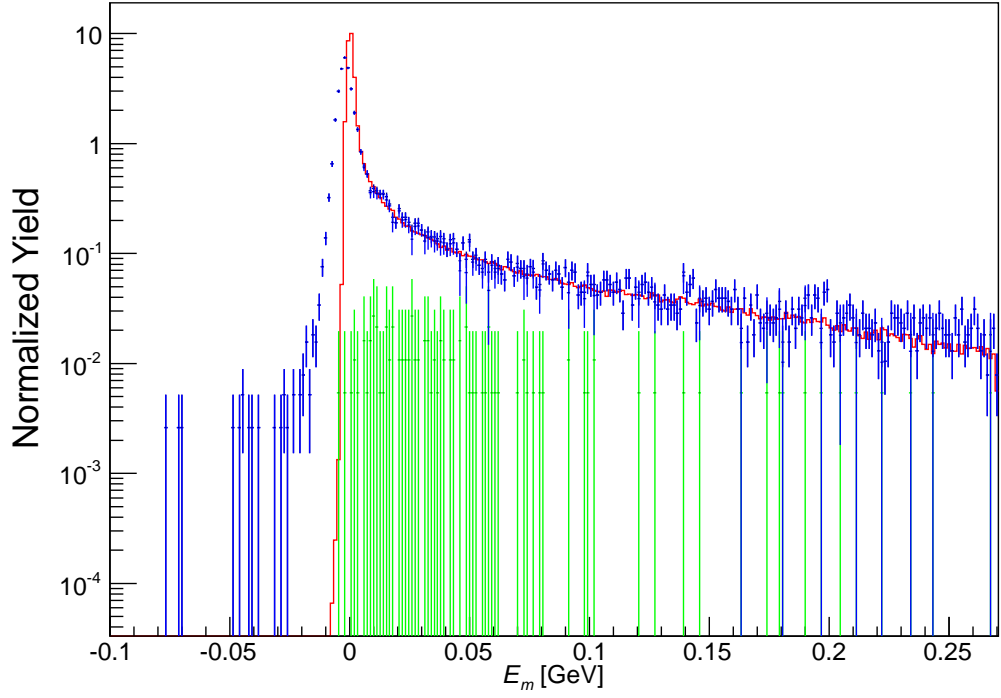}
\caption[Coincidence mode $E_{m}$ distribution for Heep $Q^2=$2.41~GeV$^2$]{Coincidence mode missing energy ($E_{m}$) distribution for Heep $Q^2=$2.41~GeV$^2$ on a logarithmic scale. Red is the simulation, blue is the dummy-subtracted data and green is the dummy. Acceptance, PID cuts, $hsbeta$-$cointime$ and missing mass cuts are applied. Random coincidence is also subtracted. \oic}
\label{fig:Heep_E_m}
\end{figure}

\begin{figure}[t]
\centering
\subfloat[][$pmoop$]{\includegraphics[width=0.33\textwidth]{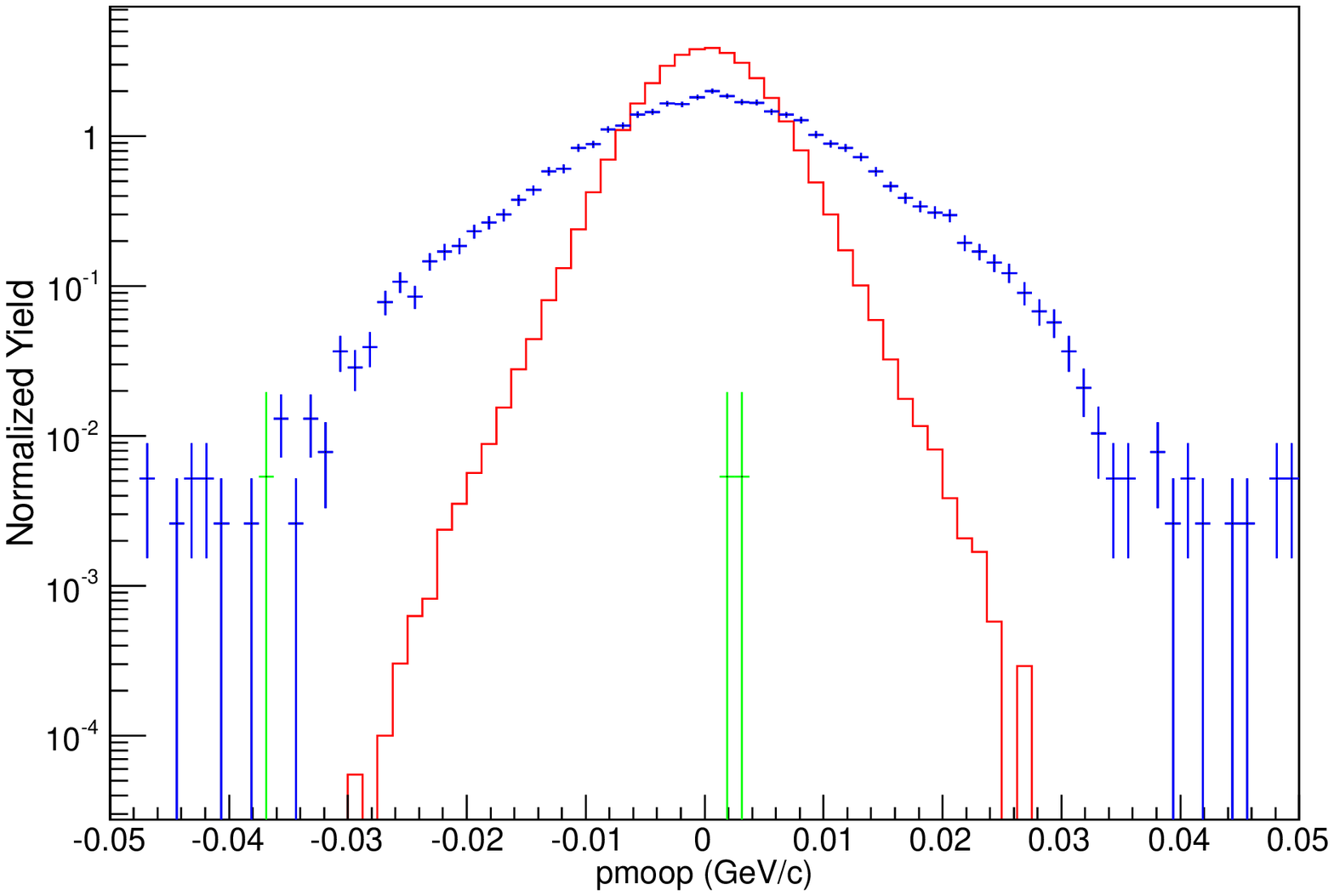}}
\subfloat[][$pmper$]{\includegraphics[width=0.33\textwidth]{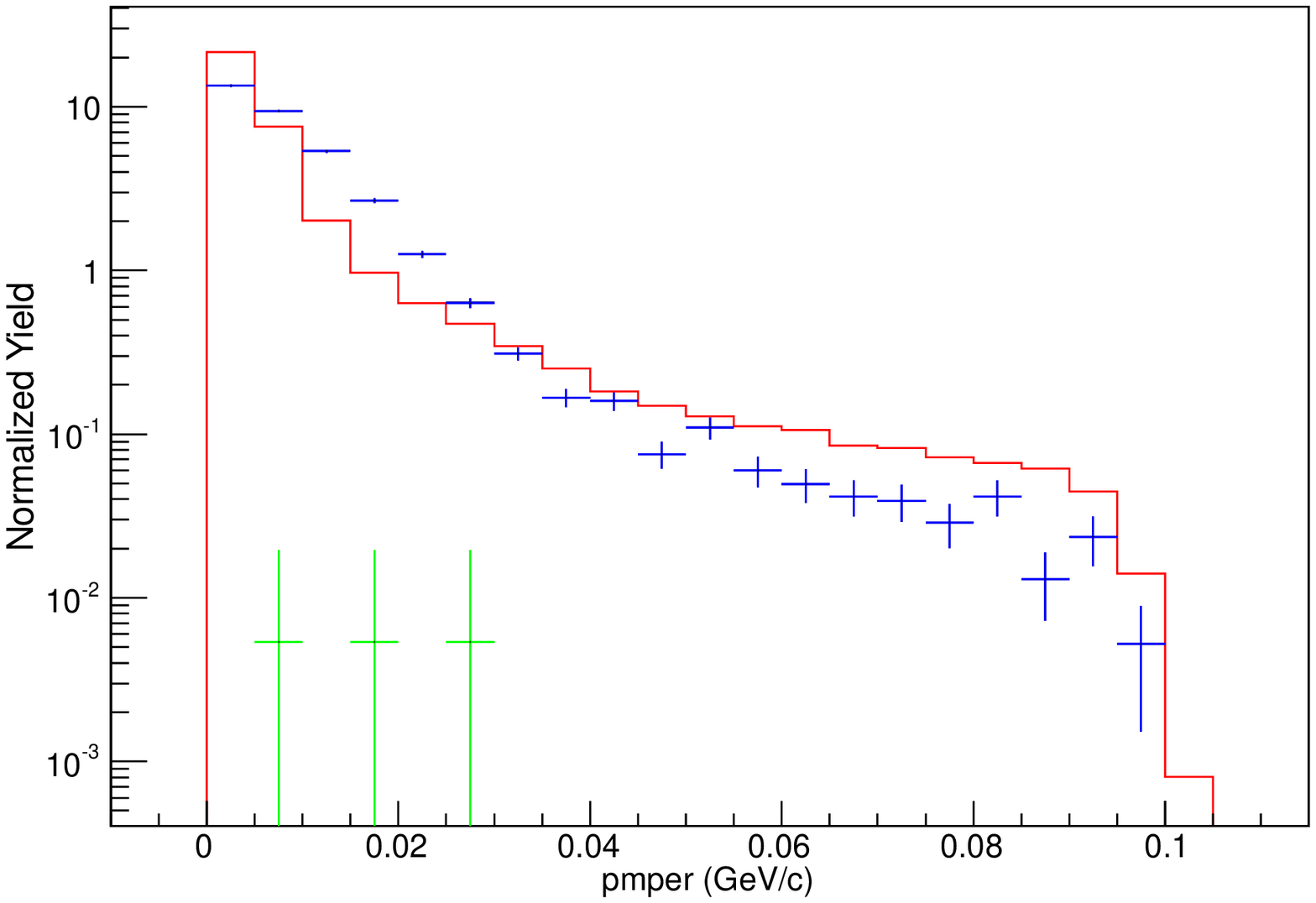}} 
\subfloat[][$pmpar$]{\includegraphics[width=0.33\textwidth]{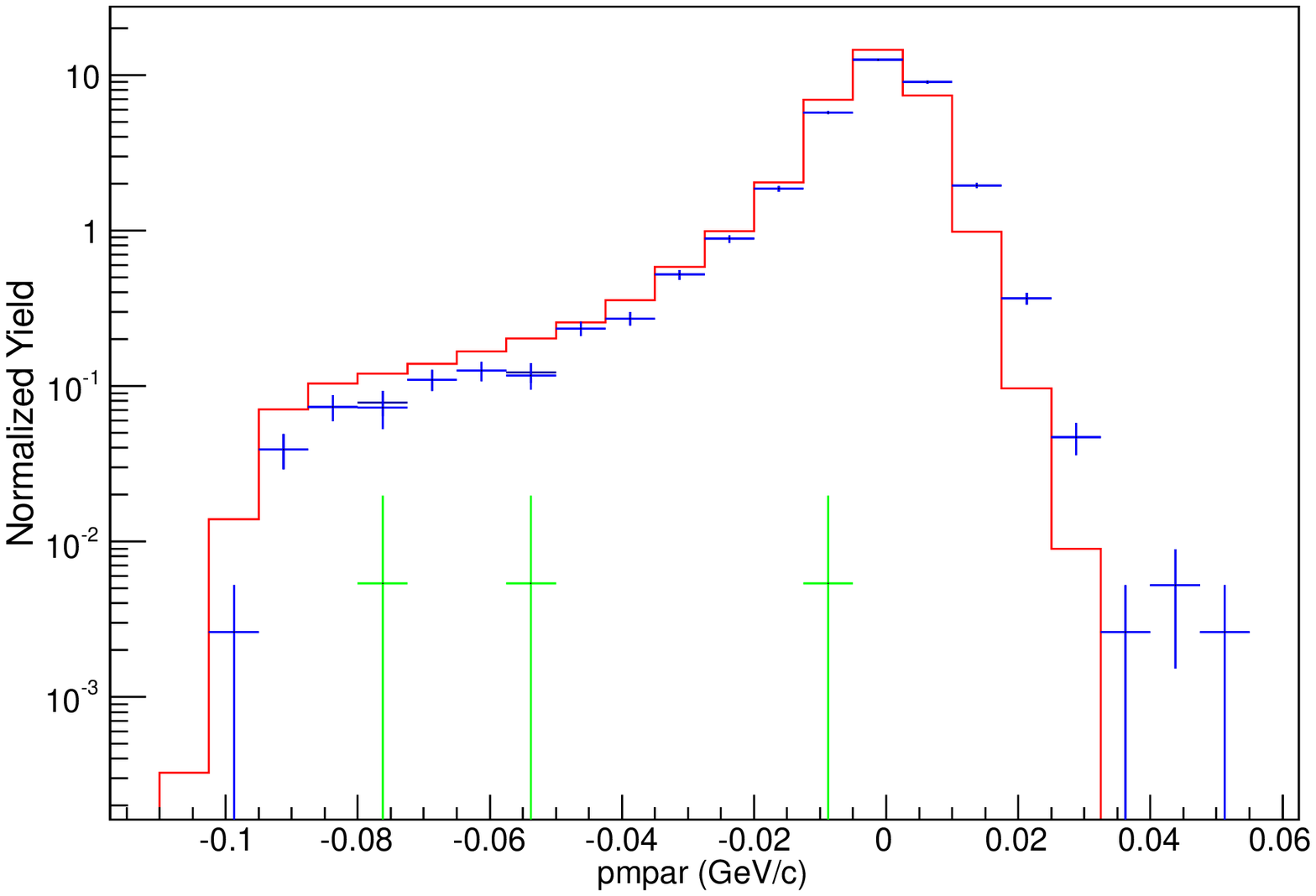}} 
\caption[Coincidence mode $\vec{p}_m$ distributions for Heep $Q^2=$2.41~GeV$^2$]{Coincidence mode missing momentum ($\vec{p}_m$) components distribution for Heep $Q^2=$2.41~GeV$^2$ on a logarithmic scale. Red is the simulation, blue is the dummy-subtracted data and green is the dummy. Acceptance, PID cuts, $hsbeta$-$cointime$ and missing mass cuts are applied. Random coincidence is also subtracted. $pmoop$ is the out-of-plane component, $pmper$ is the perpendicular and $pmpar$ is the parallel component of the $\vec{p}_m$ with respect to the $q$-vector. \oic}
\label{fig:Heep_P_m}
\end{figure}

\begin{figure}[h!]
\centering
\subfloat[][$E_m$ $<$ 0.27 GeV]{\includegraphics[width=0.5\textwidth]{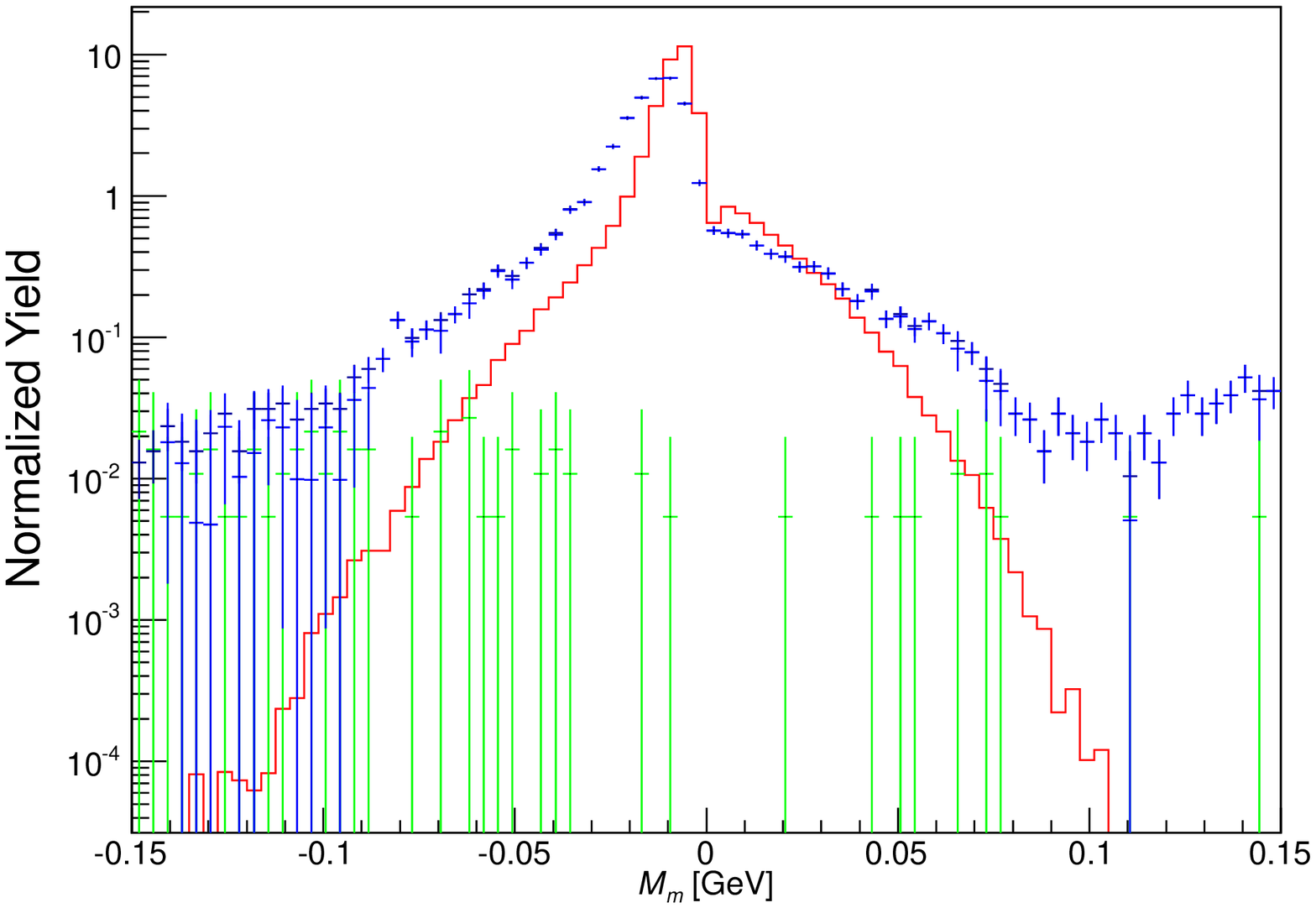}}
\subfloat[][$E_m$ $<$ 0.10 GeV]{\includegraphics[width=0.5\textwidth]{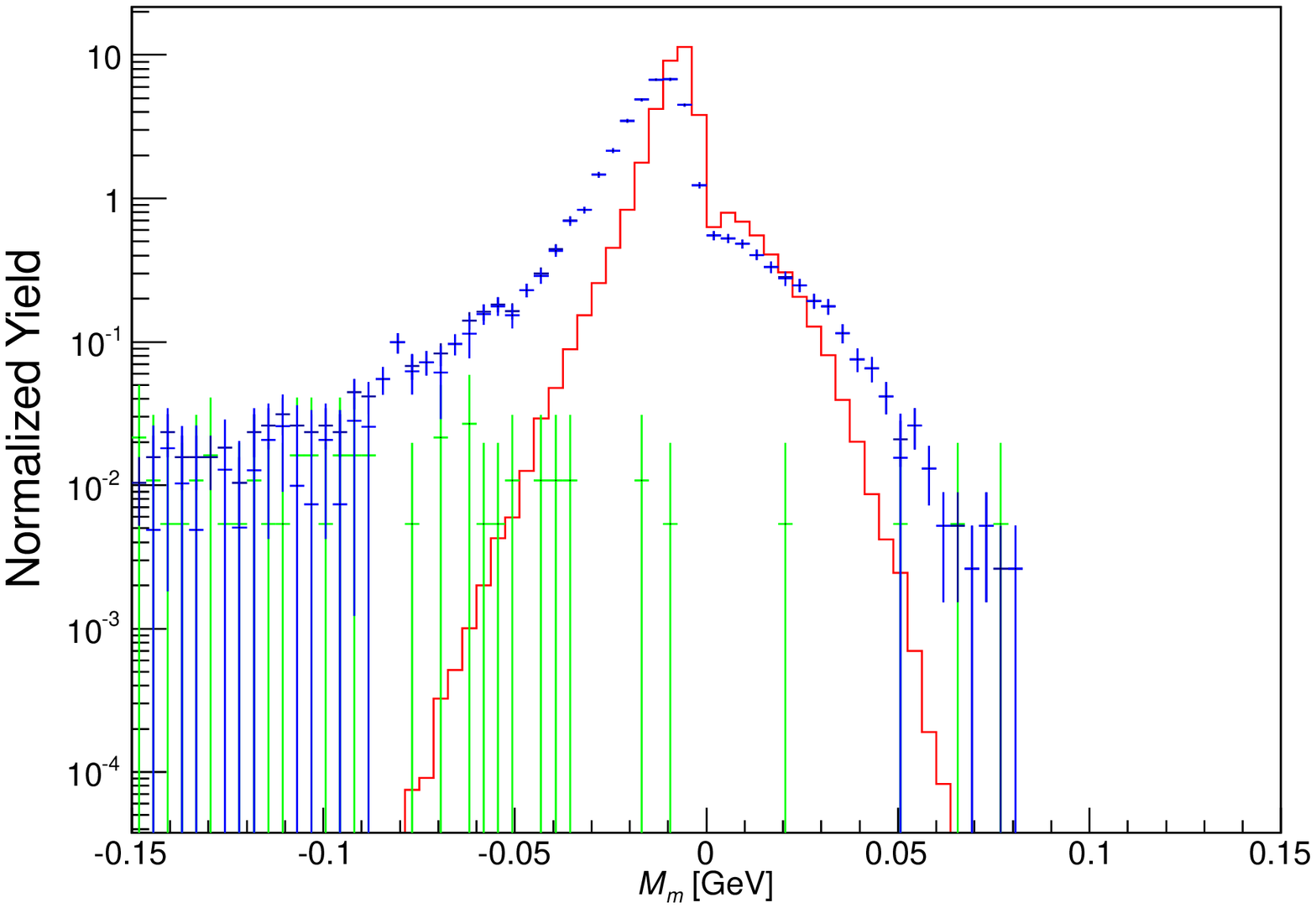}}
\caption[Coincidence mode $M_m$ distributions for Heep $Q^2=$2.41~GeV$^2$]{Coincidence mode missing mass distributions for Heep $Q^2$ = 2.41~GeV$^2$ on a logarithmic scale. Red is the simulation, blue is the dummy-subtracted data and green is the dummy. The applied $E_m$ cuts are indicated below the plots. Same acceptance, PID cuts, $hsbeta$-$cointime$ and missing mass cuts are applied. Random coincidences are also subtracted.~\oic}
\label{fig:Heep_M_m}
\end{figure}

In the coincidence Heep mode, the missing energy ($E_m$) and the missing momentum ($\vec{p}_m$\nomenclature{$\vec{p}_m$}{Missing Momentum}) of the $^1$H$(e, e^{\prime}p)$ reaction are defined as:
\begin{equation}
\begin{split}
E_{m} =& E_e - E_{e^{\prime}} - E_{p}  \,,\\
\vec{p}_{m} =& \vec{p}_e - \vec{p}_{e^{\prime}} - \vec{p}_{p} = \vec{q} - \vec{p}_p \,,
\end{split}
\end{equation}
where the $E_e$ and $\vec{p}_{e}$ are the energy and momentum of the electron beam; $E_{e^{\prime}}$ and $\vec{p}_{e^{\prime}}$ are the energy and momentum of the recoil electron; $E_{p}$ and $\vec{p}_{p}$ are the energy and momentum of detected proton; $\vec{q}$ corresponds to the three-momentum of the virtual photon. From these two quantities, one can calculate the missing mass of the system $M_m = \sqrt{E^2_m - p^2_m}$. The expected values are $M_m\sim0$ and $\vec{p}_m\sim0$ for the coincidence Heep mode.

For the electron singles mode, the Heep reaction is different: $^1$H$(e, e^{\prime})p$, since only the recoil electron is detected by the SOS. In this case, $E_m$ and the $\vec{p}_m$ are defined as:
\begin{equation}
\begin{split}
E_{m} =& E_e - E_{e^{\prime}} = E_p \,, \\  
\vec{p}_{m} =& \vec{p}_e -\vec{p}_{e^{\prime}} = \vec{p}_p \,,
\end{split}
\end{equation}
the $M_m$ is expected to be consistent with the rest mass of the proton ($M_m\sim0.938$~GeV), for the singles Heep mode.

Fig.~\ref{fig:Heep_E_m} shows a good agreement in the coincidence Heep mode missing energy ($E_m$) distribution between the dummy-subtracted data (blue) and simulation (red) up to 0.27 GeV for the nominal missing mass cut. Thus, the yield ratio is not sensitive to the $E_m$ cut. The difference is less than 0.5\% when comparing the yield ratio with $E_m$ $<$ 0.1~GeV and $E_m$ $<$ 0.27~GeV. The $E_m$ $<$ 0.1~GeV cut was chosen to narrow down the missing mass distribution, especially for $M_m$ $>$ 0~GeV. 

Fig.~\ref{fig:Heep_P_m} shows data-simulation comparisons for three components of the missing momentum $\vec{p}_m$. The dummy-subtracted data are shown in blue, dummy target data are shown in green and the simulation data are shown in simulation red. Fig.~\ref{fig:Heep_P_m} (a), (b) and (c) represent the out-of-plane, perpendicular and parallel (with respect to the $q$-vector) components of $\vec{p}_m$. Note that the average values of the data and simulations for all three components of $\vec{p}_m$ are close to the expectation ($\vec{p}_{m}$ $\sim$ 0). This validates of the momentum and angle offsets listed in Table.~\ref{tab:kin_offset}.

Fig.~\ref{fig:Heep_M_m}(a) and (b) show the missing mass distributions after applying $E_m$ $<$ 0.27~GeV and $E_m$ $<$ 0.1~GeV cuts. It is clear that for the $E_m$ $<$ 0.1~GeV missing mass distribution, the deviation between the data and simulation begins to increase significantly outside of $\pm$0.03~GeV from the peak (around $-$0.007 GeV). The nominal $M_m$ cut is defined as $\pm$0.025~GeV from the peak position, which corresponds to a cut of $-$0.032 $<$ $M_m$ $<$ 0.018~GeV.


\section{Simulating the Heep Reaction}
\label{sec:heep_model}

Elastic scattering reaction $^1$H$(e,e^{\prime}p)$ data provide a good check for spectrometer and various effects which can affect event reconstruction such as the radiative processes, multiple scattering and energy loss that are simulated in SIMC. The Monte Carlo simulates both coincidence and single arm elastic scattering (singles) events, corresponding to the $^1$H$(e,e^{\prime}p)$ and $^1$H$(e,e^{\prime})p$ reactions. The difference between coincidence and singles Heep events is further explained in Sec.~\ref{sec:heep_yield}.

In both coincidence and singles Heep modes, all kinematic quantities are calculated from the simulated in- and out-of-plane angles of the scattered electrons. In terms of the Sachs form factors, the differential cross section for elastic $ep$ scattering can be written as ~\cite{walker94},
\begin{equation}
\dfrac{d\sigma}{d \Omega} = \dfrac{\alpha^2 \cos^2\frac{\theta_e}{2}}{4~E^2\sin^4 \frac{\theta_e}{2}}  \frac{E^\prime}{E}\left( \dfrac{G^2_{E_p} + \tau G^2_{M_p}}{1 + \tau} +2~\tau ~ G^2_{M_p} \tan^2 \frac{\theta_e}{2} \right),
\end{equation}
where $\theta_e$, $E$ and $E^\prime$ represent the electron scattering angle, incident electron energy and final electron energy; $\alpha$ is the fine structure constant ($\sim$1/137);  $\tau$ is defined as $\tau = Q^2/4~M_P^2$.

The electric ($G_{E_p}$\nomenclature{$G_{E_p}$}{Electric Form Factor}) and magnetic ($G_{M_p}$\nomenclature{$G_{M_p}$}{Magnetic Form Factor}) form factors are parameterized using an empirical fit applied to the past $^1$H$(e^{\prime},ep)$ scattering data. The default parameterization for $G_{E_p}$ and $G_{M_p}$ in SIMC, is known as the Bosted parameterization~\cite{bosted95}, and is given by:
\begin{equation} \label{eq1}
\begin{split}
G_{E_p} (Q^2) &= \dfrac{1}{1 + 0.62~Q + 0.68~Q^2 + 2.8~Q^3 + 0.83~Q^4}\,, \\[5mm]
G_{M_p} (Q^2) &= \dfrac{\mu_p}{1 + 0.35~Q + 2.44~Q^2 + 0.5~Q^3 + 1.04~Q^4 + 0.34~Q^5}~\,,
\end{split}
\end{equation}
where $Q$ is the four momentum transfer between the incoming electron beam and recoil electron; $\mu_p$ is the magnetic moment of the proton and $\mu_p\approx2.793\,\mu_{\rm N}$, where $\mu_{\rm N}$\footnote{In SI Unit: $\mu_{\rm N} = 5.050783699(31) \times 10^{−27}$~J/T, in Gaussian Unit: $\mu_{\rm N} = 0.105155$~e$\cdot$fm } is the nuclear magneton constant. 

Two other parameterizations for $G_{E_p}$ and $G_{M_p}$, the AMT~\cite{AMT07} and the Brash~\cite{brash02} parameterizations, were used to study the model dependent variation in the experiment-simulation yield ratio.

The Brash parameterization~\cite{brash02} has different parameterizations of $G_{E_p}$ for different $Q^2$ regions; for $0.04<Q^2<7~$GeV$^2$ the parameterization is given by Ref.~\cite{brash02}:
\begin{equation} \label{eq2}
\begin{split}
G_{M_p} (Q^2) &= \dfrac{\mu_p}{1 + 0.1164~Q + 2.8742~Q^2 + 0.2411~Q^3 + 1.0056~Q^4 + 0.3449~Q^5} \,, \\[5mm]
G_{E_p} (Q^2) &= \left[1 - 0.130\,(Q^2 - 0.04)\right]\,\dfrac{G_{M_p}}{\mu_p}\,.
\end{split}
\end{equation}



The AMT parameterization~\cite{AMT07} is the most recent effort that used the world's data on elastic electron-proton scattering and calculations of two-photon exchange effects to extract corrected values of proton form  factors  over the full range of $Q^2$ coverage of the existing data. The effort also included the calculation of the two-photon exchange. The AMT parameterization is given as
\begin{equation} \label{eq3}
\begin{split}
G_{E_p} (Q^2) &= \dfrac{1 - 1.651~\tau + 1.287~\tau^2 - 0.185~\tau^3}{1 + 9.531~\tau + 0.591~\tau^2 + 0.0~\tau^3 + 0.0~\tau^4 + 4.994~\tau^5} \,, \\[5mm]
G_{M_p} (Q^2) &= \dfrac{\mu_p~(1 - 2.151~\tau + 4.261~\tau^2 + 0.159~\tau^3) }{1 + 8.647~\tau + 0.001~\tau^2 + 5.245~\tau^3 + 82.817~\tau^4 + 14.191~\tau^5}\,.
\end{split}
\end{equation}

The coincidence Heep experiment-simulation yield ratios are computed with the Bosted, Brash and AMT parameterizations. The results are presented in Sec.~\ref{sec:diff_para}. The singles Heep experiment-simulation yield ratio was only computed with the Bosted parameterization.


\section{Heep Study Results}
\label{sec:heep_yield}

In this section, the experiment-simulation yield ratio for the Heep study is presented. The results include the yield ratio for both coincidence mode: $^1$H$(e,e^{\prime}p)$, and electron singles mode: $^1$H$(e,e^{\prime})p$. Furthermore, different Heep parameterizations, Bosted~\cite{bosted95}, Brash~\cite{brash02} and AMT~\cite{AMT07}, were used to study the model dependent variation in the experiment-simulation yield ratio for cross-checking purposes. The detailed descriptions of all three Heep parameterizations can be found in Sec.~\ref{sec:heep_model}.

\subsection{Heep Coincidence Study}

\begin{figure}[t]
\centering
\includegraphics[width=0.8\textwidth]{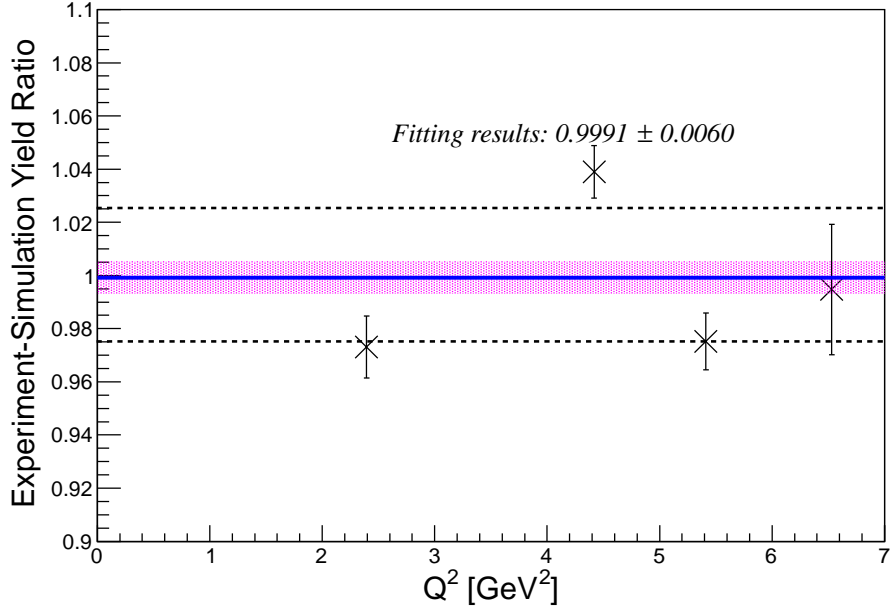}
\caption[Experimental-simulation Heep coincidence yield ratio]{Experimental-simulation Heep coincidence yield ratio. Note that only statistical error bars are shown and the uncertainty due to the missing mass cut is included. Other normalization uncertainties that are much larger than the statistical uncertainties also contribute (see first column of Table~\ref{tab:err_tab}). The weighted average fitting result: 0.9991$\pm$0.0060, is consistent with 1 within the uncertainty. The magenta band shows the fitting error and the dotted lines give $\pm$2\% point-to-point (considering error bars) uncertainty range (from the average value). \oic}
\label{fig:yield_recheck}
\end{figure}

After analysing every $Q^2$ setting, the accumulated distributions of acceptance and kinematic parameters are saved into a ROOT-file to cross-check with the Monte Carlo distributions. Fig.~\ref{fig:Heep_E_m}(b) shows the normalized missing energy distribution for the lowest $Q^2$ setting. The normalized hydrogen target (black), dummy target (green), dummy-target-subtraction (blue) and simulation (red). All four missing mass distributions are normalized to 1~mC of beam charge. As described in Sec.~\ref{sec:mm_me}, a reconstructed Heep event using the coincidence information is expected to have zero missing mass. Notice that the width of the missing mass peak is narrower (cleaner) for the simulation; this is due to the fact that the proton scattering in the target chamber and the HMS entrance/exit windows is poorly simulated (negative tail: $M_m < 0$); on the other hand, the data and Monte Carlo agree significantly better on the radiative process side (positive tail: $M_m > 0$). The radiative process (positive tail) in this context is mainly referring to the additional soft photon exchange between electron (beam) and proton (target).

Fig.~\ref{fig:yield_recheck} shows the normalized experiment-simulation yield ratio. The projected statistical error bars take into account the uncertainty due to the missing mass cut. The weighted fitting yield ratio (blue line) is consistent with 1 within the fitting error (magenta band). The point-to-point deviation (taking into account the individual error bars) of $\pm$2.5\% from the average yield ratio is plotted as the dotted line, and is used as a systematic error for the $\omega$ analysis.

The missing mass cut dependent uncertainty was determined as follows: changing the nominal missing mass cut by $\pm$0.01~GeV and reproducing three sets of yield ratios with different missing mass cuts: $-0.027<M_m<0.013$~GeV, $-0.032<M_m<0.018$~GeV (nominal cut) and $-0.037<M_m<0.023$~GeV; the average of the three yield ratios are plotted in Fig.~\ref{fig:yield_recheck} and the standard deviations are taken as the missing mass cut dependent uncertainty. The missing mass cut dependent uncertainty was added to the other statistical uncertainties in quadrature.

\subsection{Heep Singles Study}
\label{sec:singles}

The singles Heep study only uses the SOS information to reconstruct the process: $e^- + p \rightarrow e^- +X$, where the recoil $e^-$ is detected by the SOS. The SOS singles operation mode requires no coincidence information (from HMS). This allows more background events from the target cell and inelastic physics process, which results in a much higher event rate.

\begin{figure}[t]
  \centering
  \subfloat[][Heep singles invariant mass distribution]{\includegraphics[width=0.51\textwidth]{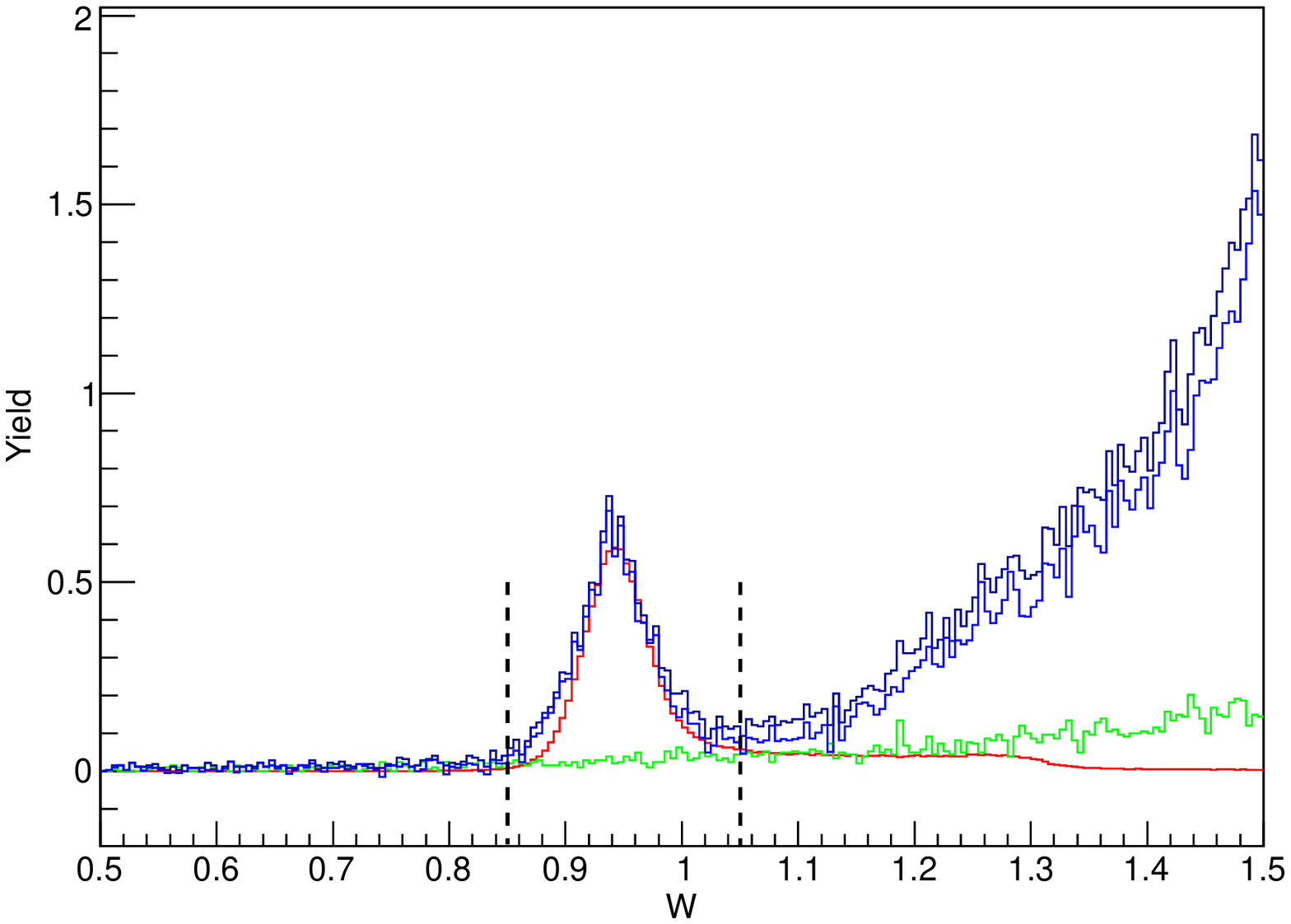}\label{single_mm}}
  \subfloat[][Heep singles elastic events yield ratio]{\includegraphics[width=0.54\textwidth]{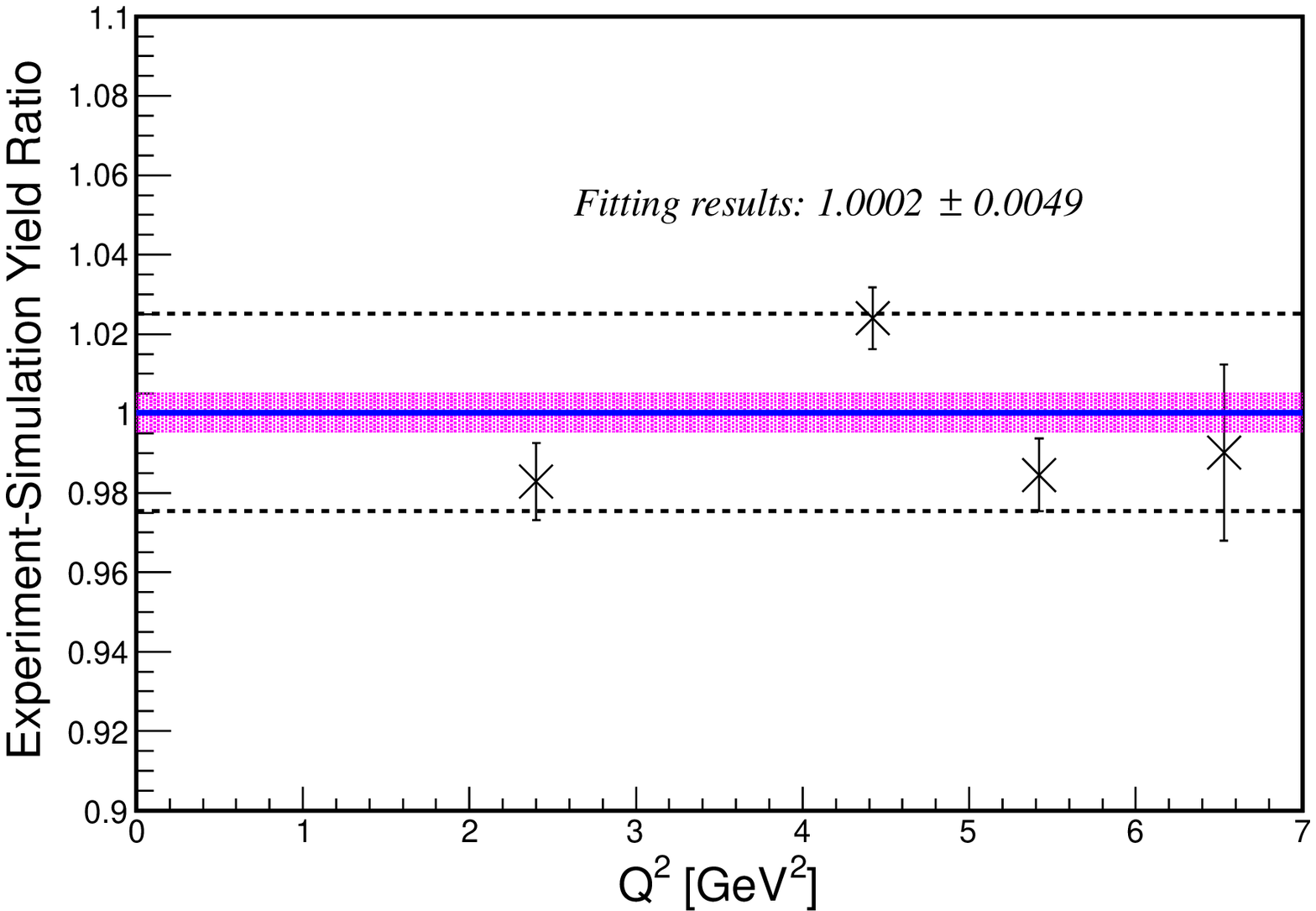}\label{single_y_ratio}}
  \caption[Heep singles invariant mass distribution and yield ratio]{(a) Example Heep SOS singles missing mass distribution at the highest $Q^2$ Setting: $Q^2=6.53$~GeV$^2$. The missing mass distribution of normalized hydrogen target is in black; dummy target in green; dummy-target-subtracted hydrogen target in blue; simulation in red. All four missing mass distributions are to 1~mC of beam charge. (b) shows the experimental-simulation Heep singles yield ratio. Note that only statistical error bars are shown. The weighted average fitting result: $1.0002\pm0.0049$. The magenta band shows the fitting error and the dotted lines indicate $\pm$2.5\% point-to-point uncertainty range. \oic}
  \label{single_final}
\end{figure}

The standard SOS acceptance and PID cuts are applied, which are defined as follows: 
\begin{description}
\item[SOS Acceptance Cut:]  $ssytar \le 1.5$ \&\& $abs(ssdelta) \le 15.$ \&\& $abs(ssxfp) \le 20.$ \&\& \\ 
 $abs(ssxptar) \le 0.04$ \&\& $abs(ssyptar) \le 0.065$.
\item[PID Cut:] $scer\_npe > 0.5$ \&\& $ssshtrk > 0.70$.
\end{description}
After applying the cuts, the normalized dummy-subtracted invariant mass ($W$) distribution is sufficiently clean around the proton mass region, as shown in Fig.~\ref{single_mm}. However, the experiment yield starts to deviate from the simulation for $W$ $>$ 1.1~GeV. This is due to the fact that the simulation doesn't take into account pion production above the elastic scattering region. A cut is enforced on the invariant mass to eliminate inelastic events:  $$0.85 < W < 1.05~\textrm{GeV}\,.$$  The cut dependent uncertainty was studied and included in the final yield ratio computation.  

Fig.~\ref{single_y_ratio} shows the normalized experiment-simulation elastic events yield ratio, which takes into account the radiative tail. Note that only statistical error bars are shown. The weighted fitting yield ratio (blue line) is consistent with 1 within the fitting error (magenta band). The $\pm$2.5\% band from the singles study is consistent with yield ratio determined in the coincidence study.

\subsection{Heep with Different Parametrizations}
\label{sec:diff_para}

\begin{figure}[t]
  \centering
  \subfloat[][Brash parameterization]{\includegraphics[width=0.52\textwidth]{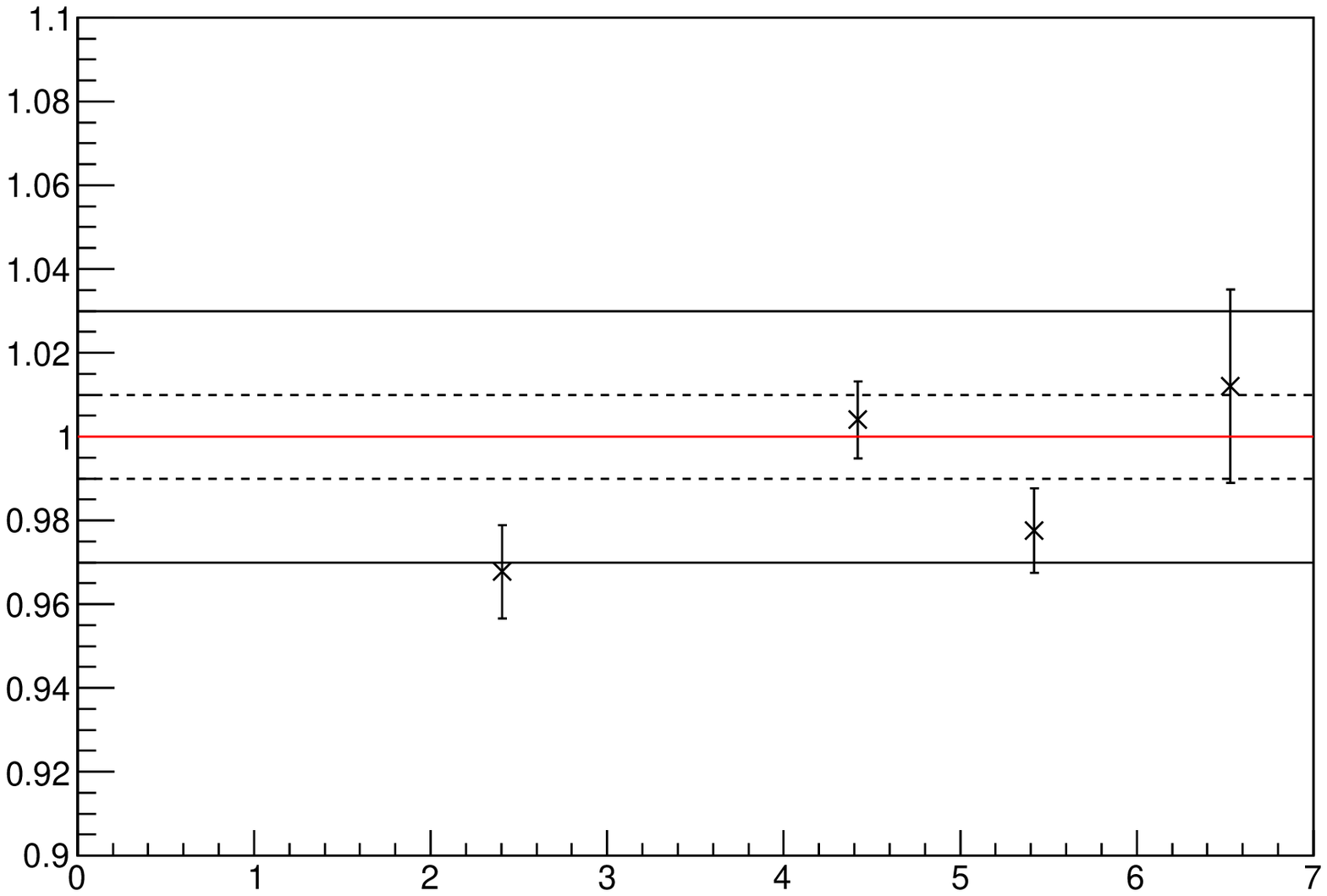}\label{brash_para}} 
  \subfloat[][AMT parameterization]{\includegraphics[width=0.52\textwidth]{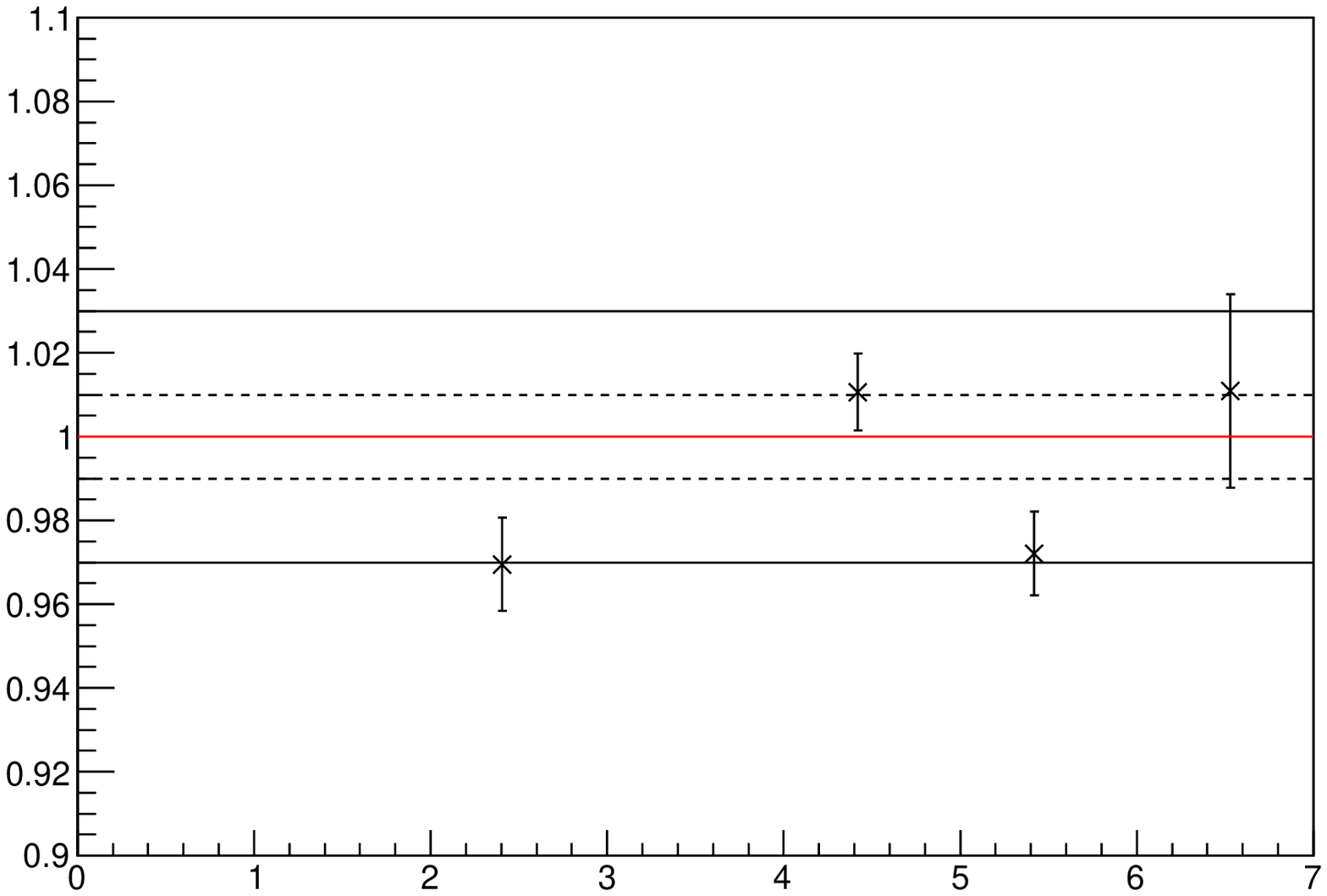}\label{amt_para}}
  \caption[Experimental-simulation yield ratio computed using the Brash and AMT parameterizations]{Experimental-simulation yield ratio computed using the Brash and AMT parameterizations defined in Sec.~\ref{sec:heep_model}. The red line indicates yield ratio of 1; the black dashed (solid) lines gives $\pm1\%$ ($\pm3\%$) band from 1. The projected error does not include the missing mass dependent uncertainty. \oic}
  \label{fig:para}
\end{figure}

In order to ensure the validity of our yield ratio results and parameterization independence, the coincidence study was repeated using the Brash and AMT parameterizations defined in Sec.~\ref{sec:heep_model}. The yield ratio results from the Brash and AMT parameterizations are shown in Fig.~\ref{fig:para} (a) and (b), respectively. Compared to the yield ratio from the Bosted parameterization, both yield ratio points for all $Q^2$ settings are slightly lowered by around 1\%. 

To further compare the yield ratio results from different parameterizations, the $\chi^2$ per degree of freedom from unity is computed using the following equation:
\begin{equation}
\frac{\chi^2}{\nu} = \frac{1}{\nu}\sum_i \left(\dfrac{x_i - 1 }{\sigma_i}\right)^2\,,
\label{eqn:chi2}
\end{equation}
where $\nu$ is the number of degrees of freedom, which is 3 (number of data points minus 1); index $i$ indicates the $Q^2$ setting; $x_i$ is the yield ratio; $\sigma_i$ is the corresponding yield ratio uncertainty. The $\chi^2/\nu=3.47$ for the Bosted parameterization~\cite{bosted95}; $\chi^2/\nu=4.59$ for the Brash parameterization~\cite{brash02}; $\chi^2/\nu=5.59$ for AMT parameterization~\cite{AMT07}. Thus, all three Heep parameterizations gave experiment-simulation yield ratios consistent with with each other, with preference for the Bosted parameterization after including systematic uncertainty of 2.5\%.

\section{Results}

From the Heep experiment-simulation yield ratio results from different modes (coincidence and singles mode) and model dependence study, Heep yield ratios are concluded to be consistent with 1 within the experimental uncertainties for all $Q^2$ setting independent of the Heep model used. This agreement between experiment and Monte Carlo (for Heep analysis) gave validation and reassurance to the data selection procedure, detector efficiencies studies, and various corrections used for the $\omega$ analysis.

\graphicspath{ {pics/6Omega/} }

\chapter{Omega Analysis}
\label{chap:omega_ana}

%
%
%

This chapter is intended to provide details regarding the analysis of the exclusive $\omega$ electroproduction $p(e,e^\prime p)\omega$ data.

\section{Overview and Introduction to the Iterative Procedure}
\label{sec:itt_pro}


\begin{table}[t]
\centering
\footnotesize
\setlength{\tabcolsep}{.62em}
\caption[A list of the kinematic values for reaction $p(e,e^{\prime}p)\omega$]{A full list of the kinematic values for experimental settings of the $^1$H$(e,e^{\prime}p)\omega$ reaction. T$_{inc}$ represents the incoming electron beam kinetic energy; P$_{\rm SOS}$ is the SOS momentum setting; $\theta^*_e$ is angle of scattered electron which defines the angle of the SOS; $\theta_q$ gives the direction of the virtual photon which corresponds to the nominal angle of HMS; $\theta_{pq}$ is the HMS angle with respective to the $q$-vector (positive angle represents rotation away from the beam line); $\theta_{\rm HMS}$ is the HMS angle with respect to the beam line; $P_p$ is the recoil proton target momentum after interaction; $P_{\rm HMS}$ is the HMS momentum setting during the experiment, note that the spectrometer momentum stays the same for all angles at the same $\epsilon$ setting.}
\label{tab:kin_tab}
\begin{tabular}{cccccccccccccc}
\toprule

T$_{inc}$ & P$_{\rm SOS}$ &  $\theta_e^*$ & $ \epsilon$  & $\theta_q$ & $\theta_{pq}$  & $\theta_{\rm HMS}$ & $P_p$ & $P_{\rm HMS}$ & $x$ & $-u$  & $-t$\\ 
MeV       & MeV     &   deg         &               &  deg     & deg            & deg           & MeV/c      & MeV/c     &  & GeV$^2$  &   GeV$^2$ \\
\midrule
\multicolumn{12}{c}{$Q^2_{\rm nominal}$ = 1.60~GeV$^2$ ~~ $W_{\rm nominal}$ = 2.21~GeV   } \\
\midrule
3772 & 785.79  & 43.09 & 0.328 & 9.53  & +1.0   & 10.53 & 2936.79 & 2927.2  & 0.2855 &  0.087  & 4.025  \\
     &         &       &       &       & +3.0   & 12.53 & 2913.20 &         &        &  0.129  & 3.983  \\
\midrule                                                                                  
4702 & 1715.79 & 25.73 & 0.593 & 13.28 &  0.0   & 13.28 & 2939.53 & 2927.2  & 0.2855 &  0.082  & 4.030  \\
     &         &       &       &       & $-$2.7 & 10.58 & 2917.79 &         &        &  0.121  & 3.991  \\
     &         &       &       &       & +3.0   & 16.28 & 2913.15 &         &        &  0.129  & 3.982  \\

\midrule
\multicolumn{12}{c}{$Q^2_{\rm nominal}$ = 2.45~GeV$^2$ ~~ $W_{\rm nominal}$ = 2.21~GeV } \\
\midrule
4210 & 770.83  & 51.48 & 0.270 & 9.19  & 1.4    & 10.59 & 3355.82 & 3331.7  & 0.3796 &  0.184  & 4.778 \\
     &         &       &       &       & 3.0    & 12.14 & 3324.12 &         &        &  0.241  & 4.721 \\
\midrule                                                                                
5248 & 1808.83 & 29.43 & 0.554 & 13.61 &  0.0   & 13.61 & 3363.86 & 3331.7  & 0.3796 &  0.169  & 4.793 \\
     &         &       &       &       & 3.0    & 16.61 & 3324.28 &         &        &  0.241  & 4.721 \\
     &         &       &       &       & $-$3.0 & 10.61 & 3324.49 &         &        &  0.240  & 4.722 \\
\bottomrule
\end{tabular}
\end{table}

\begin{figure}[p]
  \centering
  \includegraphics[width=0.97\textwidth]{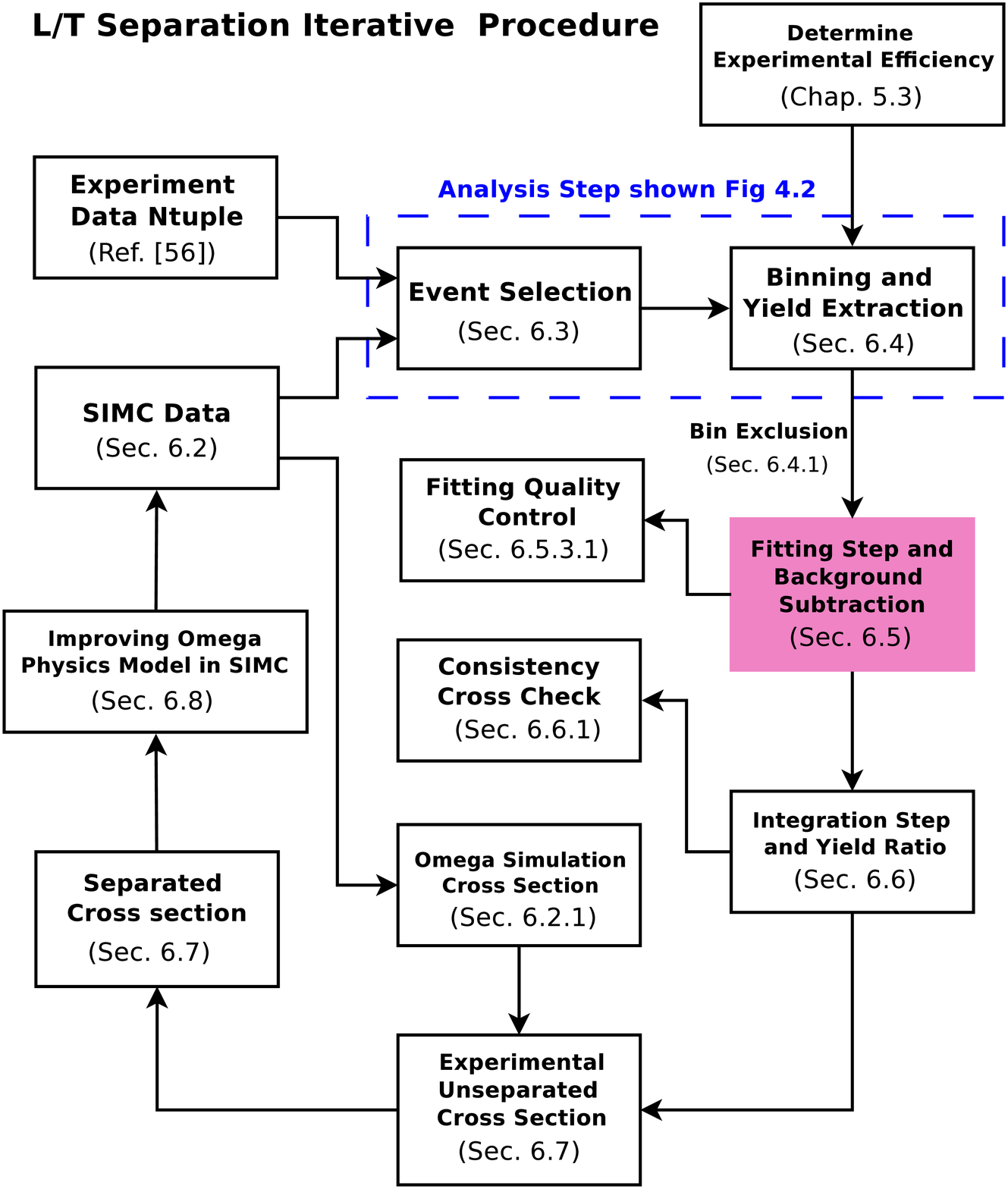}   
  \caption[Flowchart of the L/T separation iterative procedure]{Flowchart of the L/T separation iterative procedure. Note that the detail regarding each step is described in the relevant section given inside the bracket. The steps inside of the blue dashed box are performed using the new yield analysis platform introduced in Sec.~\ref{sec:ana_platform}. The procedural flowchart of this new platform is shown in Fig.~\ref{fig:flowchart}. The red shaded box indicates the most important step of the iterative procedure: the fitting step, for the physics background subtraction. ~\oic}
  \label{fig:lt_flowchart}
\end{figure}


As introduced in Sec.~\ref{sec:fpi2}, the $\omega$ data analyzed in this thesis work came from the same data set as the F$_\pi$-2-$\pi^+$ analysis. The $\omega$ data includes measurements at two $Q^2$ settings: $Q^2$ = 1.60, 2.45~GeV$^2$ at a common $W$ = 2.21~GeV. Each of the $Q^2$ settings require high and low $\epsilon$ measurements to perform a full L/T separation. To ensure maximum $\phi$ angle coverage around the $q$-vector, the $\epsilon_{high}$ data contain measurements at three different HMS angles (corresponds to a full $\phi$ coverage) and the $\epsilon_{low}$ contain measurements at two different HMS angles (corresponds to a partial $\phi$ coverage). The partial $\phi$ coverage is imposed by the physical clearance of the spectrometer and beam line components. In total, there are 10 experimental measurement settings. The nominal central kinematic values for all experimental settings are listed in Table~\ref{tab:kin_tab}.

A full L/T separation of the differential cross section is an iterative procedure which requires gradual improvement of the estimated cross section parameterization in the simulation by comparison with the data, and the improved simulation should offer an acceptable description of experimental data across the spectrometer acceptances and kinematics coverages. The iterative procedure is summarized in the flowchart shown in Fig.~\ref{fig:lt_flowchart}. As indicated by the flowchart, every step of the procedure is described in detail in a separate section (section indices are inside of the brackets).

The iterative procedure begins with generating the experimental and simulation ntuples (correlated data structure). Note that the raw experimental data calibration for the $\omega$ data was performed as part of the F$_\pi$-2-$\pi^+$ analysis, details regarding the raw data calibration and generation of the experimental data ntuples can be found in Ref.~\cite{horn}. The simulation ntuples are generated using SIMC, which was described in Sec.~\ref{sec:simc}.  Sec.~\ref{sec:sim_physics} documents the functional forms of the physics cross section models for $\omega$ and background mesons, used in SIMC to generate the simulation data.

The event selection criteria for identifying valid proton events in both experimental and simulation data including the yield computation, are described in Sec.~\ref{sec:event_select}. Details regarding binning the proton events in $u$-$\phi$ are covered in Sec.~\ref{sec:binning}.

The bin-by-bin analysis (in $Q^2$, $\epsilon$, $\theta_{pq}$, $u$, $\phi$) of subtracting the physics background and obtaining the $\omega$ events is a two step process: the fitting step and the integration step. The fitting step (described in Sec.~\ref{sec:bg_sub}) is the most critical step of the iterative procedure. It involves fitting and subtracting the physics background underneath the $\omega$ peak, then extracting the $\omega$ experimental yield. Sec.~\ref{sec:int_step} introduces the integration step, and its role is to integrate the simulation $\omega$ yield by summing the $\omega$ events among different HMS angle measurements within the common $Q^2$ and $\epsilon$ setting, then the comparing with the background subtracted experimental $\omega$ yield (obtained from the fitting step) to form yield ratios on a bin-by-bin basis ($Q^2$, $\epsilon$, $\theta_{pq}$, $u$, $\phi$). 

Sec.~\ref{sec:extracting_LT_cx} contains details regarding the computation of the experimental cross section and L/T separation. The last step of the iterative procedure is to obtain the improved parameters with the separated cross section. This is covered Sec.~\ref{sec:improving}.

Sec.~\ref{sec:issues} documents technical difficulties encountered during the analysis. The discussions related to the experimental and systemic uncertainties, as well as the overall uncertainty budget table, can be found in Sec.~\ref{sec:uncertainty}.






\section{Physics Simulation Model in SIMC}
\label{sec:sim_physics}








The first step of the iterative procedure is the generation for the simulation data of all possible contributing final states for the $^1$H$(e,e^{\prime}p)X$ interaction, where $X=\omega$, $\rho$, $2\pi$, $\eta$, $\eta^{\prime}$, and calculating the $\omega$ differential cross sections using the Monte Carlo simulation method. 




This analysis uses the standard Hall C simulation tool: SIMC, described in Sec.~\ref{sec:simc}. SIMC takes into account spectrometer acceptance and other effects such as radiative corrections and multiple scattering. In order to generate the simulation data for a specific physics process, a realistic cross section model is required as input to the SIMC. The functional form of this physics model has to offer an adequate description the data behavior in terms of the kinematics variables (such as $Q^2$ and $u$ during this analysis), whose parameters can be improved iteratively. The iterative procedure is capable of improving the input parameters obtained from the previous iteration, whereas the improved parameters can be used for the next iteration given the functional form of the model stays the same. The final parameters should reflect the optimal agreement between simulation and data for the chosen function form. 



%
%
%
%

%


\subsection{$\omega$ Production Model}
\label{sec:omega_model}

The exclusive $\omega$ electroproduction: $^1$H$(e,e^{\prime}p)\omega$, is the primary reaction of this analysis. The $\omega$ meson is a vector meson which holds the quantum numbers of $J^{\rm PC}=1^{--}$ and $I^{\rm G} = 1^+$. The valence quark content of the $\omega$ can be written as
\begin{equation}
\omega = \frac{u~\overline{u} + d~\overline{d}}{\sqrt{2}}\,.
\end{equation}
The rest mass of the $\omega$ meson is $m_{\omega}=$782.59~MeV with a narrow width of 8.49~MeV, therefore a missing mass cut of $M_m$ $>$ 0.65~GeV is included in the event selection criteria for the purpose of background rejection.


In SIMC, the function form of the $\omega$ production model depends on the Lorentz invariant quantities $Q^2$ and $u$; the components of the L/T separated differential cross section can be written as
\begin{align}
   \sigma_{\rm T}  & = \frac{t_0 + t_1 \cdot(-u)}{Q}\,,  \label{eqn:T}    \\
   \sigma_{\rm L}  & = \frac{l_0 + l_1 \cdot(-u)}{Q^4}\,,  \label{eqn:L}  \\
   \sigma_{\rm LT} & = \left[\,\frac{lt_0 + lt_1 \cdot(-u)}{Q^2}\right] \cdot \sin\theta^{*}, \\
   \sigma_{\rm TT} & = \left[\,\frac{tt_0 + tt_1 \cdot(-u)}{Q^2}\right] \cdot \sin^2\theta^{*} \label{eqn:TT}\,,
\end{align}
where $\theta^*$ corresponds to the $\omega$ emission angle (see Fig.~\ref{fig:plane}) with respect to the $q$-vector in the $\gamma^*p$ CM frame; $t_0$-$t_1$, $l_0$-$l_1$, $lt_0$-$lt_1$ and $tt_0$-$tt_1$ are the free fitting parameters whose values are improved by the iterative process. The $1/Q^n$ dependences were determined by trial and error to achieve good description of the data. The components of the differential cross section are computed in units of $\mu$b/GeV$^2$; $Q^2$ and $u$ are in GeV$^2$. For $Q^2$ = 1.6~GeV$^2$, the optimal parameterization~\footnote{Input parameterization obtained from iteration \#137} is given by:
\begin{align*}
t_0  &=  7.73587,  & t_1  &=  -7.9672    ,  \\ 
l_0  &=  13.2553 ,  & l_1  &= -47.2633   ,  \\
lt_0 &= -0.3439  ,  & lt_1 &=  5.9217    ,  \\
tt_0 &=  8.1221  ,  & tt_1 &= -139.8422  ,
\end{align*}
and the optimal parameterization for $Q^2$ = 2.45~GeV$^2$:
\begin{align*}
t_0  &=  6.16527 ,  & t_1  &=  -4.2124  ,  \\ 
l_0  &=  12.2546 ,  & l_1  &=  -29.8629 ,  \\
lt_0 &= -0.3620  ,  & lt_1 &=   3.1028  ,  \\
tt_0 &= -7.4032  ,  & tt_1 &=  63.4705  , 
\end{align*}

The separated differential cross sections are combined into the total differential cross section using the Rosenbluth Separation formula:
\begin{equation}
2 \pi \frac{d^2 \sigma}{dt ~ d\phi} = \frac{d \sigma_{\rm T}}{dt} + 
\epsilon ~ \frac{d \sigma_{\rm L}}{dt} + \sqrt{2\epsilon(1+\epsilon)}~ 
\frac{d\sigma_{\rm LT}}{dt} \cos \phi + \epsilon ~ 
\frac{d\sigma_{\rm TT}}{dt} \cos 2\phi \,,
\label{eqn:rosen_chapt_5}
\end{equation}
and then converted to the lab frame six-fold $d^6\sigma/d\Omega_{e^{\prime}}\,dE_{e^{\prime}}\,d\Omega_p\,dE_p$ via Eqns.~\ref{Xsection_6f} and \ref{Xsection_5f}.

Note that the L/T separated differential cross section given by Eqns.~\ref{eqn:T}-\ref{eqn:TT} has a $Q^2$ dependence which is included to provide a gentle correction across the acceptance of each ($Q^2$, $\epsilon$, $\theta_{pq}$, $u$, $\phi$) bin. A $W$ dependence of the form 
\begin{equation}
\frac{1}{(W^2-M^2_p)^{2}}
\label{eqn:w_dep}
\end{equation}
is directly multiplied to the total differential cross section (Eqn.~\ref{eqn:rosen_chapt_5}) for the $W$ correction. Since the $W$ coverage of the data is narrow, the $W$ dependence cannot be independently determined in this experiment; the $W$ dependence was taken the same as in Refs.~\cite{horn,volmer,bebek78}. 

The shape of the $E_m$ distribution for $\omega$ is constructed during the event generation stage of the SIMC. The following equation is used to replicate the mass distribution of the $\omega$:
\begin{equation}
M_x = m_{\omega} + 0.5 \, \Gamma_{\omega} \tan \left[\frac{(2\,r-1 )\,\pi}{2} \right],
\label{eqn:shape}
\end{equation}
where $M_x$ is the recoiling particle mass which is equaled to $M_{\omega}$; $m_\omega$ and $\Gamma_{\omega}$ represent rest mass and width of the $\omega$; $r$ indicates the randomly generated number in the range [0, 1].


The $\omega$ simulated cross section used to extract the experimental cross section (described in Sec.~\ref{sec:extracting_LT_cx}) by comparing the measured and simulated events, is also generated using the same function form and parameters.






\subsection{$\rho^0$ Production Model}
The $^1$H$(e,e^{\prime}p)\rho^0$ reaction contributes significantly to the broad physics background underneath the $\omega$ peak due to its wide rest mass distribution. The rest mass of the $\rho^{0}$ is $m_{\rho^0}$ = 775.8~MeV, which is similar to the mass of the $\omega$, but with a much wider width of $\Gamma_{rho^{0}}$ = 150.3~MeV. The $\rho^0$ is also a vector meson which holds the same $J^{\rm PC}$ quantum number as the $\omega$ meson ($J^{\rm PC}$=1$^{--}$), but with a different $I^{\rm G}=0^{-}$ quantum number.

The $\rho^0$ electroproduction model in SIMC was adopted from the one developed by the HERMES collaboration~\cite{hermes09}. The model was modified to fit the smooth background underneath the $\omega$ peak. The $\nu$ (energy of the virtual photon) and $Q^2$ dependent part of the cross section is given below~\cite{gaskell16,hermes_monte}:
\begin{equation} 
	\sigma(\nu,~Q^2) = \frac{41.263}{\nu^{0.4765}} \left[1.0 + 0.33 \, \epsilon \, \left(\frac{Q^2}{m_{\rho^0}^2}\right)^{0.61}\right]\,\left( \frac{m_{\rho^0}^2}{Q^2 + m_{\rho^0}^2} \right) ^{2.575}.
\label{eqn:rho_cx}
\end{equation}
The differential cross section which includes the $t$ dependence is given as follows:
\begin{equation}
	\frac{d^2\sigma}{dtd\phi}= \frac{\sigma(\nu,~Q^2)}{2~\pi} \, B_\rho \, e^{-B_\rho t}\,.
	\label{eqn:rho}
\end{equation}
Note that unit of the cross section in Eqn.~\ref{eqn:rho_cx} is in $\mu$b/GeV$^2$. The shape of the $t$ dependence is inspired by CLAS-6 data from Hall B at JLab~\cite{hajidakis}. The fit parameter $B_\rho$ takes different values depending on the $\Delta \tau \cdot c$ value, where $\Delta \tau$ signifies the life time of the intermediate (exchanged) particle and the $\Delta \tau \cdot c$ is equivalent to $\Delta x$ (spatial distance) according to the Heisenberg uncertainty principle, and can be determined as $$\Delta \tau \cdot c = \frac{\hbar~c}{\Delta E} = \frac{\hbar~c}{\sqrt{\nu^2 +Q^2+m^2_\rho}-\nu}\,,$$ where the Planck constant is defined as $\hbar c= 197.32697$~MeV$\cdot$fm. If $\Delta \tau \cdot c$ $<$ 2.057~fm:
$$B_\rho = -0.0941 + 3.449\,\cdot  \Delta \tau \cdot\,c,$$
and for $\Delta \tau \cdot c$ $\ge$ 2.057~fm: 
$$ B_\rho = 7.0.$$

Since the rest mass spectrum of the $\rho^0$ meson overlaps with the multiple pion production phasespace, an additional correction factor known as the Soding factor (model)~\cite{bauer78}, is required to account for the skewing of the $\rho^0$ mass distribution due to the interference between resonant and non-resonant pion pair production. In SIMC, the Soding factor is defined as
\begin{equation}
F_s=\left(\frac{M_p}{m_{\rho^0}}\right)^{-4.394+2.366 \cdot |t \cdot 10^{-6}|}\,,
\end{equation}
where $M_p$ is the mass of the proton target; $t$ is in the unit of GeV$^2$. Note that the Soding factor is directly multiplied to the total differential cross section of the $\rho^0$ (Eqn.~\ref{eqn:rho}).

The relativistic Breit-Wigner distribution is used to model the shape $M_m$ distribution of the $\rho^0$ meson~\cite{pdg}, the Breit-Wigner shape factor can be written as:
\begin{equation}
F_{BW}= \frac{(m_{\rho^0}\Gamma_{\rho^0})^2}{(M_x^2 - m_{\rho^0}^2)^2 +(m_{\rho^0} \Gamma_{\rho^0})^2},
\end{equation}
where $m_{\rho^0}$ and $\Gamma_{\rho^0}$ correspond the rest mass and width of the $\rho^0$; $M_x$ indicates the recoiling particle  mass. Since cross section is integrated over the $\omega$ peak, an additional normalization factor is required to correct the Breit-Wigner shape factor:
\begin{equation}
F^{\prime}_{BW} = F_{BW} \frac{2}{\pi\Gamma_{\rho^0}}.
\end{equation}






\subsection{Two-$\pi$ Production Phasespace Model}

The distribution of two-non resonant pion pair production reaction (phasespace): $^1$H$(e,e^{\prime}p)\pi\pi$, contributes to the broad physics background similarly to the $\rho^0$. The two-$\pi$ phasespace (later referred to as $\pi\pi$) model was derived for the Hall C $\omega$ production experiment near the resonance region  by Ambrosewicz et al.~\cite{ambrosewicz}, and can be written as  
\begin{equation}
	\frac{d^2\sigma}{d\Omega^*dM_x} = \frac{1}{32\,\pi^2} \cdot \frac{M_x}{q^{*}} \cdot \frac{p^{*}}{W^2}\,,
\end{equation}
where $M_x$ indicates the recoil mass due to the two pion production process; $q^{*}$ is the virtual photon momentum in the CM frame. Here, the unit of differential cross section is in $\mu$b/MeV/sr. The details regarding the derivation of the two pion production phasespace formalism can be found in Ref.~\cite{ambrosewicz}.

\subsection{$\eta$ and $\eta^{\prime}$ Production Models}

Comparing to the $\rho$ and two pion exchange phasespace, the contributions of $\eta$ and $\eta^{\prime}$ to the physics background underneath the $\omega$ are much less significant.

$\eta$ and $\eta^{\prime}$ are a pair of closely related pseudoscalar mesons with the common $J^{\rm PC}$ quantum number of $0^{-+}$. $\eta$ has a rest mass of $m_\eta=$ 547.86~MeV and extremely narrow width of 1.3~keV. $\eta^{\prime}$ has a rest mass of $m_{\eta^{\prime}}=$ 957.78~MeV with width of 0.3~MeV.


Based on the SU(3) symmetry of the quark model which involves the three lightest quarks, the following particle (states) are predicted: 
$$\eta_1 = \frac{u\overline{u}+d\overline{d}+s\overline{s}}{\sqrt{3}}\,$$ and $$\eta_8 = \frac{u\overline{u}+d\overline{d}-2s\overline{s}}{\sqrt{6}}\,,$$ where $\eta_1$ belongs to a singlet quark flavor state and $\eta_8$ is the octet state.

The $\eta$ and $\eta^{\prime}$ can be described as the eigenstate mixing of the $\eta_1$ and $\eta_8$ states. The linear combination of the quarks can be written as 
\begin{equation}
  \binom{\cos \theta_\textrm{P} ~~~-\sin \theta_\textrm{P}}{\sin \theta_\textrm{P} ~~~~~~~ \cos \theta_\textrm{P}} \binom{\eta_8}{\eta_1} = \binom{\eta}{\eta^{\prime}}\,,
\end{equation}
where the mixing angle $\theta_\textrm{P}=-11.5^\circ$~\cite{pdg}. The $\eta$ and $\eta^{\prime}$ quark content can be written below:  
$$\eta = \eta_8 ~ \cos\theta_{\textrm{P}} - \eta_1 ~ \cos\theta_{\textrm{P}} \approx \frac{u\overline{u}+u\overline{u}-2s\overline{s}}{\sqrt{6}}\,,$$ and $$\eta^{\prime} = \eta_8 ~ \sin\theta_{\textrm{P}} + \eta_1 ~ \sin\theta_{\textrm{P}} \approx \frac{u\overline{u}+d\overline{d}+s\overline{s}}{\sqrt{3}}\,.$$

The $^1$H$(e,e^{\prime}p)\eta$ and $^1$H$(e,e^{\prime}p)\eta^{\prime}$ reactions have small contributions to the broad physics background distribution under the $\omega$ peak. Thus, their physics models do not require complicated constraint by the kinematic variables. A simple model which gives a gentle rise in small $-u$ range is used:
\begin{equation}
    \frac{d\sigma}{dt} = a \cdot e^{-b \cdot |u|} + c\,,
\end{equation}
where $a$, $b$ and $c$ are the free fitting parameters. For $\eta$ physics model in SIMC,
\begin{equation*}
a=0.0044, ~~~~~~~  b=5, ~~~~~~~   c=0.000011,
\end{equation*}
and for the $\eta^{\prime}$ physics model,
\begin{equation*}
a=0.0088, ~~~~~~~  b=5, ~~~~~~~   c=0.000022.
\end{equation*}
The unit of the resulting cross section is in $\mu$b/GeV$^2$. The width of the $\eta$ and $\eta^{\prime}$ are constructed in the same way as the $\omega$ (Eqn.~\ref{eqn:shape}).

\section{Event Selection}
\label{sec:event_select}


Similar to the Heep analysis (described Chapter~\ref{chap:heep}), establishing the appropriate $e$-$p$ coincidence event selection criteria is extremely important. The event selection criteria used for the $\omega$ analysis for selecting experimental and simulation events are listed in Table~\ref{cut_tab}. Note that simulation data for different final states are separate, therefore they do not require PID cuts. Among the listed criteria, the spectrometer acceptance and PID are the same as those used for the Heep analysis (see Sec.~\ref{sec:PID_cuts}), and are not discussed in this chapter to avoid repetition.


The identification of $^1$H$(e,e^{\prime}p)\omega$ events depends on the correct selection of electrons and protons in the SOS and HMS spectrometers, and on the precise coincidence timing information for the separation of the true and random coincidence events. The identification of the electrons in the SOS and protons in the HMS are described in Sec.~\ref{sec:PID_cuts}. The $cointime$ spectra for the $\omega$ analysis is sufficiently different from that of the Heep analysis and is discussed in Sec.~\ref{sec:cointime_omega}. Sec.~\ref{sec:diamond} introduces the 2D selection criterion on $W$-$Q^2$ kinematics coverage, this selection criterion is specific to the full L/T separation known as the diamond cut,

\begin{table}[t]
\centering
\setlength{\tabcolsep}{0.4em}
\caption[Summary table of the event selection creteria (cuts) used for the $\omega$ analysis]{Summary table of event selection criteria (cuts) used for the $\omega$ analysis. Top section are the standard PID cuts on the PID (Cherenkov + calorimeter) detectors of the SOS and HMS spectrometer. Middle section shows the standard spectrometer acceptance cuts. $^*$ indicates the common event selection criteria used by both Heep and $\omega$ analysis.}
\label{cut_tab}

\begin{tabular}{lcccc}
\toprule
Parameter        	          & Label and cuts             & Experiment & Simulation & Reference                     \\ \toprule 
HMS Cherenkov~$^*$            & $hcer\_npe < 0.5        $  & \checkmark &            & Ref.~\cite{horn,piminus}      \\
HMS Aerogel~$^*$              & $haero\_su < 4          $  & \checkmark &            & Ref.~\cite{horn,piminus}      \\
SOS Calorimeter~$^*$          & $ssshtrk   < 0.70       $  & \checkmark &            & Ref.~\cite{horn,piminus}      \\
SOS Cherenkov~$^*$            & $scer\_npe < 0.50       $  & \checkmark &            & Ref.~\cite{horn,piminus}      \\ \midrule
HMS $|\delta|$~$^*$           & $abs(hsdelta) \le 8.0   $  & \checkmark & \checkmark & Ref.~\cite{horn,piminus}      \\ 
HMS $|y_{tar}|$~$^*$          & $abs(hsytar)  \le 1.75  $  & \checkmark & \checkmark & Ref.~\cite{horn,piminus}      \\ 
HMS $|x^{\prime}_{tar}|$~$^*$ & $abs(hsxptar) \le 0.080 $  & \checkmark & \checkmark & Ref.~\cite{horn,piminus}      \\ 
HMS $|y^{\prime}_{tar}|$~$^*$ & $abs(hsyptar) \le 0.035 $  & \checkmark & \checkmark & Ref.~\cite{horn,piminus}      \\ 
SOS $|\delta|$~$^*$           & $abs(ssdelta) \le 15.   $  & \checkmark & \checkmark & Ref.~\cite{horn,piminus}      \\ 
SOS $|y_{tar}|$~$^*$          & $ssytar \le 1.5         $  & \checkmark & \checkmark & Sec.~\ref{sec:heep_cuts}      \\ 
SOS $|x^{\prime}_{tar}|$~$^*$ & $abs(ssxptar) \le 0.04  $  & \checkmark & \checkmark & Ref.~\cite{horn,piminus}      \\ 
SOS $|y^{\prime}_{tar}|$~$^*$ & $abs(ssyptar) \le 0.065 $  & \checkmark & \checkmark & Ref.~\cite{horn,piminus}      \\ 
SOS $|xfp|$~$^*$              & $abs(ssxfp)   \le 20.   $  & \checkmark & \checkmark & Ref.~\cite{horn,piminus}      \\ \midrule
Coincidence timing (ns)       & Defined in the text        & \checkmark &            & Sec.~\ref{sec:cointime_omega} \\ \midrule
Missing mass ($M_m$)          & $missmass> 0.65         $  & \checkmark & \checkmark & Sec.~\ref{sec:sim_physics}   \\ 
Diamond ($W$ and $Q^2$) cut   & Defined in the text        & \checkmark & \checkmark & Sec.~\ref{sec:diamond}        \\ 
\bottomrule
\end{tabular}
\end{table}

\subsection{Particle Speed in the HMS vs. Coincidence Time}

\label{sec:cointime_omega}

\begin{figure}[t]
  \centering
  \includegraphics[width=0.8\textwidth]{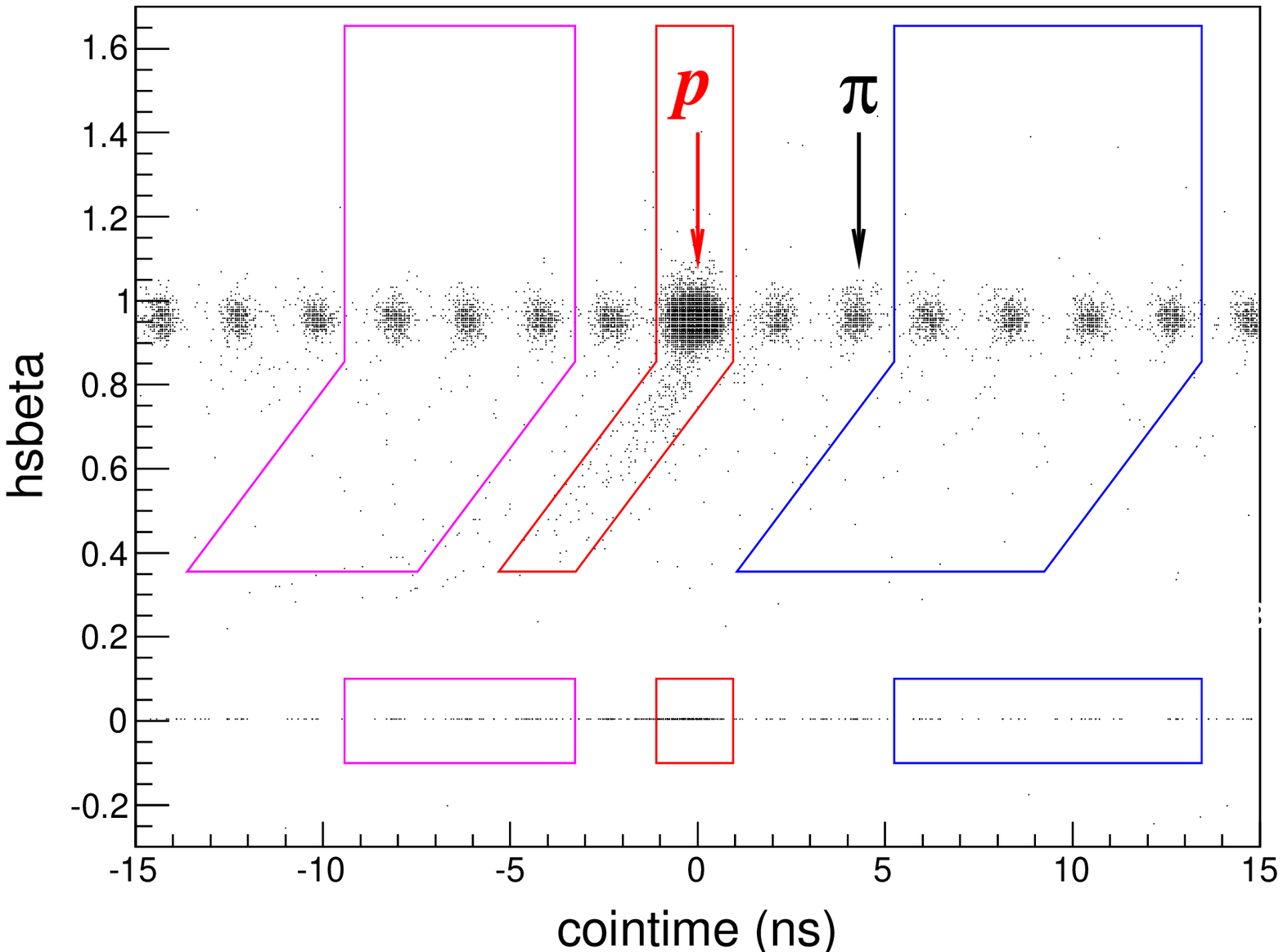}  
  \caption[$hsbeta$ versus $cointime$ for all $\omega$ runs]{$hsbeta$ versus $cointime$ distribution at $Q^2$ = 1.6~GeV$^2$, $\epsilon$ = 32, $\theta_{pq}$ = 3$^\circ$. The coincidence time offset is corrected at a run by run basis. Red box shows the real coincidence time box with width of 2.1~ns; blue box shows the early random coincidence time box with width of 8.4~ns; magenta box shows the late random coincidence time box with width of 6.3~ns. The box boundary positions are fixed across all settings. The acceptance cut and PID cut for selecting the $e$-$p$ coincidence events are applied. The black arrow indicates the region where the real $e-\pi$ coincidence events are expected, due to the applied PID cut, the $e-\pi$ events are significantly suppressed. Random coincidence window intentionally avoided the $e-\pi$ location to prevent any potential event leakage (contamination).~\oic}
  \label{fig:cointime}
\end{figure}

As described in Sec.~\ref{sec:cointime_heep}, the most effective selection criterion for the proton coincidence events is by examining the correlation between the relative particle velocity ratio inside of the HMS and the coincidence timing information ($hsbeta$-$cointime$). Conceptually, the same $hsbeta$-$cointime$ technique used for the Heep study (Sec.~\ref{sec:cointime_heep}) can be directly applied to the $\omega$ analysis. However, as shown in Fig.~\ref{fig:cointime}, the level of random coincidence background is much higher for the $\omega$ production data, therefore wider range random coincidence time windows are selected for random coincidence background subtraction. The blue boxes show the early random coincidence time window, which is 8.4~ns wide, and the magenta boxes show the late coincidence time window, which is 6.3~ns wide. The red boxes are the real coincidence windows (2.1~ns wide).

Note that after the coincidence proton events are selected, the random coincidence and dummy target contributions must be subtracted from the total experimental yield. These background subtraction procedures are identical as those described in Secs.~\ref{sec:heep_rand} and \ref{sec:heep_dummy}.

\subsection{A Diamond Cut on the $W$-$Q^2$ Coverage}
\label{sec:diamond}

The choice of kinematics for the experiment is based on maximizing the coverage in $Q^2$ at high values of the invariant mass $W$ (far above the resonance region: $W>2$~GeV), as well as differentiating the photon polarization $\epsilon$ between the two measurements. One of the measurements would be taken at a low electron beam energy (corresponds to the low $\epsilon$ value) and the other measurement would be at a high electron beam energy (corresponds to the high $\epsilon$ value). This makes the L/T separation at a given $Q^2$ setting possible, see Sec.~\ref{sec:LT_sep} for more detailed explanation regarding the experimental methodology on the L/T separation. The kinematic constraints for a given experimental measurement were imposed by the maximum achievable electron beam energy (5.7~GeV), the maximum central momentum of the SOS (1.74 GeV/c), the minimum HMS angle (10.5$^\circ$) and the minimum angle separation between the two spectrometers (30.5$^\circ$). The choice was made to keep the central value of $W$ constant for both $Q^2$ measurements. 

The nominal $W$ value for the $\omega$ sub-set of the F$_\pi$-2 data was 2.21~GeV, the nominal $Q^2$ values were 1.6 and 2.45~GeV$^2$, as shown Table~\ref{tab:kin_tab}.

\begin{figure}[t]
  \centering
  \includegraphics[width=0.9\textwidth]{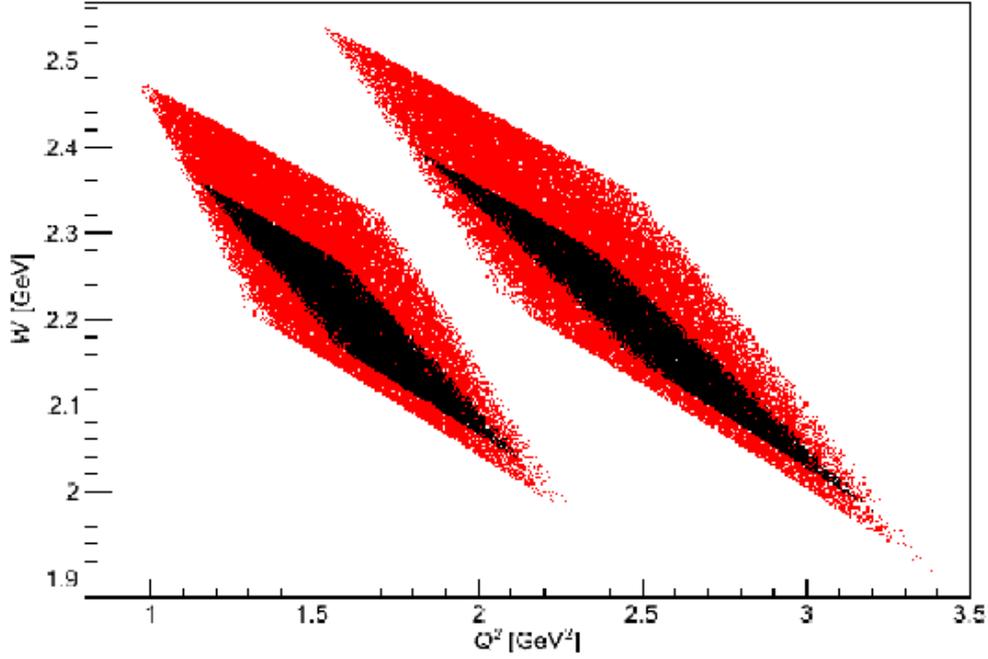}   
  \caption[$W$ versus $Q^2$ kinematics coverage for the $p(e,e^{\prime}p)X$ reaction]{$W$ versus $Q^2$ kinematics coverage for $p(e,e^{\prime}p)X$ reaction. $W$ is plotted in the y-axis; and $Q^2$ is plotted in the x-axis. The plot was generated used all experimental events which survived the acceptance and PID cuts. The left distributions are from the $Q^2=$1.60~GeV$^2$ data set; the right distributions are from the $Q^2=$2.45~GeV$^2$ data set. Black points indicate lower $\epsilon\sim0.30$ data set and red points indicate higher $\epsilon\sim0.57$ data set. ~\oic}
  \label{fig:dia_all_fig}
\end{figure}

Fig.~\ref{fig:dia_all_fig} shows the $W$ and $Q^2$ kinematic coverages for both $Q^2$ settings: $Q^2=$1.60 and 2.45~GeV$^2$. As already discussed, each $Q^2$ setting requires separate measurements at two different electron beam energies ($\epsilon$ values). The higher electron beam energy settings (corresponds to higher $\epsilon$) are shown in red, they provide larger coverages by a factor of three or four compared to those events of the lower beam energy settings (corresponds to lower $\epsilon$) shown in black. This is due to the larger SOS momentum acceptance at higher beam energy, since the percentage of the momentum acceptance remains a constant value but $P_{\rm SOS}$ is raised.

An optimal L/T separation requires the complete overlap (in $W$-$Q^2$ coverages) between the measurements at high and low electron beam energies, and the boundaries of the low beam energy settings are used as a criterion to select the high beam energy events. This data selection criterion is often referred as the `diamond cut' and any events outside of the boundaries are excluded from the analysis. In general, the experiments are designed to collect equal amounts of events within the diamond region, thus achieving comparable statistical uncertainty for the experimental yield at low and high beam energy.

\section{$u$-$\phi$ Binning and Yield Extraction}
\label{sec:binning}

\begin{figure}[p]
  \centering
  \includegraphics[width=1\textwidth]{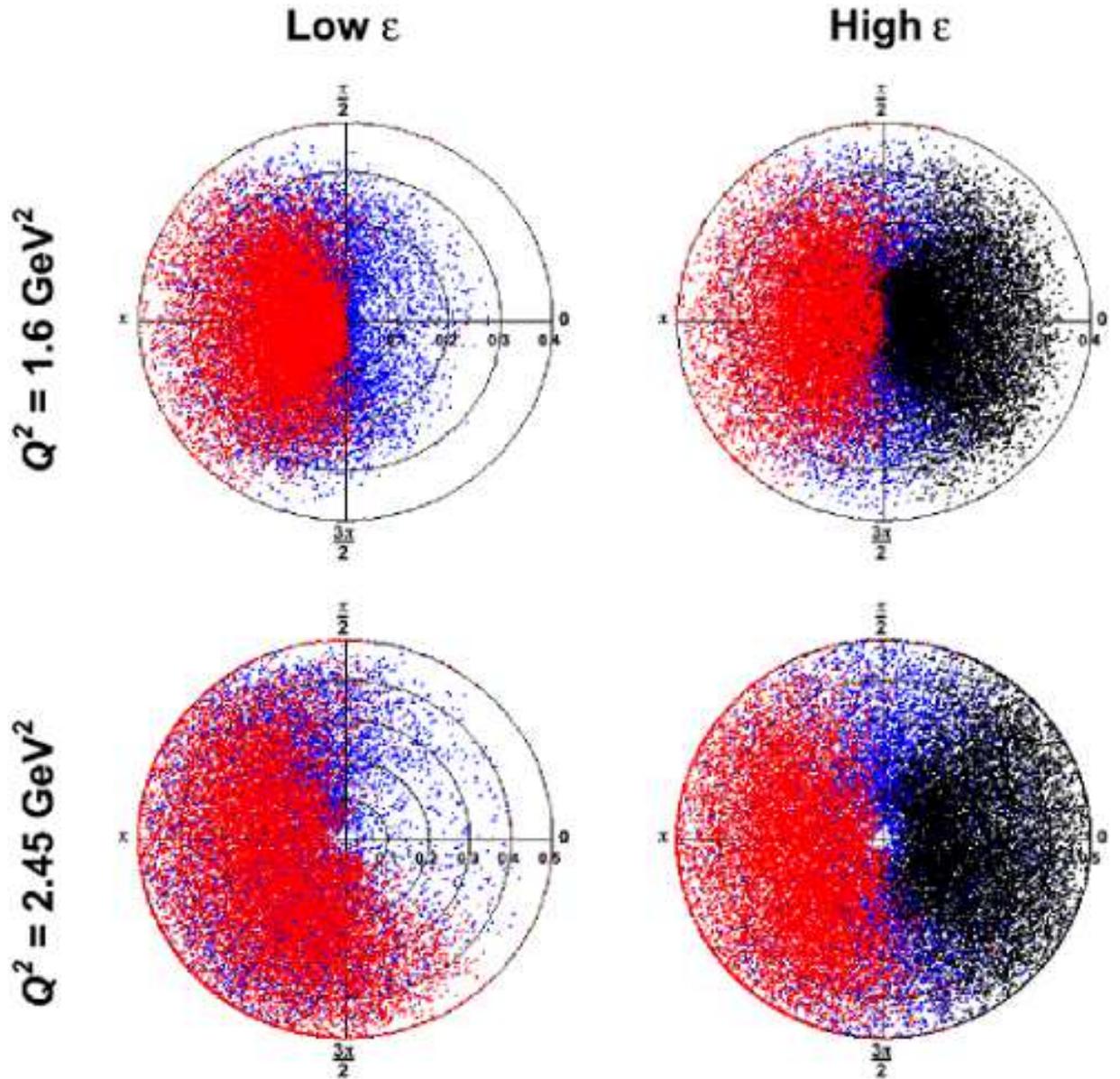}   
  \caption[$u$-$\phi$ polar distributions]{$u$-$\phi$ polar distributions for four combinations of $Q^2$ and $\epsilon$ settings. $-u$ is plotted as the radial variable and $\phi$ as the angular variable. The plots in the first row show $u$-$\phi$ distributions for $\epsilon$ = 0.32 and $\epsilon$ = 0.59 settings at $Q^2$ = 1.60~GeV$^2$; the second row plots show $u$-$\phi$ distributions for low and high $\epsilon$ = 0.27 and $\epsilon$ = 0.55 at $Q^2$ = 2.45~GeV$^2$. For the low $\epsilon$ plots, blue points represent data at $\theta_{\rm HMS}$ = 1$^\circ$ and red data points represent data at $\theta_{\rm HMS}$ = $+$3$^\circ$. For the high $\epsilon$ plots, blue points represent data at $\theta_{\rm HMS}$ = 0$^\circ$, black points represent data at $\theta_{\rm HMS}$ = $-$3$^\circ$ and red data points represent data at $\theta_{\rm HMS}$ = $+$3$^\circ$. ~\oic}
  \label{fig:bull_fig}
\end{figure}

Fig.~\ref{fig:bull_fig} shows in a polar coordinate distribution of the $u$-$\phi$ coverage for all four combinations of $Q^2$ (1.6 and 2.45 GeV$^2$) and $\epsilon$ (low and high beam energies) settings. The Mandelstam variable $-u$ is plotted as the radial component and the polar angle of the recoil proton $\phi$ is plotted as the polar component. Assuming a given setting has minimum $-u$ value of $-u_{\rm min}$ = 0, the ``bullseye'' of the distribution represents the direction of the incoming $\gamma^*$ ($q$-vector) at the $\theta_{pq}=0$ (nominal) angle setting. In this analysis, the nominal $-u_{\rm min}$ values for $Q^2$ = 1.6 and 2.45~GeV$^2$ are $-u_{\rm min}$ = 0.083 and 0.170~GeV$^2$, respectively. An intuitive demonstration of $u$, $\phi$ and $q$-vector on the scattering-reaction planes are shown in Fig.~\ref{fig:plane}.


Even though the spectrometer setting at $\theta_{pq}$ = 0$^{\circ}$ is centered with respect to the $q$-vector, which corresponds to the parallel scenario for proton (anti-parallel for $\omega$), the spectrometer acceptance of the HMS (proton arm) is not wide enough to provide uniform coverage in $\phi$ (blue events). A complete $\phi$ coverage over a full $u$ range is critical for the extraction of the interference terms (LT and TT) during the L/T separation procedure (see Sec.~\ref{sec:LT_sep}). To ensure an optimal $\phi$ coverage, additional measurements were required at the $\theta_{pq}$ = $\pm$3$^{\circ}$ HMS angles (shown as the black and red events). Constrained by the minimum HMS angle from the beam line of $\theta_{\rm HMS} = 10.5^{\circ}$,  the $\theta_{pq}$ = $0^{0}$ and $-$3$^{\circ}$ measurements were impossible at the low $\epsilon$ setting, therefore only $\theta_{pq}$ = 1$^{\circ}$ and $+$3$^{\circ}$ spectrometer angle measurements were performed. For each
setting, the $\theta_{pq}$ and $\theta_{\rm HMS}$ are shown in Table~\ref{tab:kin_tab}. Despite the lack of full $\phi$ coverage at the low $\epsilon$ settings, the full $\phi$ coverage at high $\epsilon$ and use of simulated distributions from SIMC are sufficient to determine the interference components (LT and TT) of the differential cross section.


For each $Q^2$-$\epsilon$ setting shown in Fig.~\ref{fig:bull_fig}, after populating events in $u$-$\phi$ space and obtaining a good $\phi$ coverage around the $q$-vector after combining the statistics from three (or two) HMS angle measurements, the event distribution (looks like a disk or pizza) is divided into three uneven $u$ bins (crusts), and each $u$ bin (crust) is further divided into eight even $\phi$ bins (segments) from 0 to 360$^\circ$ in 45$^{\circ}$ steps. Thus, there are 24 separate bins (divisions) for each $Q^2$-$\epsilon$ setting. 

\begin{figure}[h!]
  \centering
  \subfloat[][$Q^2=1.60$~GeV$^2$, $\epsilon=0.33$]{\includegraphics[width=0.5\textwidth]{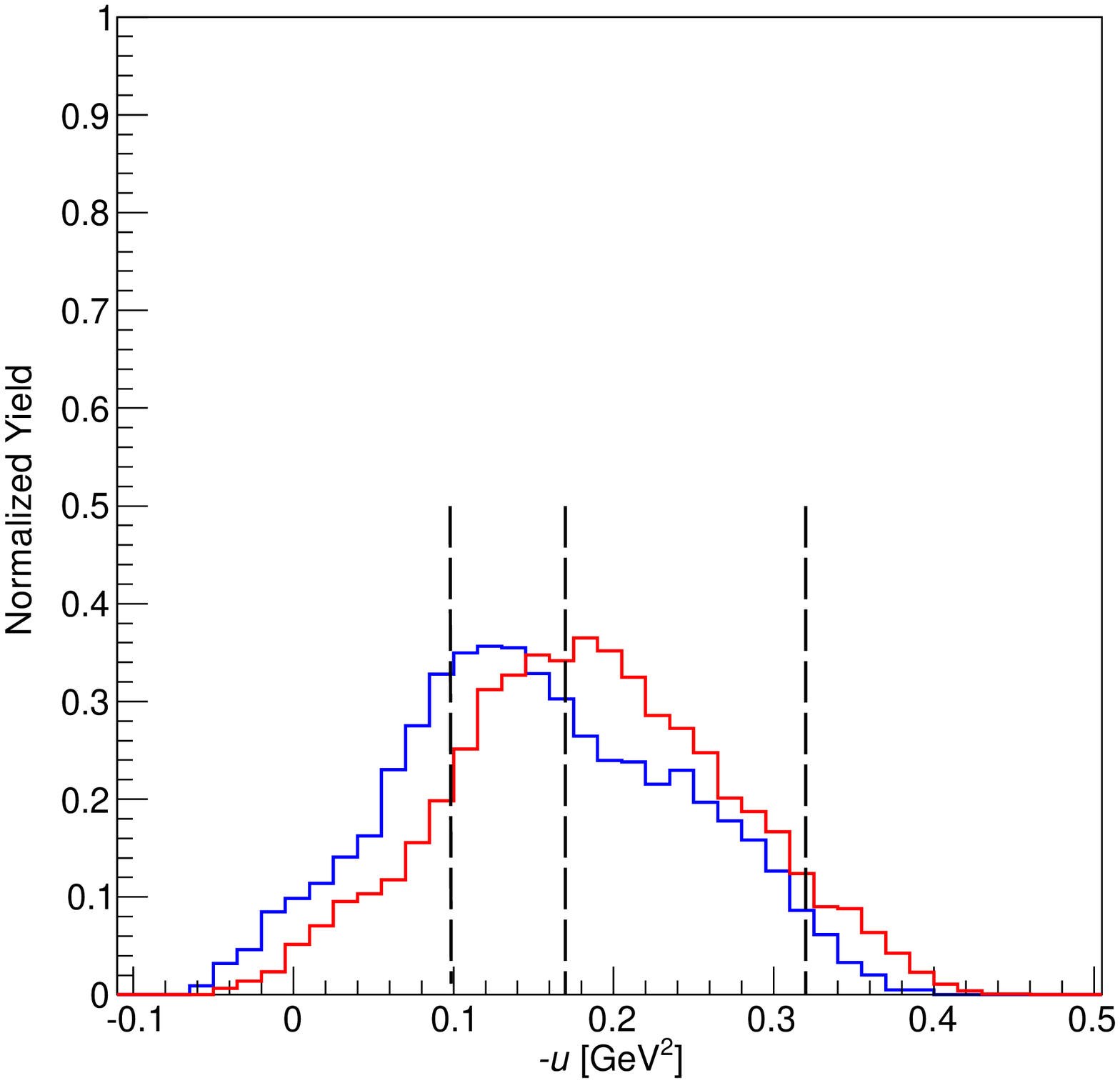}} 
  \subfloat[][$Q^2=1.60$~GeV$^2$, $\epsilon=0.59$]{\includegraphics[width=0.5\textwidth]{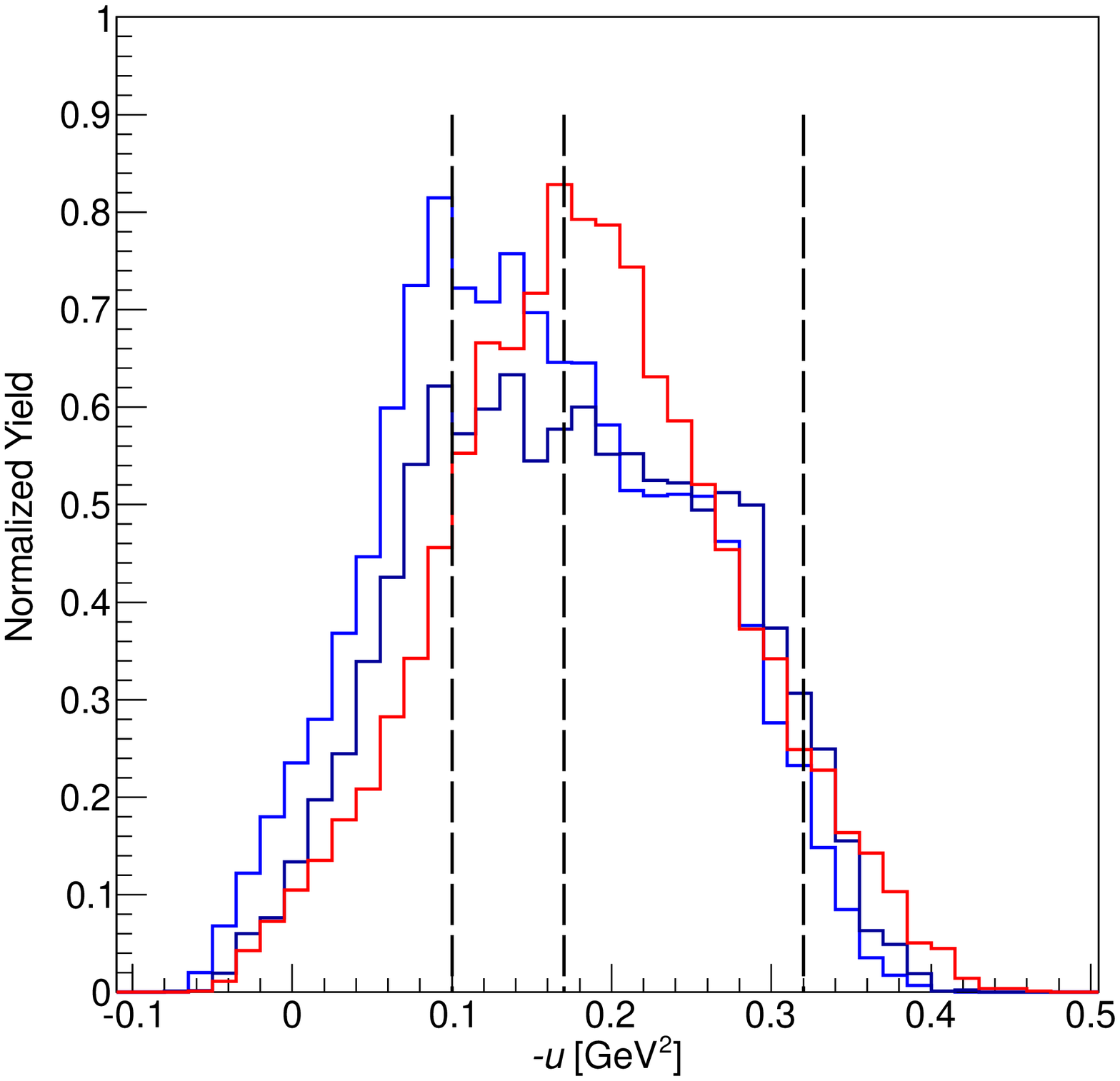}} \\
  \subfloat[][$Q^2=2.45$~GeV$^2$, $\epsilon=0.27$]{\includegraphics[width=0.5\textwidth]{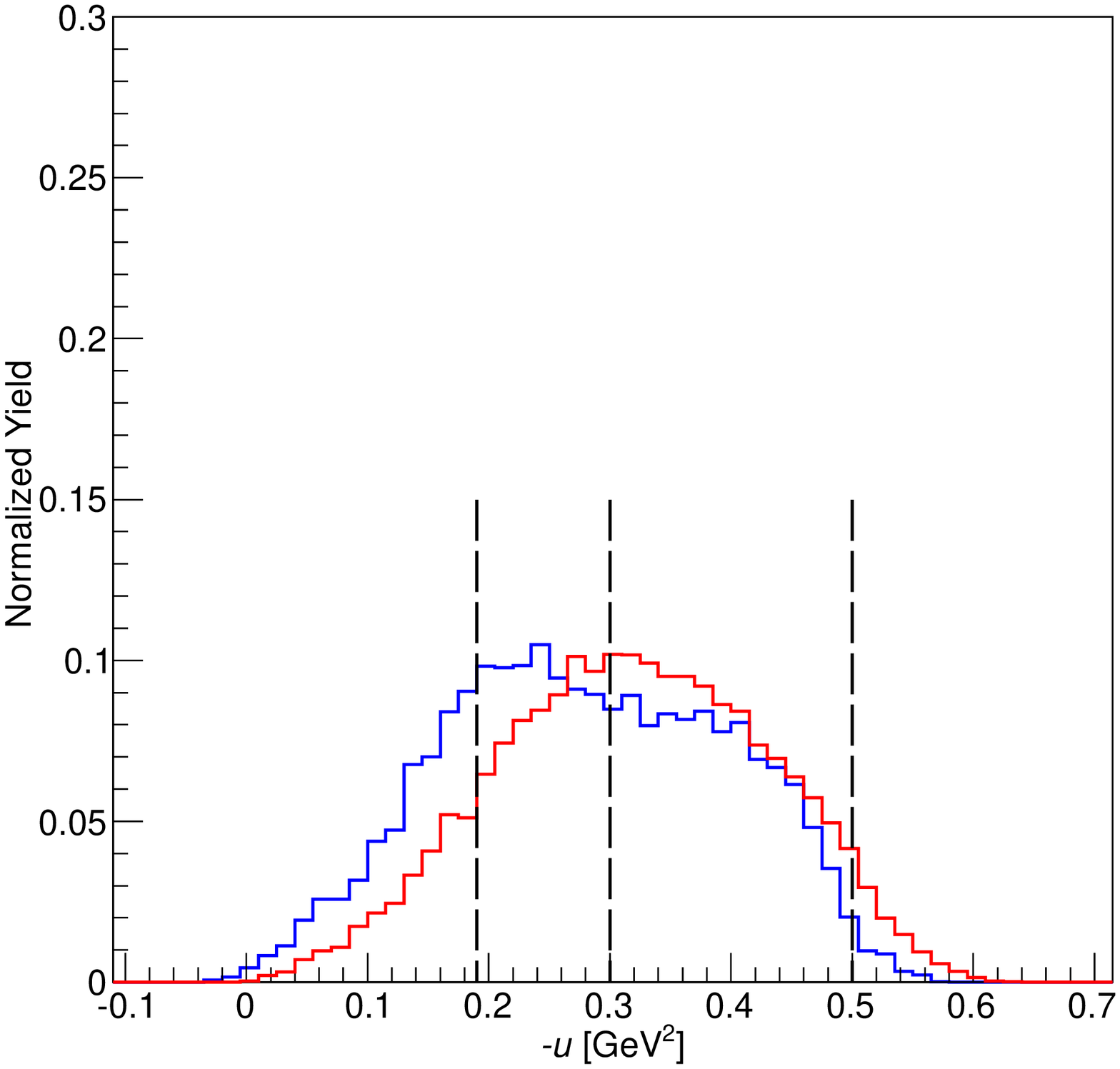}} 
  \subfloat[][$Q^2=2.45$~GeV$^2$, $\epsilon=0.55$]{\includegraphics[width=0.5\textwidth]{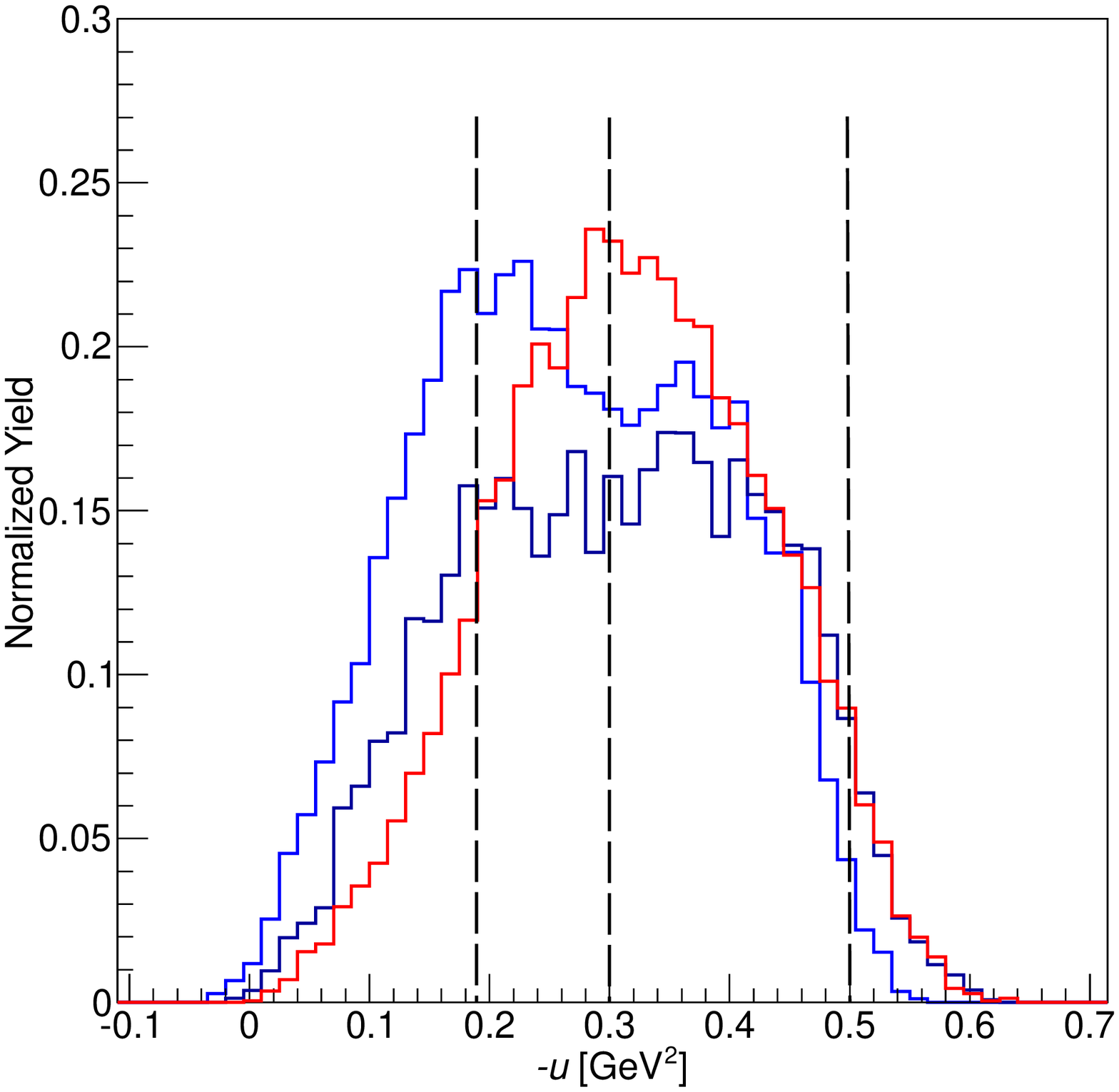}} 
  \caption[$u$ distributions and boundaries]{$u$ distributions for each $Q^2$-$\epsilon$ combinations are shown. The normalized yield (to 1~mC of beam charge) is plotted in $y$-axis. The actual values of $Q^2$ and $\epsilon$ are labeled under each plots. The same color scheme as in Fig.~\ref{fig:bull_fig} is used. The black dashed lines indicate the boundaries between the different $u$ ranges, given in Table~\ref{tab:ubin}.}
  \label{fig:u_bin_limits}
\end{figure}

The determination of the $u$ bin boundaries is based on the principle of ensuring equal statistics for the $\omega$ events among all $u$ bins. Each $Q^2$ setting has different $u$ bin coverages and the boundary values. The $u$-distribution and bin boundary limits for all $Q^2$-$\epsilon$ combinations are shown in Fig.~\ref{fig:u_bin_limits} and the boundary values for the $u$ bins are given in Table \ref{tab:ubin}. Note that events exceeding the upper limit of the third $u$-bin ($-u$ $>$ 0.32 for $Q^2$ = 1.6~GeV$^2$ and $-u$ $>$ 0.50 for $Q^2$ = 2.45~GeV$^2$) are excluded from the analysis for background rejection purpose, since the edge of the simulated $\omega$ distribution is far below these limits. Note that the same treatment was applied to the both experimental and the simulation data (same software). Each ($Q^2$, $\epsilon$, $\theta_{pq}$, $u$, $\phi$) bin requires independent analysis, involving reconstruction of the $M_m$ distribution and computation of normalized yield.

\begin{table}[t]
\centering
\footnotesize
\setlength{\tabcolsep}{2em}
\caption[$-u$ Bin boundaries for the $\omega$ analysis]{$-u$ Bin boundaries for the $\omega$ analysis. The central value for each $u$ bin is shown in the square bracket. $^*$ During the analysis, the first $-u$ bin limit are set to be $-u$ $<$ 0.1 and $u$ $<$ 0.19 to include to include events with $-u$ $<$ 0. The reason for this change is further elaborated in Sec.~\ref{sec:u_less_zero}.}
\label{tab:ubin}
\begin{tabular}{cccc}
\toprule
Q$^2$       &  \multicolumn{3}{c}{$-u$ Bin Boundary}                                     \\ 
GeV$^2$     & 1st Bin (GeV$^2$)         & 2nd Bin (GeV$^2$)     &  3rd Bin (GeV$^2$)     \\ \midrule
1.60        & 0.00-0.10 $^*$ ~ [0.050]   & 0.10-0.17 ~ [0.135]   &  0.17-0.32  ~ [0.245]  \\
2.45        & 0.00-0.19 $^*$ ~ [0.110]   & 0.19-0.30 ~ [0.245]   &  0.30-0.50  ~ [0.400]  \\
\bottomrule
\end{tabular}
\end{table}

The normalized experimental and simulation yield (to 1~mC beam charge) for every ($Q^2$, $\epsilon$, $\theta_{pq}$, $u$, $\phi$) bin needs to be accurately determined. The normalized yields were obtained using the same methodology as in Sec.~\ref{sec:yield_cal}. Note that obtaining an accurate normalized yield ratio requires a good understanding of the overall experimental efficiencies; these efficiencies were determined based on the studies described in Sec.~\ref{sec:heep_eff} and further discussed in Sec.~\ref{sec:uncertainty}.

At this stage, the normalized experimental yield ratio includes the events not only from the $\omega$ production, but also for all possible $^1$H$(e, e^{\prime}p)X$ final states. The physics background subtraction (Sec.~\ref{sec:bg_sub}) is required to extract the $\omega$ events.

\section{The Fitting Step and Physics Background Subtraction}
\label{sec:bg_sub}

\begin{figure}[t]
  \centering
  \includegraphics[width=0.75\textwidth]{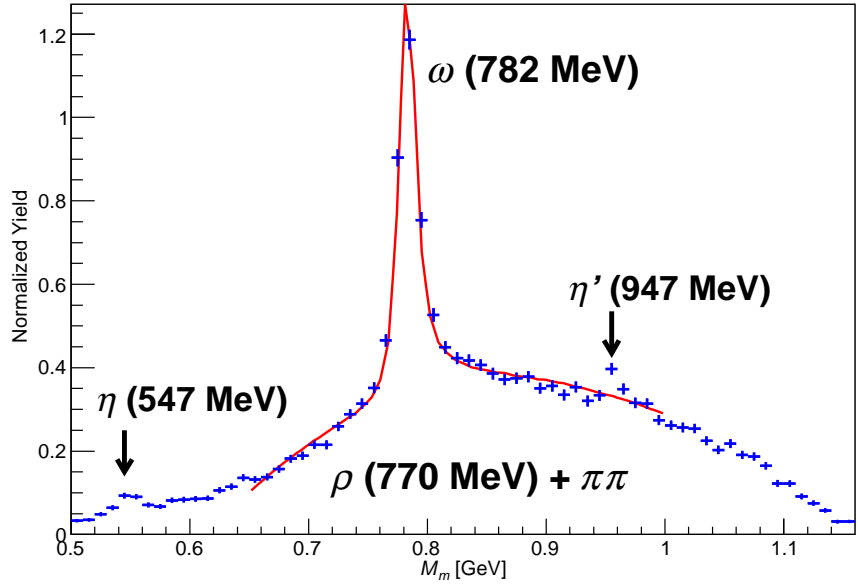}   
  \caption[Missing mass spectrum for $^1$H$(e, e^{\prime}p)X$]{Missing mass spectrum for $^1$H$(e, e^{\prime}p)X$ reaction at $Q^2$ = 2.45~GeV, $\epsilon$ = 0.55, $\theta_{pq}$ = $-$3$^\circ$. The spectrum includes all events from the setting over all 24 ($Q^2$, $\epsilon$, $\theta_{pq}$, $u$, $\phi$) bins. The normalized yield (to 1~mC of electron beam change) is plotted on the $y$-axis and $M_m$ is on the $x$-axis. The main feature of the distribution includes distinctive narrow peak for $\omega$ and a broad background underneath the $\omega$ which contains $\rho$ and two-pion production phase space ($\pi\pi$). Note that $\eta$ and $\eta^{\prime}$ peaks can also be seen in the data distribution in the correct $M_m$ regions. The corresponding rest mass values for these final state particles are shown in the brackets. The red line shows the description of the data by the polynomial fitting method, which involves the $\omega$ simulation and a second second order polynomial. }
  \label{fig:mm_tot}
\end{figure}

The primary reaction of the F$_\pi$-2-$\pi^+$ analysis was exclusive $\pi^+$ production: $^1$H$(e,e^\prime \pi^{+})n$, the reconstructed $M_m$ distribution (centered at the rest mass of the neutron) is distinct and clean with no physics background underneath. In comparison, the reconstructed $M_m$ peak of the $\omega$ electroproduction reaction: $^1$H$(e,e^\prime p)\omega$, has a sharp peak with physics backgrounds underneath the $M_m$ peak, see Fig.~\ref{fig:mm_tot}.

In the $^1$H$(e,e^\prime p)X$ meson production reaction, the final state particle $X$ can be a variety of mesons including: $\omega$, $\rho$, $\eta$, $\eta^{\prime}$ and two-pion production phasespace ($\pi\pi$). The advantage to fit the $M_m$ distribution is the convenience of using the narrow $\omega$ width to establish an effective integration range around the its rest mass, while avoiding over constraining the fitting algorithm by fitting additional physics or kinematic variables such as $P_m$, $W$ and $Q^2$.

Fig.~\ref{fig:mm_tot} shows an example of the reconstructed $M_m$ distribution of the reaction: $^1$H$(e,e^\prime p)X$, which shows the physics background under the primary $\omega$ peak. The selected $M_m$ spectrum is for setting $Q^2=$2.45~GeV, $\epsilon$ = 0.55, $\theta_{pq}$ = $-$3$^\circ$. A sharp peak corresponding to the $\omega$ is at 782~MeV, as expected. As parts of the physics background, the pseudoscalar mesons $\eta$ (547~MeV) and $\eta^{\prime}$ (947~MeV) are visible at their corresponding missing mass ranges. Underneath the $\omega$ peak, a broad background containing the contributions from vector meson $\rho^0$ and two $\pi$ production phasespace is observed.

This section describes the methodology used to subtract the physics background underneath the $\omega$ peak in the $M_m$ distribution and obtain the experimental yield of the $\omega$: $\rm{Y}_{\omega\,{Exp}}$ (defined in Sec.~\ref{sec:bin-by-bin}). Two different fitting methods were attempted to give a description of the broad physics background, both methods are described in Sec.~\ref{sec:fitting_methods}. The bin-by-bin background subtraction is handled by a procedure referred as the fitting step, which is discussed the Sec.~\ref{sec:bin-by-bin}.

\subsection{Bin Exclusion}

\label{sec:bin_exclude}



\begin{figure}[t]
  \centering
  \subfloat[][Example of low statistics bin]{\includegraphics[width=0.49\textwidth]{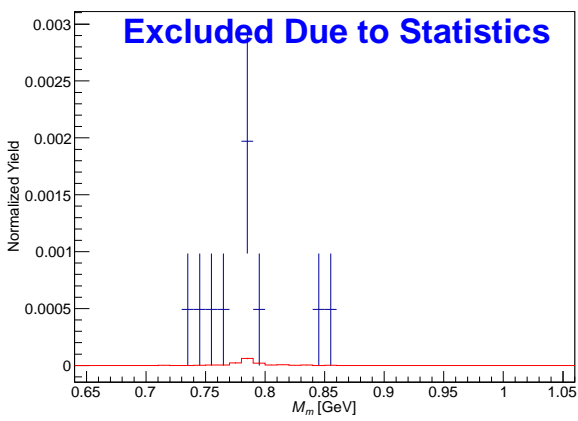}}  ~~
  \subfloat[][Example of excessive radiative tail bin]{\includegraphics[width=0.49\textwidth]{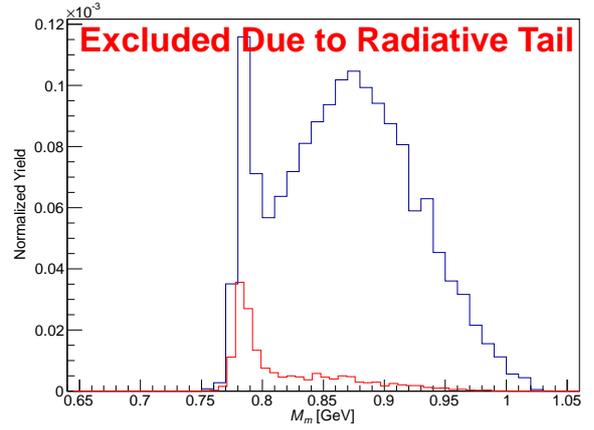}} 
  \caption[Bins excluded from the analysis]{(a) and (b) show examples of excluded bins due to low statistics and excessive amount of radiative tail, respectively. In figure (b), the simulations with and without radiative correction are shown in blue and red distributions, respectively.}
  \label{fig:exclusion}
\end{figure}

Prior to performing the bin-by-bin missing mass distribution fitting procedure, two kinds of bins need to be identified and excluded from the analysis. There are 240 bins (2$Q^2$ $\times$ 5$\theta_{pq}$ $\times$ 3$u$ $\times$ 8$\phi$) in total for the $\omega$ analysis, 149 of them are valid bins with 91 bins excluded from the analysis. The criteria for excluding a given ($Q^2$, $\epsilon$, $\theta_{pq}$, $u$, $\phi$) bin are defined as follows:
\begin{description}

\item[Low statistics:] For a given bin, the raw experimental yield is less than 70 counts after the random and dummy target subtraction. In this case, the $M_m$ distribution cannot be reliably fitted to extract any meaningful scale factors. An example of a low statistics $u$-$\phi$ bin is shown in Fig.~\ref{fig:exclusion} (a). There are 70 bins excluded from the analysis due to low statistics. 
  
\item[Excessive radiative tail:] For a given bin, the simulated $\omega$ peak contains excessive radiative tail which contributes more than 60\% of the overall $M_m$ distribution. The cause of the radiative tail is due to the additional photon emitted by the scattered electron and recoil proton immediately after the primary interaction (described in Sec.~\ref{sec:rad_process}) and the center of the $E_m$ distribution shifts to greater than 0.9 GeV. Due to the uncertainties associated with the radiative correction in the SIMC, the simulation description to experimental data becomes less accurate as the radiative tail grows. An example $u$-$\phi$ bin for the excessive radiative tail is shown in Fig.~\ref{fig:exclusion} (b). There are 21 bins excluded from the analysis due to the excessive radiative tail. 

\end{description}


\subsection{Fitting Methods}
\label{sec:fitting_methods}

\subsubsection{A Failed Attempt: Polynomial Fitting Method}



The most challenging aspect of the $\omega$ analysis is to reliably subtract the physics backgrounds underneath the $\omega$ peak. Conventionally, the polynomial fitting method is sufficient to describe the combined physics background in the $M_m$ distribution, as shown in Fig.~\ref{fig:mm_tot}. The red line shows the fitting result combining the $\omega$ simulation and a smooth second order polynomial function of the form: $$y= a + b~x + c~x^2\,,$$ where $a$, $b$ and $c$ are the free fitting parameters. Despite the fact that the polynomial fit gives a good description for the physics background over a setting, it fails to consistently describe the $M_m$ distributions for every ($Q^2$, $\epsilon$, $\theta_{pq}$, $u$, $\phi$) bin.

\begin{figure}[t]
   \subfloat[][Background at the right side]{\includegraphics[width=0.333\textwidth]{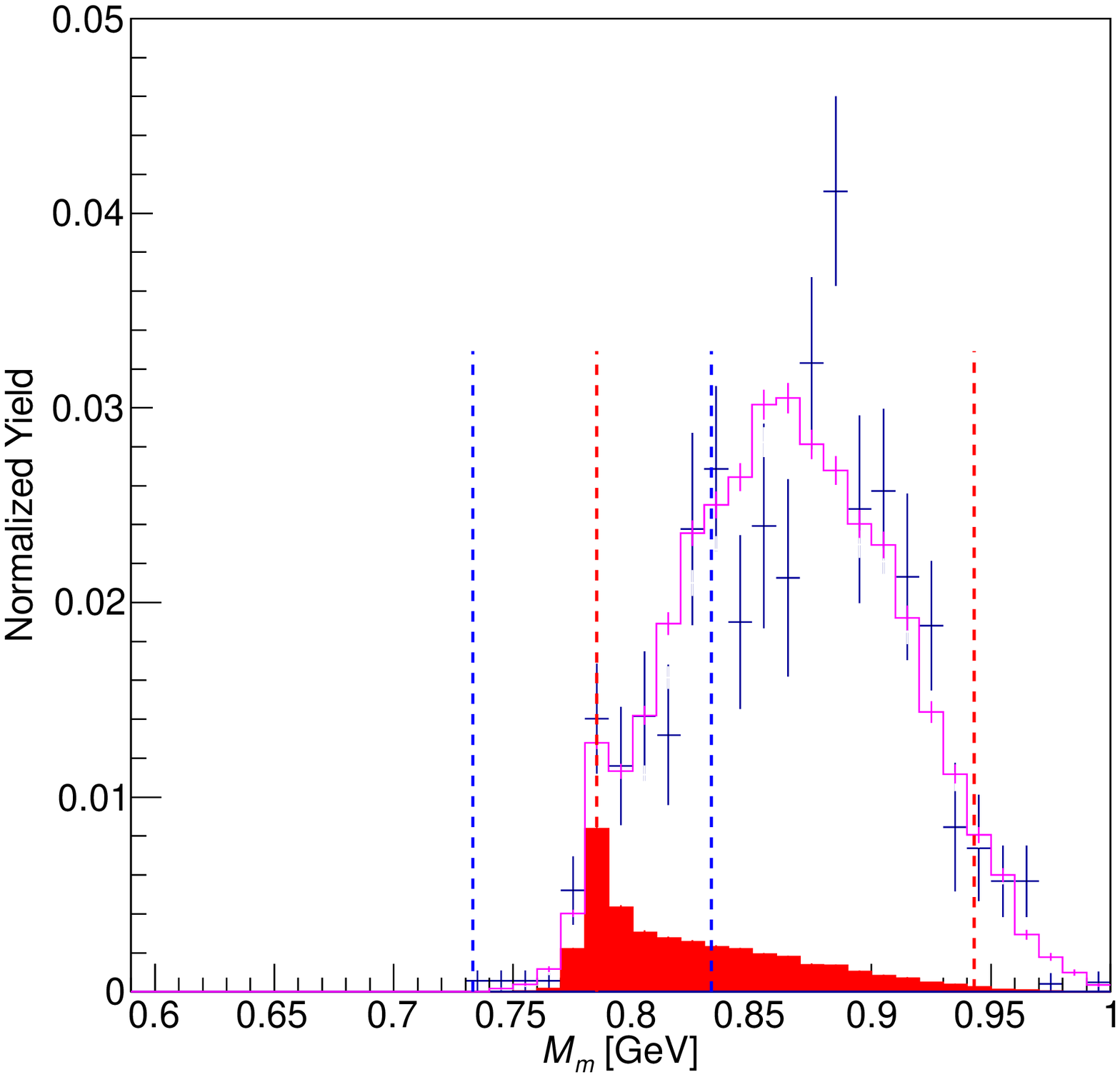}} 
   \subfloat[][Background at the center]{\includegraphics[width=0.333\textwidth]{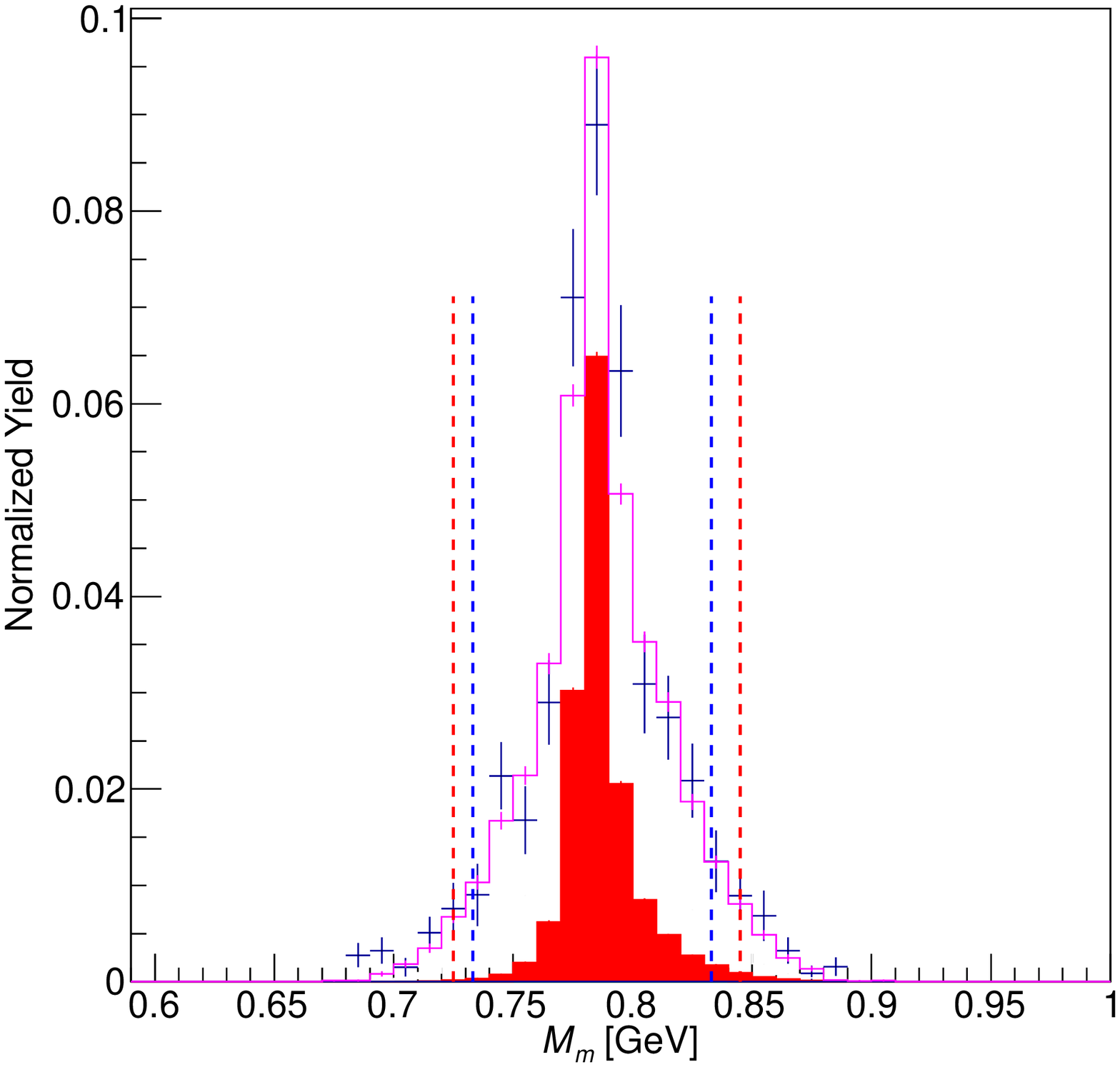}}
   \subfloat[][Background at the left side]{\includegraphics[width=0.333\textwidth]{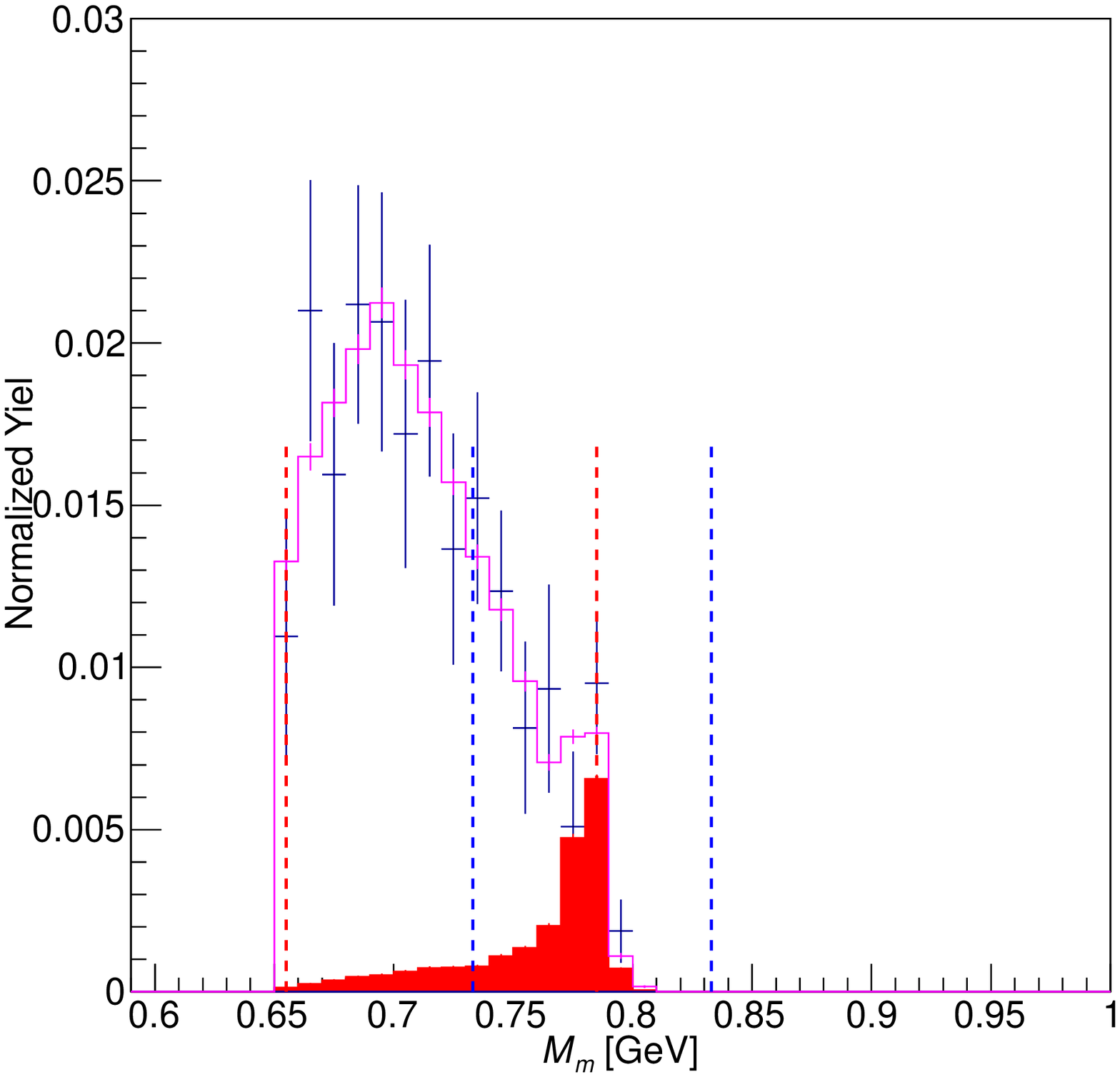}} 
   \caption[Shifting background behavior in M$_m$ distribution]{A set of binned (in $Q^2$, $\epsilon$, $\theta_{pq}$, $u$, $\phi$) missing mass distributions (a), (b) and (c), which demonstrate the shifting behavior of the background at the left, center and right side of the $\omega$ peak, respectively. The normalized yield (to 1~mC of beam charge) is plotted in $y$-axis. The blue crosses indicate data and the relative systemic error. The red shaded distribution is the $\omega$ simulation after the fitting scale factor is applied after. The magenta distribution is the sum of the all scaled simulated distributions. Note that $\rho^0$ and other background distributions are not shown in the figure. All three $M_m$ distributions are at $Q^2$ = 1.6~GeV$^2$ and $\epsilon$ = 0.32. Other relevant kinematics variables for (a): $\theta_{pq}$ = 0$^\circ$, $-u$ = 0.245~GeV$^2$, $\phi$ = 112.5$^{\circ}$; for (b): $\theta_{pq}$ = 0$^\circ$, $-u$ = 0.135~GeV$^2$, $\phi$ = 337.5$^\circ$; for (c): $\theta_{pq}$ = $+$3$^\circ$, $-u$ = 0.050~GeV$^2$, $\phi$ = 157.5$^\circ$.~\oic}
  \label{fig:mm_example}
\end{figure}

Fig.~\ref{fig:mm_example} shows three typical selected $M_m$ distribution examples after the ($Q^2$, $\epsilon$, $\theta_{pq}$, $u$, $\phi$) binning. In all three examples, the position of the $\omega$ peak stays close to its expected its rest mass value, however, the broad physics background shifts around the $\omega$ peak depending on the $u$ coverage of the bin and the $\theta_{\rm HMS}$ setting. The unstable appearance of the background position would significantly vary the quality of the polynomial fit; particularly when the $\omega$ peak is close to the edges of the overall distribution, the polynomial fitting method fails completely.

%

\subsubsection{Simulation Fitting Method}

In order to reliably describe the physics background and extract the $\omega$ events in every ($Q^2$, $\epsilon$, $\theta_{pq}$, $u$, $\phi$) bin for a given setting, a different fitting method is required to utilize the simulated distributions of all possible final states particles of $^1$H$(e,e^{\prime}p)X$ reactions.

Recall there are five different possible final states for $^1$H$(e,e^{\prime}p)X$, where $X=\omega$, $\rho$, $\pi\pi$, $\eta$ and $\eta^{\prime}$. For a given $u$-$\phi$ bin, the total normalized simulation yield can be represented as the sum of the individual normalized simulation yields from five possible final states after appropriate scaling, and can be written as 
\begin{equation}
\begin{split}
\textrm{Y}_{\rm SIMC} & = {\textrm{Y}_{\rm \omega \, SIMC}} + \textrm{Y}_{\rho^0\, {\rm SIMC}} + \textrm{Y}_{\pi\pi \, {\rm SIMC}} + \textrm{Y}_{\eta \, {\rm SIMC}} + \textrm{Y}_{\eta^{\prime} \, {\rm SIMC}}   \\
& = a \cdot \mathcal{Y}_{\omega\,{\rm SIMC}} + b \cdot \mathcal{Y}_{\rho^0\,{\rm SIMC}} + c \cdot \mathcal{Y}_{\pi\pi\,{\rm SIMC}} + d \cdot \mathcal{Y}_{\eta\,{\rm SIMC}} + e \cdot \mathcal{Y}_{\eta^{\prime}\,{\rm SIMC}} \,,
\label{eqn:sim_yield}
\end{split}
\end{equation}
where $\mathcal{Y}_{\omega\,{\rm SIMC}}$ is the normalized simulation yield for the $^1$H$(e,e^{\prime}p)\omega$; $\mathcal{Y}_{\rho^0\,{\rm SIMC}}$ is for $^1$H$(e,e^{\prime}p)\rho^0$; $\mathcal{Y}_{\pi\pi\,{\rm SIMC}}$ is for $^1$H$(e,e^{\prime}p)\pi\pi$; $\mathcal{Y}_{\eta\,{\rm SIMC}}$ is for $^1$H$(e,e^{\prime}p)\eta$; $\mathcal{Y}_{\eta^{\prime}\,{\rm SIMC}}$ is for $^1$H$(e,e^{\prime}p)\eta^{\prime}$; $a$-$e$ are the corresponding scale factors (i.e., $a$ for $\omega$) determined by the fitting algorithm; the normalized simulation yield after the scaling ${\textrm{Y}_{\rm \omega \, SIMC}}$, $\textrm{Y}_{\rho^0\, {\rm SIMC}}$, $\textrm{Y}_{\pi\pi \, {\rm SIMC}}$,  $\textrm{Y}_{\eta \, {\rm SIMC}}$ and $\textrm{Y}_{\eta^{\prime} \, {\rm SIMC}}$, are the products of the corresponding individual simulation yield ($\mathcal{Y}_{\omega\,\textrm{SIMC}}$, ...,$\mathcal{Y}_{\eta^{\prime}\,\textrm{SIMC}}$.) and the corresponding scale factor ($a$, ...,$e$.).

Compared to the polynomial fitting method, the simulation fitting method describes the experimental data by adjusting the relative height of the individual simulation distribution through the usage of the scale factors. Therefore, the shape of the simulation distributions are not changed. A significant advantage of the simulation fitting method is its capability of adapting to the kinematics and optical acceptance for each individual bin, and capturing the any shifting of the distribution (demonstrated in Sec.~\ref{sec:q_cross_checks}). By fitting the experimental $M_m$ distribution with five simulated distributions, five scale factors are extracted as described in Eqn.~\ref{eqn:sim_yield}.

The effectiveness of the simulation fitting method greatly relies on the good spectrometer resolution and the quality of the Monte Carlo simulation (SIMC). Both of these characteristics for $e$-$p$ coincidence experiments using the HMS-SOS setup are demonstrated by the experiment-simulation agreement in the Heep analysis, in particular, the reconstructed physics parameters (shown in Figs.~\ref{fig:Heep_E_m}, \ref{fig:Heep_P_m} and \ref{fig:Heep_M_m}) and the yield ratio result (shown in Fig.~\ref{fig:yield_recheck}).


The simulation fitting method offers a bin-by-bin data description from the scaled simulations, and is based on the principle of treating all ($Q^2$, $\epsilon$, $\theta_{pq}$, $u$, $\phi$) bins equally. This means the algorithm applies the same general criteria for all bins and gives no customized accommodation to any given bin. This generalized bin-by-bin fitting algorithm is further described in the next subsection.




\subsection{Fitting Step: A Bin-by-Bin Fitting Algorithm}
\label{sec:bin-by-bin}

\begin{figure}[t!]
  \centering
  \subfloat[][Overall comparison]{\includegraphics[width=0.5\textwidth]{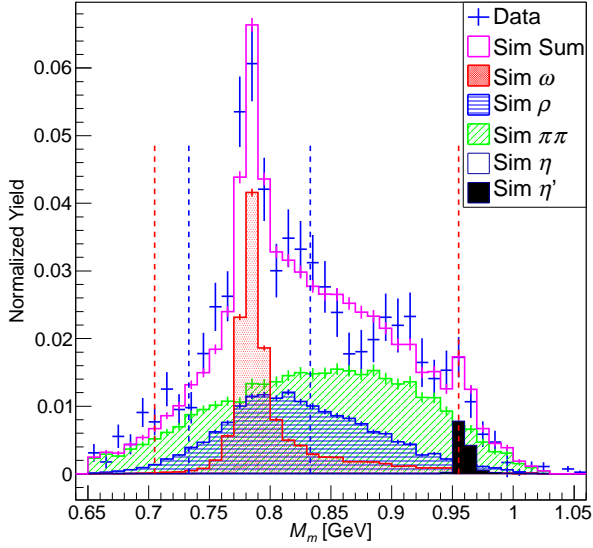}} 
  \subfloat[][Comparison of $\omega$]{\includegraphics[width=0.5\textwidth]{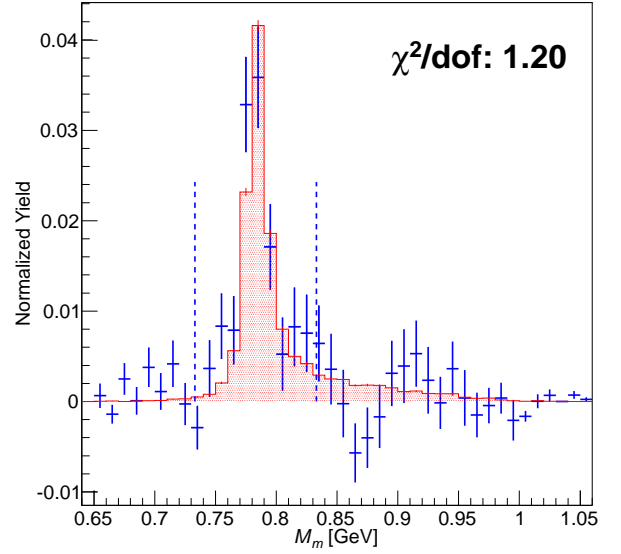}} \\
  \subfloat[][Comparison of background]{\includegraphics[width=0.5\textwidth]{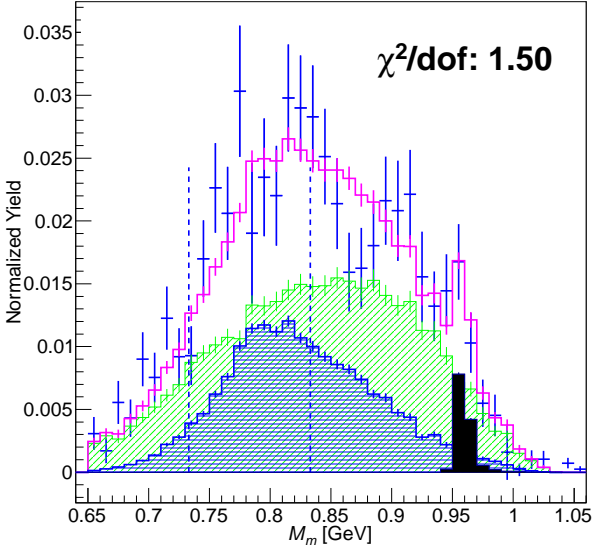}} 
  \subfloat[][Comparison of zero]{\includegraphics[width=0.5\textwidth]{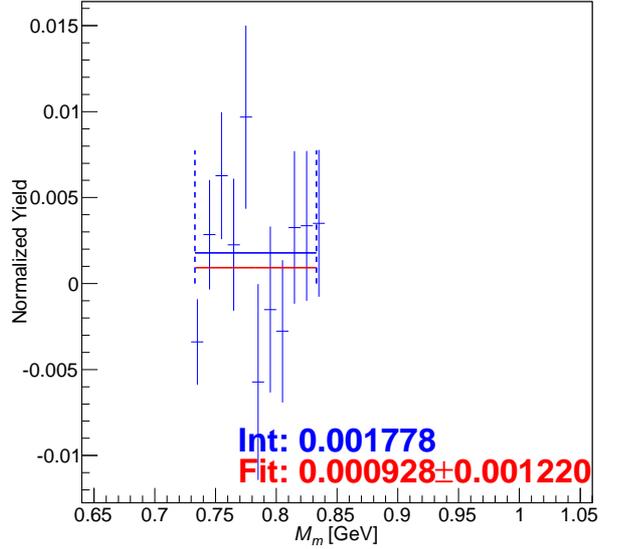}} 
  \caption[The $M_m$ and the cross check distributions for a selected bin]{The $M_m$ and the cross check distributions for a pseudo-randomly selected ($Q^2$, $\epsilon$, $\theta_{pq}$, $u$, $\phi$) bin.  The normalized yield (to 1~mC of beam charge) is plotted in $y$-axis. The chosen bin corresponds to $Q^2$ = 2.45~GeV$^2$, $\epsilon$ = 0.55, $\theta_{pq}$ = $-$3$^\circ$, $-u$ = 0.4~GeV$^2$, $\phi$ = 22.5$^\circ$. Same color scheme applies to all four panels.~\oic}
  \label{fig:bin_by_bin_mm}
\end{figure}


As described in the earlier text, the fitting step is the most critical step in the iterative analysis procedure. Its main purpose is to obtain an adequate bin-by-bin description of the broad background using scaled simulated (four background) distributions and obtain the experimental yield for $\omega$ (${\rm Y}_{\omega~{\rm Exp}}$) through the background subtraction.   

The $M_m$ distribution for a typical ($Q^2$, $\epsilon$, $\theta_{pq}$, $u$, $\phi$) bin is shown in Fig.~\ref{fig:bin_by_bin_mm} (a). The experimental data are shown as the blue crosses. The simulation distributions of $\omega$ (red), $\rho^0$ (blue), $\pi\pi$ phasespace (green), $\eta^{\prime}$ (black) are appropriately scaled and summed to construct the total simulation distribution shown in magenta. From a qualitative visual comparison, the simulation sum gives a good description of the data.  

Note that there two boundary regions indicated by the red and blue dashed lines which are essential to the iterative procedure. The red dashed lines define the fitting range, and the fitting algorithm would only fit the experimental data within this range. The integration range is defined by the blue dashed lines, and is an important component for the integration step, which is introduced in Sec.~\ref{sec:int_step}.


As shown in Figs.~\ref{fig:mm_example} (a), (b) and (c), the behavior (position and shape) of $\omega$ and background peaks vary significantly, depending on the nominal kinematic values for a given bin. Therefore, it is not possible to choose a static fitting limit for the fitting algorithm to describe the data for all ($Q^2$, $\epsilon$, $\theta_{pq}$, $u$, $\phi$) bins simultaneously. After trial and error, a dynamical determination of the fitting limit is implemented in the fitting algorithm, which takes into account the shape and position variation of the data distribution. For a given bin, the fitting algorithm would automatically exclude 4\% from either end of the data distribution, and fit the middle 92\% of the distribution to determine the scale factors ($a$-$e$) defined in Eqn.~\ref{eqn:sim_yield}. The uncertainty associated with the percentage of the excluded distribution from the edge is discussed in Sec.~\ref{sec:uncertainty}.

Fitting the entire (100\%) data distribution was also attempted, however, the sharp drop of the statistics near the edge of the distribution in some ($Q^2$, $\epsilon$, $\theta_{pq}$, $u$, $\phi$) bins would cause the fitting algorithm to fail.


In addition, the fitting algorithm uses different sets of simulation distributions to fit the data, depending on the $M_{m}$ coverage. For $M_m < 0.783$~GeV, the radiative tail from $\eta$ plays a significant role; for $M_m$ around $\eta^{\prime}$ peak ($M_m\sim0.947$)~GeV, the $\eta^{\prime}$ contribution must be taken into account. After some trial and error, the best simultaneous fitting results were achieved to include either $\eta$ or $\eta^{\prime}$ in the fitting, but not both. The determination of whether to include $\eta^{\prime}$ depends on if the integral of data distribution for $M_m > 0.947$~GeV exceeds 10\% of the overall distribution. If the $\eta^{\prime}$ is included in the fitting, the scale factor $e$ for $\eta$ distribution is set to 0, and Eqn.~\ref{eqn:sim_yield} becomes
$$\textrm{Y}_{\rm SIMC} = {\textrm{Y}_{\rm \omega \, SIMC}} + \textrm{Y}_{\rho^0\, {\rm SIMC}} + \textrm{Y}_{\pi\pi \, {\rm SIMC}} + \textrm{Y}_{\eta^{\prime} \, {\rm SIMC}}.$$ If $\eta^{\prime}$ is not included in the fitting algorithm, the $\eta$ would be included instead, and Eqn.~\ref{eqn:sim_yield} becomes $$\textrm{Y}_{\rm SIMC} = {\textrm{Y}_{\rm \omega \, SIMC}} + \textrm{Y}_{\rho^0\, {\rm SIMC}} + \textrm{Y}_{\pi\pi \, {\rm SIMC}} + \textrm{Y}_{\eta \, {\rm SIMC}}.$$
Note that in the example $M_m$ distribution from Fig.~\ref{fig:bin_by_bin_mm} (a), $\eta^{\prime}$ is included in the fitting algorithm for this particular $u$-$\phi$ bin.




\subsubsection{Fitting Quality Control}
\label{sec:fit_quality_control}


It is important to monitor the behavior of all five simulated distributions (and fitted scaled factors), and to check if the total simulated distributions are consistent with the experimental $M_m$ distributions on a bin-by-bin basis.

In order to ensure the sum of the simulation distributions (obtained from the fitting algorithm) correctly describes the data, a number of cross-checks were introduced to examine the agreement between the experimental and reconstructed simulated distributions. These cross-checks (comparisons) are shown in Figs.~\ref{fig:bin_by_bin_mm} (b), (c) and (d).

Fig.~\ref{fig:bin_by_bin_mm} (b) shows the comparison between the scaled $\omega$ simulation ($\rm{Y}_{\omega \, {\rm SIMC}}$) and $\omega$ experimental distribution $\rm{Y}_{\omega~\textrm{Exp}}$, which is defined as the data distribution after the background physics distributions are subtracted 
\begin{equation}
\textrm{Y}_{\omega~\textrm{Exp}} = \textrm{Y}_{\rm Data} - \textrm{Y}_{\rho^0 \, {\rm SIMC}} - \textrm{Y}_{\pi\pi \, {\rm SIMC}} - \textrm{Y}_{\eta^{\prime} \, {\rm SIMC}}.
\label{eqn:exp_yield}
\end{equation}
This comparison is referred as the `comparison of $\omega$' in the later part of this section. As part of the quantitative comparison, the $\chi^2/{\rm dof}=1.20$ is computed, using Eqn.~\ref{eqn:chi2}.

Fig.~\ref{fig:bin_by_bin_mm} (c) shows the comparison between the sum of the simulation background distributions ($\textrm{Y}_{\rho^0 \, {\rm SIMC}} + \textrm{Y}_{\pi\pi \, {\rm SIMC}} + \textrm{Y}_{\eta^{\prime} \, {\rm SIMC}}$) and data distribution after the $\omega$ distribution is subtracted ($\textrm{Y}_{\rm Data} - \textrm{Y}_{\omega \, {\rm SIMC} }$). This comparison is referred as the `comparison of background' in the later part of the section. The $\chi^2/{\rm dof}=1.50$ is computed, using Eqn.~\ref{eqn:chi2}.


The subtracted difference between the data and simulation sum within the integration range is shown in Fig.~\ref{fig:bin_by_bin_mm} (d). This comparison is referred as the `comparison of zero' in the later part of the section. Adequate agreement between simulation and data would should yield a distribution consistent with zero within the statistical uncertainty. The blue horizontal line (0.001778) indicates the sum of the distribution, where the red horizontal line (0.000928$\pm$0.00122) is the error weighted fitting result of the scattered points. The two sets of `zero' values agree with each other and 0 within uncertainties. It is also important to make sure there is no systematic structure for the scattered zero distribution. Note that the zero comparison is only performed within the integration range (blue dashed lines).

\begin{figure}[t!]
  \centering
  \subfloat[][Comparison of $\omega$]{\includegraphics[width=0.5\textwidth]{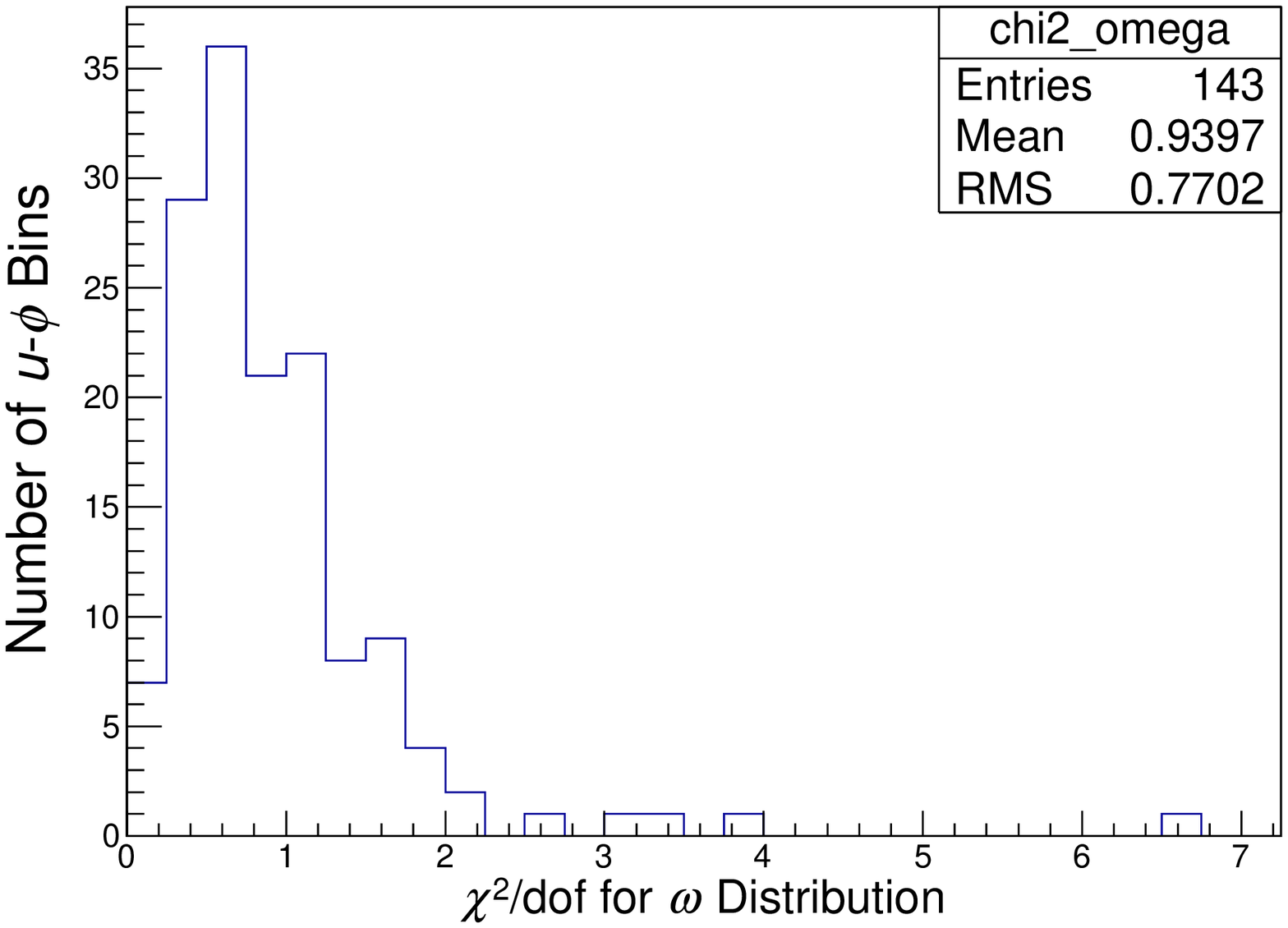}} 
  \subfloat[][Comparison of background]{\includegraphics[width=0.5\textwidth]{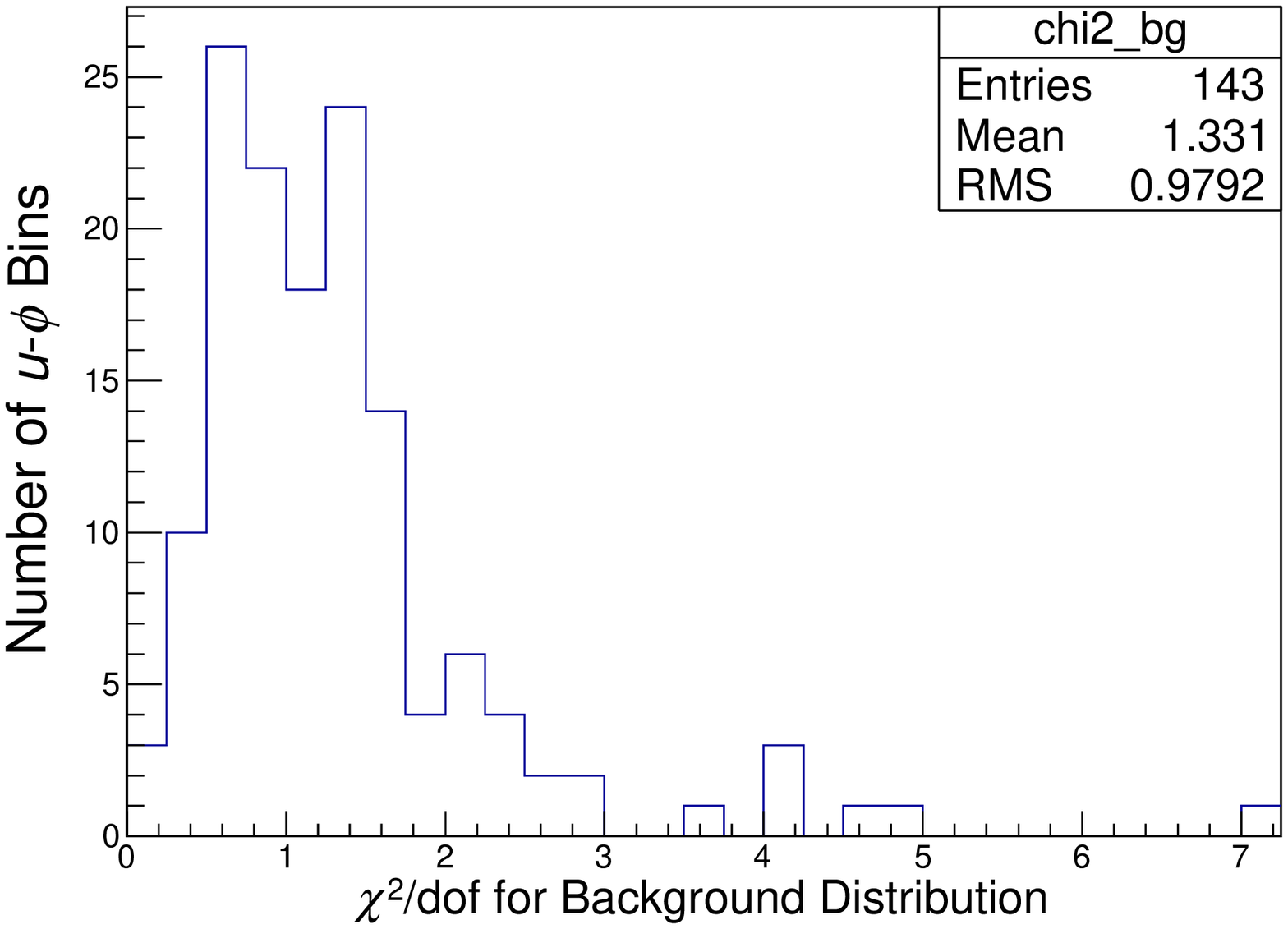}}
  \caption[Chi2 study]{(a) shows the $\chi^2/$dof distribution for comparing $\omega$ simulation ($\textrm{Y}_{\omega \, {\rm SIMC}}$) and data distribution after background subtraction ($\textrm{Y}_{\rm Data} - \textrm{Y}_{\rho^0 \, {\rm SIMC}} - \textrm{Y}_{\pi\pi \, {\rm SIMC}} - \textrm{Y}_{\eta^{\prime} \, {\rm SIMC}}$) for all analyzed $u$-$\phi$ bins. An example of the comparison is shown in Fig.~\ref{fig:bin_by_bin_mm} (b). The $\chi^2/$dof results comparing between the background distributions using sum of the all the physics background simulations ($\textrm{Y}_{\rho^0 \, {\rm SIMC}} + \textrm{Y}_{\pi\pi \, {\rm SIMC}} + \textrm{Y}_{\eta^{\prime} \, {\rm SIMC}}$) and data distribution after subtraction the $\omega$ distribution ($\textrm{Y}_{\rm Data} - \textrm{Y}_{\omega \, {\rm SIMC}}$) for all analyzed ($Q^2$, $\epsilon$, $\theta_{pq}$, $u$, $\phi$) bins. The example of the comparison is shown in Fig.~\ref{fig:bin_by_bin_mm} (c). As indicated in the statistics box, the total number of valid bins for
the entire $\omega$ analysis is 134.}
  \label{fig:chi2_study}
\end{figure}

Figs.~\ref{fig:chi2_study} (a) and (b) show the $\chi^2/{\rm dof}$ distributions of $\omega$ comparison and background compression over all valid ($Q^2$, $\epsilon$, $\theta_{pq}$, $u$, $\phi$) bins in the analysis, respectively. There are 240 bins in total for the $\omega$ analysis and 149 of them are valid bins with 91 bins excluded from the analysis by the bin exclusion criteria introduced in Sec.~\ref{sec:bin_exclude}. The global average of the $\chi^2/{\rm dof}$ values for $\omega$ comparison is 0.940, with a standard deviation of 0.770; The global average of the $\chi^2/{\rm dof}$ values for background comparison among the valid bins is 1.331, with a standard deviation of 0.979.

Based on the global averages of $\chi^2/{\rm dof}$ for both comparisons, the selected example $M_{m}$, shown in Fig.~\ref{fig:bin_by_bin_mm}, has $\chi^2/{\rm dof}$ values of 1.20 and 1.5. This would rank this particular bin slightly below the average in terms of fitting quality. Furthermore, ($Q^2$, $\epsilon$, $\theta_{pq}$, $u$, $\phi$) bins with high $\chi^2/{\rm dof}$ are typically low statistics bins and the $\omega$ is near the edge of the distribution. The general shapes of both global $\chi^2/{\rm dof}$ distributions are consistent with the Poisson statistical distribution with mean value around 1, the rare occurrences of high $\chi^2/{\rm dof}$ value bins are consistent with the statistical expectation.

The fitting algorithm has a built-in refit functionality, which is capable of repeating the fitting algorithm with narrower fitting limits (i.e. fitting 90\% of the total distribution instead of 92\%). The refit criteria are based on the fitting status (i.e. failure to converge) and $\chi^2/{\rm dof}$ values of both comparisons (surpass certain threshold). Note that the two $\chi^2/{\rm dof}$ are correlated, and are not independent measures of the overall fitting quality. This refit functionality is not used during the $\omega$ analysis, since fitting for all bins were successful and the both global $\chi^2/{\rm dof}$ distributions follow the statistical expectation.

\begin{figure}[t]
  \centering
  \includegraphics[width=0.9\textwidth]{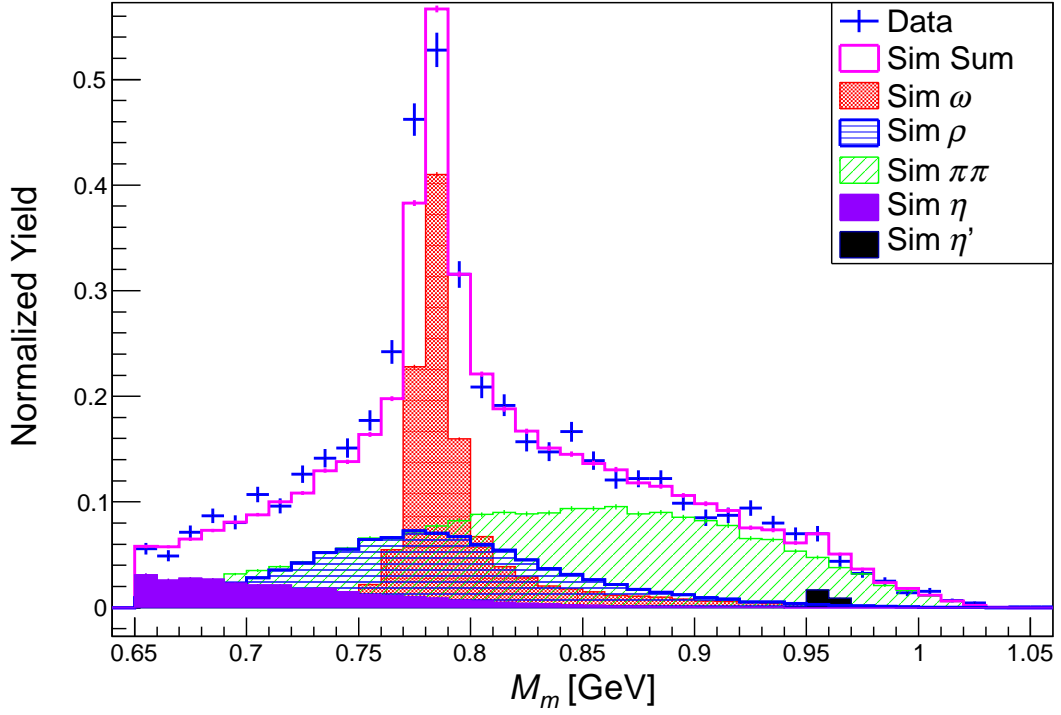}   
  \caption[Simulation method of extracting the background]{Simulation method of extracting the background. Blue distribution represents the overall experimental data; magenta distribution is the total simulated data (sum of $\rho$, $\omega$, $\eta$, $\eta^{\prime}$ and two-pion exchange simulation). Kinematics of the shown example plot is $Q^2$ = 2.45~GeV$^2$, $\epsilon$ = $0.55$ and $\theta_{pq}$ = $-$3$^{\circ}$.~\oic}
  \label{fig:all_mm_sim} 
\end{figure}

One additional validation of the fitting method comes from a comparison of the reconstructed $M_m$ distribution with data over all 24 ($Q^2$, $\epsilon$, $\theta_{pq}$, $u$) bins (i.e. summed over $\phi$). In Fig.~\ref{fig:all_mm_sim}, the sum of the simulation distributions, over all 24 bins after the fitting step is completed, are shown in magenta and the sum of the experimental data points are shown in blue crosses. The colored distributions represent the sum of corresponding simulated distributions (see legend). From the comparison, an excellent overall agreement between simulation and data is achieved. Furthermore, the contribution from each physics background can be identified directly.

\subsubsection{Fitting Step Remarks}

Recall, the main objective of the background fitting step is to determine the physics background underneath the $\omega$ peak. In addition, there are two important remarks regarding usage of the fitted simulation distributions during this step: 
\begin{itemize}

\item The $\omega$ distributions obtained during the background fitting step are for consistency check and fitting quality control only, and are not used to compute the experimental cross sections.

\item For a new iteration, this background fitting step is not required to be repeated. In fact, the background fitting results are kept constant intentionally, to maintain the stability of the extracted cross section during the iterative procedure. The iteration to iteration fluctuation of the background is further discussed in Sec.~\ref{sec:uncertainty}.     
\end{itemize}

\section{Integration Step and Yield Ratio}
\label{sec:int_step}

The goal of the integration step is to integrate and sum the $\omega$ events in both experimental and simulated distributions (from the current iteration) within the integration range (blue dashed lines), in order to determine the simulation $\omega$ yield (${\rm Y}_{\omega~{\rm SIMC}}$) on a bin-by-bin basis. 


Different from the fitting range, whose boundary locations change depending on shape and position of the data distribution, the integration range is fixed for all $u$-$\phi$ bins. The integration range is always centered at the $\omega$ rest mass, $M_m = 0.783$~GeV, and the boundary lines are located $\pm$40~MeV from the center. Any events outside of the integration range are excluded from the analysis.   



\begin{figure}[t]
  \centering
  \includegraphics[width=0.8\textwidth]{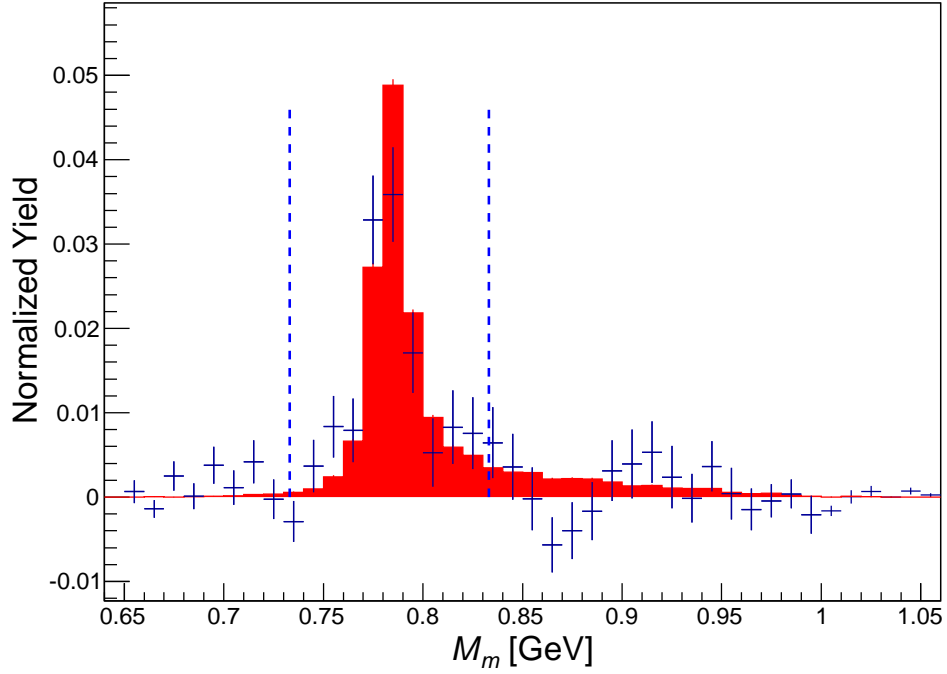}   
  \caption[Integrating the $\omega$ experimental and simulated distribution]{Integrating the $\omega$ experimental and simulated distribution. For clarity, the same bin is chosen as in Fig.~\ref{fig:mm_example}. The integration limits are shown as the blue dashed lines. The background subtracted $\omega$ distribution (blue crosses) is determined from the fitting step, and is the same data distribution shown in Fig.~\ref{fig:mm_example} (b).~\oic}
  \label{fig:omega_int} 
\end{figure}

\begin{figure}[t]
  \centering
  \includegraphics[width=1\textwidth]{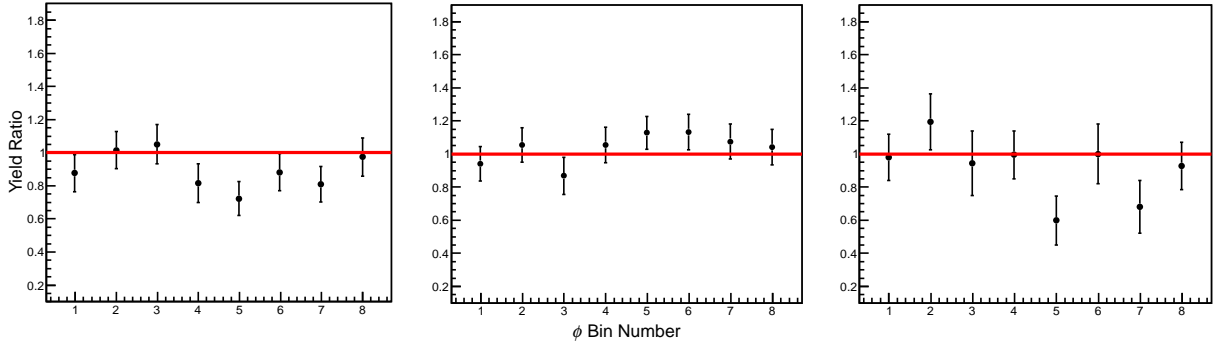}   
  \caption[Experiment-simulation yield ratio versus $\phi$ bin number]{The figure shows the experiment to simulation $\omega$ yield ratio (taking into account two or three $\theta_{pq}$ angles) versus $\phi$ bin number from the lowest to highest $-u$ bin (left to right plots), respectively. The chosen example is for $Q^2$ = 1.6~GeV$^2$, $\epsilon$ = 0.59. The $-u$ coverage for the left plot: $-u$ $\le$ 0.1~GeV$^2$; for the center plot: 0.10 $<$ $-u$ $\le$ 0.17~GeV$^2$; for the right plot: 0.17 $<$ $-u$ $\le$ 0.32~GeV$^2$. The first $\phi$ bin corresponds to $\phi$ = 22.5$^\circ$ with increment of 45$^\circ$. The red line indicates the desired yield ratio of 1.~\oic}
  \label{fig:yield_ratio} 
\end{figure}

\begin{figure}[h!]
  \centering
  \includegraphics[width=0.8\textwidth]{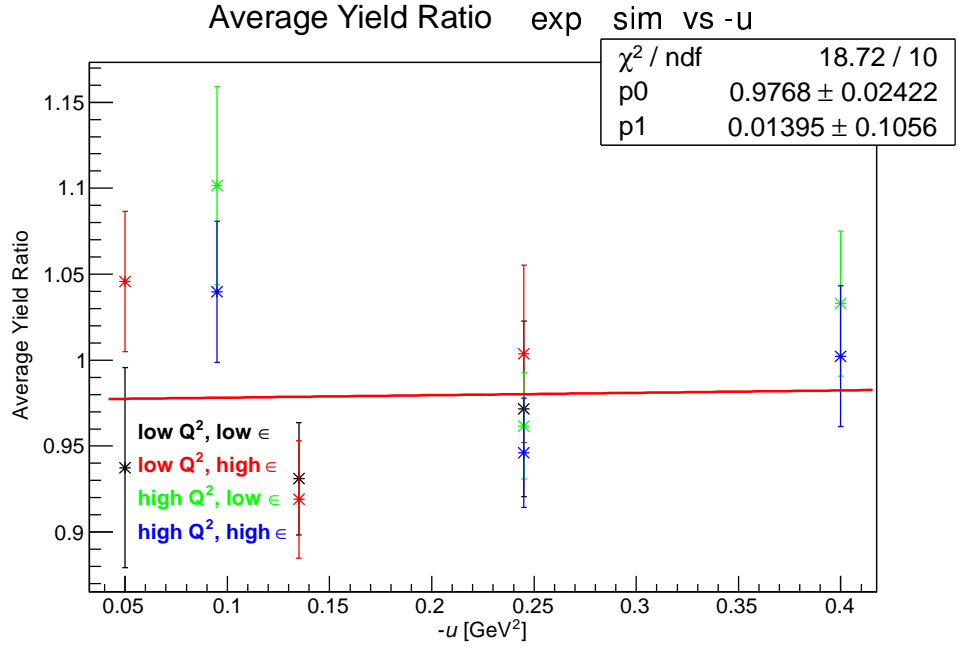}   
  \caption[A global view of the average yield ratio $R$ vs $-u$]{ A global view of the average yield ratio $R$ vs $-u$. Note that each point represents an $R$ value averaged over available $\phi$ bins for a given $-u$ bin. ~\oic}
  \label{fig:yield_ratio_check} 
\end{figure}

Fig.~\ref{fig:omega_int} shows an example $\omega$ distribution for the integration step. Note that the simulated distribution (in red) is not scaled to match the data distribution (shown as blue crosses). Meaning, the $\omega$ distribution is directly obtained from the simulation and it is not rescaled based on the fitting result. This is to be distinguished from the similar plot shown in Fig.~\ref{fig:bin_by_bin_mm} (b), whose $\omega$ distribution is rescaled based on the fitting result.


For a given ($Q^2$, $\epsilon$, $u$, $\phi$) bin of a single $\theta_{pq}$ (directly related to $\theta_{\rm HMS}$) setting, the experiment-simulation yield ratio is defined as the ratio between the background subtracted experimental $\omega$ yield ${\rm Y}_{\omega~{\rm Exp}}$ (defined in Eqn.~\ref{eqn:exp_yield}), and un-scaled simulated $\omega$ yield $\mathcal{Y}_{\omega~{\rm SIMC}}$. The yields are obtained by integrating corresponding distributions over the integration limits. The yield ratio is written as 
\begin{equation}
\begin{split}
R & = \frac{\textrm{Y}_{\omega\,{\rm Exp}}}{\mathcal{Y}_{\omega\,{\rm SIMC}}} \\[5mm]
  & = \frac{\textrm{Y}_{\rm Data} - \textrm{Y}_{\rho^0 \, {\rm SIMC}} - \textrm{Y}_{\pi\pi \, {\rm SIMC}} - \textrm{Y}_{\eta \, {\rm SIMC}} - \textrm{Y}_{\eta^{\prime} \, {\rm SIMC}}}{\mathcal{Y}_{\rm \omega \, SIMC}},
\label{eqn:yield_ratio}
\end{split}
\end{equation} 
where $\textrm{Y}_{\omega\,{\rm Exp}}$ represents the experimental data distribution. The simulated final states yields: $\textrm{Y}_{\rm \omega \, SIMC}$, $\textrm{Y}_{\rho^0 \, {\rm SIMC}}$, $\textrm{Y}_{\pi\pi \, {\rm SIMC}}$, $\textrm{Y}_{\eta \, {\rm SIMC}}$ and $\textrm{Y}_{\eta^{\prime} \, {\rm SIMC}}$, are scaled simulation distributions for the corresponding final states what were defined in Eqn.~\ref{eqn:sim_yield}




There are multiple $\theta_{\rm HMS}$ measurements per $\epsilon$ setting as shown Table~\ref{tab:kin_tab}. The yield ratio for a given ($Q^2$, $\epsilon$, $u$, $\phi$) bin, at high $\epsilon$ setting (three $\theta_{pq}$ angle measurements) can be written as
%
%
\begin{equation}
R   = \frac{\textrm{Y}_{\omega\,{\rm Exp}} (\theta_{pq}= -3^{\circ})   + \textrm{Y}_{\omega\,{\rm Exp}} (\theta_{pq}= 0^{\circ}) +  \textrm{Y}_{\omega\,{\rm Exp}} (\theta_{pq}= +3^{\circ})}{ \mathcal{Y}_{\omega\,{\rm SIMC}}(\theta_{pq}= -3^{\circ}) + \mathcal{Y}_{\omega\,{\rm SIMC}}(\theta_{pq}= 0^{\circ}) + \mathcal{Y}_{\omega\,{\rm SIMC}}(\theta_{pq}= +3^{\circ}) },
\label{eqn:Ratio_hi}
\end{equation}
and  at low $\epsilon$ setting (two $\theta_{\rm HMS}$ angles):
\begin{equation}
R   = \frac{\textrm{Y}_{\omega\,{\rm Exp}} (\theta_{pq}= 0^{\circ}) +  \textrm{Y}_{\omega\,{\rm Exp}} (\theta_{pq}= +3^{\circ})}{ \mathcal{Y}_{\omega\,{\rm SIMC}}(\theta_{pq}= 0^{\circ}) + \mathcal{Y}_{\omega\,{\rm SIMC}}(\theta_{pq}= +3^{\circ})}\,.
\label{eqn:Ratio_lo}
\end{equation}
The yield summation over different HMS angles consolidates the bin structure from ($Q^2$, $\epsilon$, $\theta_{pq}$, $u$, $\phi$) to ($Q^2$, $\epsilon$, $u$, $\phi$), where bins having less number of events can be compensated by the same bin from another HMS angle.


It is important to note that when performing a new iteration, the background extraction (through the fitting step algorithm) is not required, therefore the simulated background distributions and ${\rm Y}_{\omega~{\rm Exp}}$ remains constant. In order to help extract a more accurate parameterization, the background fit and ${\rm Y}_{\omega~{\rm Exp}}$ calculation (fitting step) is repeated after five to seven iterations.

Fig.~\ref{fig:yield_ratio} shows the yield ratio $R$ versus $\phi$ bin number for the $Q^2$ = 1.60~GeV$^2$, $\epsilon$ = 0.59 setting. The panel on the left shows $R$ for the lowest $-u$ bin, and the right plot shows $R$ for the highest $-u$ bin. In this particular setting, the $R$ values for most of the $\phi$ bins are within 1-2 $\sigma$ of unity, thus indicating good agreement between data and simulation.

A global view of the yield ratio is shown in Fig.~\ref{fig:yield_ratio_check}. The plot shows the averaged $R$ versus the nominal $-u$ value. Here, each average $R$ point is the average over the eight $\phi$ bins, i.e. Fig.~\ref{fig:yield_ratio} contributed 3 points shown in red. The fitted line is for demonstration purpose only, to show the general tend of the averaged $R$ values for the final iteration.







%
%
%
%
%


\subsection{Consistency Cross Checks}
\label{sec:q_cross_checks}




\begin{figure}[hp]
  \centering
  \subfloat[][Reconstructed missing energy $E_m$ distribution]{\includegraphics[width=0.5\textwidth]{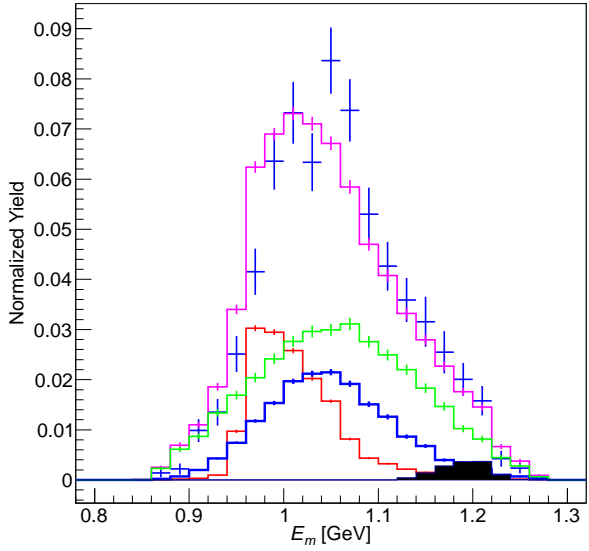}} 
  \subfloat[][Reconstructed missing momentum $P_m$ distribution]{\includegraphics[width=0.5\textwidth]{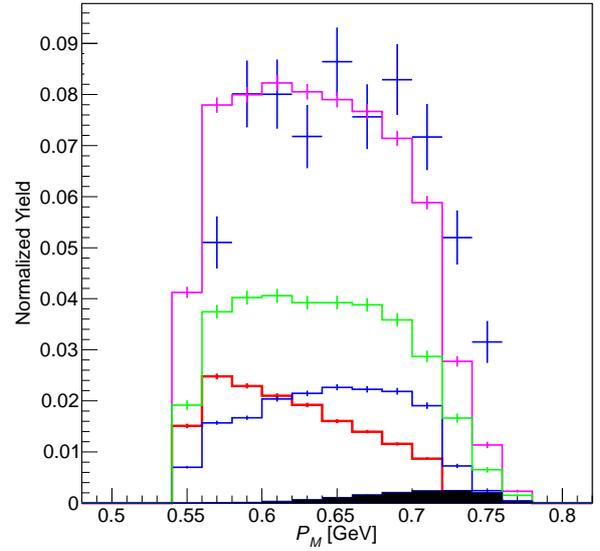}} \\
  \subfloat[][Reconstructed $hsdelta$ distribution]{\includegraphics[width=0.5\textwidth]{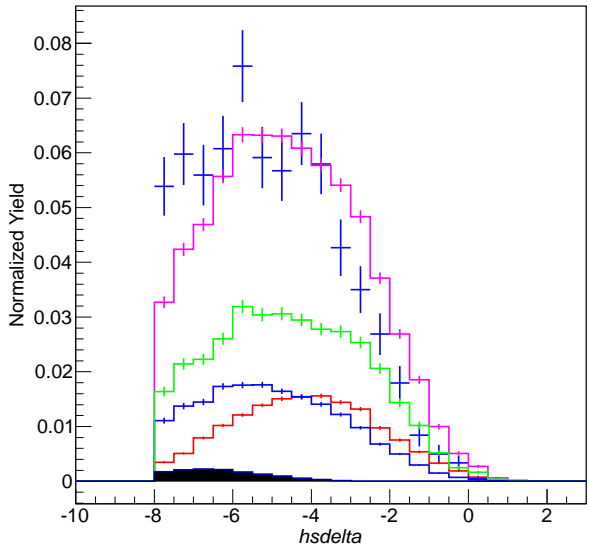}} 
  \subfloat[][Reconstructed $hsyptar$ distribution]{\includegraphics[width=0.5\textwidth]{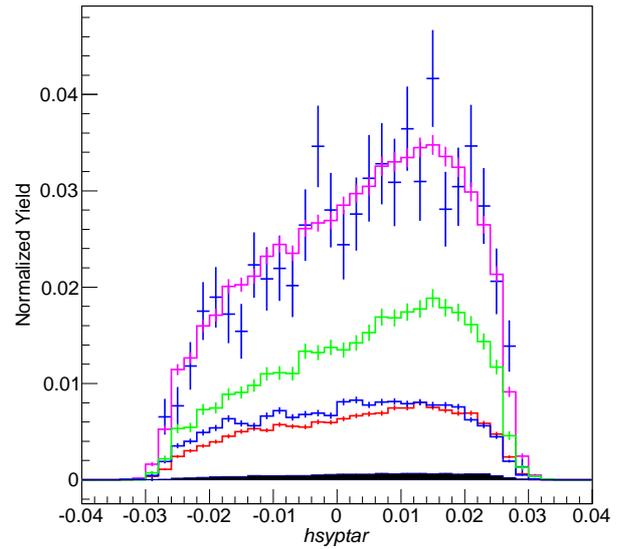}} 
  \caption[Physics and optics variable reconstructions]{Reconstructed distributions for physics ($E_m$ and $P_m$) and optics parameters ($hsdelta$ and $hsyptar$). The normalized yield (to 1~mC of beam charge) is plotted in $y$-axis. $hsdelta$ gives the percentage difference in particle momentum compared to the nominal momentum ($(p_{particle}-p_{norminal})/p_{norminal}$), and $hsyptar$ defines vertical angle of the particle entering the spectrometer from the target station. For clarity, same example ($Q^2$, $\epsilon$, $\theta_{pq}$, $u$, $\phi$) bin as the one shown in Fig.~\ref{fig:bin_by_bin_mm} is selected. The figures use the same color as in Fig.~\ref{fig:bin_by_bin_mm}. ~\oic}
  \label{fig:var_recon}
\end{figure}

Similar to the fitting quality control cross check after the fitting step, it is also important to verify the agreement between the data and simulation distributions after the integration step is completed. As a reminder, during the integration step, the $\omega$ simulation distribution is the direct output from the simulation (unscaled), and the background simulations are scaled by the fitting step and therefore are kept constant.

The cross checks after the integration step involves reconstructions of various of critical physics parameters such as $P_{m}$ and $E_{m}$, and acceptance parameters such as $hsdelta$ and $hsyptar$. Note that the simulation sum (magenta distribution) has a different definition than the one presented in the fitting step. Here, the simulation sum takes into account four scaled background distributions and an unscaled $\omega$ simulation distribution, and can be written as,
\begin{align}
\textrm{Y}_{\rm SIMC} & = Y_{\rm \omega \, SIMC} + \textrm{Y}_{\rho^0\, {\rm SIMC}} + \textrm{Y}_{\pi\pi \, {\rm SIMC}} + \textrm{Y}_{\eta \, {\rm SIMC}} + \textrm{Y}_{\eta^{\prime} \, {\rm SIMC}}  \nonumber \\
& = \mathcal{Y}_{\omega\,{\rm SIMC}} + b \cdot \mathcal{Y}_{\rho^0\,{\rm SIMC}} + c \cdot \mathcal{Y}_{\pi\pi\,{\rm SIMC}} + d \cdot \mathcal{Y}_{\eta\,{\rm SIMC}} + e \cdot \mathcal{Y}_{\eta^{\prime}\,{\rm SIMC}}.
\end{align}


For clarity, the same example bin is chosen as the one used for the fitting step (shown in Fig.~\ref{fig:bin_by_bin_mm}) with slightly worse than average fitting quality. The reconstructed physics parameter distributions $P_{m}$ and $E_{m}$ of the example bin are shown in Figs.~\ref{fig:var_recon} (a) and (b), respectively. The experimental data points are shown as the blue crosses. The scale factors ($b$-$e$) used to scale the background simulation distributions are obtained from the fitting step for each ($Q^2$, $\epsilon$, $u$, $\phi$) bin.



In addition to the $P_m$ and $E_m$ distributions, the distributions of three critical spectrometer acceptance parameters (described in Sec.~\ref{sec:secptrometer_acc}): $hsdelta$, $hsxptar$ and $hsyptar$ are also reconstructed on a bin-by-bin basis and compared to the experimental data. Figs.~\ref{fig:var_recon} (b) and (c) show the reconstructed $hsdelta$ and $hsyptar$ distributions, respectively. The reconstructed $hsxptar$ distribution shows similar agreement as the $hsyptar$ comparison, and therefore is not shown. 

The reconstructed physics and acceptance parameters are in good agreement with the data, particularly in terms of the coverage and cut-off of the distributions. This implies the kinematics and spectrometer acceptance offsets are simulated accurately. The $hsdelta$ show slight disagreement in terms of the height at one end of the distribution. For the optical parameters, it is critical to match the coverage of the distribution to ensure the spectrometer acceptance of the simulated and experimental data is identical, the distribution height is less important particularly on a bin-by-bin basis. Discrepancies are also observed in the reconstructed peak of the $E_m$ and $P_m$ distributions. As shown in Heep analysis, small differences are also observed for the missing $E_m$, $P_m$ and $M_m$ distributions, shown in Figs.~\ref{fig:Heep_E_m}, \ref{fig:Heep_P_m} and \ref{fig:Heep_M_m}, respectively. This is due to the fact that the proton scattering in the target chamber and the HMS entrance/exit windows is poorly simulated.

In addition, it is impossible to parameterize the physics model to replicate the behavior of the experimental data for every reconstructed parameter and every $u$-$\phi$ bin. The kinematics coverage in terms of $Q^2$, $W$ and $-u$ for the real data are slightly different in each bin, and the generation of the simulated data requires the experimental parameters such the spectrometer angles and momentum settings as input, where these input can only represent the nominal kinematic values. Since the role of the SIMC is to achieve the best possible overall agreement between the data and the simulation, this would inevitably create small difference between the simulation and the experimental data in certain bins.

The conclusion, based on the $\chi^2/{\rm dof}$ values from the fitting quality control (in Sec.~\ref{sec:fit_quality_control}) is that the selected example ($Q^2$, $\epsilon$, $u$, $\phi$) bin has lower than average fitting quality. Therefore, this particular bin is a good representation of an average bin in terms of fitting quality and acceptable agreement for the reconstructed parameters.

These qualitative comparisons of the reconstructed variables are not used to determine the background fitting quality on a bin-by bin-basis, they are only used as a consistency check to validate the sum between the unscaled $\omega$ simulation distribution and the scaled background simulation distributions. For bins having large $\chi^2/{\rm dof}$ values (greater 3) for both $\omega$ and background comparisons, the reconstructed distributions are expected to be worse. The disagreement for any bin will not cause the refit of the background.

Any significant disagreement between reconstructed simulation and data observed in a large number of bins for multiple parameters would indicate a serious issue, such as hidden spectrometer offsets, insufficient fitting and integration limits, over or under estimated uncertainties, potential coding error in the analyzer and a number of other potential errors. The reconstructed parameters are a useful diagnostics tool to help with locating and revolving the errors.



\section{Experimental Cross Section and L/T Separation}
\label{sec:extracting_LT_cx}

\begin{figure}[p]
  \centering

  \subfloat[][$-u$ $\le$ 0.1~GeV$^2$]{\includegraphics[width=0.333\textwidth]{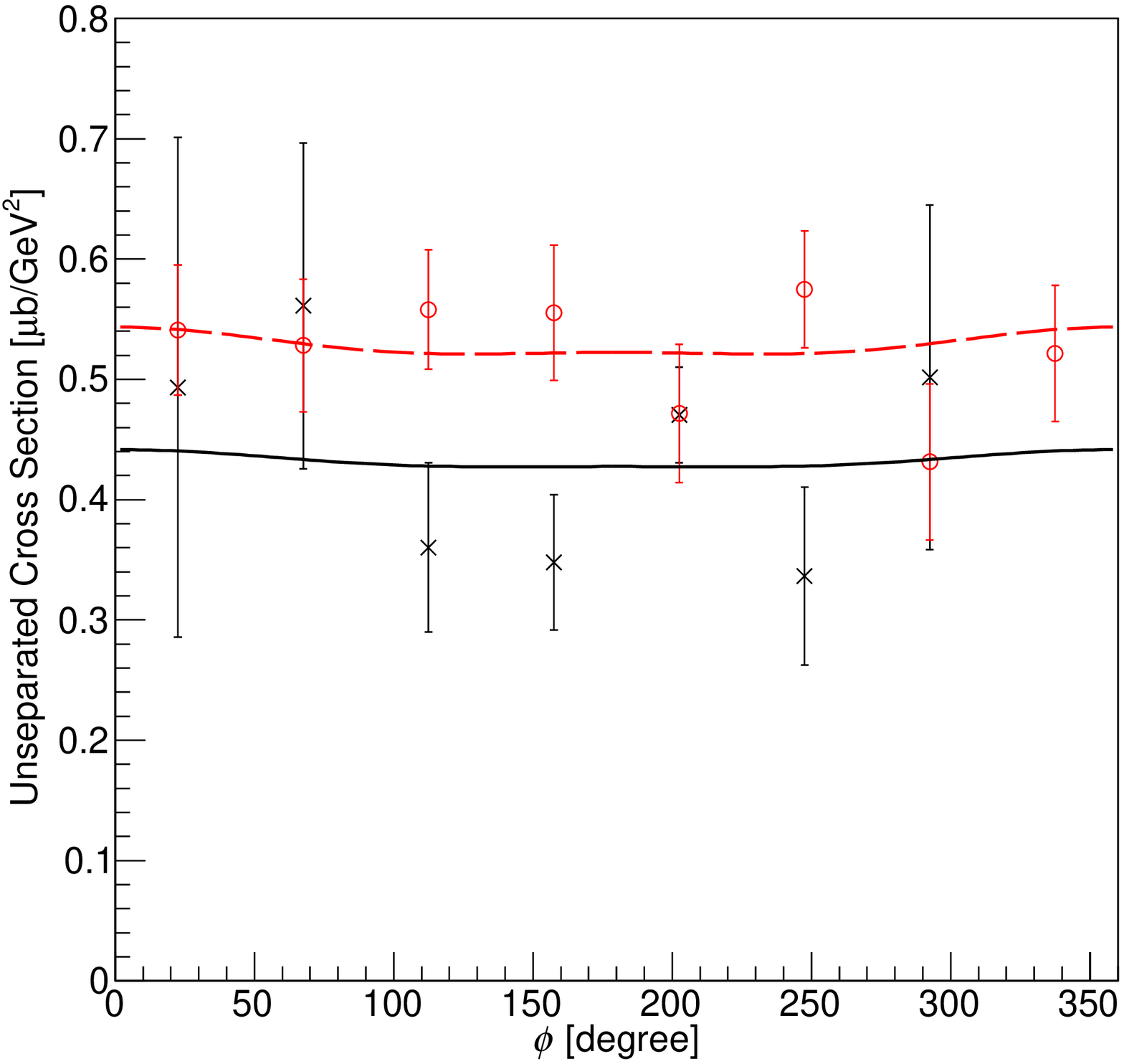}} 
  \subfloat[][0.10 $<$ $-u$ $\le$ 0.17~GeV$^2$]{\includegraphics[width=0.333\textwidth]{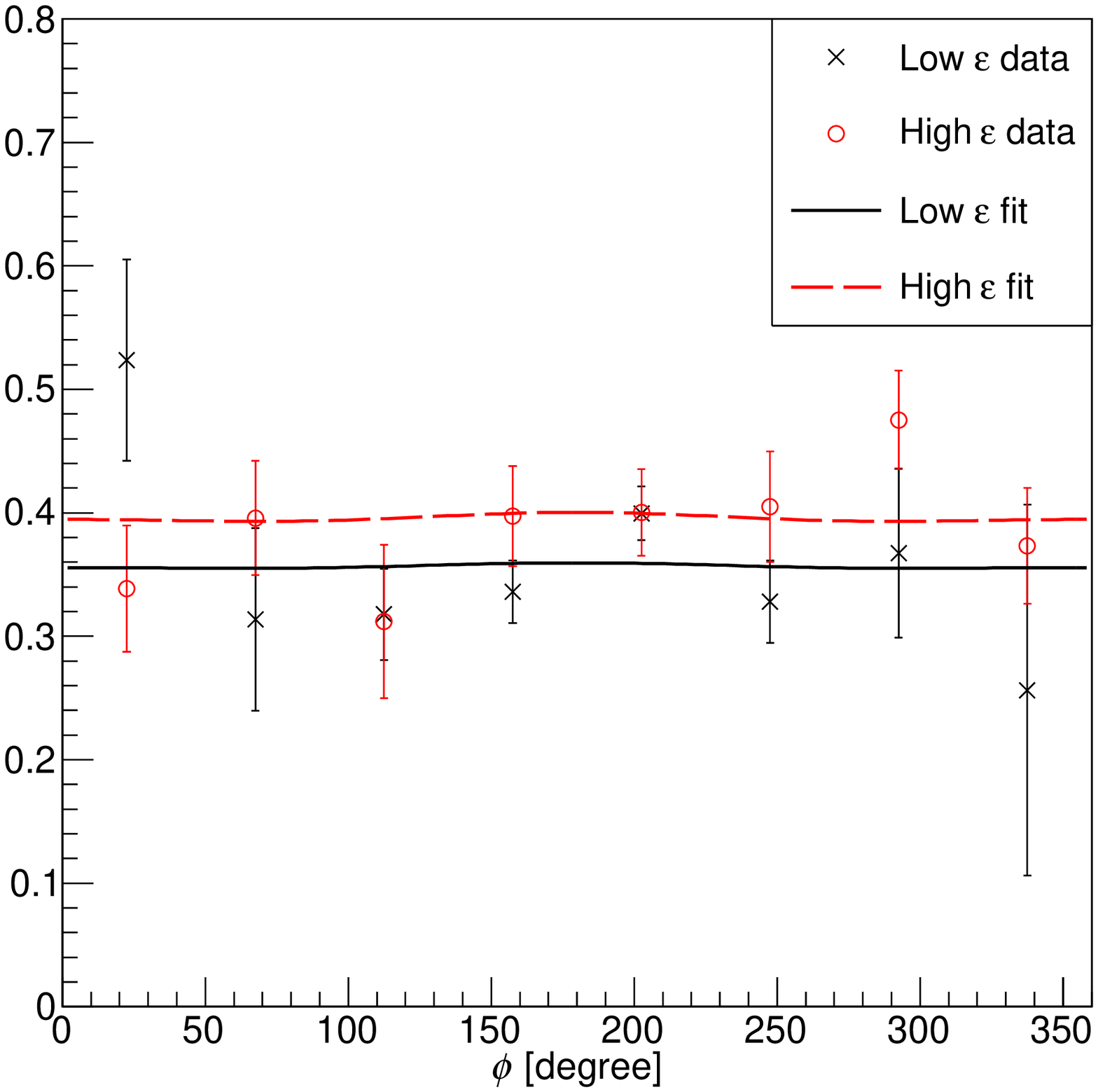}} 
  \subfloat[][0.17 $<$ $-u$ $\le$ 0.32~GeV$^2$]{\includegraphics[width=0.333\textwidth]{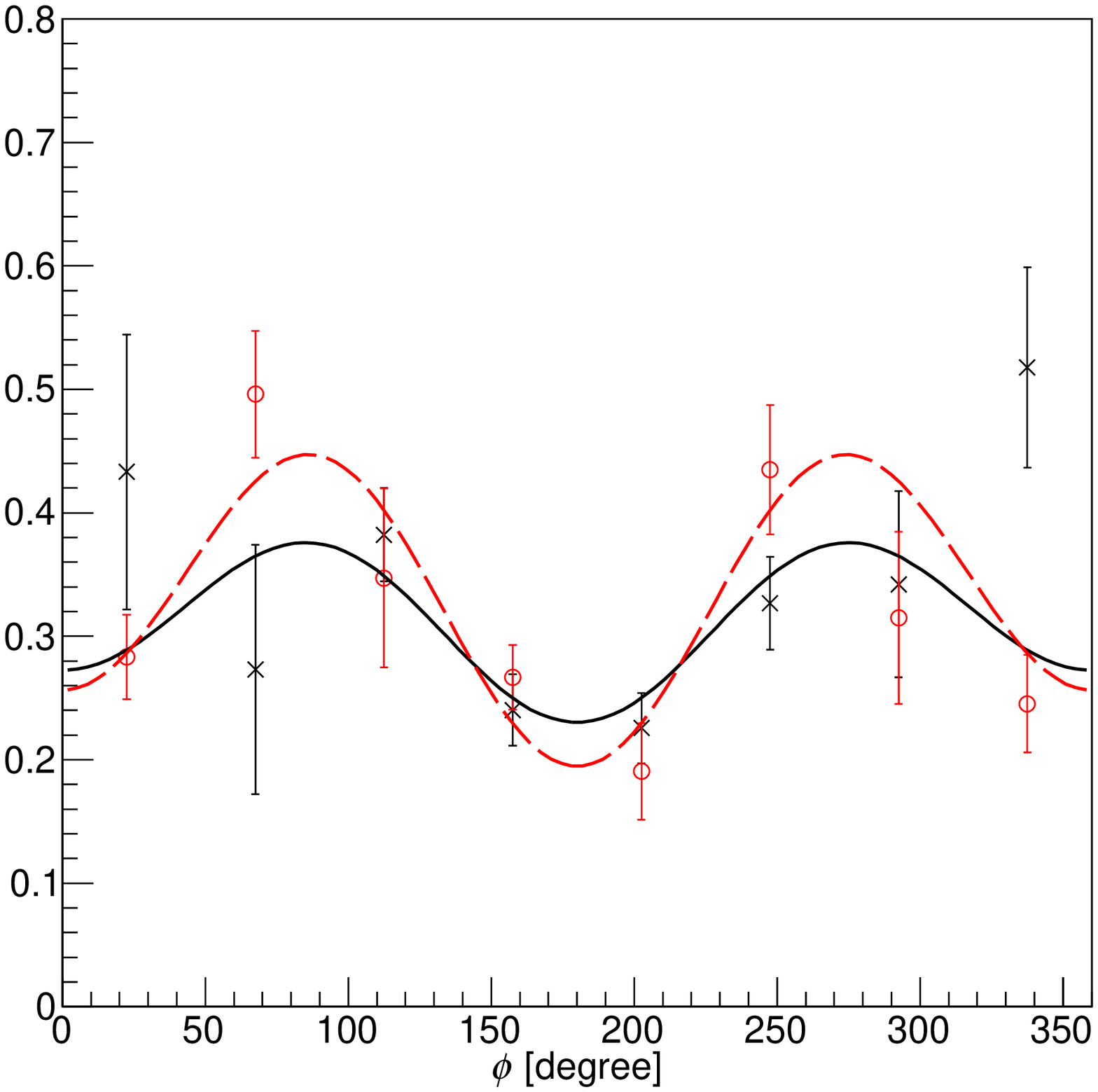}}
  \caption[Unseparated differential cross section for $Q^2$ = 1.60~GeV$^2$]{Unseparated differential cross section $\sigma_u$ versus $\phi$ for all three $u$ ranges at $Q^2$ = 1.60~GeV$^2$. Here, $\sigma_u$ is the short form for $2\pi\cdot d^2\sigma/dtd\phi$. The black and red data points indicate the unseparated $\sigma_{\omega}$ at low and high $\epsilon$, respectively. The black dashed and red solid curves show the fitting results using Eqn.~\ref{eqn:rosen_chapt_5} for low and high $\epsilon$, respectively. The $-u$ coverage is indicated below each plot. Statistical uncertainties are shown in the plot.~\oic}
  \label{fig:sig_exp_160} 

  \subfloat[][$-u$ $\le$ 0.19~GeV$^2$]{\includegraphics[width=0.333\textwidth]{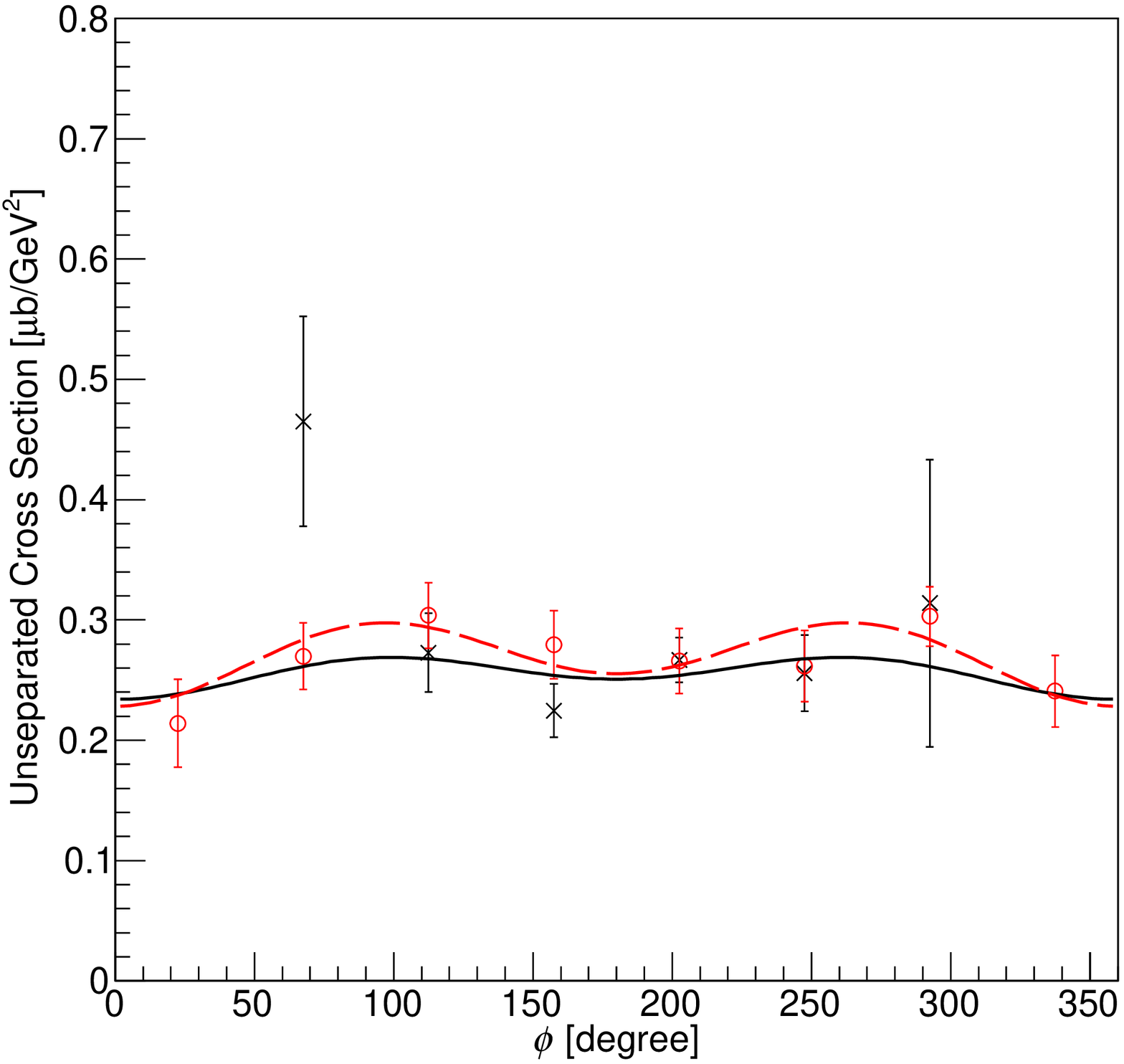}} 
  \subfloat[][0.19 $<$ $-u$ $\le$ 0.30~GeV$^2$]{\includegraphics[width=0.333\textwidth]{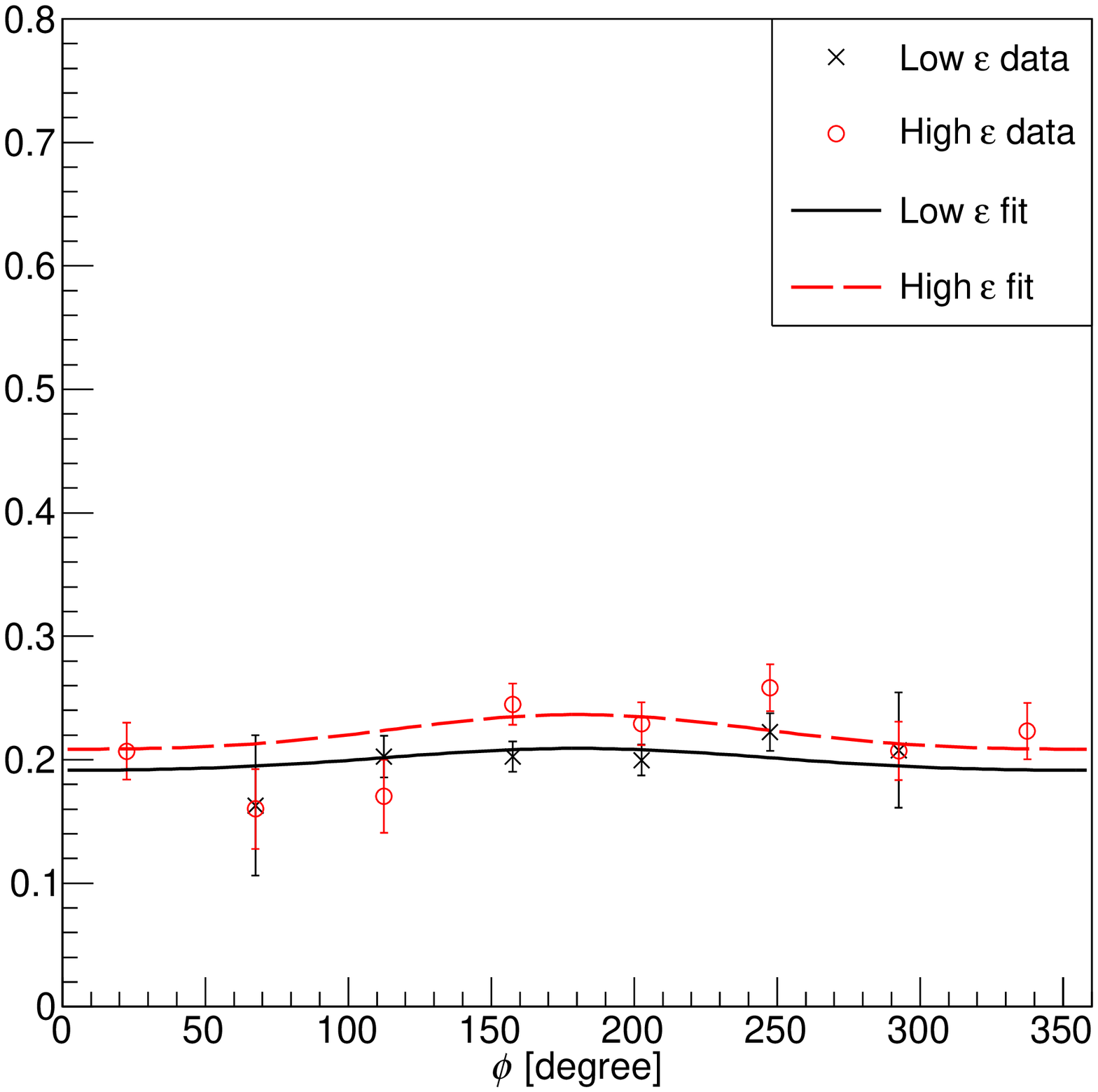}} 
  \subfloat[][0.30 $<$ $-u$ $\le$ 0.50~GeV$^2$]{\includegraphics[width=0.333\textwidth]{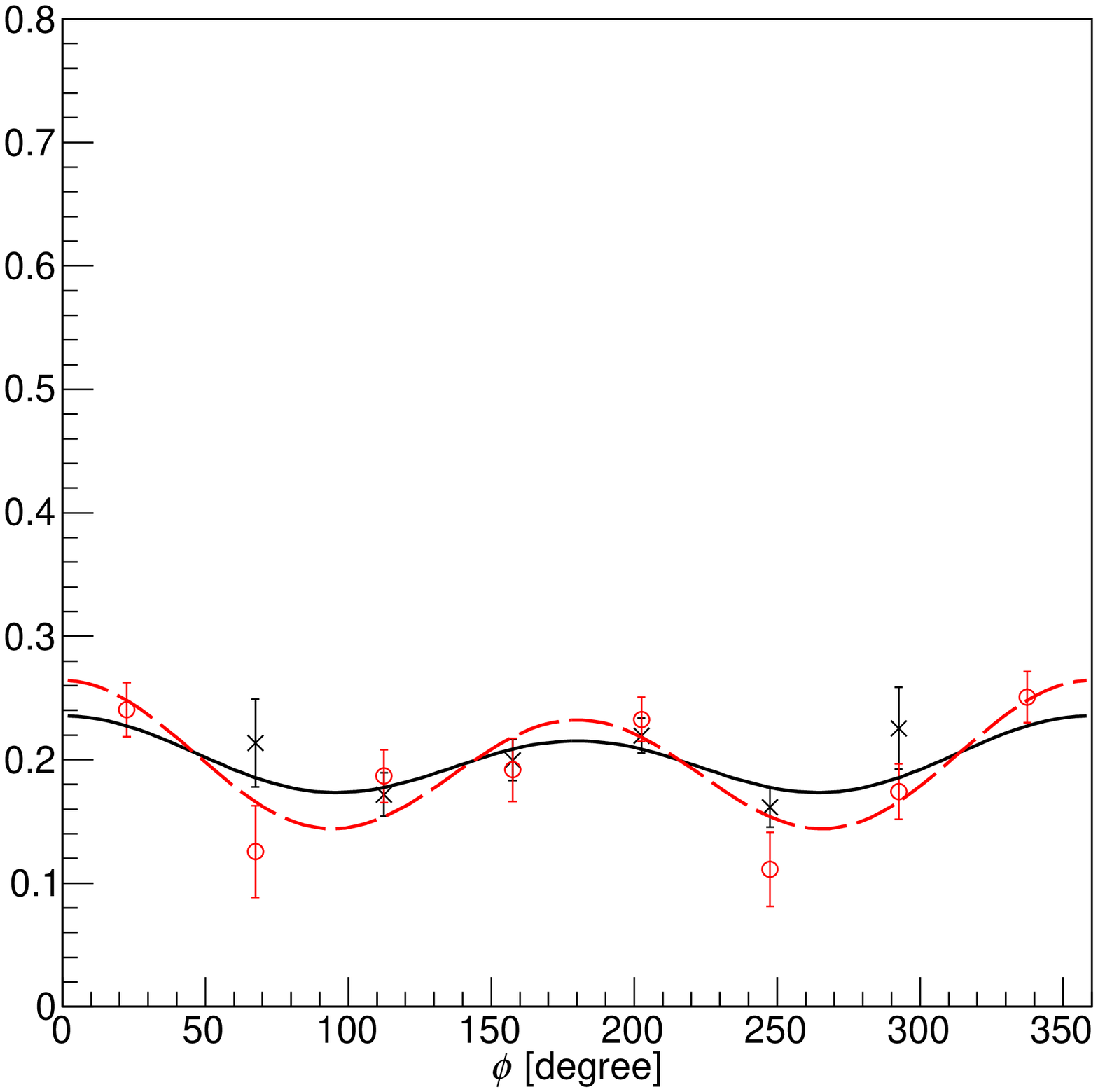}} 
  \caption[Unseparated differential cross section for $Q^2$ = 2.45~GeV$^2$]{Unseparated differential cross section $\sigma_u$ versus $\phi$ for all three $u$ ranges at $Q^2$ $=$ 2.45~GeV$^2$. Same marker and color schemes are used as in Fig.~\ref{fig:sig_exp_160}. The $-u$ coverage is indicated below each plot. ~\oic}
  \label{fig:sig_exp_245} 


\end{figure}

The extraction of the experimental cross section is complicated due to correlations between the kinematic variables and the nonuniform angular acceptance. In order to evaluate the experimental cross section at a specific point within the bin acceptance, the dependence of the cross section on all kinematic variables has to be well understood. From the previous work~\cite{horn,volmer}, the cross section model may depend on kinematic variables including: $t$, $Q^2$, $W$, $\theta^*$ and $\phi$.

The experimental cross sections are determined by comparing the experimental yields to the SIMC simulated yields. If the simulation describes the experimental data properly, the experimental cross section can be extracted by iterating the model input cross section until the best agreement between the data and Monte Carlo is achieved. If the model input cross section describes the dependence on all kinematic variables ($W$, $Q^2$, $u$, $\theta^*$, $\phi$) correctly, the experimental cross section can be extracted
\begin{equation}
\begin{split}
\sigma_{\omega\,{\rm Exp}}  & = \frac{\textrm{Y}_{\omega\,{\rm Exp}}}{\textrm{Y}_{\omega\,{\rm SIMC}}} \cdot \sigma _{\omega\,{\rm SIMC}} \\[5mm]
& = R \, \sigma_{\omega\,{\rm SIMC}} \,,
\label{eqn:xsection}
\end{split}
\end{equation} 
where the yield ratio $R$ is given in Eqn.~\ref{eqn:yield_ratio}. The $\sigma$ represents the total differential cross section $d^2\sigma/dtd\phi$.

Using Eqn.~\ref{eqn:xsection}, the $\omega$ experimental differential cross section $\sigma_{\omega~\rm{Exp}}$ can be calculated for $Q^2$ = 1.60 and 2.45~GeV$^2$, and plotted versus $\phi$ in Figs.~\ref{fig:sig_exp_160} and \ref{fig:sig_exp_245}, respectively.


For each $Q^2$-$u$ bin, the Rosenbluth formula (given by Eqn.~\ref{eqn:rosen_chapt_5}) is used to simultaneously fit the high (red) and low {back} data points to extract $\sigma_{\rm T}$, $\sigma_{\rm L}$, $\sigma_{\rm LT}$, $\sigma_{\rm TT}$ for a given ($Q^2$, $u$) bin.

Note, there are 24 $\sigma_{\omega~\rm{Exp}}$, corresponding to eight $\phi$ angles at each $\epsilon$. However, due to excluded bins, $\phi$ bins without $\sigma_{\omega~\rm{Exp}}$ are expected, particularly for low $\epsilon$ measurements, where there is no $\theta_{pq}$ $<$ 0$^{\circ}$ setting.

The unseparated differential cross sections $\sigma_{u}$ and the kinematics values of each $-u$ ($\braket{W}$ and $\braket{Q^2}$) bin are listed in Table~\ref{tab:unxsec_raw}. Here, the $\sigma_{u}$ is taken as the integral of the fitted function curve shown in Figs.~\ref{fig:sig_exp_160} and \ref{fig:sig_exp_245}, divided by $2\pi$. $\braket{W}$ and $\braket{Q^2}$ correspond to the averaged $W$ and $Q^2$ values of the $\omega$ simulation distribution within the kinematics acceptance of each bin; the $\braket{-u}$ of each bin is taken as the central value of the $-u$ limit of that bin; $\braket{\epsilon}$ values are calculated from the averaged $\braket{Q^2}$ and $\braket{W}$ values.

\begin{table}[t]
\centering
\setlength{\tabcolsep}{1em}
\caption[Unseparated cross section $\sigma_u$ for $Q^2$ = 1.60 and 2.45 GeV$^2$]{Unseparated cross section $\sigma_u$ for $Q^2$ = 1.60 and 2.45 GeV$^2$. The determination of the total statistical uncertainty $\delta\sigma_{u}$ (includes statistical and uncorrelated point-to-point error) and systematic systematic $\Delta\sigma_{u}$ are discussed in Sec.~\ref{sec:uncertainty}. The meaning of the variables are defined in the text.}
\label{tab:unxsec_raw}
\begin{tabular}{ccccc}
\toprule
$\braket{-u}$ & $\braket{W}$ & $\braket{Q^2}$ & $\braket{\epsilon}$ & $\sigma_{u}$ $\pm$ $\delta\sigma_{u}$ $\pm$ $\Delta\sigma_{u}$      \\
$\mu$b/GeV$^2$  & GeV  & GeV$^2$   &            & $\mu$b/GeV$^2$                                                      \\ \midrule
\multicolumn{5}{c}{$W_{\rm nominal}$ = 2.21~GeV, $Q^2_{\rm nominal}$ = 1.60~GeV$^2$, $\epsilon_{\rm nominal}$ = 0.32} \\ \midrule
0.058           & 2.26 & 1.47 &  0.316   &  0.432 $\pm$ 0.027 $\pm$ 0.014                                             \\
0.135           & 2.22 & 1.58 &  0.327   &  0.357 $\pm$ 0.014 $\pm$ 0.011                                             \\
0.245           & 2.19 & 1.67 &  0.334   &  0.313 $\pm$ 0.016 $\pm$ 0.011                                             \\ \midrule
\multicolumn{5}{c}{$W_{\rm nominal}$ = 2.21~GeV, $Q^2_{\rm nominal}$ = 1.60~GeV$^2$, $\epsilon_{\rm nominal}$ = 0.59} \\ \midrule
0.058           & 2.26 & 1.47 &  0.586   &  0.527 $\pm$ 0.020 $\pm$ 0.017                                             \\
0.135           & 2.22 & 1.58 &  0.593   &  0.396 $\pm$ 0.016 $\pm$ 0.013                                             \\
0.245           & 2.19 & 1.67 &  0.597   &  0.336 $\pm$ 0.015 $\pm$ 0.011                                             \\ \midrule
\multicolumn{5}{c}{$W_{\rm nominal}$ = 2.21~GeV, $Q^2_{\rm nominal}$ = 2.45~GeV$^2$, $\epsilon_{\rm nominal}$ = 0.27} \\ \midrule
0.117           & 2.28 & 2.23 &  0.258   &  0.256 $\pm$ 0.013 $\pm$ 0.008                                             \\
0.245           & 2.23 & 2.39 &  0.268   &  0.199 $\pm$ 0.007 $\pm$ 0.006                                             \\
0.400           & 2.18 & 2.52 &  0.277   &  0.197 $\pm$ 0.008 $\pm$ 0.006                                             \\ \midrule
\multicolumn{5}{c}{$W_{\rm nominal}$ = 2.21~GeV, $Q^2_{\rm nominal}$ = 2.45~GeV$^2$, $\epsilon_{\rm nominal}$ = 0.55} \\ \midrule
0.117           & 2.28 & 2.23 &  0.547   &  0.269 $\pm$ 0.010 $\pm$ 0.009                                             \\
0.245           & 2.23 & 2.39 &  0.553   &  0.220 $\pm$ 0.008 $\pm$ 0.007                                             \\
0.400           & 2.18 & 2.52 &  0.559   &  0.194 $\pm$ 0.008 $\pm$ 0.006                                             \\
\bottomrule
\end{tabular}
\end{table}

\section{Improving the Physics Model in SIMC}
\label{sec:improving}

\begin{figure}[t]
  \centering
  \subfloat[][Full dependence]{\includegraphics[width=0.5\textwidth]{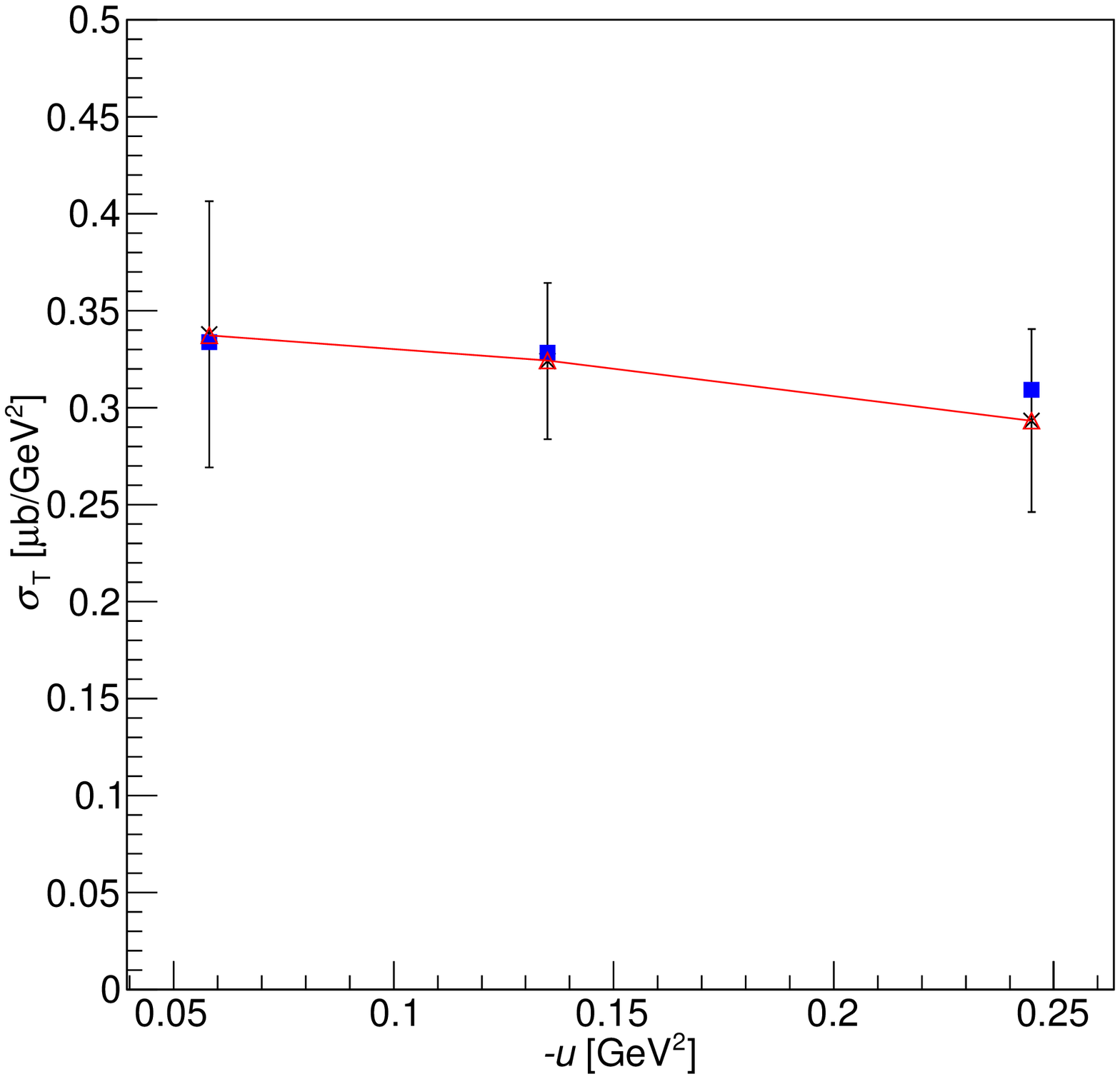}} 
  \subfloat[][Without $Q^2$ and $W$ dependence]{\includegraphics[width=0.5\textwidth]{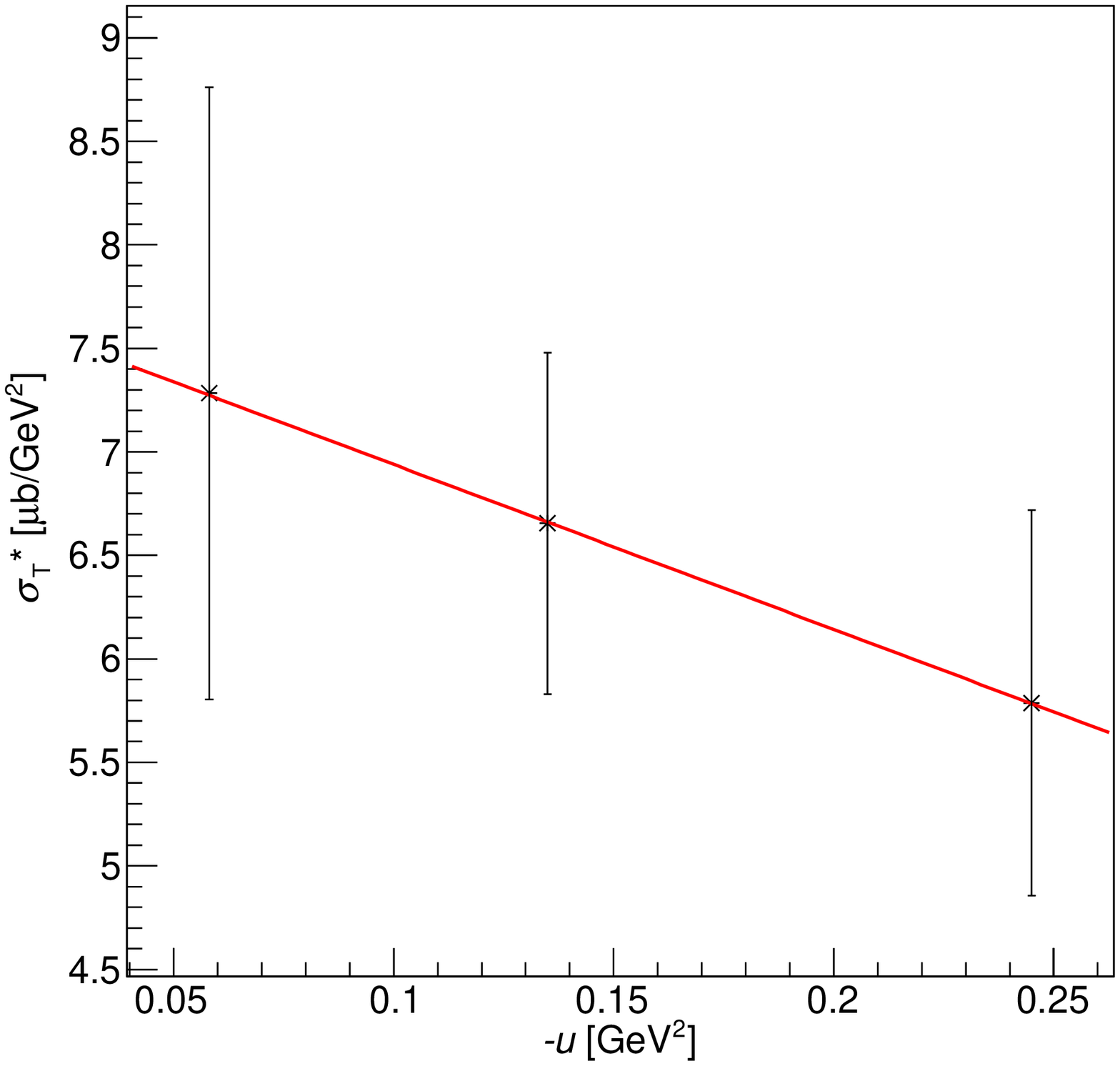}} 
  \caption[Improving the physics model in SIMC]{(a) shows the extracted transverse differential cross section $\sigma_{\rm T}$ (short form of $d\sigma_{\rm T}/dt$) versus $-u$ at $Q^2$ = 1.6~GeV$^2$. The black crosses are the extracted data points; the red triangles are the reconstructed points based on the fitting results from (b); the blue square boxes are the projected results computed using the parameterization from previous iteration. Figure (b) shows the $\sigma^*_{\rm T}$ versus $-u$, where $\sigma^*_{\rm T}$ is equivalent to $\sigma_{\rm T}$ with the $Q^2$ and $W$ dependence removed (see Sec.~\ref{sec:wqscaling}). A linear fit using Eqn.~\ref{eqn:T} is performed to demonstrate the $-u$ dependence and is shown in red solid line.~\oic}
  \label{fig:improving_model}
\end{figure}

In this section, the last step of the iterative procedure is described. Since the actual L/T separated cross sections are documented in Sec.~\ref{sec:lt_sep_res}, this section only covers the concept of using the extracted L/T separated cross sections to improve the free fitting parameters which dictate the $u$ and $Q^2$ dependences in the $\omega$ physics model, defined in Eqns.~\ref{eqn:T}-\ref{eqn:TT}. The improved parameters are then used as the input to generate the $\omega$ simulation in SIMC for the new iteration.

Fig.~\ref{fig:improving_model} (a) shows the transverse differential cross section $d\sigma_{\rm T}/dt$ (abbreviated as $\sigma_{\rm T}$) versus $u$ at $Q^2$ = 1.6~GeV$^2$. The $\sigma_{\rm T}$ is extracted from the total differential cross section $2\pi~d^2\sigma/dtd\phi$ (abbreviated as $\sigma_{\omega {\rm Exp}}$) using the method described in the previous section. 

The $\sigma_{\rm T}$ shown in black crosses cannot be used for fitting directly to extract $-u$ dependence, due to the presence of the bin-to-bin $\braket{Q^2}$ and $\braket{W}$ dependences. The $Q^2$ and $W$ dependences are described by Eqns.~\ref{eqn:T} and ~\ref{eqn:w_dep}, respectively, and they must be stripped away from the $\sigma_{\rm T}$ in order to study the behavior of the data with respect to $-u$.

The $\sigma_{\rm T}$ after removing the $Q^2$ and $W$ dependences, using Eqns.~\ref{eqn:T} and \ref{eqn:w_dep}, are plotted in Fig.~\ref{fig:improving_model} (b) as $\sigma^*_{\rm T}$. A fitting is performed using the functional form of the transverse component of the differential cross section given in Eqn.~\ref{eqn:T}; the shown red curve shows results of the fitting of the $u$ dependence. Note that there are two parameters, $t_0$ and $t_1$, whose values are determined through the fitting.


The new parameters, $t_0$ and $t_1$, extracted from the fitting of the $\sigma^*_{\rm T}$, are used to construct the $\sigma_{\rm T}$ values by reintroducing the $Q^2$ and $W$ dependences and form the red triangles in Fig.~\ref{fig:improving_model} (a), so that the improved parameterization can be compared with the previous parameterization shown in blue squares. 


The identical procedure is performed to extract parameters for the longitudinal and interference differential cross sections: $\sigma_{\rm L}$, $\sigma_{\rm LT}$ and $\sigma_{\rm TT}$, except, there is an additional $\sin\theta^*$ dependence for $\sigma_{\rm LT}$ and $\sigma_{\rm TT}$, which is handled in the same way as the $Q^2$ and $W$ dependences.

The final step involves including the improved parameters in SIMC as the input for the next iteration, thus completing the iterative process.

\section{Encountered Issues}
\label{sec:issues}

\subsection{Events at $-u<0$~GeV$^2$}
\label{sec:u_less_zero}

Based on the conservation of momentum and energy, the minimum $-u$ value for exclusive $\omega$ production is required to be greater than zero and any event below zero would violate the physical limits of maximum possible backward-angle of $\theta_\omega$ = 180$^{\circ}$. Due to the broad physics background and imperfect spectrometer resolution near the edge of the distribution, it can be seen that the experimental data distribution extends into the unphysical region of $-u$ $<$ 0~GeV$^2$ in Fig.~\ref{fig:u_bin_limits}, particularly for the $Q^2$ = $1.6$~GeV$^2$ settings.

After consultation with other experts~\cite{huber17, gaskell16}, the $-u$ $<$ 0 events are included in the analysis since they are too close to the $\omega$ simulation to be cut off. As the result, there has been a slight improvement in the fitting quality.


\subsection{Kinematics Shift and $M_m$ Distribution Cut-off}
\label{sec:miss_focusing}
%

Through careful examination of the $M_m$ distributions of the first (lowest) $u$ bin, specially at $\theta_{\rm HMS}$ = $+$3$^{\circ}$ setting, the simulated $\omega$ peak (located at the right edge of the background) seems to contain a distinctive tail towards the lower missing mass range. An example is shown in Fig.~\ref{fig:mm_example} (c). This tail is different from the tail caused by the radiative process, since the direction of the radiative tail is towards the higher missing mass range (as shown in Fig.~\ref{fig:exclusion} (b)).

The tail toward the lower missing mass (lower $M_m$ tail) only exists in ($Q^2$, $\epsilon$, $\theta_{pq}$,  $u$, $\phi$) bins with the lowest $-u$ value in both $Q^2$ setting. They are caused by the $M_m$ distribution cut-off in combination with the skewed field of view at $\theta_{pq}$ = $\pm$3$^\circ$ measurement. As it was described in the earlier text, every $Q^2$-$\epsilon$ setting requires at least two measurement at different HMS angles to populate the full $\phi$ coverage around the $q$-vector, as shown in Fig.~\ref{fig:bull_fig}. In this setup, the $q$-vector is always defined by the $\theta_{pq}=0^\circ$. When measurements at $\theta_{pq}$ = $\pm$3$^\circ$ angles are performed, the proton events for the lowest $-u$ value are off the center of the focal plane, particularly for $\theta_{pq}$ = $+$3$^{^\circ}$ it only occupies a small section of the focal plane; for the $\theta_{pq}$ = $-$3$^\circ$ setting, protons occupy half of the focal plane, which gives a more complete
 $M_m$ distribution for $\omega$. In short, the lower $M_m$ tail is not a spectrometer acceptance effect, but rather that due to four momentum conservation, the low $-u$ region starts to become forbidden for certain proton momenta.

Thanks to the multiple HMS angle measurements, the lack of statistics in certain bins within a HMS angle setting can be compensated by the same bin from another angle setting. This, in combination with the SIMC, makes the $M_m$ cut-off less significant in terms of impact to the overall quality of the cross section extraction.

\section{Uncertainty Budget}
\label{sec:uncertainty}

The uncertainty in the extraction of the experimental yield consists of both statistical and systematic uncertainties. In this section, the statistical uncertainties are expressed by the $\delta$ symbol, and systematic errors used $\Delta$ symbol. Furthermore, the percentage uncertainty are indicated as $\delta (\%)$ and the absolute uncertainties as $\delta ({\rm abs})$.

\subsection{Statistical Uncertainty}

The statistical uncertainty is determined by the uncertainty in the number of good $\omega$ events and in detector efficiencies as well as the beam charge. The combined efficiency (taking into account detectors, event tracking and DAQ) and its uncertainty is determined on a run-by-run basis. For every run, the combined efficiency uncertainties and charge uncertainties are added in quadrature, then multiplied by the accumulated beam charge (product of the combined efficiency and beam charge):
\begin{equation}
\delta_\textrm{run}^2 ({\rm abs.})= (\textrm{efficiency} \times \textrm{charge})^2 \times (\delta_\textrm{efficiency}^2 + \delta_\textrm{charge}^2 )\,.
\end{equation}
For each setting, the normalized uncertainty can be obtained as:
\begin{equation}
\delta_{setting}^2 (\%) =  \dfrac{\displaystyle\sum_{\textrm{run}} \delta_\textrm{run}^2}{\displaystyle\left(\sum_{\text{run}} \textrm{efficiency} \times \textrm{charge}\right)^2 } ,
\end{equation}
The experimental yield uncertainty (percentage) is computed by adding the $\delta_{setting}$ and statistical uncertainty of the selected events ($\sqrt{N}$) in quadrature,
\begin{equation}
\delta {\rm Y}_{\rm Data} (\%) = \sqrt{\delta_{setting}^2 + \left(\frac{\sqrt{N}}{N}\right)^2},
\end{equation}
where $N$ is the total number of $^1$H$(e,e^{\prime}p)X$ events surviving the event selection criteria for the setting. 

Since the determination of the good $\omega$ events requires background fit and subtraction, therefore the yield ratio (defined in Eqn.~\ref{eqn:yield_ratio}) uncertainty is computed by adding the total uncertainty of the experimental yield and scaled simulation yield in quadrature, 
\begin{equation}
\delta R (\%) =  \sqrt{ \delta {\rm Y}_{\rm Data} ^2 +  \delta \mathcal{Y}_{\omega\,{\rm SIMC}}^2 + b^2 \delta \mathcal{Y}_{\rho^0\,{\rm SIMC}}^2 + c^2 \delta \mathcal{Y}_{\pi\pi\,{\rm SIMC}}^2 + d^2 \delta \mathcal{Y}_{\eta\,{\rm SIMC}}^2 + e^2 \mathcal{Y}_{\eta^{\prime}\,{\rm SIMC}}^2},
\end{equation}
where $\delta \mathcal{Y}_{\omega\,{\rm SIMC}}$, $\delta \mathcal{Y}_{\rho^0\,{\rm SIMC}}$, $\delta \mathcal{Y}_{\pi\pi\,{\rm SIMC}}$, $\delta \mathcal{Y}_{\eta\,{\rm SIMC}}$ and $\mathcal{Y}_{\eta^{\prime}\,{\rm SIMC}}$ are statistical uncertainties of the $\omega$ and background distributions, which are computed as the square root of the corresponding simulation yield (i.e. $\delta \mathcal{Y}_{\rho^0\,{\rm SIMC}} = \sqrt{\mathcal{Y}_{\rho^0\,{\rm SIMC}}}$); $b$-$e$ are scale factors determined from the fitting algorithm. The unseparated cross section for a given ($Q^2$, $\epsilon$, $\theta_{pq}$, $u$, $\phi$) bin has the same percentage statistical error as the yield ratio.

The statistical uncertainty of the unseparated cross section for a given $u$ bin (sum over 8 $\phi$ bins) is computed using the uncertainty of the weighted average~\cite{taylor97},
\begin{equation} 
\delta{\sigma_{u}} ({\rm abs}) = \frac{1}{\sqrt{\Sigma w_i}}
\end{equation}
where $i$ = 1-8, which corresponds to the number of valid $\phi$ bins; $w_i$ is the weight factor for each $\phi$ bin and is defined as $$w_i = \frac{1}{\delta \sigma_{i}^2},$$ where $\sigma_{i}$ is the absolute uncertainty of unseparated cross section in each $\phi$ bin.


\begin{table}[p]
\centering

\caption[Summary of uncertainties for the $\omega$ analysis]{Summary of uncertainties for the F$_{\pi}$-2-$\omega$ analysis. Where two values are given, they corresponds to the two $Q^2$ points. When a range is specified, it corresponds to the range in $u$. The systematic uncertainties in each row are added quadratically to obtain the total systematic uncertainty shown in the last row.}
\label{tab:err_tab}

\begin{tabular}{lcccc}

\toprule
Correction        	  & Uncorrelated &  $\epsilon$ uncorr. & Correlated &     Section    \\
                  	  & (Pt-to-Pt)   &  $u$ corr.   &  (scale)   &                \\
               	      & (\%)   &  (\%)     & (\%)   &                \\ \toprule 
HMS Cherenkov         &        &           &  0.02  & Sec.~\ref{sec:gas_cherenkov} \\
HMS Aerogel           &        &           &  0.04  & Sec.~\ref{sec:aero}          \\
SOS Calorimeter       &        &           &  0.17  & Sec.~\ref{sec:calorimeter}   \\
SOS Cherenkov         &        &           &  0.02  & Sec.~\ref{sec:gas_cherenkov} \\
HMS beta              & 0.4    &           &        & Sec.~\ref{sec:cointime_heep} \\ \midrule 
HMS Tracking          &        &  0.4      &  1.0   & Sec.~\ref{sec:tr_eff}        \\
SOS Tracking          &        &  0.2      &  0.5   & Sec.~\ref{sec:tr_eff}        \\
HMS Trigger           &        &  0.1      &        & Sec.~\ref{sec:trigger}       \\
SOS Trigger           &        &  0.1      &        & Sec.~\ref{sec:trigger}       \\
Target Thickness      &        &  0.3      &  1.0   & Secs.~\ref{sec:target_thick_mess}, \ref{sec:target_boil_study}\\ 
CPU LT                &        &  0.2      &        & Sec.~\ref{sec:CLT}           \\
Electronic LT         &        &  0.1      &        & Sec.~\ref{sec:ELT}           \\
Coincidence Blocking  &        &           &  0.1   & Sec.~\ref{sec:coin_blocking} \\ \midrule 
$d\theta$             &  0.1   &  0.7-1.1  &        & Ref.~\cite{blok08}           \\
$dE_{\rm Beam}$       &  0.1   &  0.2-0.3  &        & Ref.~\cite{blok08}           \\
$dp_e$                &  0.1   &  0.1-0.3  &        & Ref.~\cite{blok08}           \\
$d\theta_p$           &  0.1   &  0.2-0.3  &        & Ref.~\cite{blok08}           \\ \midrule  
PID                   &        &  0.2      &        & Sec.~\ref{sec:PID_cuts}      \\
Beam Charge           &        &  0.3      &  0.5   & Sec.~\ref{sec:beam}          \\
Radiative Correction  &        &  0.3      &  1.5   & Sec.~\ref{sec:rad_process}   \\
Acceptance            &  1.0   &  0.6      &  1.0   & Sec.~\ref{sec:secptrometer_acc} \\
Proton Interaction    &        &           &  0.7   & Sec.~\ref{sec:p_int_corr}    \\ \midrule 
Background Fitting Limit   &  2.0   &  0.8      &  0.8   & Secs.~\ref{sec:bin-by-bin}, \ref{sec:sys_err}  \\  
$\omega$ Integration Limit &  1.7   &  1.0      &  0.3   & Secs.~\ref{sec:int_step}, \ref{sec:sys_err}    \\  
Model Dependence           &  0.7   &           &        & Secs.~\ref{sec:omega_model}, \ref{sec:sys_err} \\ \midrule 
Total                      &  2.9   &  1.7-2.0  &  2.6   &                              \\
\bottomrule
\end{tabular}

\end{table}

\subsection{Systematic Uncertainty}

\label{sec:sys_err}

The systematic uncertainties can be subdivided into the correlated and uncorrelated contributions. The correlated uncertainties, i.e. those that are the same for both $\epsilon$ points, such as the target thickness corrections and beam charge variation, contribute directly to the separated cross section. The uncorrelated uncertainties are attributed to the unseparated cross sections, which inflates the uncertainties in the separated of $\sigma_{\rm L}$ and $\sigma_{\rm T}$.

All systematic uncertainties of this analysis are listed in Table~\ref{tab:err_tab}. They are added in quadrature to obtain the total systematic uncertainties. Further details regarding the uncertainty estimations related to the instrumental and acceptance can be found in Refs.~\cite{blok08,horn,ArringtonThesis,christy04,tvaskis}.  The influence of the uncertainties in the offsets in spectrometer variables, such as beam energy, momentum and angles, were determined by changing the variables by their statistical uncertainty and evaluating the resultant changes in the unseparated cross section. These well established uncertainties were studied by previous Hall C analyses such the F$_\pi$-2-$\pi^+$~\cite{horn}, therefore were not re-determined in this analysis. The $\epsilon$-uncorrelated uncertainties can be subdivided into uncertainties that are the same for all $u$ values at a given $\epsilon$ value, and ones that are also uncorrelated in $u$.

The largest contributions of the point-to-point uncorrelated uncertainty come from the fitting and integration limits, which are critical components for the fitting step (Sec.~\ref{sec:bin-by-bin}) and integration step (Sec.~\ref{sec:int_step}).





\subsubsection*{Background Fitting Limit Uncertainties}

In order to fully understand the each component (uncorrelated, $\epsilon$ uncorrelated $u$ correlated, and correlated) of systematic uncertainty due to the fitting limit, a study was performed to monitor the deviations of the unseparated cross sections computed from three separated analyses. In each of the three analyses, different fitting limits were used, which corresponds to 90\%, 92\% (nominal limit), 95\% of the central $M_m$ distributions. Note that there are 12 sets of unseparated $\sigma$ (2$Q^2$ $\times$ 2$\epsilon$ $\times$ 3$u$ bins) for each of the three analyses. Note that during this study, no iteration was done after changing the fitting limit.

The 12 sets of $\sigma$ from 90\% fitting limit analysis, and 12 sets of $\sigma$ of 95\% analysis, are separately compared with the 12 sets of $\sigma$ from 92\% (nominal) analysis. Then the average percentage difference (in 12 sets of $\sigma$) and standard deviation obtained for comparing 90\% and 92\% analyses are denoted as $aver(90\%)$ and $std(90\%)$; correspondingly, the average percentage difference (in 12 sets of $\sigma$) and standard deviation by comparing 95\% and 92\% analyses are $aver(95\%)$ and $std(95\%)$. The correlated systematic uncorrelated uncertainty is computed using $aver(90\%)$ and $aver(95\%)$,
\begin{equation}
\textrm{Correlated~Error} = \frac{|aver(90\%)| +|aver(95\%)|} {2};
\end{equation}
and the point-to-point (uncorrelated) error is computed using $aver(90\%)$ and $aver(95\%)$,
\begin{equation}
\textrm{Point-to-Point} = \frac{|std(90\%)| +|std(95\%)|} {2}.
\end{equation}
The determination of the $\epsilon$ uncorrelated $u$ correlated uncertainty also requires the 12 sets of the percentage difference between 90\% and 92\% analyses, and 12 sets of percentage difference between 95\% and 92\% analyses. These 24 sets of percentage differences are used to compute three separate (percentage difference) averages according to the $u$ range, i.e. the first average is calculated among eight lowest $-u$ bins, second average is for eight mid $u$ bins and third average is for the eight highest $u$ bins. The standard deviation among the three average values is taken as the $\epsilon$ uncorrelated $u$ correlated uncertainty.

\subsubsection*{$\omega$ Integration Limit}

The correlated and uncorrelated systematic uncertainties due to the integration limits are estimated using a similar methodology as the one used for estimating the fitting limits uncertainties. The $\sigma_u$ are computed using three different integration limits: $M_{\omega}\pm$30~MeV, $\pm$40~MeV (nominal), $M_{\omega}\pm$50~MeV. The estimated uncertainties are similar in size as the fitting limit uncertainties, and are listed in Table~\ref{tab:err_tab}.

\subsubsection*{Model Dependence Uncertainties}

There were two studies performed regarding the model dependence contribution to the $\sigma_{u}$ uncertainties. In the first study, the $\omega$ physics model parameters given in Sec.~\ref{sec:omega_model} were scaled up by 5\%, this resulted a negligible difference ($<$ 0.1\%) in $\sigma_u$. 

In the second model dependence uncertainty study, the LT and TT components of the differential cross section were turned off, the percentage difference in $\sigma_u$ seems to suggest a point-to-point uncertainty of 0.7\% (standard deviation). In addition, when the physics background fit and subtraction is performed for a new iteration (instead of using the background fit and subtraction from previous iteration), the deviation in $\sigma_u$ is less than 0.1\%. Note that during this study, no iteration was done after changing the fitting limit.

Note that the unseparated experimental cross section should not be dramatically sensitive to any small tweak in the functional form of the physics model. As a consistency test, three iterations were performed with $1/Q^8$ dependence for $\sigma_{\rm L}$ (instead of $1/Q^{4}$ in Eqn.~\ref{eqn:L}). The unseparated extraction cross section variation is within the model dependence uncertainty shown in Table~\ref{tab:err_tab}. The agreement for separated cross sections are well within the statistical uncertainty.



\subsubsection*{Uncertainty Propagation for $\sigma_{\rm T}$ and $\sigma_{\rm L}$}


The unseparated cross sections at low ($\epsilon_1$) and high $\epsilon_2$ values (shown in Figs.~\ref{fig:sig_exp_160} and \ref{fig:sig_exp_245}), can be expressed in terms of the separated cross sections $\sigma_{\rm L}$ and  $\sigma_{\rm T}$,
\begin{align}
\sigma_1 &= \sigma_{\rm T} + \epsilon_1 \, \sigma_{\rm L} = \sigma_{\rm T} ~ (1+\frac{\epsilon_1}{R}) \label{eqn:sig_1}, \\
\sigma_2 &= \sigma_{\rm T} + \epsilon_2 \, \sigma_{\rm L} = \sigma_{\rm T} ~ (1+\frac{\epsilon_2}{R}) \label{eqn:sig_2},
\end{align}
where $\sigma_1$ and $\sigma_2$ represent unseparated cross sections at $\epsilon_1$ and $\epsilon_2$, respectively; $R$ is the transverse-longitudinal (T-L) ratio defined as
$$R = \frac{\sigma_{\rm T}}{\sigma_{\rm L}}.$$ Through substitution and manipulation of Eqns.~\ref{eqn:sig_1} and \ref{eqn:sig_2}, $\sigma_{\rm T}$ and $\sigma_{\rm L}$ can be expressed in terms of $\sigma_1$ and $\sigma_2$, 
\begin{align}
\sigma_{\rm L} &= \frac{\sigma_1-\sigma_2}{(\epsilon_1 - \epsilon_2)}, \\
\sigma_{\rm T} &= \frac{\sigma_2\,\epsilon_1-\sigma_1\,\epsilon_2}{(\epsilon_1 - \epsilon_2)}.
\end{align}
By differentiating $\sigma_{\rm L}$ and $\sigma_{\rm T}$, their percentage errors can be expressed as, 
\begin{equation}
\frac{\delta \sigma_{\rm T}}{\sigma_{\rm T}} (\%) = \frac{1}{(\epsilon_1 - \epsilon_2)} \sqrt{ \epsilon^2_1 \left(\frac{\delta\sigma_1}{\sigma_1}\right)^2\left(1 + \frac{\epsilon_2}{R} \right)^2 + \epsilon^2_2  \left(\frac{\delta\sigma_2}{\sigma_2} \right)^2 \left(1 + \frac{\epsilon_1}{R}\right)^2 },
\label{eqn:sigt_err}
\end{equation}
\begin{equation}
\frac{\delta \sigma_{\rm L}}{\sigma_{\rm L}} (\%) = \frac{1}{(\epsilon_1 - \epsilon_2)} \sqrt{ \left(\frac{\delta \sigma_1}{\sigma_1}\right)^2 (R + \epsilon_1)^2 + \left(\frac{\delta \sigma_2}{\sigma_2}\right)^2 (R+\epsilon_2)^2 },
\label{eqn:sigl_err}
\end{equation}
where $\delta\sigma_1$ and $\delta\sigma_2$ are the total statistical uncertainties (quadratic sum of statistical and point-to-point uncorrelated systematic uncertainties) of the $\sigma_1$ and $\sigma_2$, respectively. The inflation factor is approximately $1/(\epsilon_{2} - \epsilon_{1})$, which is $\sim3$. The calculated percentage uncertainties of $\frac{\delta \sigma_{\rm T}}{\sigma_{\rm T}}$ and $\frac{\delta \sigma_{\rm L}}{\sigma_{\rm L}}$ values are shown in $(\delta \sigma_{\rm T})_{\rm Form.}$ and $(\delta \sigma_{\rm L})_{\rm Form.}$ columns of Table~\ref{tab:LT_error_tab}. These can be directly compared with the uncertainties generated by the fitting function shown in $(\delta \sigma_{\rm T})_{\rm Fit}$ and $(\delta \sigma_{\rm L})_{\rm Fit}$ columns. Note that when these fitting errors are generated, they include contributions from both statistical and uncorrelated systematic uncertainties of 2.9\% (shown in Table~\ref{tab:err_tab}). The propagated uncertainties
$(\delta\sigma_{\rm T})_{\rm Form.}$ and $(\delta\sigma_{\rm L})_{\rm Form.}$ are generally comparable to the
 fitting uncertainties $(\delta\sigma_{\rm T})_{\rm Fit}$ and $(\delta\sigma_{\rm L})_{\rm Fit}$. This is an important indicator showing the error correlation for the measurements are small, since the uncertainty propagation formulas (Eqn.~\ref{eqn:sigt_err} and \ref{eqn:sigl_err}) assume uncorrelated errors, whereas the fitting errors include the correlated errors.

Note that the fitting errors, i.e. $(\delta \sigma_{\rm L})_{\rm Fit}$ and $(\delta \sigma_{\rm L})_{\rm Fit}$, are used as the total statistical uncertainties for the official separated cross section results.







\begin{table}[t]
\centering
\small
\setlength{\tabcolsep}{0.5 em}

\caption[Total statistical and systematic uncertainties for $\sigma_{\rm T}$ and $\sigma_{\rm L}$]{Summary table of the relevant parameters for the estimation of the total statistical uncertainties $(\delta\sigma_{\rm L})_{\rm Form.}$ and $(\delta \sigma_{\rm T})_{\rm Form.}$ using Eqns.~\ref{eqn:sigt_err} and ~\ref{eqn:sigl_err}. Note that these total statistical uncertainties include a contribution from the point-to-point systematic uncertainty.  Fitting uncertainties for $\sigma_{\rm L}$ and $\sigma_{\rm T}$ obtained from the simultaneous fit of $\sigma_u$ at $\epsilon_{low}$ and $\epsilon_{high}$  (shown in Fi.g~\ref{fig:sig_exp_160}), are listed in columns $(\delta \sigma_{\rm L})_{\rm Fit}$ and $(\delta \sigma_{\rm T})_{\rm Fit}$. Note that the $\sigma_{\rm T}/\sigma_{\rm L}$ shown in the table does not fully take into account the all uncertainties, and should not be considered as part of the final results.}
\label{tab:LT_error_tab}
\begin{tabular}{cccccccccc}
\toprule
$u$ & $\sigma_{\rm T}/\sigma_{\rm L}$ & $\epsilon_1$ & $\epsilon_2$ & $\delta\sigma_1/\sigma_1$ & $\delta\sigma_2/\sigma_2$ & $(\delta\sigma_{\rm T})_{\rm Form.}$ & $(\delta \sigma_{\rm T})_{\rm Fit}$ & $(\delta \sigma_{\rm L})_{\rm Form.}$ & $\delta (\sigma_{\rm L})_{\rm Fit}$ \\
GeV$^2$ &         &          &     	&  \%    &  \%    &   \%      & \%        &     \%    &  \%       \\ \toprule
\multicolumn{10}{c}{$W_{\rm nominal}=2.21$~GeV, ~~ $Q^2_{\rm nominal}=$2.45~GeV$^2$}                  \\ \toprule
0.058   & 0.90    &  0.316   &  0.586  &  6.43  &  3.61    &  16.30  & 20.92   &  36.52    &   36.57  \\ 
0.135   & 2.10    &  0.327   &  0.593  &  3.64  &  3.93    &  11.65  & 13.03   &  52.96    &   57.17  \\
0.245   & 3.29    &  0.334   &  0.597  &  4.96  &  4.30    &  13.12  & 15.05   &  94.03    &   101.52 \\ \midrule
\multicolumn{10}{c}{$W_{\rm nominal}=2.21$~GeV, ~~ $Q^2_{\rm nominal}=$2.45~GeV$^2$}                  \\ \midrule
0.117   & 5.04    &  0.267   &  0.552  &  4.85  &  3.71    &  9.09   & 11.62   &  116.27   &   125.05 \\
0.245   & 2.44    &  0.275   &  0.557  &  3.39  &  3.42    &  8.55   & 9.87    &  49.92    &   54.14  \\
0.400   & -17.76  &  0.285   &  0.563  &  3.79  &  4.13    &  9.06   & 9.65    &  349.83   &   383.60 \\  
\bottomrule
\end{tabular}

\end{table}

\subsubsection*{Estimation of the Systematic Scale Error}

There are three components contributing to the total systematic scale error:
\begin{itemize}
\item Correlated scale systematic error of 2.6\% (shown in Table~\ref{tab:err_tab});
\item Unseparated cross section scale error, due to the $\epsilon$ uncorrelated $u$ correlated systematic error (also shown in Table~\ref{tab:err_tab});
\item Separated cross section scale error, due to the choice of physics model parameterization (Eqns.~\ref{eqn:T}-\ref{eqn:TT}), and the binning limits in $\phi$ and $u$;
\end{itemize}
The total scale error (for each bin) are calculated as the quadratic sum of all three scale error components.

In order to quantify the contribution of unseparated cross section scale error, the $\epsilon$ uncorrelated $u$ correlated errors (1.7\%, 2.0\% for $Q^2$=1.6, 2.45 GeV$^2$ shown in Table~\ref{tab:err_tab}) are used to study the variations in the separated cross sections. There are four scenarios studied: 
\begin{itemize}
\item $\epsilon_{low}$ is shifted up by the $\epsilon$ uncorrelated $u$ correlated error while $\epsilon_{high}$ is fixed.
\item $\epsilon_{low}$ is shifted down by the $\epsilon$ uncorrelated $u$ correlated error while $\epsilon_{high}$ is fixed.
\item $\epsilon_{low}$ is fixed while $\epsilon_{high}$ is shifted up by the $\epsilon$ uncorrelated $u$ correlated error.
\item $\epsilon_{low}$ is fixed while $\epsilon_{high}$ is shifted down by the $\epsilon$ uncorrelated $u$ correlated error.
\end{itemize}
The absolute percentage difference (compared to the official separated cross sections) from the first and second scenarios are averaged, and same for the third and forth scenarios. The two sets of averaged absolute percentage differences are then added in quadrature to obtain the unseparated cross section scale error.


\begin{table}[t]
\centering
\setlength{\tabcolsep}{1 em}
\caption[Total systematic scale uncertainties]{Total systematic scale uncertainties for the separated cross sections.}

\label{tab:scale_error_tab}
\begin{tabular}{ccccc}
\toprule
$u$     & $\Delta\sigma_{\rm T}$ & $\Delta\sigma_{\rm L}$ &  $\Delta\sigma_{\rm LT}$ & $\Delta\sigma_{\rm TT}$  \\
GeV$^2$ &    \%     &  \%      &   \%      & \%                 \\ \toprule 
\multicolumn{5}{c}{$W_{\rm nominal}=2.21$~GeV, ~~ $Q^2_{\rm nominal}=$1.60~GeV$^2$} \\ \toprule
0.058   &  6.592  & 12.879   &   92.238	 &  729.139   \\ 
0.135   &  5.805  & 23.767   &  239.567	 &  828.214   \\
0.245   &  6.036  & 42.813   &   27.732	 &   49.298   \\ \midrule
\multicolumn{5}{c}{$W_{\rm nominal}=2.21$~GeV, ~~ $Q^2_{\rm nominal}=$2.45~GeV$^2$} \\ \midrule
0.117   &  8.764  &  93.974  &   78.334  &   44.779   \\
0.245   &  8.143  &  41.078  &   16.709  &  286.531   \\
0.400   & 15.342  & 391.930  &  118.128  &   60.462   \\  
\bottomrule
\end{tabular}

\end{table}

Two independent re-analyses were performed to investigate the separated cross section scale error:
\begin{itemize}
\item Re-analysis with the initial parameter of $\sigma_{\rm LT}$ and $\sigma_{\rm TT}$ set to 0.
\item Re-analysis with 10$^\circ$ offset to the center of $\phi$ bins.
\end{itemize}
The percentage difference (compared to the official separated cross sections) for each of the re-analyses is calculated and added in quadrature to give the separated cross section scale error. The scale error due to the $u$ bin limits is considered to be small compared to the contributions from the $\phi$ binning and initial parameterization. The separated cross section scale error will be revised to include the $u$ bin limits contribution before the final publication.

The total scale errors are calculated as the quadratic sum of all three components of the scale error, and are listed in Table~\ref{tab:scale_error_tab}.

\graphicspath{{pics/7results/} }
\label{chap:results}

\chapter{Results and Discussion}

In Sec.~\ref{sec:lt_sep_res}, the separated differential cross section results are presented and the $Q^2$ behavior of the L and T differential cross sections, as well as the $\sigma_{\rm L}/\sigma_{\rm T}$ and $\sigma_{\rm TT}/\sigma_{\rm T}$ ratios, are also discussed. 

Sec.~\ref{sec:u_channel_peak} presents scaled F$_\pi$-2 data points on the same $-t$ axis as the Morand data from CLAS~\cite{morand05}, which shows a potential $u$-channel peak in the exclusive $\omega$ electroproduction at $Q^2$~= 1.60 and 2.45~GeV$^2$.

In Sec.~\ref{sec:Xsection}, the extracted transverse component of the differential cross section $d\sigma_{\rm T}/dt$ is compared to the theoretical predictions made by the TDA framework at both $Q^2$ settings.

\section{The L/T Separated Extracted Differential Cross Sections}
\label{sec:lt_sep_res}

\subsection{Separated Cross Sections and General Remarks}

\begin{table}[p]
\centering
\setlength{\abovetopsep}{1ex}
\small
\setlength{\tabcolsep}{0.38em}
\caption[Separated differential cross sections of $^1$H$(e,e^{\prime}p)\omega$]{Separated differential cross sections for exclusive $\omega$ production: $p(e,e^{\prime}p)\omega$, at $Q^2_{\rm nominal}$ = 1.60 and 2.45~GeV$^2$, $W_{\rm nominal}$ = 2.21~GeV. For each contribution, the $\delta$ represents total statistical uncertainty which includes the fitting error, shown in Table~\ref{tab:LT_error_tab}; $\Delta$ is for the systematic (scale) uncertainty, listed in Table~\ref{tab:scale_error_tab}. $\braket{-u}$, $\braket{u_{\rm min}}$, $x$, $\braket{W}$ and $\braket{Q^2}$ represent the corresponding kinematics values of each bin; $\theta^{*}$ is the $\omega$ emission angle in the CM frame.}
\label{tab:xsec_raw}
\begin{tabular}{ccccccc|c|c}
\toprule
$\braket{-u}$  & $\braket{-u_{\rm min}}$ &  $u^{\prime}$ & $\braket{W}$ & $\braket{Q^2}$ &  $\theta^*$  &  $\braket{x}$ & $\sigma_{\rm T}\pm\delta\sigma_{\rm T}\pm\Delta\sigma_{\rm T}$ & $\sigma_{\rm L}\pm\delta\sigma_{\rm L}\pm\Delta\sigma_{\rm L}$ \\
GeV$^2$ & GeV$^2$ & GeV$^2$ & GeV  & GeV$^2$      &  $^\circ$ &  & $\mu$b/GeV$^2$   & $\mu$b/GeV$^2$    \\ \midrule
\multicolumn{9}{c}{$W_{\rm nominal}=2.21$~GeV, ~~ $Q^2_{\rm nominal}=$1.60~GeV$^2$} \\ \midrule
0.058 & 0.058 & 0.000 & 2.26 & 1.47 &  180 & 0.26 & 0.320 $\pm$ 0.067 $\pm$ 0.021 &  0.356 $\pm$ 0.130 $\pm$ 0.046 \\
0.135 & 0.078 & 0.057 & 2.22 & 1.58 &  166 & 0.28 & 0.309 $\pm$ 0.040 $\pm$ 0.018 &  0.147 $\pm$ 0.083 $\pm$ 0.035 \\
0.245 & 0.097 & 0.148 & 2.19 & 1.67 &  157 & 0.23 & 0.284 $\pm$ 0.043 $\pm$ 0.017 &  0.087 $\pm$ 0.089 $\pm$ 0.037 \\ \midrule
\multicolumn{9}{c}{$W_{\rm nominal}=2.21$~GeV, ~~ $Q^2_{\rm nominal}=$2.45~GeV$^2$} \\ \midrule
0.117 & 0.117 & 0.000 & 2.28 & 2.23 &  180 & 0.34 & 0.243 $\pm$ 0.028 $\pm$ 0.021 &  0.048 $\pm$ 0.060 $\pm$ 0.045 \\
0.245 & 0.188 & 0.091 & 2.23 & 2.39 &  164 & 0.37 & 0.179 $\pm$ 0.017 $\pm$ 0.014 &  0.073 $\pm$ 0.040 $\pm$ 0.030 \\
0.400 & 0.252 & 0.207 & 2.18 & 2.52 &  155 & 0.39 & 0.203 $\pm$ 0.019 $\pm$ 0.031 & -0.011 $\pm$ 0.044 $\pm$ 0.045 \\ 
\bottomrule
\end{tabular}

\vspace{20mm}

\caption[Separated differential cross sections of $^1$H$(e,e^{\prime}p)\omega$ (continued)]{Continuation of Table~\ref{tab:xsec_raw}. }
\label{tab:xsec_raw_LT_TT}
\begin{tabular}{cccc}
\toprule
$\sigma_{\rm LT}\pm\delta\sigma_{\rm LT}\pm\Delta\sigma_{\rm LT}$ & $\sigma_{\rm TT}\pm\delta\sigma_{\rm TT}\pm\Delta\sigma_{\rm TT}$ & $\sigma_{\rm L}/\sigma_{\rm T} \pm \delta \pm \Delta$ & $\sigma_{\rm TT}/\sigma_{\rm T} \pm \delta \pm \Delta$\\
$\mu$b/GeV$^2$ & $\mu$b/GeV$^2$ & $\mu$b/GeV$^2$ & $\mu$b/GeV$^2$ \\ \midrule
\multicolumn{4}{c}{$W_{\rm nominal}=2.21$~GeV, ~~ $Q^2_{\rm nominal}=$1.60~GeV$^2$} \\ \midrule
  0.008 $\pm$ 0.020 $\pm$ 0.007  &  0.007 $\pm$ 0.045 $\pm$ 0.054 &  1.114 $\pm$ 0.469 $\pm$ 0.070 &  0.023 $\pm$ 0.142 $\pm$ 0.166 \\
 -0.002 $\pm$ 0.015 $\pm$ 0.005  &  0.003 $\pm$ 0.034 $\pm$ 0.027 &  0.476 $\pm$ 0.279 $\pm$ 0.086 &  0.011 $\pm$ 0.112 $\pm$ 0.088 \\
  0.022 $\pm$ 0.012 $\pm$ 0.006  & -0.185 $\pm$ 0.036 $\pm$ 0.091 &  0.304 $\pm$ 0.312 $\pm$ 0.112 & -0.651 $\pm$ 0.159 $\pm$ 0.282 \\ \midrule
\multicolumn{4}{c}{$W_{\rm nominal}=2.21$~GeV, ~~ $Q^2_{\rm nominal}=$2.45~GeV$^2$} \\ \midrule
 -0.010 $\pm$ 0.012 $\pm$ 0.008  & -0.050 $\pm$ 0.026 $\pm$ 0.023 &  0.199 $\pm$ 0.249 $\pm$ 0.169 & -0.207 $\pm$ 0.109 $\pm$ 0.075  \\
 -0.011 $\pm$ 0.008 $\pm$ 0.002  &  0.005 $\pm$ 0.019 $\pm$ 0.013 &  0.410 $\pm$ 0.225 $\pm$ 0.135 &  0.025 $\pm$ 0.106 $\pm$ 0.070  \\
  0.012 $\pm$ 0.008 $\pm$ 0.014  &  0.093 $\pm$ 0.019 $\pm$ 0.056 & -0.056 $\pm$ 0.216 $\pm$ 0.212 &  0.457 $\pm$ 0.105 $\pm$ 0.206  \\ 
\bottomrule
\end{tabular}
\end{table}

\begin{figure}[th!]
\centering
\includegraphics[width=1.0\textwidth]{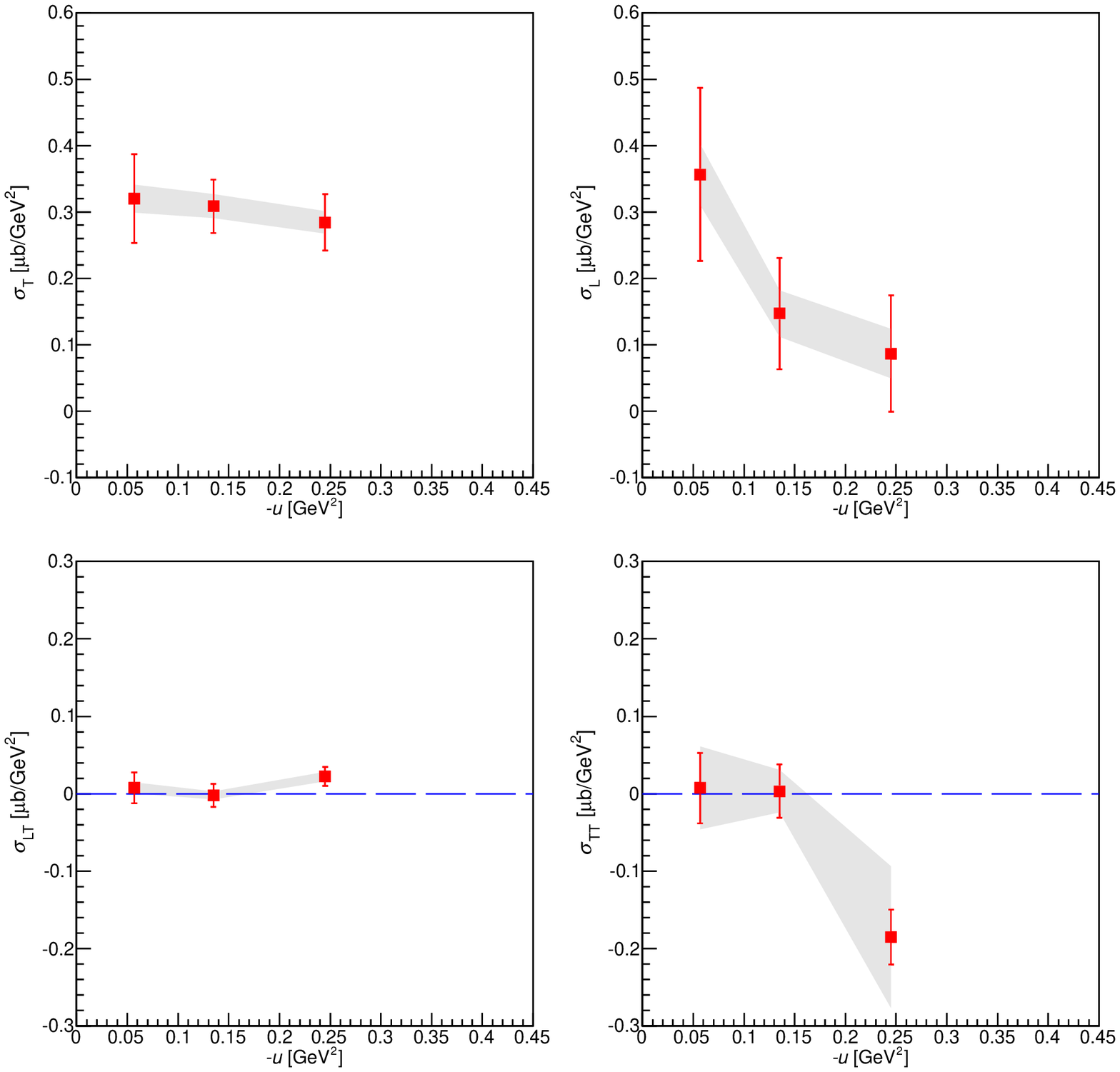}
\caption[Separated cross sections for $W$ = 2.21~GeV and $Q^2$ = 1.60~GeV$^2$]{Extracted $\sigma_{\rm T}$, $\sigma_{\rm L}$, $\sigma_{\rm LT}$ and $\sigma_{\rm TT}$ versus $-u$ for $W$ = 2.21~GeV and $Q^2$ = 1.60~GeV$^2$. These data points are not scaled to the common $Q^2$ and $W$ value. Grey bands indicate the systematic errors. The blue dashed lines indicate zero for the interference cross sections. The numerical values are listed in Tables~\ref{tab:xsec_raw} and \ref{tab:xsec_raw_LT_TT}.~\oic}
\label{fig:sigma_160}
\end{figure}

\begin{figure}[th!]
\centering
\includegraphics[width=1.0\textwidth]{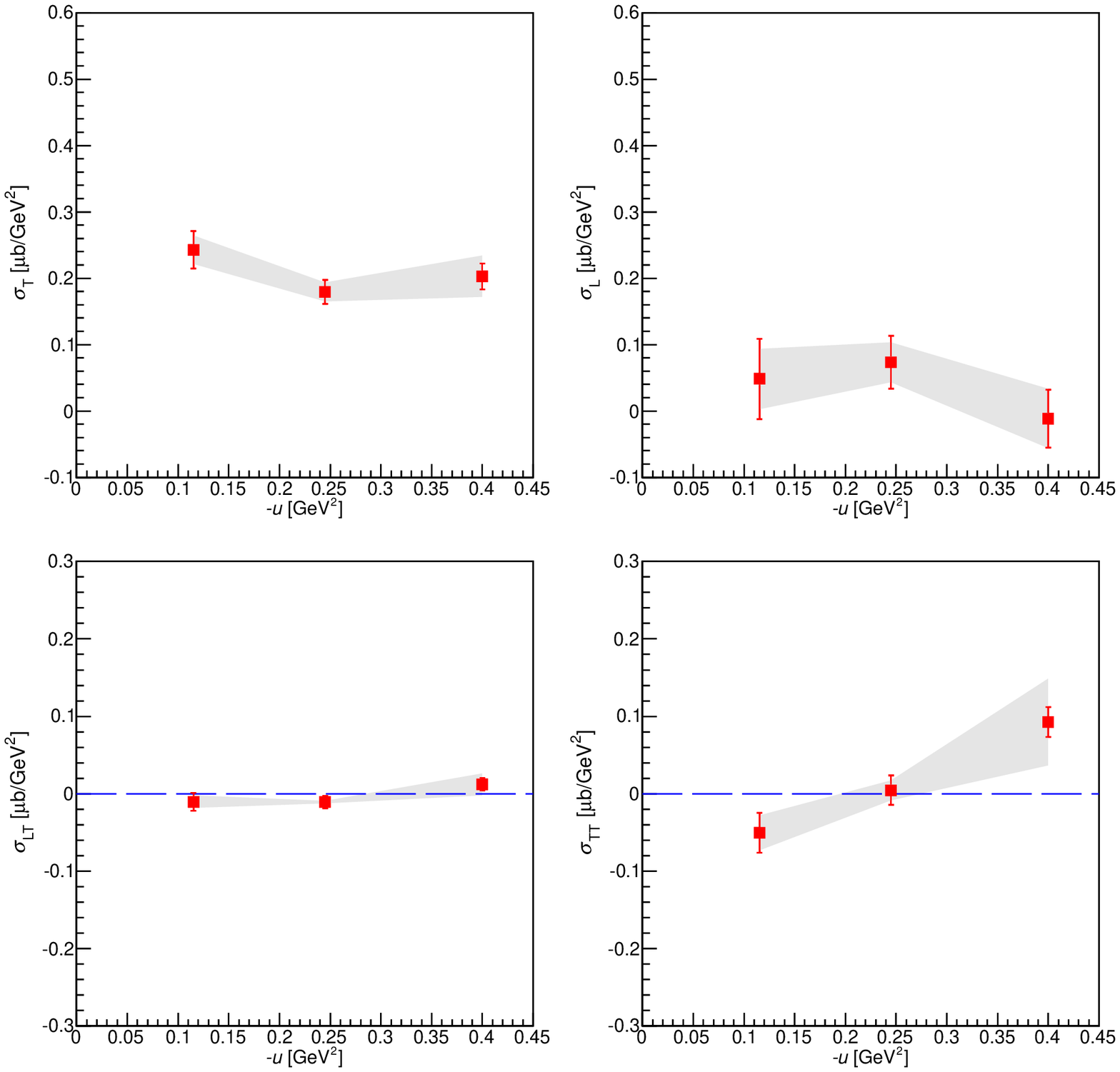}
\caption[Separated cross sections for $W$ = 2.21~GeV and $Q^2$ = 2.45~GeV$^2$]{Extracted $\sigma_{\rm T}$, $\sigma_{\rm L}$, $\sigma_{\rm LT}$ and $\sigma_{\rm TT}$ versus $-u$ for $W$ = 2.21~GeV and $Q^2$ = 2.45~GeV$^2$. These data points are not scaled to the common $Q^2$ and $W$ value. Grey bands indicate the systematic errors. The blue dashed lines indicate zero for the interference cross sections. The numerical values are listed in Tables~\ref{tab:xsec_raw} and \ref{tab:xsec_raw_LT_TT}. ~\oic}
\label{fig:sigma_245}
\end{figure}

The differential cross sections presented here have been extracted using the Monte Carlo simulation and the relation described in Eqn.~\ref{eqn:xsection}. The differential cross sections for the low and high $\epsilon$ measurements at both $Q^2$ settings are shown in Figs.~\ref{fig:sig_exp_160} and~\ref{fig:sig_exp_245},  and these numerical values are listed in Table~\ref{tab:unxsec_raw}. The separated differential cross sections $d\sigma_{\rm T}/dt$ and $d\sigma_{\rm L}/dt$ for all three $-u$ bins are listed in Table~\ref{tab:xsec_raw}, whereas the $d\sigma_{\rm LT}/dt$ and $d\sigma_{\rm TT}/dt$ are listed in Table~\ref{tab:xsec_raw_LT_TT}. The statistical uncertainties for the seperated cross sections come from the function fitting, which includes contributions from the statistical and point-to-point uncorrelated systematic uncertainties. Estimations of the systematic scale errors were discussed in Sec.~\ref{sec:sys_err}. From this point onwards, the separated differential cross sections, such as $d\sigma_{\rm T}/dt$ and $d\sigma_{\rm L}/dt$, are written as $\sigma_{\rm T}$ and $\sigma_{\rm L}$ for simplicity purposes.

Figs.~\ref{fig:sigma_160} and \ref{fig:sigma_245} show $\sigma_{\rm T}$, $\sigma_{\rm L}$, $\sigma_{\rm LT}$ and $\sigma_{\rm TT}$ as functions of $-u$ for $Q^2$ = 1.60 and 2.45~GeV$^2$, respectively. It is important to note that these data points are extracted according to the individual kinematics coverage ($\braket{W}$ and $\braket{Q^2}$) and are not scaled to the common $W$ and $Q^2$ values, therefore should not be used for the $u$ dependence study. 

From the general trends of the L/T separated differential cross sections, some qualitative remarks can be drawn:
\begin{itemize}

\item The behavior of $\sigma_{\rm T}$ and $\sigma_{\rm L}$ is similar at both $Q^2$ values: $\sigma_{\rm T}$ shows weak dependence with respect to $-u$, where $\sigma_{\rm L}$ falls more quickly.

\item For $Q^2$ = 2.45~GeV$^2$, $-u$ = 0.4~GeV$^2$, $\sigma_{\rm L}$ is nearly consistent with zero.

\item L-T interference contribution $\sigma_{\rm LT}$ is consistent with zero, even at large $-u$ $>$ 0.2 at both $Q^2$ settings. This is consistent with the observed diminishing of $\sigma_{\rm L}$ at large $-u$.

\item In contrast, the data show a more significant $\sigma_{\rm TT}$ contribution, particularly at large $-u$ values. Furthermore, $\sigma_{\rm TT}$ has a different dependence at different $Q^2$ settings, i.e. $\sigma_{\rm TT} (u=0.25)$ $<$ 0 at $Q^2$ = 1.6~GeV$^2$, and $\sigma_{\rm TT} (u=0.40)$ $>$ 0 at $Q^2$ = 2.45~GeV$^2$.

\end{itemize}

\subsection{Cross Section $\sigma_{\rm L}/\sigma_{\rm T}$ Ratio Studies}

\begin{figure}[p]
\centering
\includegraphics[width=0.8\textwidth]{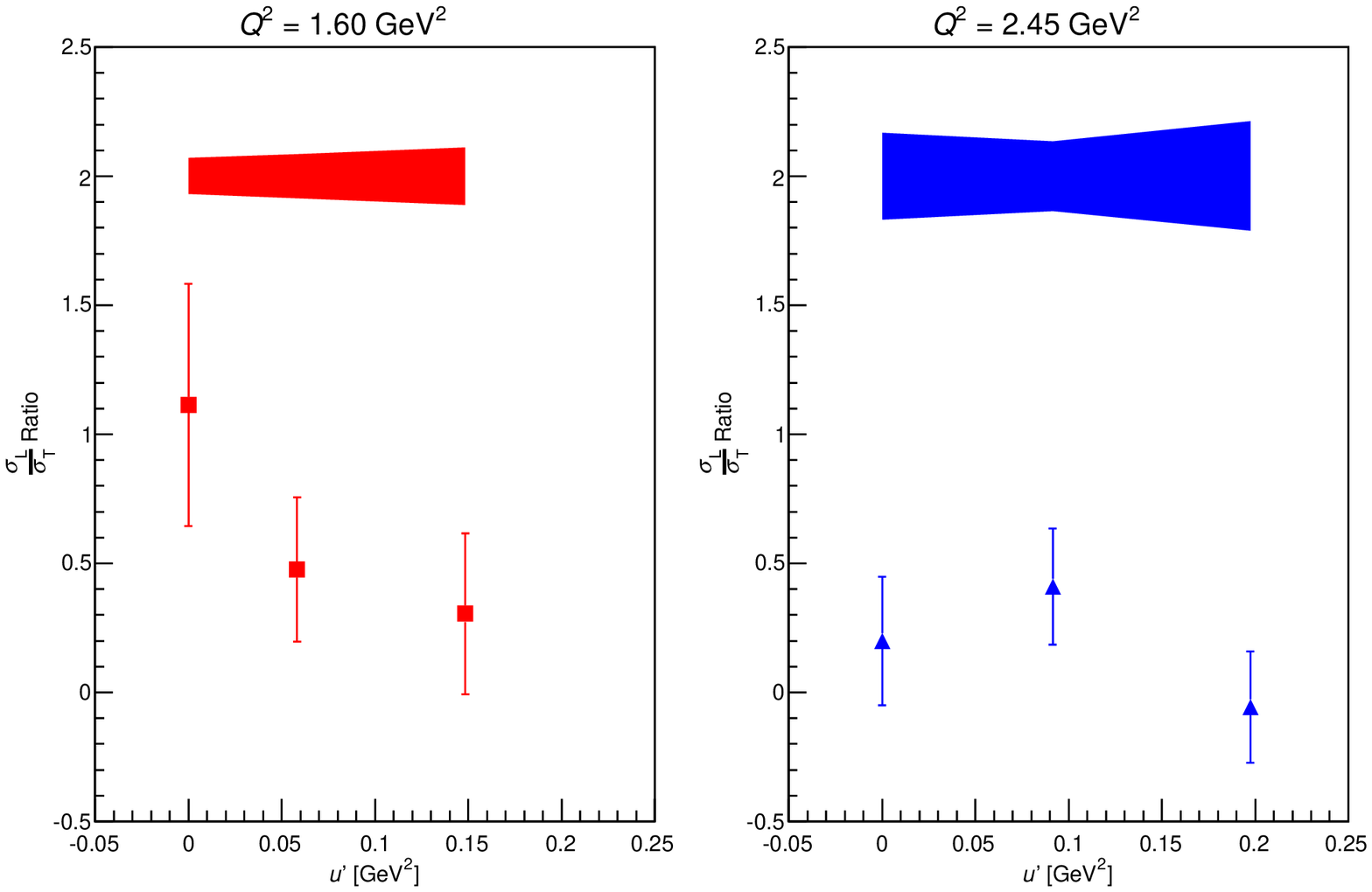}
\caption[$\sigma_{\rm L}/\sigma_{\rm T}$ ratio versus $u^{\prime}$]{$\sigma_{\rm L}/\sigma_{\rm T}$ versus $u^{\prime}$ for $Q^2$ = 1.60 and 2.45 GeV$^2$. These data points are not scaled to the common $W$ and $Q^2$ values. The numerical values are listed in Tables~\ref{tab:xsec_raw} and \ref{tab:xsec_raw_LT_TT}.~\oic}
\label{fig:t_l_ratio_uprime}
\centering
\includegraphics[width=0.8\textwidth]{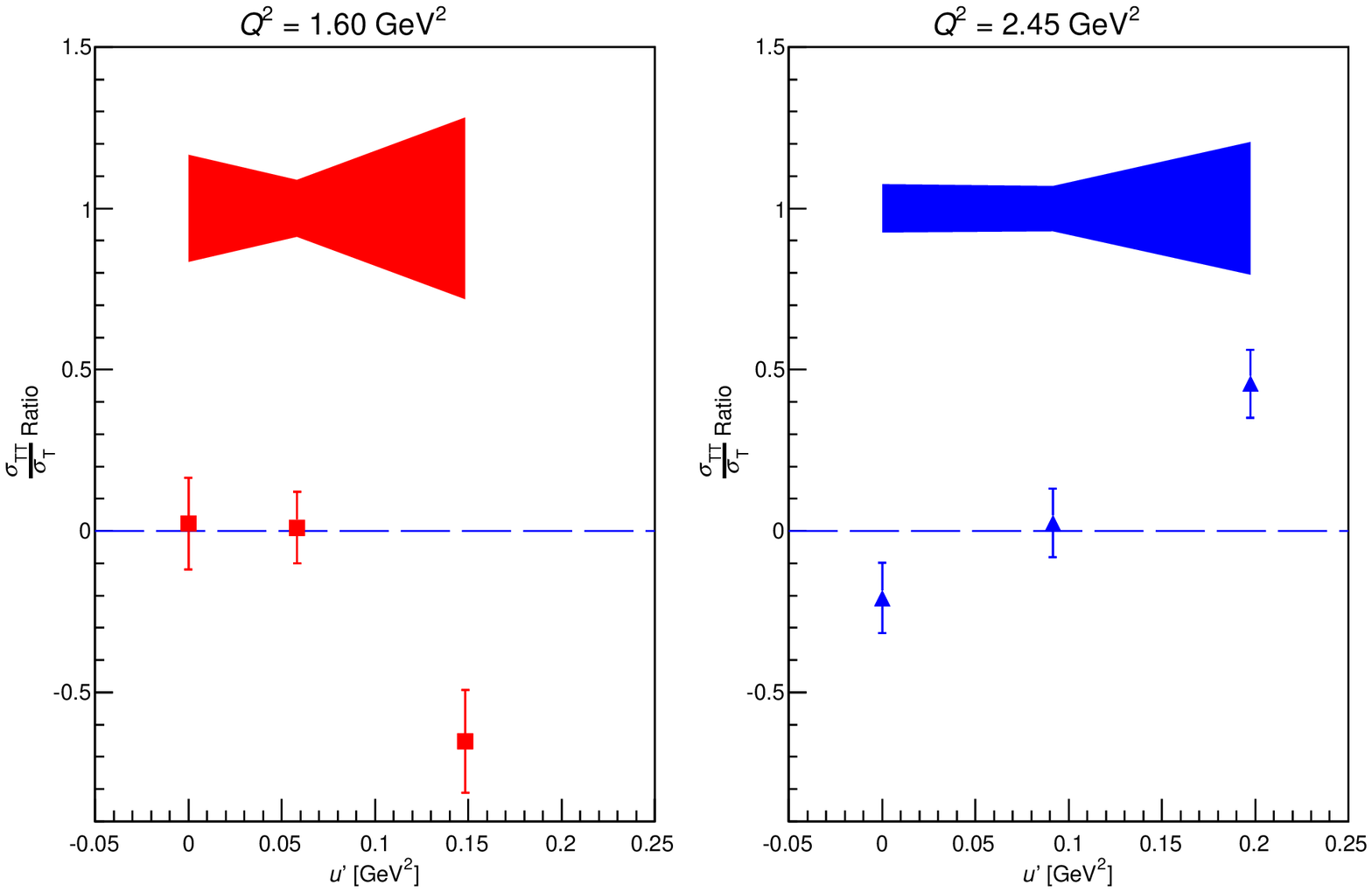}
\caption[$\sigma_{\rm TT}/\sigma_{\rm T}$ ratio versus $u^{\prime}$]{$\sigma_{\rm TT}/\sigma_{\rm T}$ versus $u^{\prime}$ for $Q^2$ = 1.60 and 2.45 GeV$^2$. These data points are not scaled to the common $W$ and $Q^2$ values. The systematic uncertainty shown is from the contribution from $\sigma_{\rm T}$. The blue dashed lines indicate zero. The numerical values are listed in Tables~\ref{tab:xsec_raw} and \ref{tab:xsec_raw_LT_TT}.~\oic}
\label{fig:t_tt_ratio}
\end{figure}

Figs.~\ref{fig:t_l_ratio_uprime} shows the differential cross section ratios $\sigma_{\rm L}/\sigma_{\rm T}$ versus $u^{\prime}$ at $Q^2$ = 1.60 and 2.45~GeV$^{2}$. Here $u^{\prime} = |u-u_{\rm min}|$, whose values are listed in Table~\ref{tab:xsec_raw}. These data points are not scaled to the common $W$ and $Q^2$ values. At both $Q^2$ settings, the ratios appear to drop as $u^{\prime}$ increases. With large statistical uncertainty, data show $\sigma_{\rm L}$ being more suppressed $Q^2$ = 2.45~GeV$^2$ compared $Q^2$ = 1.60~GeV$^2$ for $u^{\prime}$ $<$ 0.05~GeV$^2$. Note that the ratio is consistent with zero for $Q^2$ = 2.45~GeV$^2$, $u^{\prime}$ = 0.2~GeV$^2$.

The cross section $\sigma_{\rm TT}/\sigma_{\rm T}$ ratios versus $u^{\prime}$ for $Q^2$ = 1.60 and 2.45~GeV$^{2}$ are plotted in Fig.~\ref{fig:t_tt_ratio}. These data points are not scaled to the common $W$ and $Q^2$ values. At $u^{\prime}$ $<$ 0.1, the $\sigma_{\rm TT}$ ratios at both $Q^2$ settings are consistent with zero, since $\sigma_{\rm TT} \rightarrow 0$ are required by the physical constraints imposed by the antiparallel kinematics (described in Sec.~\ref{sec:LT_sep}). Furthermore, the ratios deviate from zero at $u^{\prime}$ $>$ 0.1 for both $Q^2$ settings.

\begin{figure}[t]
\centering
\includegraphics[width=0.8\textwidth]{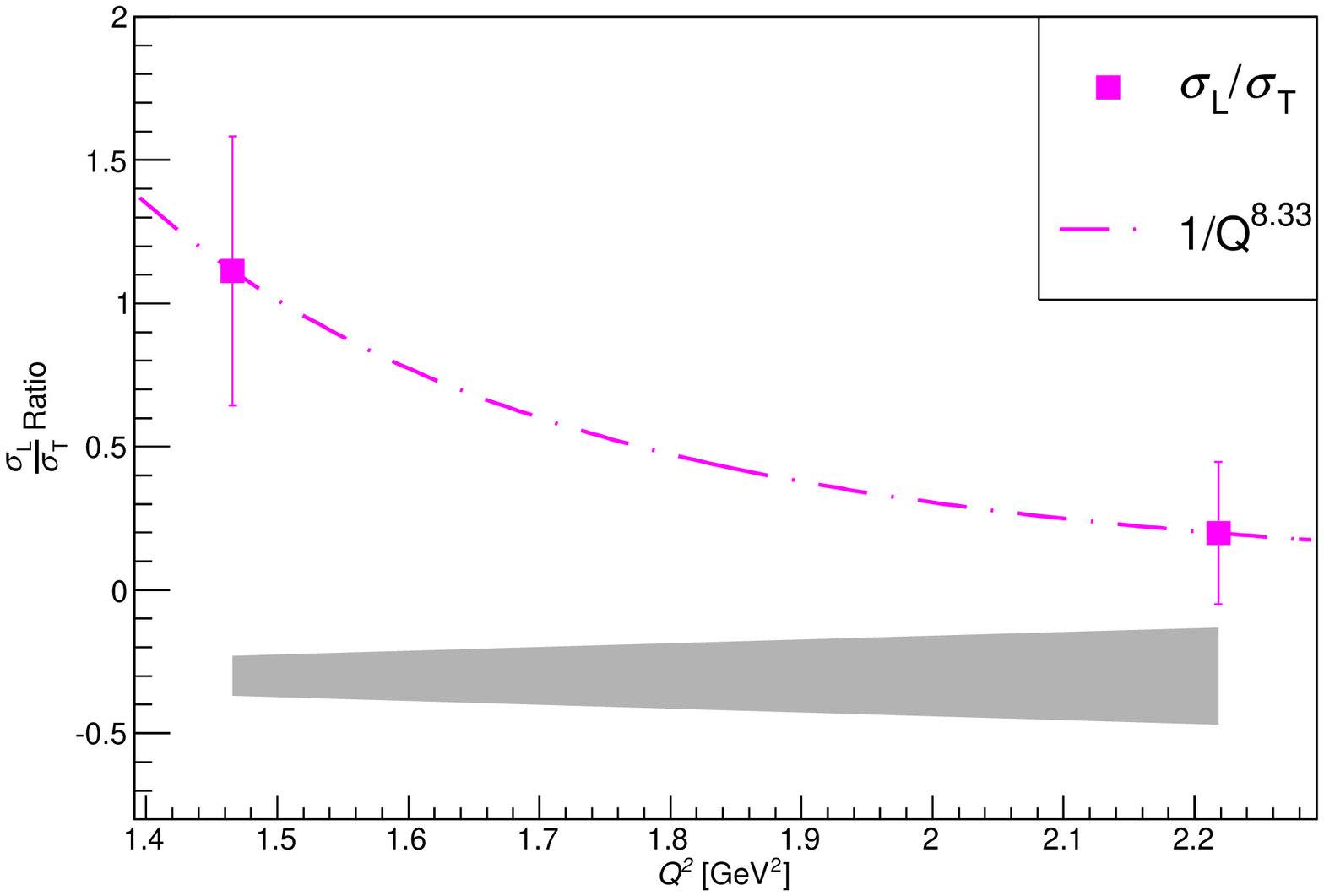}
\caption[$\sigma_{\rm L}/\sigma_{\rm T}$ ratio versus $Q^2$]{$\sigma_{\rm L}/\sigma_{\rm T}$ ratio versus $Q^2$ at $u^{\prime}$ = 0 GeV$^2$. All curves are normalized to the data point at $Q^2 = 1.47$ GeV$^2$. The $1/Q^n$ dependence is fitted using Eqn.~\ref{eqn:Qfit}. The fitted $n=8.33\pm6.38$. These data points are not scaled to the common $W$ and $Q^2$ values. The numerical values are listed in Tables~\ref{tab:xsec_raw} and \ref{tab:xsec_raw_LT_TT}.

~\oic}
\label{fig:q2_ratio}
\end{figure}

In addition to the $u^{\prime}$ dependence studies, the $Q^2$ dependence of the cross section ratio ($\sigma_{\rm L}/\sigma_{\rm T}$) at the lowest $-u$ bin is also studied ($u^{\prime}$ = 0 GeV$^2$). The $\sigma_{\rm L}/\sigma_{\rm T}$ ratio versus $Q^2$ is shown Fig.~\ref{fig:q2_ratio}. In order to extract the $Q^2$ dependence, the following equation is used to fit the cross section ratio:
\begin{equation}
C = \frac{A}{Q^{2B}} = \frac{A}{Q^n}
\label{eqn:Qfit}
\end{equation}
where $C$ is $\sigma_{\rm L}/\sigma_{\rm T}$, $\sigma_{\rm T}$ or $\sigma_{\rm L}$ (later sections); $A$ and $B$ are free parameters; $n = 2\cdot B$. The fitting result suggests a $1/Q^{8.33 \pm 6.38}$ dependence for the $\sigma_{\rm L}/\sigma_{\rm T}$ ratio. The full fitting results are listed in Table~\ref{tab:ratio_fit}.

\subsection{$W$ and $Q^2$ Scaling Factors}

\label{sec:wqscaling}

\begin{figure}[t]
\centering
\includegraphics[width=0.8\textwidth]{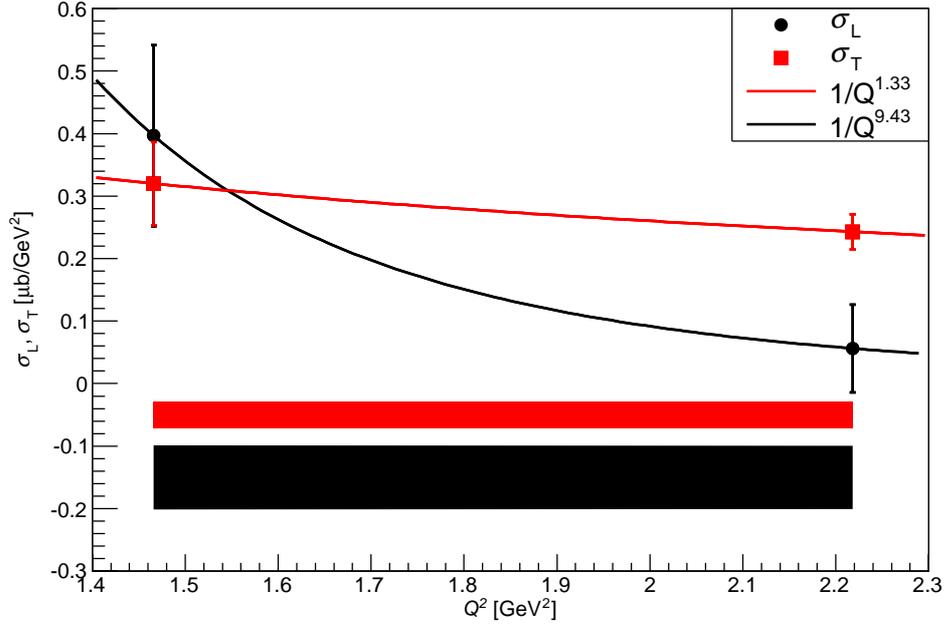}
\caption[$\sigma_{\rm L}$ and $\sigma_{\rm T}$ versus $Q^2$]{$\sigma_{\rm L}$ and $\sigma_{\rm T}$ versus $Q^2$ for the lowest $-u$ bin ($u^\prime$ = 0 GeV$^2$). These the data are scaled to the common $W$ = 2.21 GeV. The $1/Q^n$ dependence is fitted using Eqn.~\ref{eqn:Qfit}. The fitted $n=1.33\pm1.12$ for $\sigma_{\rm T}$ and $n=9.43\pm6.28$ for $\sigma_{\rm L}$.~\oic}
\label{fig:lt_q2}
\end{figure}

As described in Sec.~\ref{sec:extracting_LT_cx}, each ($Q^2$, $u$) bin has slightly different $\braket{Q^2}$ and $\braket{W}$ values that deviate from the nominal $Q^2_{\rm nom}$ and $W_{\rm nom}$ values. Therefore, a small $Q^2$ and $W$ correction is required to adjust these small deviations in order to perform a quantitative comparison between the separated cross sections to theoretical predictions or measurements from other experiments.

In terms of the $W$ dependence scaling, adjusting $\braket{W}$ to a given $W_{\rm nom}$, the following expression can be used,
\begin{equation}
\frac{(\braket{W}^2-M_p^2)^2}{(W_{\rm nom}^2-M_p^2)^2}.
\label{eqn:w_scale}
\end{equation}
This expression is based on Eqn.~\ref{eqn:w_dep}, and is our best estimate for the $W$ correction~\cite{brauel79}, and is small since the $W$ and $W_{\rm nom}$ are close.

\begin{table}[t]
\centering
\caption[$Q^2$ dependence for $\sigma_{\rm T}$, $\sigma_{\rm L}$ and $\sigma_{\rm L}/\sigma_{\rm T}$]{Fitting results of the $Q^2$ dependence for $\sigma_{T}$, $\sigma_{\rm L}$ and $\sigma_{\rm L}/\sigma_{\rm T}$. Free parameter $A$ and $B$ are defined in Eqn.~\ref{eqn:Qfit}. $n = 2\cdot B$ and quantify the $1/Q^n$ dependence.}
\label{tab:ratio_fit}
\setlength{\tabcolsep}{1.5em}
\begin{tabular}{cccc}
\toprule
                                 &  $A$  $\pm$ $\delta A$ &  $B$  $\pm$ $\delta B$ & $n$ $\pm$ $\delta n$ \\ \midrule 
$\sigma_{\rm T}$                 &  0.41 $\pm$ 0.18       &  0.65 $\pm$ 0.58       & 1.33 $\pm$ 1.12      \\ 
$\sigma_{\rm L}$                 &  2.41 $\pm$ 3.25       &  4.72 $\pm$ 3.14       & 9.43 $\pm$ 6.28      \\ 
$\sigma_{\rm L}/\sigma_{\rm T}$  &  5.47 $\pm$ 7.72       &  4.16 $\pm$ 3.19       & 8.33 $\pm$ 6.38      \\ 
\bottomrule
\end{tabular}
\end{table}

In order to determine the appropriate $Q^2$ scaling factor to adjust $\braket{Q^2}$ to a given $Q^2_{\rm nom}$ value, $\sigma_{\rm L}$ and $\sigma_{\rm T}$ at the lowest $-u$ bin versus $\braket{Q^2}$ is shown in Fig.~\ref{fig:lt_q2}. The fitting result (using Eqn.~\ref{eqn:Qfit}) suggest a flat $1/Q^{1.33 \pm 1.21}$ dependence for the $\sigma_{\rm T}$, and a stronger $1/Q^{9.43 \pm 6.28}$ dependence for the $\sigma_{\rm L}$. The fitting results are listed in Table~\ref{tab:ratio_fit}. Note that these data are scaled to the common value of $W_{\rm nom}$ = 2.21~GeV from their $\braket{W}$ values using Eqn.~\ref{eqn:w_scale}, and the points are plotted at their actual $\braket{Q^2}$ values. Note that the $1/Q$ dependence for $\sigma_{\rm T}$ and $1/Q^4$ dependence for $\sigma_{\rm L}$ are used to parameterize the $\omega$ physics model in SIMC (shown in Eqns.~\ref{eqn:T} and \ref{eqn:L}).

Considering the large uncertainties for the fitted $n$ values, conservative $1/Q$ and $1/Q^8$ dependences are chosen for $\sigma_{\rm T}$ and $\sigma_{\rm L}$, respectively, to perform $Q^2$ corrections. The expression of the $Q^2$ scaling factor is similar to the $W$ scaling factor, and is given as
\begin{equation}
\frac{\braket{Q^2}^m}{(Q^2_{\rm nom})^m},
\end{equation}
where $m$ = 0.5 for $\sigma_{\rm T}$; $m$ = 4 for $\sigma_{\rm L}$.

In addition to the $W$ and $Q^2$ scaling factors, the $\braket{-u_{\rm min}}$ values listed in Table~\ref{tab:xsec_raw} (minimum possible $-u$ value corresponding to $\theta$ = 180$^\circ$) for each bin are slightly different from the nominal $-u_{\rm min, nom}$ values due to the variations in $\braket{Q^2}$ and $\braket{W}$ values. The difference between $\braket{u}$ and $\braket{u_{\rm min}}$ is written as $u^\prime$, and is defined as 
\begin{equation}
u^{\prime} = |\braket{u} - \braket{u_{\rm min}}|.
\end{equation}
$u^{\prime}$ is a good intermediate parameter to shift cross sections measured in $\braket{-u}$ space to $-u$ space with a nominal $u_{\rm nom, min}$ offset. The $-u$ value can be calculated as 
\begin{equation}
-u = -u_{\rm nom, min} + u^{\prime}.
\label{eqn:u_cor}
\end{equation}
This $-u$ value adjustment technique is an adequate methodology to correct the cross sections from three separate $u$ bins to a common $u_{\rm min}$ offset and is used for all result comparisons in this chapter.

\section{The $u$-Channel Peak}
\label{sec:u_channel_peak}

The charged pion photoproduction ($\gamma p \rightarrow n \pi^+$) data~\cite{anderson69, anderson76, boyarski68}, shown in Fig.~\ref{fig:vgl_data} from Sec.~\ref{sec:regge_model_intro}, contains a strong $t$-channel (forward-angle) peak and a $u$-channel (backward-angle) peak. The dominant contributions of these peaks were explained by the Regge trajectory based VGL model~\cite{vgl96, guidal97}, as the saturations of the exchanged meson (Regge) trajectories (in $t$-channel) and of the exchanged baryon trajectories (in $u$-channel).

\begin{figure}[p]
\centering
\includegraphics[width=0.85\textwidth]{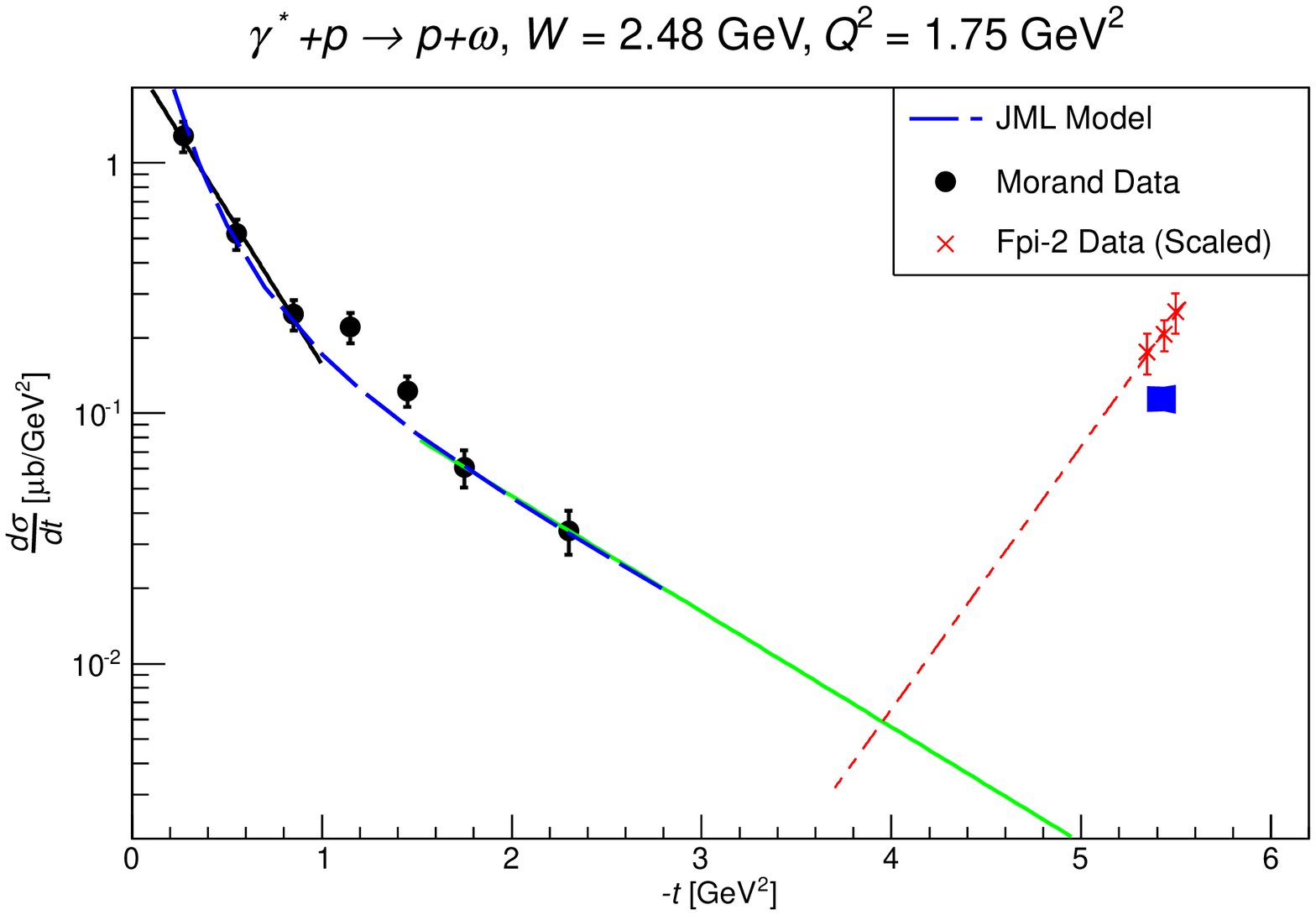}
\includegraphics[width=0.85\textwidth]{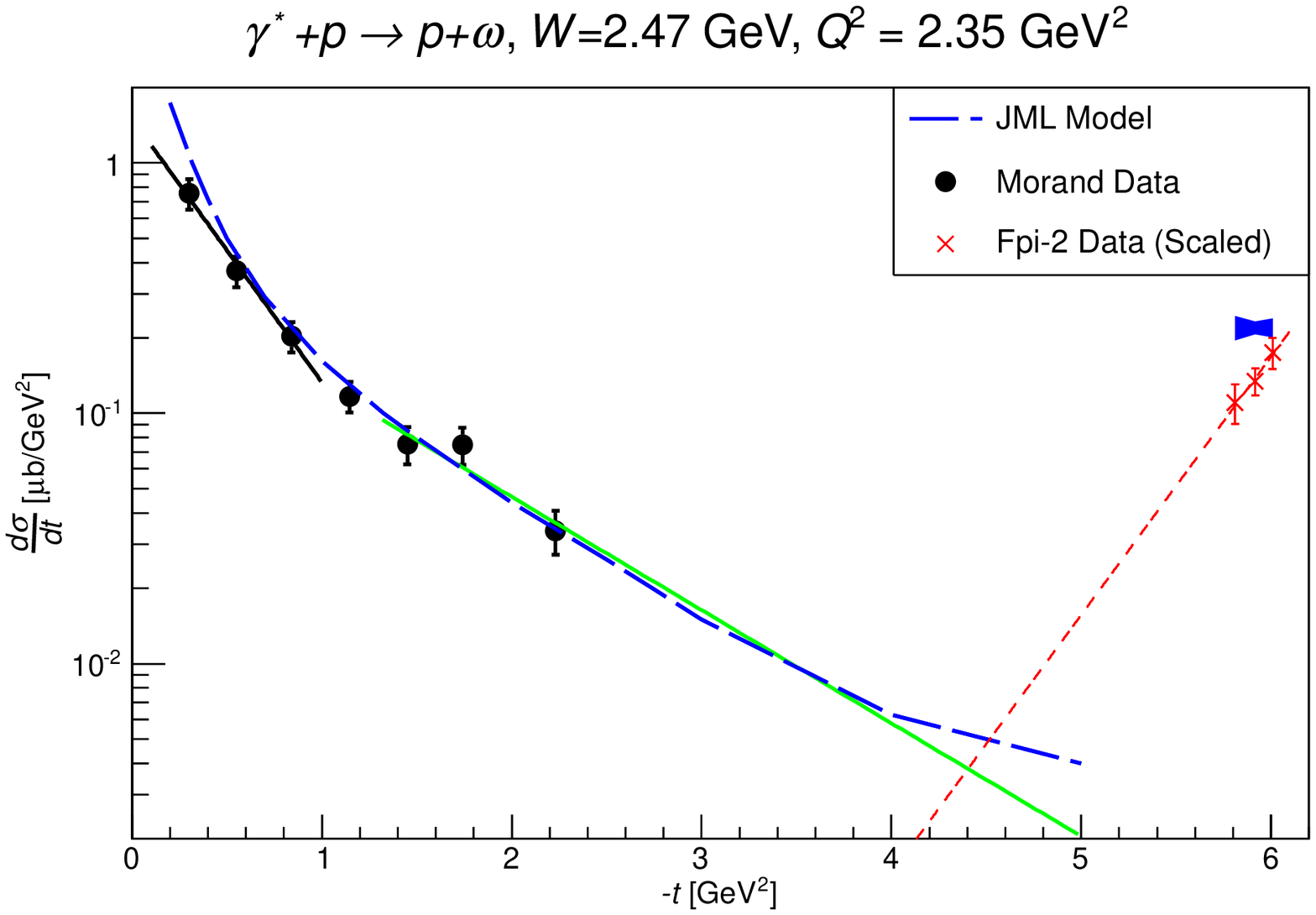}
\caption[Combined plots of $\sigma_{u}$ versus $-t$ from CLAS and F$_{\pi}$-2]{ $\sigma_{u}$ versus $-t$ for $W$ = 2.47~GeV, $Q^2$ = 1.75~GeV (top) and $W$ = 2.47~GeV, $Q^2$ = 1.75~GeV (bottom). The black dots show published CLAS results~\cite{morand05}. The red crosses show the reconstructed $\sigma_u$ (Eqn.~\ref{eqn:cross_sec_rec}) using the scaled $\sigma_{\rm T}$ and $\sigma_{\rm L}$ from this analysis, the systematic error bands are shown in the blue. The blue dashed lines represent the predictions by the Regge trajectory based JML model~\cite{laget04}. The black lines are the fitted curve showing the contribution of the forward-angle softer process (meson exchange); the green solid lines are the fitted curve showing a flatter $-t$ dependence due to the interaction with harder parton structure (harder process); red dashed line are the fitted curves which might indicate the contribution due to the softer baryon exchange in the backward-angle. The numerical values are listed in Table~\ref{tab:xsec_regge}. \oic}
\label{fig:sigma_t}
\end{figure}

For the exclusive $\omega$ electroproduction process: $\gamma^* p \rightarrow \omega p$ above the resonance region ($W>2$~GeV), a strong $t$-channel peak has been observed and reported by the CLAS collaboration~\cite{morand05} at $W$ = 2.48~GeV, $Q^2$ = 1.75~GeV$^2$ and at $W$ = 2.47~GeV, $Q^2$ = 2.35~GeV$^2$. The differential cross sections (measured in $\mu$b/GeV$^2$) of the CLAS data versus $-t$ are shown in Fig.~\ref{fig:sigma_t} as the black dots. The blue dashed lines are the JML model predictions~\cite{laget04, laget00, laget02}, which include the meson exchange Regge trajectories (dominant contribution) and other contributing effects (particularly at $-t$ $>$ 1~GeV). The model predictions seem to give an excellent description to the CLAS data at both settings even at $-t$ $\sim$ 2 GeV$^2$.

The nominal $Q^2$ and $W$ values of the F$_\pi$-2 experimental data are different from those of the CLAS data. In order to compare the two data sets, the separated differential cross sections ($\sigma_{\rm T}$ and $\sigma_{\rm L}$) of F$_\pi$-2 must be corrected to match $W$ and $Q^2$ values of the CLAS data before computing the total differential cross section with the $\epsilon$ value from the CLAS data.

The $W$ = 2.21 GeV, $Q^2$ = 1.60 GeV$^2$ data from F$_{\pi}$-2 are scaled to $W$ = 2.48 GeV, $Q^2$ = 1.75 GeV$^2$; $W$ = 2.21 GeV, $Q^2$ = 2.45 GeV$^2$ data from F$_{\pi}$-2 are scaled to $W$ = 2.47 GeV, $Q^2$ = 2.35 GeV$^2$. The extrapolations of the $\sigma_{\rm T}$ and $\sigma_{\rm L}$ from $\braket{Q^2}$ and $\braket{W}$ to a new set of nominal $Q^2$ and $W$ values requires the following expressions,
\begin{equation}
\sigma_{\rm T} (W_{\rm nom}, Q^2_{\rm nom}) = \frac{\sqrt{\braket{Q^2}}}{Q_{\rm nom}} \frac{(\braket{W}^2-M_p^2)^2}{(W_{\rm nom}^2-M_p^2)^2} ~ \sigma_{\rm T}(\braket{W},\braket{Q^2}),
\label{eqn:sigt_scale}
\end{equation}
and
\begin{equation}
\sigma_{\rm L} (W_{\rm nom}, Q^2_{\rm nom}) =\frac{(\braket{Q^2})^4}{Q^8_{\rm nom}} \frac{(\braket{W}^2-M_p^2)^2}{(W_{\rm nom}^2-M_p^2)^2} ~ \sigma_{\rm L}(\braket{W},\braket{Q^2}),
\label{eqn:sig_scale}
\end{equation}
Finally, the unseparated differential cross section ($\sigma_u$) is computed using the corrected $\sigma_{\rm L}$ and $\sigma_{\rm T}$,
\begin{equation}
\sigma_u = \sigma_{\rm T} + \epsilon~\sigma_{\rm L},
\label{eqn:cross_sec_rec}
\end{equation}
where $\epsilon$ = 0.59 and 0.50 for the lower $Q^2$ and higher $Q^2$ settings of the CLAS data. The conversion between the $-u$ space to the $-t$ space is done using Eqn.~\ref{eqn:constraints}. The calculated  differential cross section $\sigma_u$, the scaled separated differential cross sections ($\sigma_{\rm T}$, $\sigma_{\rm L}$), and associated kinematics variables (such as $-u$ and $-t$) are listed in Table~\ref{tab:xsec_regge}.

Using the definition of the $-u$ and $-t$ limits described in Sec.~\ref{sec:u_t_limits} and shown in Fig.~\ref{fig:scattering_u_t}, the $-t$ coverage in Fig.~\ref{fig:sigma_t} can be divided into four different regions,
\begin{description}
\centering
\item[Low $-t$ Region:] 0.03 $<-t<$ 1 GeV$^2$
\item[Low $-u$ Region:]  5 $<-t<$ 6 GeV$^2$
\item[High $-t$ Region:] 1 $<-t<$ 3 GeV$^2$ 
\item[High $-u$ Region:] 3 $<-t<$ 5 GeV$^2$
\end{description}
Note the high $-t$ region can be combined with the high $-u$ region, to form the Large Emission Angle (LEA) region.

In Fig.~\ref{fig:sigma_t}, the extrapolated $\sigma_u$ from F$_\pi$-2 versus $-t$ are plotted as the red crosses. In both $Q^2$ settings, the F$_{\pi}$-2 data points show a strong $u$-channel peak for the low $-u$ region ($-t$ $>$ 5~GeV$^2$). The general trend of $\sigma_{u}$ as a function of $-t$, at both $Q^2$ settings, shows a gradual increase in $\sigma_{\rm T}$ as $-t$ increases. This observation offers experimental evidence for the existence of the backward-angle peak for the differential meson cross section of the $\omega$ electroproduction. The statistical and systematic scale errors associated with the $Q^2$ extrapolation will be revised before the final publication.

Equivalent to the $u$-channel peak observed in the charged pion photoproduction data (shown in Fig.~\ref{fig:vgl_data}), the $u$-channel peak in the $\omega$ electroproduction can potentially be described by the VGL and JML models, which take into account the saturation of exchange baryon trajectories; examples of baryon trajectory are shown in Fig.~\ref{fig:regge_trajectory} (b). The leading candidates for the $u$-channel vector mesons ($\omega$, $\rho^0$ and $\phi$) electroproduction are shown in Table~\ref{tab:regge_table}.

\begin{table}[t]
\centering
\footnotesize
\setlength{\tabcolsep}{0.8em}
\caption[Scaled $p(e,e^{\prime}p)\omega$ data from the F$_{\pi}$-2 experiment]{Scaled $p(e,e^{\prime}p)\omega$ data from the F$_{\pi}$-2 experiment for comparison with unseparated CLAS-6 data. $-u$ is corrected using Eqn.~\ref{eqn:u_cor} with the listed $-u^\prime$ values and $-u_{\rm nom, min}$, $\sigma_{\rm L}$ and $\sigma_{\rm T}$ are scaled to the corresponding nominal $W$ and $Q^2$ values of the CLAS data settings, and $\sigma_u$ is calculated using Eqn.~\ref{eqn:cross_sec_rec}.}
\label{tab:xsec_regge}
\begin{tabular}{ccc|c|c|c}
\toprule
$-u$     &  $-u^\prime$  &  $-t$     & $\sigma_{u}\pm\delta\sigma_{u}\pm\Delta\sigma_{u}$ & $\sigma_{\rm T}\pm\delta\sigma_{\rm T}\pm\Delta\sigma_{\rm T}$   & $\sigma_{\rm L}\pm\delta\sigma_{\rm L}\pm\Delta\sigma_{\rm L}$ \\
GeV$^2$  &  GeV$^2$      &  GeV$^2$  & $\mu$b/GeV$^2$                       & $\mu$b/GeV$^2$              & $\mu$b/GeV$^2$           \\ \midrule
\multicolumn{6}{c}{$W$ = 2.48~GeV, $Q^2$ = 1.75~GeV$^2$, $\epsilon$ = 0.59, $x$ = 0.25, $-u_{\rm min, nom}$ = 0.031 GeV$^2$}          \\ \midrule
0.031    &  0.000        &  5.496    & 0.255 $\pm$ 0.046 $\pm$ 0.015 & 0.188 $\pm$ 0.039 $\pm$ 0.012 & 0.113 $\pm$ 0.041 $\pm$ 0.015 \\
0.088    &  0.057        &  5.440    & 0.206 $\pm$ 0.029 $\pm$ 0.013 & 0.173 $\pm$ 0.023 $\pm$ 0.010 & 0.057 $\pm$ 0.032 $\pm$ 0.013 \\
0.179    &  0.148        &  5.348    & 0.176 $\pm$ 0.032 $\pm$ 0.013 & 0.153 $\pm$ 0.023 $\pm$ 0.009 & 0.039 $\pm$ 0.040 $\pm$ 0.017 \\ \midrule
\multicolumn{6}{c}{$W$ = 2.47~GeV, $Q^2$ = 2.35~GeV$^2$, $\epsilon$ = 0.50, $x$ = 0.31, $-u_{\rm nom, min}$ = 0.069 GeV$^2$}          \\ \midrule
0.069    &  0.000        &  6.009    & 0.175 $\pm$ 0.025 $\pm$ 0.019 & 0.162 $\pm$ 0.019 $\pm$ 0.014 & 0.026 $\pm$ 0.033 $\pm$ 0.024 \\
0.160    &  0.091        &  5.918    & 0.135 $\pm$ 0.017 $\pm$ 0.013 & 0.111 $\pm$ 0.011 $\pm$ 0.009 & 0.048 $\pm$ 0.026 $\pm$ 0.020 \\
0.276    &  0.207        &  5.802    & 0.110 $\pm$ 0.020 $\pm$ 0.024 & 0.115 $\pm$ 0.011 $\pm$ 0.018 &-0.009 $\pm$ 0.033 $\pm$ 0.034 \\ \bottomrule
\end{tabular}
\end{table}

It is obvious that the CLAS data at low $-t$ ($-t<1$~GeV$^2$), CLAS data at high $-t$ ($-t > 1$~GeV$^2$) and F$_\pi$-2 data low $-u$ ($<0.6$) region have different $-t$ dependences. Here, one can apply the standard technique~\cite{aaron08,abramowicz95} used in high energy physics to extract the (exponential) slope of the $-t$ dependence, by fitting the $d\sigma/dt$ with the following function:  
\begin{equation}
\frac{d\sigma}{dt} = a~e^{-b~(-t)}\,
\label{eqn:t_dep}
\end{equation}
where $a$ and $b$ are free parameters. The $b$ parameter can be linked (through phenomenological models)~\cite{aaron08,abramowicz95} to the interaction radius of between $\gamma^*$ and $p$ target (inside of the proton structure) through 
\begin{equation}
R_{\rm int} = \sqrt{|b|} ~ \hbar c,
\end{equation}
where $\hbar c$ = 0.197~GeV$\cdot$fm.

\begin{table}
\centering
\caption[$R_{\rm int}$ interaction length]{The fitted $b$ parameter values and calculated $R_{\rm int}$ for all three $-t$ regions at both $Q^2$ settings.}
\label{tab:R_int}
\setlength{\tabcolsep}{1.5em}
\begin{tabular}{cccc}
\toprule
$-t$ Region  &  $-t$ (Fitting) Range  &  $|b| \pm \delta b$  &  $R_{\rm int}$  \\ 
             &  GeV$^2$               &  GeV$^{-2}$          &  (fm)           \\ \midrule
\multicolumn{4}{c}{$W$ = 2.48~GeV, $Q^2$ = 1.75~GeV$^2$, $\epsilon$ = 0.59, $x$ = 0.25} \\ \midrule
Low  $-t$    &  $0<-t<1$     & 2.818 $\pm$ 0.362  &  0.331 $\pm$ 0.042  \\ 
High $-t$    &  $1.5<-t<2.5$ & 1.063 $\pm$ 0.477  &  0.203 $\pm$ 0.091  \\ 
Low  $-u$    &  $5<-t<6$     & 2.420 $\pm$ 1.818  &  0.306 $\pm$ 0.230  \\ \midrule
\multicolumn{4}{c}{$W$ = 2.47~GeV, $Q^2$ = 2.45~GeV$^2$, $\epsilon$ = 0.50, $x$ = 0.31} \\ \midrule
Low  $-t$    &  $0<-t<1$     & 2.424 $\pm$ 0.388  &  0.307 $\pm$ 0.049  \\ 
High $-t$    &  $1<-t<2.5$   & 1.040 $\pm$ 0.301  &  0.201 $\pm$ 0.058  \\ 
Low  $-u$    &  $5<-t<6$     & 2.374 $\pm$ 1.200  &  0.304 $\pm$ 0.153  \\ 
\bottomrule
\end{tabular}
\end{table}

In Fig.~\ref{fig:sigma_t}, the fitting results are shown in black solid, green solid and red dotted lines for $-t$ $<$ 1~GeV$^2$, 1 $<$ $-t$ $<$ 2.5 and $-t$ $>$ 4.5~GeV$^2$, respectively. For the top plot, the green curve fitting range is 1.5 $<$ $-t$ $<$ 2.5 GeV$^2$ and for the bottom plot, 1 $<$ $-t$ $<$ 2.5 GeV$^2$ (shown in Table~\ref{tab:R_int}).

The F-$_{\pi}$-2 data points in the $-t$ $>$ 4.5~GeV$^2$ show an increasing trend, therefore Eqn.\ref{eqn:t_dep} needs to be modified,
\begin{equation}
\sigma = a~e^{b~(-t + t_{\rm max} )},
\label{eqn:t_dep_mod}
\end{equation}
where $t_{\rm \max}$ represents the maximum possible $-t$ value at the scaled $Q^2$ and $W$ values. An alternate fitting in terms of $u$ was attempted, the same result was obtained. 

The fitted $b$ parameters and calculated interaction radius ($R_{\rm int}$) are listed in Table~\ref{tab:R_int} for both $Q^2$ settings. Base on the listed numerical results, some general observations are as follows: 
\begin{itemize}

\item At the low $-t$ region ($-t$ $<$ 1~GeV$^2$), $R_{\rm int}$ = 0.331 $\pm$ 0.042~fm at $Q^2$ = 1.75~GeV$^2$ and 0.307 $\pm$ 0.049~fm at $Q^2$ = 2.45~GeV$^2$. The distinctive peak in this region corresponds to the exchange of mesons (softer exchange process).

\item At the high $-t$ region (1 $<$ $-t$ $<$ 2.2~GeV$^2$), $R_{\rm int}$ = 0.203 $\pm$ 0.091~fm at both $Q^2$ settings. This indicates the virtual photon couples more directly to parton structure which is smaller than meson (harder in terms of the structure), therefore we observed a harder $t$-dependence.

\item At the low $-u$ region ($-t$ $>$ 5~GeV$^2$), $R_{\rm int}$ = 0.306 $\pm$ 0.23~fm at the lower $Q^2$ and 0.304 $\pm$ 0.153~fm at the higher $Q^2$ setting, which likely corresponds to baryon exchange that is also considered as a softer process.

\item The calculated $R_{\rm int}$ values at low $-t$ and low $-u$ regions are comparable at lower $Q^2$ setting; similarly, $R_{\rm int}$ at low $-t$ and low $-u$ regions are comparable at higher $Q^2$ setting. Note that the uncertainties of the $R_{\rm int}$ values in low $-u$ are much greater than the low $-t$ region. The fact that $R_{\rm int}$ at $Q^2$=1.75 GeV$^2$ is larger than at higher $Q^2$ in both the low $-t$ and region $-u$ region, weakly supports the classic interpretation of wavelength of the virtual photon (directly related to the interaction radius) inversely proportional to the $Q^2$. The $R_{\rm int}$ at high $-t$ shows sign of $Q^2$-independent behavior that has been reported in Ref.~\cite{morand05}. Clearly more data over a wider range of $-u$ would be helpful to confirm this interpretation.

\end{itemize}

It is noteworthy that the $-t$ evolution from softer process (meson exchange) at low $-t$, to a harder process at high $-t$, then back to the softer process (baryon exchange) is similar to the $-t$ evolution observed in charged pion photoproduction as shown in Fig.~\ref{fig:vgl_data}.

Currently, JML has not made any specific calculations regarding backward-angle vector meson electroproduction. It is hoped that more theoretical interest will be generated by the completion of this thesis work on the subject. In addition, the JML model has the capability of generating the L/T separated differential cross section~\cite{laget04}. The comparison between the L/T separated cross section extracted from this analysis and the JML model prediction should be an important and exciting study to challenge the limitations of the Regge-based model, particularly at the higher $Q^2$ values ($Q^2$ = 2.45 GeV$^2$).

The intersections of the red and green curves from both $Q^2$ settings in Fig.~\ref{fig:sigma_t}, seem to suggest a minimum $\sigma_{u}$ occurs at $-t$ $\sim$ 4~GeV$^2$. Based on the result of this analysis, $\sigma_{\rm L}$ drops significantly with respect to $-u$ which would result in a vanishing $\sigma_{\rm LT}$. Therefore, in the high $-u$ region (3 $<$ $-t$ $<$ 5 or 1 $<$ $-u$ $<$ 3 GeV$^2$), the differential cross section should only contain $\sigma_{\rm T}$ and $\sigma_{\rm TT}$. In addition, a smooth and shallow (almost flat) behavior of $\sigma_u$ similar to the one observed in the hard process of the photoproduction (shown in Fig.~\ref{fig:vgl_data}) is expected in the high $-u$ region.

Currently, there is no known experimental methodology to access the cross sections at the high $-u$ region (3 $<$ $-t$ $<$ 5~GeV$^2$). A feasibility should be performed using similar technique as that of the high $-t$ measurements presented in the recent F$_\pi$-2-$\pi^+$-high-$t$ analysis~\cite{samip17}, to investigate the possibility of accessing the $\omega$ production data in this region.

\section{$d\sigma_{\rm T}/dt$ Comparison to the TDA Calculations}
\label{sec:Xsection}

\begin{figure}[p]
\centering
\includegraphics[width=0.85\textwidth]{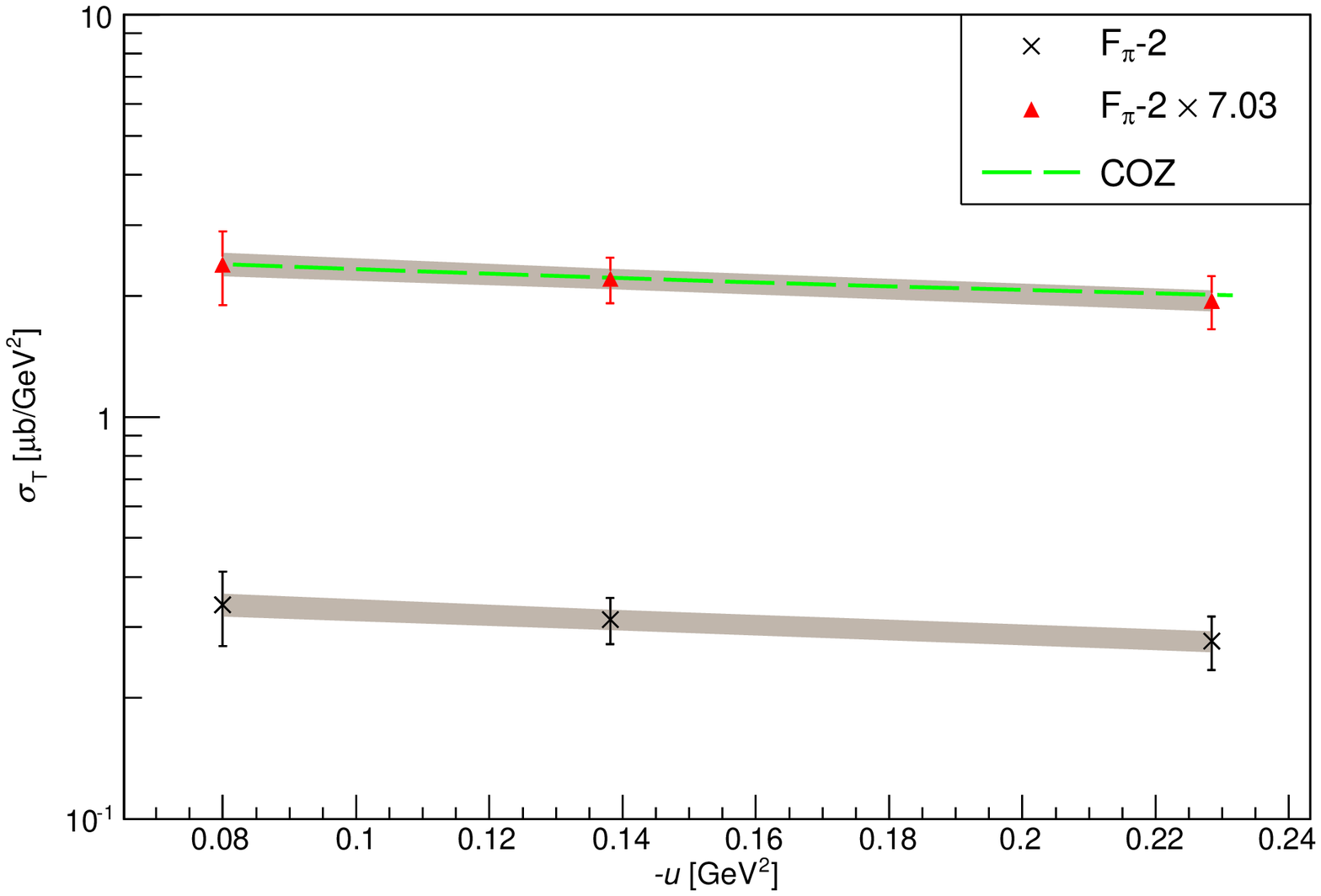}
\caption[$\sigma_{\rm T}$ versus $-u$ for $Q^2$ = 1.60~GeV$^2$]{$\sigma_{\rm T}$ versus $-u$ for $Q^2$=1.60~GeV$^2$. These data are scaled to $W$ = 2.21 GeV and $Q^2$ = 1.60~GeV$^2$. The green dashed line represents the TDA prediction using the COZ model. Red triangular data points represent the normalized F$_{\pi}$-2 data points to the prediction at $-u$ = 0.5 GeV$^2$. The numerical values are listed in Table~\ref{tab:xsec_regge}.~\oic}
\label{fig:tda_compare_160}
\includegraphics[width=0.85\textwidth]{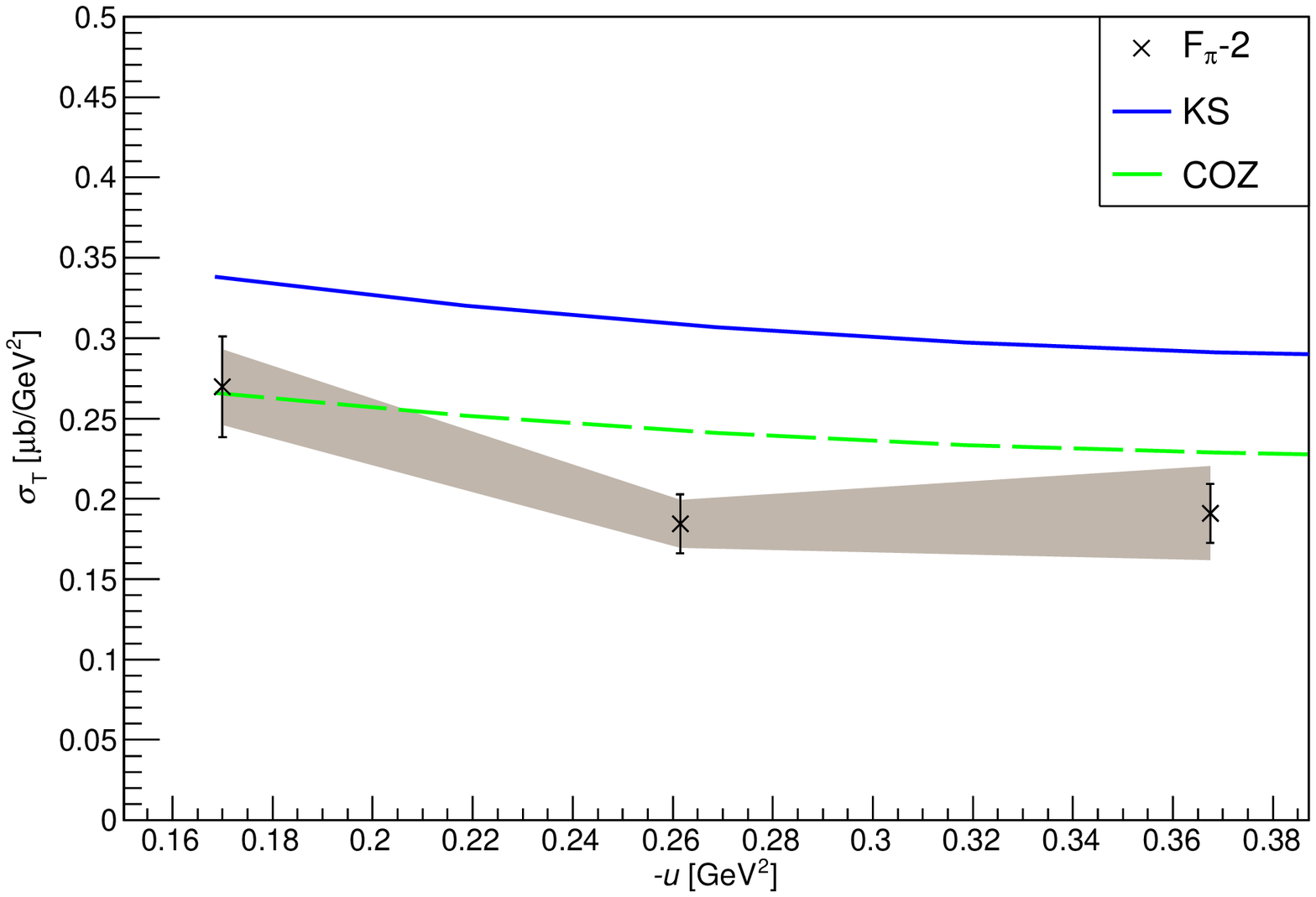}
\caption[$\sigma_{\rm T}$ versus $-u$ for $Q^2$ = 2.45~GeV$^2$]{$\sigma_{\rm T}$ versus $-u$ for $Q^2$=2.45~GeV$^2$. These data are scaled to $W$ = 2.21 GeV and $Q^2$ = 2.45~GeV$^2$. The blue solid line and green dashed line represent the TDA predictions using the KS and COZ nucleon DA models, respectively. The numerical values are listed in Table~\ref{tab:xsec_regge}.~\oic}
\label{fig:tda_compare_245}
\end{figure}

Currently, the only existing theoretical prediction on the $u$-channel exclusive electroproduction of $\omega$ comes from the TDA framework~\cite{pire05, pire15}, described in Sec.~\ref{sec:TDA}. 

From Fig.~\ref{fig:lt_q2}, the general trend of the $\sigma_{\rm T}$ seems to have a weak $Q^2$ dependence, where the difference in $\sigma_{\rm T}$ values between $\braket{Q^2}$ = 1.47 and 2.23~GeV$^2$ is 10-15\%. This is significantly different from the TDA predicted $1/Q^8$ scaling at fixed $x$. Although there are no data on the $x$-dependences of this process at fixed $Q^2$, it seems unlikely that the $x$-dependence is sufficient to explain this discrepancy. The TDA formalism is not applicable at low $Q^2$ values, as TDA collinear factorization requires at least $Q^2$ =10~GeV$^2$. Measurements at much larger $Q^2$ are needed to properly verify the $Q^2$ scaling prediction. Comparing to $\sigma_{\rm T}$, which only mildly depends on $Q^2$, the general trend of $\sigma_{\rm L}$ may suggest a stronger $Q^2$ dependence.

The theoretically predicted and experimentally extracted transverse differential cross section $d\sigma_{\rm T}/dt$ (or $\sigma_{\rm T}$) versus $-u$ dependences at $Q^2$ = 1.60 and 2.45~GeV$^2$ are shown in Figs.~\ref{fig:tda_compare_160} and \ref{fig:tda_compare_245}, respectively. The TDA model predictions using the COZ and KS nucleon DA models (shown in Fig.~\ref{fig:NDA}) are drawn in blue solid line and green dash line, respectively. Note that at $Q^2$ = 1.6~GeV$^2$, only one TDA prediction (with COZ) is available, since this $Q^2$ value is too low compared to the optimal $Q^2$ range of the TDA framework.

\begin{table}[t]
\centering
\setlength{\tabcolsep}{1em}
\caption[$p(e,e^{\prime}p)\omega$ kinematics for the F$_{\pi}$-2 experiment.]{Scaled $p(e,e^{\prime}p)\omega$ data from the F$_{\pi}$-2 experiment to $Q^2_{\rm nominal}$ = 1.60 GeV$^2$ and 2.45 GeV$^2$, $W_{\rm nominal}$ = 2.21 GeV.}
\label{tab:xsec_TDA}
\begin{tabular}{ccc}
\toprule
$-u$ & $-t$   &  $\sigma_{\rm T}\pm\delta\sigma_{\rm T}\pm\Delta\sigma_{\rm T}$    \\
GeV$^2$ & GeV$^2$  &  $\mu$b/GeV$^2$                                   \\ \midrule
\multicolumn{3}{c}{$W$ = 2.21~GeV, $Q^2$ = 1.60~GeV$^2$, $x$ = 0.28}   \\ \midrule
0.080  & 4.031  & 0.341 $\pm$ 0.071 $\pm$ 0.022                        \\
0.137  & 3.974  & 0.313 $\pm$ 0.041 $\pm$ 0.018                        \\
0.228  & 3.883  & 0.277 $\pm$ 0.042 $\pm$ 0.017                        \\ \midrule
\multicolumn{3}{c}{$W$ = 2.21~GeV, $Q^2$ = 2.45~GeV$^2$, $x$ = 0.37}   \\ \midrule
0.170  & 4.791  & 0.270 $\pm$ 0.031 $\pm$ 0.024                        \\
0.261  & 4.700  & 0.184 $\pm$ 0.018 $\pm$ 0.015                        \\
0.377  & 4.584  & 0.191 $\pm$ 0.018 $\pm$ 0.029                        \\ 
\bottomrule
\end{tabular}
\end{table}

It seems that at $Q^2$ = 1.60~GeV$^2$ (Fig.~\ref{fig:tda_compare_160}), the TDA prediction with COZ $N$ DA correctly predicted the flat $-u$ dependence for the $\sigma_{\rm T}$, but over predicted its strength by a factor of 7.04. The TDA predictions at $Q^2$ = 2.45~GeV$^2$ with COZ and KS $N$ DA are shown in Fig.~\ref{fig:tda_compare_245}. The predicted $\sigma_{\rm T}$ strength is much closer to the data (compared to the $Q^2$ = 1.60~GeV$^2$ prediction), the TDA prediction with COZ is consistent with the data within the experimental uncertainties.

Considering the optimal $Q^2$ range of the TDA model is $Q^2$ $>$ 10~GeV$^2$ and that these predictions were made without any previous experimental constraints, the TDA model predictions are able to capture the main features of the data. 
Specially, at $Q^2$ = 2.45~GeV$^2$ setting, the TDA model works surprisingly well in describing the experimental data, which demonstrates the predictive power of this parton-based model. It is extremely important to perform more backward-angle experiments in support of developing this promising model in the 12~GeV era of Jefferson lab.

\section{Conclusion and Closing Remarks}

\subsection{Conclusion}

This thesis work has demonstrated that the missing mass reconstruction technique, in combination with the high precision spectrometers in coincidence mode at Hall C, can be used to reliably extract the backward-angle $\omega$ cross section through the exclusive reaction $^1$H$(e, e^{\prime}p)\omega$, while performing a full L/T separation. Since the missing mass reconstruction method does not require the detection of the produced meson, this allows physicists the possibility to extend experimental kinematics coverage that was considered to be inaccessible through the standard direct detection method. The backward-angle interactions, which have been previously ignored, are anticipated to play an important role and offer complementary information on nucleon structure. Additionally, any future $u$-channel physics studies at Hall C will benefit from the knowledge gained during this thesis work.

Through studying the general trends of the separated differential cross sections of the exclusive $^1$H$(e,e^{\prime}p)\omega$ reaction, the transverse component $\sigma_{\rm T}$ appears to have a flat $\sim1/Q^{1.33 \pm 1.21}$ dependence, whereas $\sigma_{\rm L}$ with large statistical uncertainty has a stronger $1/Q^{9.43 \pm 6.28}$ dependence in the extreme backward-angle kinematics. With $\sim$90\% confidence level, the $\sigma_{\rm L}/\sigma_{\rm T}$ ratio indicates the dominance of $\sigma_{\rm T}$, at $Q^2$ = 2.45~GeV$^2$.

After translating the F$_\pi$-2 data from $-u$ to the $-t$ space of the CLAS-6 data, the cross sections show evidence of a backward-angle peak for the $\omega$ exclusive electroproduction at both $Q^2$ setting. These features, including a forward-angle ($t$-channel) peak shown by the CLAS-6 data and a possible backward-angle ($u$-channel) peak shown by the F$_\pi$-2 data, are consistent with those observed in the $\pi$ photoproduction data. Since the $\pi$ photoproduction peaks were successfully described by a Regge trajectory based model, the observed backward-angle peak in $\omega$ electroproduction calls for the resurrection of the $u$-channel studies through the Regge trajectory based model, such as the JML model. Additionally, the transition from the soft physics region (low $-t$ or low $-u$) to the hard physics region (large angle emission region) at a higher $Q^2$ value would be an interesting topic for future studies.

The $\sigma_{\rm T}$ are compared to the TDA model prediction. At $Q^2$ = 2.45 GeV$^2$, the TDA model predictions are within one to two $\sigma$ band of the data, depending on whether COZ or KS DA are used. In addition, the indication of $\sigma_{\rm T}$ dominance over $\sigma_{\rm L}$ at $Q^2$ = 2.45 GeV$^2$, seems to agree with the postulated TDA factorization condition. On the other hand, the TDA prediction at $Q^2$ = 1.6~GeV$^2$ missed the data by a factor of 7, indicating the TDA factorization doesn't apply for this setting. As the JLab 12~GeV experiments offer experimental data much closer to the TDA preferred $Q^2$ range of $Q^2$ $>$ 10 GeV$^2$, the TDA formalism should be carefully studied and tested. 

\subsection{Closing Remarks}

It is anticipated that as $Q^2$ is extended towards the optimal range of the TDA model ($Q^2$ $>$ 10~GeV$^2$), the Regge-based model might become less effective due to the transition between hadronic and partonic degrees of freedom within the nucleon. Studying the ``crossing point'' in terms of model effectiveness between the JML (exchanges of mesons and baryons) and TDA (exchanges of quarks and gluons) models, is equivalent to studying the nucleon structure transition, which is the grand goal stated in the Chap.~\ref{chap:intro} of this thesis work.

It is the author's wish that this analysis effort can encourage more experimental and theoretical interest on backward-angle physics during the 12 GeV era of JLab. The ultimate scenario would be to perform collaborative measurements using different equipments in different Halls (described in Sec~\ref{sec:u_12}), and combine data sets at similar kinematics to map out the complete $-t$ (or $-u$) evolution and $Q^2$ scaling for a given meson production process. These valuable results will then be used to constrain and develop (hadronic) models such as the JML, and study early insight to (partonic) models such as the TDA.

In the distant future, the Electron Ion Collider\footnote{Electron Ion Collider is the next generation particle accelerator that is currently in the planning stage.} (EIC\nomenclature{EIC}{Electron Ion Collider})~\cite{accardi12} can greatly extend the maximum accessible beam energy and $Q^2$ limit. The measured cross sections in the forward and backward meson production, particularly the L/T separated cross sections, are the ultimate tools to study the effectiveness and limitations of the JML and TDA models, and eventually establish the ``crossing point'' of this transition process.

\graphicspath{ {pics/8outlook/} }

\chapter{Future Outlook}

This chapter gives a brief summary (based on the author's best knowledge), on the backward-angle ($u$-channel) and large emission angle (high $t$-channel) meson production experiments at JLab and other research facilities in the near future. Some of these of experiments use completely different experimental techniques than the one described in this thesis. Table~\ref{tab:merit_tab} lists the possible mesons that can be studied by the described experiments and the availability of the theory predictions.

\begin{table}[t]
\centering
\footnotesize
\setlength{\tabcolsep}{0.9em}

\caption[Opportunities of studying backward meson productions and theory prediction availability]{Table of merit of potential opportunities of studying backward and large emission angle meson production and theory prediction availability~\cite{clas12_phi,panda}.  $^{*}$ indicates large emission angle (high $-t$) meson production experiments.}
\label{tab:merit_tab}

\begin{tabular}{lccccc|cc}
\toprule
             & F$_{\pi}$-2 & F$_{\pi}$-12   & Hall C $\pi^0$ &  E12-12-007$^{*}$  & $\overline{\textrm{P}}$ANDA        &    Regge      &  TDA          \\ \midrule
$\pi^{0}$      &             &              & \checkmark     &              & \checkmark   &               &               \\ 
$\eta$         &             &              &                &  \checkmark  &        	   &               &               \\
$\rho$         &             &              &                &              &        	   &               &               \\
$\omega$       & \checkmark  & \checkmark   &                &              &        	   &               & \checkmark    \\
$\eta^{\prime}$&             &              &                &              &        	   &               &               \\
$\phi$         &             & \checkmark   &                &  \checkmark  &        	   &               & \checkmark    \\ \midrule
Facility       & JLab Hall C & JLab Hall C  & JLab Hall C    & JLab Hall B  & GSI    	   &               &               \\ 
\bottomrule
\end{tabular}
\end{table}

\section{Backward and Large Angle Meson Production at JLab}
\label{sec:u_12}

Upon the successful completion of the JLab 12~GeV upgrade, physicists are presented with opportunities to extend nucleon structure studies with virtual and real photons through $s$- and $t$-channel interactions. 

Thanks to the recent hardware upgrades, JLab acquired the optimized equipment to pursue $u$-channel physics at a more preferred energy range for both Regge theory and the TDA theoretical framework. In JLab Hall C, the standard SHMS-HMS setup is the optimal experimental apparatus to perform high luminosity parallel (low $-t$) and anti-parallel (low $-u$) meson electroproduction studies and perform L/T separations; whereas the CLAS-12 detector, with its high precision and large solid angle acceptance, is optimized to simultaneously study meson electroproduction at high $-t$ (high $-u$) region and $Q^2$ scaling. The GlueX detector at Hall D, with its high intensity real photon beam, would be the ideal place to study the $t$ evolution of meson photoproduction. A few related physics programs are chosen as examples, and are discussed in the following subsections.

\subsection{Backward Angle $\omega$ and $\phi$ Electroproduction from the F$_\pi$-12 Experiment at Hall C}

\begin{figure}[t]
  \centering
  \includegraphics[width=0.55\textwidth]{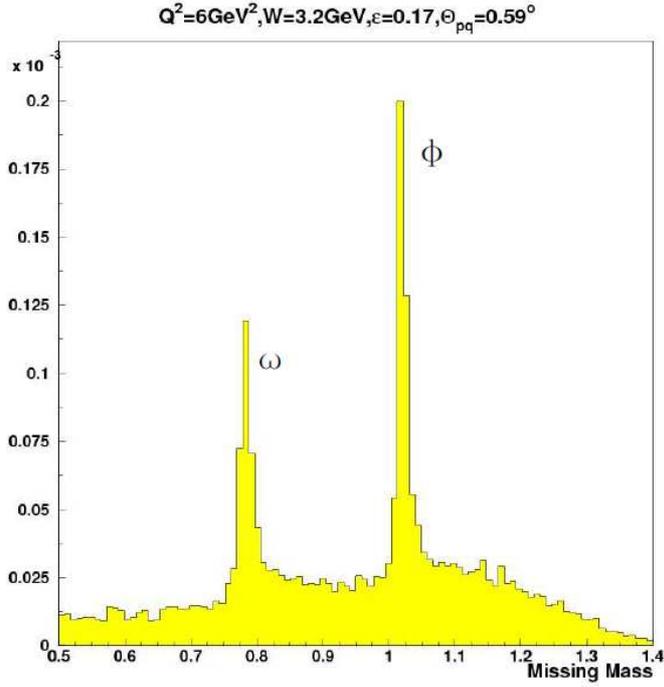}   
  \caption[Simulated $M_m$ distribution for $\omega$ and $\phi$ mesons for the F$_\pi$-12 experiment]{Simulated $M_m$ distribution for $\omega$ and $\phi$ mesons for the F$_\pi$-12 experiment. Note that the narrow width of the $\phi$ is comparable to the width of the $\omega$, the same background simulation (subtraction) technique developed from this thesis  will be tried to extract differential cross sections for both mesons. The relative heights of the distributions are estimated based on the $\omega$-$\phi$ production ratio in the $t$-channel process. Plot was created by G. Huber~\cite{huber17}.~\oic}
  \label{fig:omega_phi_fpi12}
\end{figure}

Similar to the F$_\pi$-2 experiment, which fortuitously projected the coincidence protons (from the backward $\omega$ production) in the center of the SOS+HMS spectrometer acceptance, preliminary studies~\cite{huber17} have shown that the $\phi$ and $\omega$ mesons are near the center the HMS+SHMS acceptance for the F$_\pi$-12\footnote{Hall C experiment E12-06-101}\nomenclature{F$_\pi$-12}{Third charged pion form factor experiment (E12-06-101)} experiment~\cite{fpi12} (third charged pion form factor experiment). Note the F$_\pi$-12 experiment applies the same experimental methodology and detects the same physics observables as the F$_\pi$-2 experiment at a higher and wider range of kinematic variables ($Q^2$ and $W$). The F$_\pi$-12 $\omega$ and $\phi$ electroproduction settings are also at the extreme backward-angle (low $-u$ region, also referred as the soft physics region by Regge theory terminology) region. The reconstructed missing mass distribution for $\phi$ has a narrow and distinctive peak which is similar to the $\omega$, which allows a reliable cross section extraction. An example $M_m$ distribution of $\omega$ and $\phi$ mesons for the F$_{\pi}$-12 experiment is shown in Fig.~\ref{fig:omega_phi_fpi12}.

As the natural continuation of this Ph.D. work, the $u$-channel $\omega$ data from the F$_\pi$-12 experiment will be able to further extend the separated $\omega$ differential cross section to higher $Q^2$ range, which can test the predictions from the JML and TDA models.

There has been significant theory interest on the backward-angle $\omega$ and $\phi$ productions, particularly for $\phi$. The TDA calculation for $\omega$ and $\phi$ electroproduction has already been made for the F$_\pi$-12 kinematics~\cite{pire15}. 

Furthermore, the $\phi$ has a unique $s\overline{s}$ quark structure. Currently, the $u$-channel $\phi$ electroproduction mechanism is unclear~\cite{laget04}, since the backward-angle $\phi NN$ coupling constant is an unconstrained quantity. The F$_\pi$-12 $u$-channel $\phi$ cross section can contribute to the determination the $\phi NN$ coupling constant and quantify the model dependent $s$ quark contributions of the nucleon.

\subsection{A New Proposal: Backward Angle $\pi^0$ Electroproduction at Hall C}

The experience and knowledge gathered from this Ph.D. work has initiated a new experimental proposal by the author and collaborators, to measure the backward-angle ($u$-channel) differential cross section of the neutral pion electroproduction reaction, $^1$H$(e,e^\prime p)\pi^0$, and perform a full L/T separation. This experiment is proposed to use the 11 GeV electron beam and take place at Hall~C of JLab. The measurement will be taken above the resonance region $W>2.0$~GeV and at a variety of $Q^2$ values.

In comparison to the $\omega$, $\pi^0$ electroproduction has much less physics background from other mesons. However, the contribution of the backward-angle photon production needs to be studied in the detail for $\pi^0$. A tight missing mass cut around the $\pi^0$ rest mass would significantly eliminate the random background, therefore it is expected to have smaller overall uncertainties than the $\omega$ analysis. All these features, combined with a wider kinematic coverage offered by a higher energy electron beam, would offer high quality experimental results in a more favorable range of the theory predictions.

A letter of intent on this new backward-angle $\pi^0$ proposal will be submitted by the author to the PAC in summer 2018, and the full proposal is expected to be submitted in summer 2019.

\subsection{Large Angle $\phi$ Meson Electroproduction at Hall B}

Beyond the $u$-channel study opportunities in the extreme backward-angle at Hall C, Hall B aim to study meson electroproduction in the large-emission-angle region (high $-t$ region or harder region). The approved Hall B experiment E12-12-007~\cite{guidal12} will measure exclusive $\phi$ meson electroproduction, $e+p \rightarrow e^{\prime} + p^{\prime} + \phi$, with the CLAS-12 detector. The kinematic range extends in $W$ from 2-5 GeV, $Q^2$ from 1-12 GeV$^2$, and $t^{\prime} = |t-t_{min}|$ from near zero to $\sim$4~GeV$^2$. The $\phi$ will be detected through the $K^+ K^-$ decay channel. Differential cross sections and beam spin asymmetries will be measured as a function of the $\phi\rightarrow K^+K^-$ decay angles, $\theta$ and $\phi$, to extract $\sigma_{\rm T}$, $\sigma_{\rm L}$, $\sigma_{\rm TT}$ and $\sigma_{\rm LT}$.

Exclusive $\phi$ electroproduction at $Q^2 \sim$ few GeV$^2$ is of special significance as a probe of the gluon GPDs of a nucleon. The purpose of the study is similar to the expected $\phi$ production data from the F$_{\pi}$-12 experiment, which is to provide information on potential intrinsic strangeness in the nucleon in the soft region (high $-t$ or high $-u$ region).

\section{$\pi^0$ Production from $\overline{\bf P}$ANDA at FAIR (GSI)}
\label{sec:panda}

Beside the JLab 12~GeV backward-angle and large emission angle programs, other nuclear physics research facilities have also started to explore the possibility to establish experimental access to study this relatively unknown field.

An example is the study of the backward-angle $\pi^0$ meson production~\cite{panda14} by the $\overline{\textrm{P}}$ANDA experiment~\cite{bettoni07,panda09} at FAIR\footnote{Facility for Antiproton and Ion Research GSI, Planckstrasse 1, 64291 Darmstadt, Germany. https://www.gsi.de/}. The FAIR accelerator complex is currently under construction at the GSI Helmholtz Centre for Heavy Ion Research in Darmstadt, Germany. The experimental setup requires a proton beam to be accelerated to an energy of 29~GeV before being directed at an antiproton production target.

Interest in the backward-angle reaction involves the $\pi^0$ production channel: $$\overline{p}+p\rightarrow l^+ + l^- + \pi^0,$$ where $\overline{p}$ is the incoming antiproton beam and $p$ is the proton target; $l^+$ and $l^-$ represent detected lepton and antileption, respectively. This experimental channel can be accessed through observables including $\overline{p}+p\rightarrow \gamma^* + \pi^0 \rightarrow e^+ e^- + \pi^0$ and $\overline{p}+p\rightarrow J/\psi + \pi^0 \rightarrow e^+ e^- + \pi^0$~\cite{panda14}.

The measurement is planed for two kinematical settings: $s=W^2=5$~GeV$^2$, $3<q^2<5$~GeV$^2$ and for $s=W^2=10$~GeV$^2$, $5<q^2<10$~GeV$^2$. Note that the four-momentum transfer squared, $q^2=-Q^2$, is positive for the time-like virtual photon exchange process; $Q^2$ is positive for the space-like virtual photon exchange process.

The main objective of the $\overline{\bf P}$ANDA $u$-channel study is to test the QCD collinear factorization through time-like $q^2$ scaling behavior. Recall the purpose of the proposed backward-angle $\pi^0$ is to study the QCD collinear factorization through space-like $Q^2$ scaling behavior. The combination of both time-like and space-like measurements would give significant experimental constraints and allow theoretical physicists to develop an accurate and complete picture of the quark-gluon spatial distribution inside of the nucleon in $u$-channel physics.




\addcontentsline{toc}{chapter}{Bibliography}

\InputIfFileExists{refs.bbl}


\newpage
\addcontentsline{toc}{chapter}{List of Abbreviations}
\printnomenclature[10ex]

\end{document}